\documentclass[12pt,a4paper]{book}
\usepackage{german}
\usepackage{graphicx}
\usepackage{tabularx}
\usepackage{exscale}

\renewcommand{\baselinestretch}{1.2}
\newcommand{\hidden}[1]{}
\newcommand{\hi}[1]{{#1\index{#1}}}
\newcommand{\hii}[2]{{#1\index{#2}}}

\newcommand{\name}[2]{{\sc #1\index{Name: #1, #2}}} 
\newcommand{\captionstyle}{
  \renewcommand{\normalsize}{\small \bf}
  \renewcommand{\baselinestretch}{1.0}
  }
\newcommand{\mycaption}[2]{
  \captionstyle
  \caption[#1]{\rm #2}
  }

\newcommand{\farbe}{}

\newcommand{\vp}[2]{\left<#1,#2\right>}
\newcommand{\mx}[1]{{\mbox{\scriptsize #1}}}
\newcommand{\meter}[1]{{#1} m}
\newcommand{\qmeter}[1]{{#1} m$^{\bf 2}$}
\newcommand{\metersec}[1]{{#1} m/s}
\newcommand{\pqmeter}[1]{{#1} P/m$^{\bf 2}$}
\newcommand{\pmetersec}[1]{{#1} P/ms}
\newcommand{\gradc}[1]{{#1}$^{\circ}$~C}

\newcommand{\Ra}{\vec{r}_\alpha} \newcommand{\Rb}{\vec{r}_\beta}
\newcommand{\Va}{\vec{v}_\alpha} \newcommand{\Vb}{\vec{v}_\beta}
\newcommand{\Vaprime}{\vec{v}_{\alpha^\prime}}
\newcommand{\Eanull}{\vec{e}\,^0_{\!\alpha}}
\newcommand{\Esanull}{\vec{e}\,^\perp_{\!\alpha}}
\newcommand{\Ebnull}{\vec{e}\,^0_{\!\beta}}

\newcommand{\Fanull}{\vec{f}\,^0_{\!\alpha}}
\newcommand{\Fbnull}{\vec{f}\,^0_{\!\beta}}
\newcommand{\vanull}{v^0_\alpha}
\newcommand{\vamax}{v^{max}_\alpha}

\newcommand{\Vanull}{v^0_\alpha\,\vec{e}\,^0_{\!\alpha}}
\newcommand{\Vbnull}{v^0_\beta\,\vec{e}\,^0_{\!\beta}}
\newcommand{\rp}{r^{\prime}}
\newcommand{\rpp}{r^{\prime\prime}}
\newcommand{\Utr}{U_\mx{tr}}
\newcommand{\Ftr}{\vec{f}_{\!\mx{tr}}}
  
\newcommand{\Pac}{P_{\!\alpha \bf C}}
\newcommand{\PPac}{P^\prime_{\!\alpha \bf C}}
\newcommand{\Pacp}{P_{\!{\alpha^\prime} \bf C}}
\newcommand{\Pacprime}{P_{\!\alpha \bf C^\prime}}

\newcommand{\lanull}{l_\alpha^{\,0}}
\newcommand{\ave}[1]{\left<#1\right>}

\newcommand{\mucommalam}{(\mu,\lambda)}
\newcommand{\mupluslam}{(\mu\!+\!\lambda)}
\newcommand{\sq}[1]{{#1}^{\bf 2}}
\newcommand{\Z}{{\cal Z}}
\newcommand{\N}{{\cal N}}

\newcommand{\fussgaenger}{Fu"s"-g"an"-ger}
\newcommand{\einfuehrung}{Ein"-f"uhr"-ung}
\newcommand{\naechsten}{n"ach"-sten}
\newcommand{\fussgaengerstroeme}{Fu"s"-g"an"-ger"-str"o"-me}
\newcommand{\Zusaetzlich}{Zu"-s"atz"-lich}
\newcommand{\Binaer}{{Bi"-n"ar}}
\newcommand{\Reihenfolgenunabhaengigkeit}
{Reihen"-fol"-gen"-un"-ab"-h"an"-gig"-keit}
\newcommand{\Reifo}
{Reihen"-fol"-gen"-un"-ab"-h"an"-gig"-keit}
\makeindex
\begin{document}
\pagestyle{empty}
\begin{titlepage}
  \renewcommand{\baselinestretch}{1.2}
  \begin{center}
    \vfill
    {\LARGE \bf Modellierung und Simulation  der Dynamik von
      {\fussgaenger}str"omen} 
    \par
    \vfill
    Von der Fakult"at Physik der Universit"at Stuttgart 
    zur Erlangung der W"urde eines Doktors der 
    Naturwissenschaften (Dr.~rer.~nat.) genehmigte Abhandlung
    \par
    \vfill
    Vorgelegt von P\'eter Moln\'ar aus G"ottingen
    \par
    \vfill
    Hauptberichter: Prof.~Dr.~Dr.~h.c.~W. Weidlich \\
    Mitberichter: Apl.~Prof.~Dr.~habil.~A. Wunderlin \\
    Tag der m"undlichen Pr"ufung: 15.~Dezember 1995 \\
    \par
    \vfill
    II.~Institut~f"ur~Theoretische~Physik der Universit"at~Stuttgart
    \par
    \vfill
    1995
    \vfill
  \end{center}
\end{titlepage}
\setcounter{page}{5}
\pagestyle{headings}
\chapter*{Vorwort}
\markboth{VORWORT}{}
Die vorliegende Arbeit stellt ein mikroskopisches Modell zur Beschreibung von
{\fussgaengerstroeme}n vor, das auf der Basis der Soziale-Kr"afte-Theorie
entwickelt wurde. 

Die Arbeit verfolgt zwei Ziele: 
\begin{itemize}
\item Entwicklung eines realit"atsnahen Modells, das sich als Werkzeug zum Entwurf
  bedarfsgerechter {\fussgaenger}anlagen eignet.
\item Die Verifizierung einer sozialwissenschaftlichen Theorie durch ein 
  Modell, f"ur das eine ausreichende Menge an Daten zur Verf"ugung steht.
\end{itemize}
 
Die Untersuchung des {\fussgaenger}modells zeigten, da"s sich  trotz 
einfacher  Verhaltensmuster der Individuen durch das  
Zusammenspiel der Wechselwirkungen in {\fussgaengerstroeme}n komplexe
r"aumliche und zeitliche Strukturen ausbilden. So kann die Emergenz
kollektiven Verhaltens in {\fussgaenger}mengen gezeigt werden, in denen die 
einzelnen Individuen nur zwei Verhaltensregeln kennen: 1. Sie wollen ihr
vorgegebenes Ziel auf direktem Wege mit einer gewissen Wunschgeschwindigkeit
erreichen. 2. Sie sind bem"uht, untereinander  und gegen"uber
Hindernissen einen gewissen Abstand zu halten.
Das selbstorganisierte kollektive Verhalten tritt in Form von Spuren auf,
die sich durch die Menge ziehen und nur {\fussgaenger} einer Bewegungsrichtung
enthalten. Als weiteres Ph"anomen ist die Oszillation der Bewegungsrichtung an
Durchg"angen oder Kreuzungen zu beobachten.    

Weiterhin werden starke Abh"angigkeiten der Eigenschaft der
{\fussgaengerstroeme} von geometrischen Formen der Geb"aude gezeigt und der
Einflu"s geometrische Ver"anderungen auf die Leistungsmerkmale der Str"ome
untersucht. In einem Beispiel lie"s sich durch eine Verkleinerung der
begehbaren Fl"ache eine Effizienzsteigerung erreichen. 

Dieser Effekt l"a"st sich zur Optimierung von {\fussgaenger}anlage einsetzen,
indem die
Grundrisse einem evolution"aren Proze"s unterzogen werden.
Dazu wird die Implementierung eines evolution"aren Programms vorgestellt,
das auf den urspr"unglichen Ans"atzen der genetischen
Algorithmen und Evolutionsstrategien aufbaut.

In einer Erweiterung des {\fussgaenger}modells wird das Soziale-Kr"afte-Modell
durch ein Entscheidungsmodell erg"anzt, um  das  Auswahlverhalten
alternativer Ziele zu beschreiben. Weiterhin wird die Anpassungs- und
Lernf"ahigkeit von {\fussgaenger}n in das Modell aufgenommen. Damit k"onnen 
{\fussgaenger} ihr Ausweichverhalten und ihre Entscheidungsstrategien
anhand ihrer gesammelten Erfahrungen verbessern. Hierzu werden die
Modellparameter einem modifizierten evolution"aren Proze"s unterzogen. 

In gr"o"seren {\fussgaenger}anlagen ist das Orientierungsverm"ogen und die
Wegewahl der Passanten ausschlaggebend f"ur die Belastung einzelner
Abschnitte des Geb"audes. Zur Bestimmung von Belastungen einzelner
Teilstrecken in einem Wegesystem wird ein Verfahren entwickelt, das die
Belastung der Teilstrecken des gesamten Wegesystems unter Ber"ucksichtigung
subjektiver Auswahlkriterien der {\fussgaenger} ermittelt.
Das Verfahren kann auch regelm"a"sige Netzwerke korrekt behandeln und
Toleranzen bei der Wegewahl zulassen.

Schlie"slich ensteht auf der Basis des  Soziale-Kr"afte-Modells der
{\fussgaenger}dynamik 
ein Modell, das die selbst"andige Entwicklung von Wegesystemen mit
minimalen Umwegen beschreibt, wie sie in der Natur h"aufig anzutreffen sind. 
Bei diesen Transportwegesystemen stehen die Gesamtl"ange eines Netzwerkes und
die (Material-)Kosten f"ur die Wegstrecken in einem optimalen Verh"altnis.
In dem vorgestellten Modell bilden {\fussgaenger}
Trampelpfade aus, die durch h"aufige Nutzung aufrecht erhalten werden und
bei Nichtbenutzung wieder verschwinden. In der Simulation produziert das
Modell aus einem vollst"andig verkn"upften
Netzwerk ein minimales Umwegesystem.  

Zur Berechnung der Modelle wurde eine eigene Simulations-Software entwickelt,
die dem Anwender eine komfortable Modellspezifikation in einer speziellen
Beschreibungssprache erm"oglicht.  Die Steuerung und Beobachtung der
Simulationsabl"aufe, sowie die Darstellung der Ergebnisse erfolgen "uber eine
grafische Benutzeroberfl"ache.

Abschlie"send werden Themen aus der Simulationstheorie behandelt, wie etwa das
Konzept der objekt-orientierten Modellspezifikation- und Implementation sowie
die Problematik der Gleichzeitigkeit in Simulationen.

F"ur die anregenden Diskussionen und Ratschl"age, f"ur das
Korrekturlesen und f"ur die Unterst"utzung bei der Verwirklichung dieser Arbeit
m"ochte ich an dieser Stelle 
\newcommand{\nn}[2]{#2 #1}
\nn{Eifert}{Wolfram},
\nn{Helbing}{Dr.\ Dirk},
\nn{Keltsch}{Joachim},
\nn{R"oder}{Regina},
\nn{Saam}{Dr.\ Nicole J.\ },
\nn{Schenk}{Martin},
\nn{Schweitzer}{Dr.\ Dr.\ Frank}, 
\nn{Starke}{Jens},
\nn{Steinhauser}{Anja},
\nn{Wedekind}{Jens},
\nn{Weidlich}{Prof.\ Dr.\ Dr.\ h.c.\ Wolfgang}, den Mitarbeitern des
II. Instituts f"ur Theoretische Physik und ganz besonders meiner Familie
danken.
\par
Stuttgart, Dezember 1995 \hfill {\em P\'eter Moln\'ar}

\tableofcontents
\listoffigures
\listoftables

\chapter{Einleitung}
Bis in die f"unfziger Jahre waren unsere St"adte haupts"achlich durch den
{\fussgaenger}verkehr gepr"agt. Mit zunehmender Verbreitung von PKW
ver"anderten sich jedoch die urbanen Strukturen und wurden immer
fu"s"-g"an"-ger"-unfreundlicher. Dies l"oste einen Proze"s der  Trennung von
{\fussgaenger}- und (Kraft-)Fahrzeugverkehr aus. In der heutigen Zeit wird
der Entstehung von {\fussgaenger}zonen, Einkaufszentren und Freizeitanlagen 
eine immer gr"o"serer Bedeutung beigemessen. 

Eine bedeutende Rolle in der Planung von bedarfsgerechten
{\fussgaenger}anlagen spielt dabei die Modellierung und Simulation der
Personenstr"ome. 
So k"onnen Entw"urfe von Anlagen bereits im Planungsstadium auf die
erwarteten Anforderungen hin gepr"uft werden. Engp"asse oder ungenutzte
Bereiche in Geb"auden lassen sich vermeiden. 

Der {\fussgaenger}verkehr ist ma"sgeblich durch die sozial-psychologischen
Interaktionen zwischen den Passanten und durch deren Reaktionen gegen"uber ihrer
Umgebung bestimmt. Technische Aspekte wie etwa Beschleunigungszeit, Bremsweg
oder Einschr"ankungen beim Richtungswechsel, die den (Kraft-)Fahrzeugverkehr
ausmachen, spielen dagegen eine untergeordnete Rolle.
  
Daher stellen {\fussgaenger} ein interessantes soziales System dar, das
sich aufgrund seiner leichten Beobachtbarkeit zur Verifizierung
sozio-dynamischer Modelle anbietet: Das Verhalten der
{\fussgaenger} wird durch Geschwindigkeits"anderungen sichtbar und kann 
in physikalischen Gr"o"sen gemessen werden. Die
Ereignisse der Interaktionen zwischen Individuen treten in gro"ser H"aufigkeit
auf, so da"s sich statistische Aussagen machen lassen.

Ferner l"a"st sich das {\fussgaenger}verhalten in Ebenen
unterschiedlicher Komplexit"at betrachten, die durch einzelne Modellierungen
beschrieben werden k"onnen.

Unter den beiden Aspekten wurde ein mikroskopisches Modelle der
{\fussgaenger}dynamik entwickelt, das in der vorliegenden Arbeit vorgestellt
wird.  

Die hier vorgestellte Methode der Modellierung und Simulation von
individuellem Verhalten (in den {\fussgaengerstroeme}n) 
ist sowohl auf dem Gebiet der Stadtplanung, als auch in den
Sozialwissenschaften neu.

Um die Ans"atze interdisziplin"arer Forschung unter ein
breites (und teilweise skeptisches) Publikum zu verbreiten, ist eine
ansprechende und eing"angige 
Gestaltung der Ergebnisse sowie die Bereitstellung eines Werkzeugs zur
Modellierung und Simulation sehr hilfreich. Aus diesem Grund wurde eine
Simulation-Software entwickelt, die dem Anwender erm"oglicht eigene Modelle zu
spezifizieren, zu simulieren und auszuwerten.

\subsubsection{Entwurf bedarfsgerechter {\fussgaenger}anlagen}
Bereits seit den sechziger Jahren sind {\fussgaengerstroeme} Gegenstand der 
Verkehrsforschung. Dabei wurden in zahlreichen Beobachtungen sowohl das
Orientierungsverhalten von einzelnen {\fussgaenger}n, als auch das Verhalten
von {\fussgaenger}mengen in \hii{urbanen Strukturen}{urbane Strukturen}
untersucht (Abschn.~\ref{sec:bewegungsverhalten}).
Aus Messungen der Geschwindigkeit und der Dichte in {\fussgaengerstroeme}n
wurden dann Kriterien zur Bemessung von 
{\fussgaenger}anlagen abgeleitet (Abschn.~\ref{sec:bemessung}).  

In den letzten Jahren entstanden neben den empirischen Untersuchungen f"ur
verschiedene Anwendungszwecke auch makro- und mikroskopische Modelle zur
Simulation von {\fussgaenger}mengen  (Abschn. \ref{sec:modellierung}).

Mikrosimulationen erlauben eine detaillierte 
Untersuchung von {\fussgaengerstroeme}n in urbaner Umgebung. Dadurch lassen
sich bereits in der Planungsphase einer Anlage aussagekr"aftige Untersuchungen
am Geb"aude durchf"uhren und eventuelle Schwachstellen im Grundri"s aufdecken.
Die Eigenschaften der simulierten {\fussgaenger} k"onnen dabei leicht an den
Zweck des Geb"audes und die erwartete Zusammensetzung des
{\fussgaenger}verkehrs angepa"st werden.

Durch die mathematische Beschreibung der {\fussgaenger} (Kapitel
\ref{cha:pedestrianmodell}) k"onnen zudem
Bewertungskriterien herangezogen werden, die in der empirischen Beobachtung
nur mit gro"sem  Aufwand  zu erheben sind. Dazu geh"oren Gr"o"sen wie 
der Grad des Wohlbefindens und der psychischen Anspannung, sowie die
situationsbedingte Wahrnehmungsf"ahigkeit von "au"seren Eindr"ucken, zum
Beispiel die Warenangebote in einem Kaufhaus.   

Mit den herk"ommlichen sowie den neuen Bewertungskriterien lassen sich
Anlagengrundrisse in der Simulation bewerten und optimieren.
Dies kann mit Hilfe der Modellierung evolution"arer Prozesse
durchgef"uhrt werden. Kapitel \ref{cha:formoptimierung} behandelt diese 
seit den siebziger Jahren auf vielf"altige Weise zur technischen Optimierung
angewendete Methode und stellt eine
Implementation f"ur die Optimierung von {\fussgaenger}anlagen vor.

Betrachtet man gr"o"sere {\fussgaenger}anlagen, so mu"s auch das
Orientierungsverm"ogen von {\fussgaenger}n modelliert werden.
Mit der Anlage vertraute {\fussgaenger} haben eine genaue
Vorstellung von den Verbindungswegen  und deren Beschaffenheit.
Die Wegewahl nach diesen \hi{kognitiven Karten}  erfolgt
dabei nicht nur nach Kriterien wie Streckenl"ange oder Anstrengung, sondern
auch nach anderen subjektiven Empfindungen, die h"aufig auch durch das
{\fussgaenger}aufkommen in den einzelnen Streckenabschnitten bestimmt werden
(Abschnitte \ref{sec:bemessung} und \ref{sec:bewertungskriterien}).

Das {\fussgaenger}aufkommen in den einzelnen Stre"ckenabschnitten h"angt von
der 
Produktionsrate der Eintrittspunkte ab, ferner von der Attraktivit"at der
Zielknoten sowie von den Wegen, die die {\fussgaenger} benutzen.
Suchalgorithmen, die die Wegewahl der {\fussgaenger} in Bezug auf deren
pers"onlichen Bewertungskriterien durchf"uhren, bestimmen dabei sogenannte
Belastungsfrequenzen f"ur die
einzelnen Streckenabschnitte eines Wegesystems.

Auf diese Weise lassen sich bereits aus dem {\fussgaenger}aufkommen, den
Bed"urfnissen der {\fussgaenger} und den Eigenschaften einzelner Teilst"ucke
einer {\fussgaenger}anlage eventuell auftretende Probleme entdecken, 
die durch "uberlastete
Strecken oder Durchg"ange enstehen k"onnen. 
Auch besonders stark frequentierte und
deshalb f"ur Verkaufsfl"achen interessante Stellen werden dadurch sichtbar. 

Die charakteristischen Eigenschaften, die die Teilst"ucke einer Anlage f"ur
bestimmte {\fussgaenger}mengen und -zusammensetzungen aufweisen, k"onnen dabei
durch die Mikrosimulation des Soziale-Kr"afte-Modells
(Kap.\ \ref{cha:sozialekraeftemodell}) bestimmt werden. 

In {\fussgaenger}anlagen sind sogenannte Systeme minimaler Umwege von gro"sem
Interesse, da einerseits Verbindungswege in ausreichendem Umfang zur
Verf"ugung gestellt werden sollen, andererseits nicht die ganze Fl"ache der
Anlage durch Fu"swege zerschnitten werden darf. 
In der Natur weisen  viele Transportsysteme diese Eigenschaften auf.
Dazu geh"oren etwa Versorgungssysteme in Pflanzen und Verkehrswegenetze von
Tieren und Menschen. Die Entstehung solcher Wegesysteme, die sich den
Bed"urfnissen des  {\fussgaenger}verkehrs anpassen, lassen sich ebenfalls in
der Theorie 
der sozialen Kr"afte  modellieren (Kapitel \ref{cha:trailformation}). 

\subsubsection{Soziales Verhalten r"aumlich verteilter Individuen}
In der Psychologie und in den Sozialwissenschaften wurden verschiedene Ans"atze
unternommen, die Beziehung zwischen r"aumlich verteilten Individuen zu
beschreiben. Zwei davon werden in Kapitel \ref{cha:theorien}
vorgestellt. Sie bilden die Grundlage f"ur das Soziale-Kr"afte-Modell der
{\fussgaenger}dynamik.  

Die einfache Aufgabe eines {\fussgaenger}s von einem Ort zum anderen zu gehen,
erfordert im Gegensatz zur Auswahl der Zielorte keine gro"sen "Uberlegungen.
Hat ein {\fussgaenger} seine Richtung bestimmt, geht er fast automatisch auf
sein Ziel zu. Automatisch in dem Sinne, da"s er Hindernissen und
entgegenkommenden Passanten ausweicht, ohne weiter dar"uber nachzudenken.
Dabei hilft ihm seine Erfahrung, optimal auf ein Ereignis zu reagieren.

Die Beobachtungen zeigen, da"s sich die Bewegungsabl"aufe einzelner
{\fussgaenger} stark gleichen. Dies erlaubt eine mathematische
Beschreibung durch die Theorie der sozialen Kr"afte.

Das Verhalten von {\fussgaenger}n kann in verschiedenen Ebenen
unterschiedlicher Komplexit"at betrachtet werden. Die Gehbewegung, durch die die
{\fussgaenger} zu ihrem Ziel gelangen, legt dabei der Grundstock zu weiteren
Modellierungsebenen. 

"Uber die Ebene des Soziale-Kr"afte-Modells der {\fussgaenger}dynamik l"a"st
sich etwa ein Entscheidungsmodell der Zielwahl ansiedeln. 
Dabei stehen die Ebenen in gegenseitiger Abh"angigkeit ihrer Modellzust"ande:
So 
wird die Entscheidung f"ur eine Zielrichtung durch die Gehbewegung
ausgel"ost. Das Ergebnis der Entscheidungsfindung gibt wiederum die neue
Zielrichtung vor. Das mikroskopische Modell der {\fussgaenger} kann mit nahezu
beliebig vielen komplexen Verhaltensregeln ausgebaut werden.

\subsubsection{Anwenderfreundliche Simulations-Software}
Aufgrund dieser zwei Aspekte der Modellierung von {\fussgaengerstroeme}n wurde
eine Si"-mu"-la"-ti"-ons-Software entwickelt, die zum einen als Bestandteil von
CAD-Programmen\footnote{Computer-Aided-Design-Programme zur Erstellung von
  (Bau-)Zeichnungen und Pl"anen} f"ur Architekten und St"adteplanern
dienen kann, zum anderen aber auch ein eigenst"andiges Anwendungsprogramm
darstellt, mit dem sich Soziale-Kr"afte-Modelle entwerfen und simulieren
lassen.

In Hinblick auf die Verbreitung der Theorie und m"oglicher Anwender des
Simulators in unterschiedlichen (weniger computer-begeisterten) Disziplinen
wurde die Software mit einer leicht zu bedienenden, grafischen
Benutzeroberfl"ache ausgestattet.
Damit bietet sie dem Benutzer M"oglichkeiten zur Beobachtung und
Steuerung des Ablaufs der Simulation. Die Simulationsergebnisse lassen sich als
Animation auf Video aufzeichnen oder als grafische Darstellungen auf Papier
bringen.

Die Modellspezifikation geschieht in einer eigens daf"ur entwickelten
Beschreibungssprache. Eine kurze Einf"uhrung in diese Sprache, sowie die
Erkl"arung der Komponenten und des objekt-orientierten Konzeptes,
das dem Simulator zu Grunde liegt, werden in Kapitel \ref{cha:simulatioprogram}
behandelt.

\chapter{Untersuchung des {\fussgaenger}verkehrs}
\label{cha:untersuchung}
\label{cha:fussg_observ}
Bereits seit den sechziger Jahren sind {\fussgaengerstroeme} Gegenstand der 
Verkehrsforschung. Dabei wurden in zahlreichen Beobachtungen sowohl das
Orientierungsverhalten von einzelnen {\fussgaenger}n, als auch das Verhalten
von {\fussgaenger}mengen in \hii{urbanen Strukturen}{urbane Strukturen}
untersucht 
(Abschn.~\ref{sec:bewegungsverhalten}).
Aus Messungen der Geschwindigkeit und der Dichte in {\fussgaengerstroeme}n
wurden dann Kriterien zur Bemessung von 
{\fussgaenger}anlagen abgeleitet (Abschn.~\ref{sec:bemessung}).  

In den letzten Jahren entstanden neben den empirischen Untersuchungen f"ur
verschiedene Anwendungszwecke auch makro- und mikroskopische Modelle zur
Simulation von {\fussgaenger}mengen  (Abschn. \ref{sec:modellierung}).

\section{Bewegungsverhalten\hidden{ von {\fussgaengern}}}
\label{sec:bewegungsverhalten}
\subsubsection{Gehbewegung}
Der Durchschnitt der in der verkehrstechnischen Literatur angegebenen
{\fussgaenger}geschwindigkeit liegt bei \metersec{1.34}. 
Er ist bei M"annern mit \metersec{1.41} um ca.\ 10~\% h"oher als bei
Frauen (\metersec{1.27}). Die h"ochste Gehgeschwindigkeit wird im Alter von
etwa 20 Jahren erreicht. Oberhalb von 50 Jahren geht sie deutlich
zur"uck. Die Geschwindigkeiten in einer {\fussgaenger}menge sind 
dabei normalverteilt. Bei dem Mittelwert von  \metersec{1.34} wird in
\cite{Weidmann:1993} eine Standardabweichung von \metersec{0.26} angegeben.

Auch die Begleitumst"ande spielen beim Gehen eine gro"se
Rolle \cite{Weidmann:1993}:  
\begin{itemize}
\item der {\fussgaenger}verkehr kann in vier Kategorien mit jeweils
  unterschiedlichen mittleren Geschwindigkeiten unterteilt werden:
  Im \hi{Nutz- und Werksverkehr} laufen die {\fussgaenger} mit
  \metersec{1.61} am schnellsten. Gefolgt vom \hi{Pendlerverkehr} mit
  \metersec{1.49}. Beim Einkaufen \index{Einkaufsverkehr} und in ihrer
  Freizeit \index{Freizeitverkehr} schlendern die {\fussgaenger} im Schnitt
  nur noch mit \metersec{1.16} bzw.\ mit \metersec{1.10}.
\item Zwischen der \hi{Tageszeit} und der {\fussgaenger}geschwindigkeit besteht
  eine Abh"angigkeit. Dabei ist die Geschwindigkeit am Morgen am 
  h"ochsten. W"ahrend des Vormittags und am fr"uhen Nachmittag sind
  deutliche Einbr"uche zu beobachten. Nur zur Mittagzeit und am Abend steigt
  die mittlere {\fussgaenger}geschwindigkeit vor"ubergehend an.
\item auch die \hi{Umgebungstemperatur} beeinflu"st die
  {\fussgaenger}geschwindigkeit. Sie betr"agt bei \gradc{25} nur noch 92~\%
  des Mittelwertes, steigt dagegen bei \gradc{0} auf etwa 109~\% an.
\item ein Einflu"s der \hi{Fu"swegl"ange} ist auf ebenen Fl"achen innerhalb
  des beobachtbaren L"angenbereichs nicht zu erkennen, wohl aber
  auf Treppen.
\end{itemize} 

\subsubsection{Dynamischer Platzbedarf}
Ein stehender {\fussgaenger} ben"otigt eine Fl"ache von mindestens
\qmeter{0.15}. Das entspricht einem sehr starken Gedr"ange, in dem die Personen
keine Bewegungsm"oglichkeit haben. Der dynamische Platzbedarf f"ur das Gehen
ist deutlich gr"o"ser. In seitlicher Richtung m"ussen die  Breitenverteilung
der {\fussgaenger} inklusive mitgef"uhrtem Gep"ack, die Schwankungen des
K"orpers bei Bewegung auf idealer Bahn und  die Abweichungen von dieser
idealen Bahn ber"ucksichtigt werden. 

In Bewegungsrichtung ben"otigen die {\fussgaenger}  Freiraum
f"ur ihre n"achsten Schritte. {\Zusaetzlich} mu"s 
ein gewisser Sicherheitsabstand zu den vorderen Personen eingehalten werden.
Eine mittlere Breite von \meter{0.71} f"ur {\fussgaenger} ohne Gep"ack
und eine Schrittl"ange von \meter{0.63} bei einer Geschwindigkeit von
\metersec{1.34} ergeben  einen 
minimalen \hi{dynamischen Platzbedarf} von \qmeter{0.45} und eine maximal
m"ogliche {\fussgaenger}dichte von \pqmeter{2.2} (Personen pro Quadratmeter).
H"ohere Dichten f"uhren bei
konstanter Spurbreite zu k"urzeren Schrittweiten, und damit zu geringeren
Geschwindigkeiten. Bei einer {\fussgaenger}dichte von etwa \pqmeter{5.4} kommt
die Gehbewegung zum Stillstand \cite{Weidmann:1993}.

\subsubsection{Abstandhalten}
Die Angaben "uber die Abst"ande, die {\fussgaenger} zu den W"anden und
Begrenzungen halten, fallen in der Literatur sehr unterschiedlich
aus. Einheitlich ist jedoch, da"s der gehaltene Abstand stark von der
Beschaffenheit einer Begrenzung bestimmt wird. Als Richtgr"o"sen werden in
\cite{Weidmann:1993} f"ur Abst"ande in Korridoren mit Betonwand \meter{0.25}
und mit Metallwand \meter{0.20} sowie f"ur Abst"ande auf Gehwegen zu
Hausw"anden \meter{0.45}, Gartenz"aunen \meter{0.35} und zur Fahrbahn
\meter{0.35} vorgeschlagen.
 
\subsubsection{Orientierung}
\label{sec:orientierung}
{\fussgaenger} laufen in der Regel auf dem k"urzestem Weg in Richtung auf ihr
Ziel. Ist das Ziel in Sichtweite der {\fussgaenger} und der Weg dorthin frei,
so laufen sie m"oglichst in gerader Linie. 
Kompliziertere Strecken k"onnen durch einen Polygonzug mit mehreren
Zwischenstationen dargestellt werden.  

Die {\fussgaenger} orientieren sich dabei an markanten, visuell auf"-f"alligen
Punkten, auf die sie zusteuern. Solche Orientierungspunkte k"onnen zum
Beispiel die Lichtsignale an {\fussgaenger}"uberwegen sein. Durch eine
Anordnung der Signale f"ur jede "Uberquerungsrichtung auf der
gegen"uberliegenden {Stra"sen}seite rechts (oder links) hat man versucht, die
entgegenlaufenden {\fussgaengerstroeme} zu trennen \cite[S. 32]{Schubert:1967}.

Auf l"angeren Strecken kann es zu Richtungskorrekturen kommen,
weil beim N"aherkommen neue Hindernisse oder
Orientierungspunkte auftauchen.

\begin{figure}[t]
  \begin{center}
    \leavevmode \includegraphics[width=0.8\textwidth,angle=-90,clip]{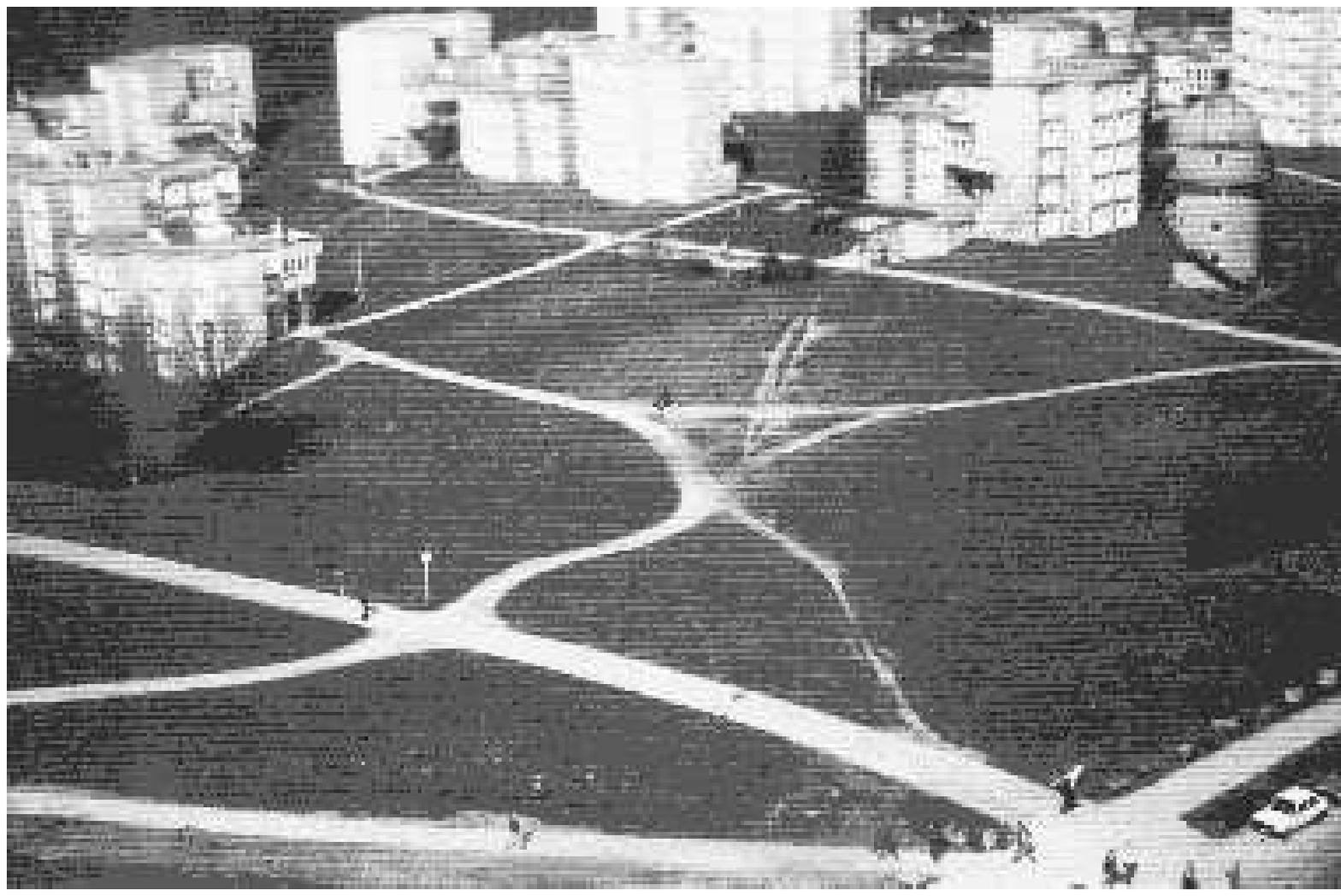}
  \end{center}
  \mycaption{Entwicklung von Trampelpfaden auf dem Campus}
  {Campus Vaihingen der
    Universit"at Stuttgart. Neben den angelegten 
    Fu"swegen haben sich auch einige Trampelpfade entwickelt. Im unteren Teil
    ist ein fast horizontal verlaufender Pflasterweg zu erkennen. Er wurde
    angelegt, nachdem dort ein stark benutzter Trampelpfad entstanden war.}
  \label{fig:trail_vaih}
\end{figure}

Fu"swege bieten den {\fussgaenger}n durch ihren Verlauf eine
Orientierungshilfe. Wenn die {\fussgaenger} jedoch durch eine ung"unstige
F"uhrung der vorgegebenen Fu"swege zu gr"o"seren Umwegen gezwungen werden,
bahnen sie sich, soweit m"oglich, einen eigenen Weg. Von diesem
Verhalten zeugen die zahlreichen Trampelpfade "uber die Rasenfl"achen in
Parkanlagen (vgl. Abbildung \ref{fig:trail_vaih}).

Dabei ist ein zweites Ph"anomen gut zu beobachten: {\fussgaenger} neigen
durchaus dazu, auf vorgegebenen Wegen zu laufen. Das ist einfacher und weniger
anstrengend als Orientierungspunkte anzupeilen und eigene Pfade anzulegen.
Gewisse Hemmungen, Rasenfl"achen durch Trampelpfade zu zerst"oren, d"urften 
die Tendenz existierende Pfade zu benutzen noch verst"arken.

Die Nutzung existierender Pfade f"uhrt zu den charakteristischen Eigenschaften
von ungeplanten Wegesystem, die bei \name{Schaur}{Eda} auch f"ur andere
Transportwegesysteme und selbstbildende Strukturen der Natur gefunden wurden
\cite{Schaur:1994}: 
\begin{enumerate}
\item Zwei aufeinandertreffende Wege schmiegen sich aneinander. Der Verlauf
  der Pfade ist durch leichte Kurven gepr"agt. 
\item Die meisten Knoten werden von drei Wegen gebildet.
\end{enumerate}
In Abbildung \ref{fig:trail_vaih} sind die
Unterschiede zwischen geplanten Wegen (unterer Bildteil) und
ungeplanten Trampelpfaden gut zu erkennen. Die Modellierung der
Entwicklung ungeplanter Wegesysteme auf der Basis des Soziale-Kr"afte-Modells
der {\fussgaenger} (Kap.\ \ref{cha:fussgaengermodell}) wird in Kapitel
\ref{cha:entwicklungvontrampelpfaden} behandelt. 

Eine Untersuchung von \name{Schenk}{Schenk, Martin} \cite{Schenk:1995} zeigt
die beiden Effekte der {\fussgaenger}orientierung:
\begin{figure}[p]
  \begin{center}
    \includegraphics[width=13cm]{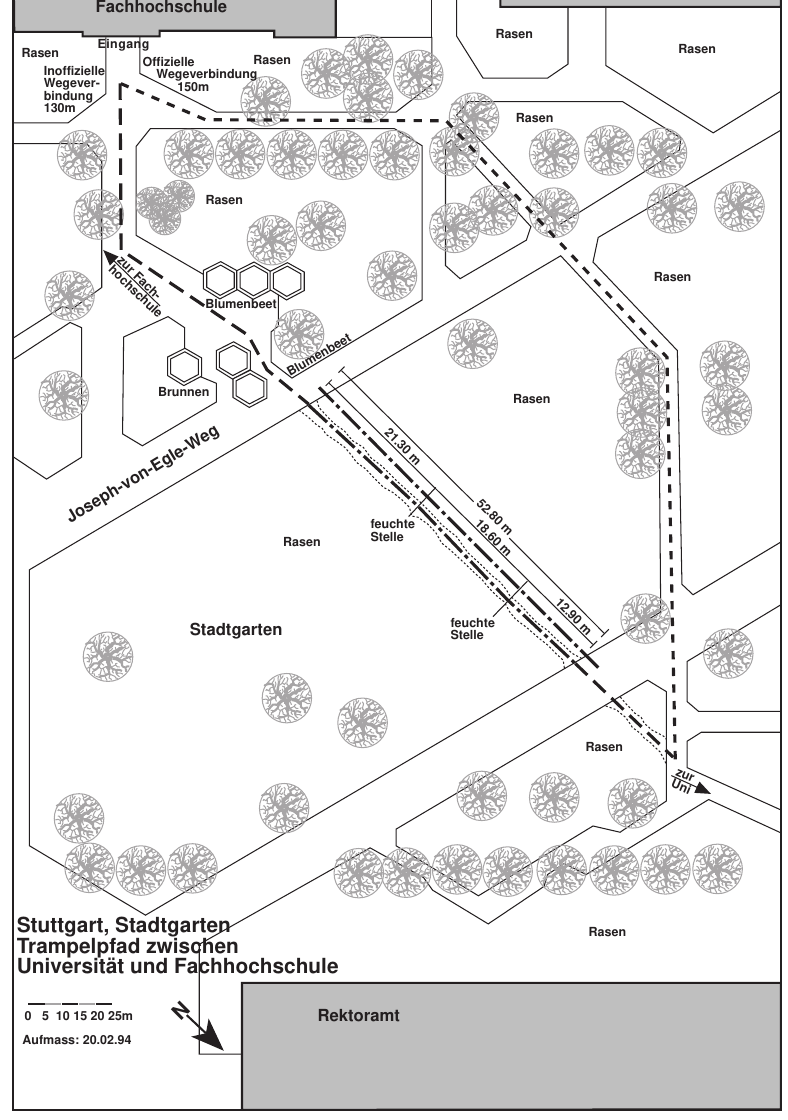}  
  \end{center}
  \mycaption{Trampelpfad durch den Stuttgarter Stadtgarten}
  {Stuttgarter Stadtgarten zwischen Universit"at und 
    Fachhochschule. Der Trampelpfad bildet einen fast gerade Verbindungsweg
    durch den Park \protect\cite{Schenk:1995}.}
  \label{fig:stadtgarten}
\end{figure}
Ein Trampelpfad im Stuttgarter Stadtgarten stellt eine wichtige
Wegebeziehung zwischen Universit"atsgeb"auden und der Fachhochschule her. Der
direkte Weg, der mit \meter{130} um \meter{20} k"urzer als der Weg auf den
angelegten Fu"swegen ist, f"uhrt als fast direkte Linie "uber die Rasenfl"ache
des Parks.  Trotz mehrmaliger Zerst"orung des Weges, d.h. Wiederherstellung
des Rasens, hat sich dieser Pfad jedesmal wieder neu entwickelt. Bei n"aherer
Betrachtung werden zwei Knickstellen im Pfad auf"-f"allig.

Das Orientierungverhalten der von den Universit"atsgeb"auden
kommenden {\fussgaenger} erkl"art die Ursache f"ur diese beiden
Kurskorrekturen. Am Anfang des Weges (unten rechts in Abbildung
\ref{fig:stadtgarten}) ist der Eingang der Fachhochschule durch B"aume
verdeckt. Nur ein St"uck des Fu"sweges zum Eingang ist zwischen den B"aumen zu
erkennen und dient als erster Orientierungspunkt. Beim N"aherkommen entdecken
die {\fussgaenger}, da"s sie auf eines der Brunnenbecken sto"sen werden und
korrigieren ihren Kurs, um genau zwischen den Brunnen und dem Baum
durchzukommen. Sie k"onnen an dieser Stelle noch nicht sehen, da"s sie damit
schnurstracks auf das Blumenbeet zulaufen und deshalb ihren  Kurs noch
ein zweites Mal "andern m"ussen. 

Die von der Fachhochschule zur Universit"at
laufenden {\fussgaenger} haben dieses Problem nicht. Ihr Orientierungspunkt ist
von weitem gut zu erkennen, und es stehen keine Hindernisse im Weg. 
Der Umstand, da"s sie trotzdem zweimal ihren Kurs "andern, ist nur durch die
Orientierung an dem bereits existierenden Pfad zu erkl"aren.  

\subsubsection{Umgebung und Beschaffenheit des Untergrunds}
\label{sec:umgebung}
{\fussgaenger} lassen sich au"serdem von der Umgebung und der Beschaffenheit
des Untergrunds leiten. Je nach Wetterlage werden eher
offene oder "uberdachte Teile der Verkehrsfl"ache bevorzugt.

Viele {\fussgaenger} scheinen freie,
menschenleere Fl"achen zu meiden. Auf weitl"aufigen Pl"at"-zen ist zu
beobachten, wie sie, Umwege in Kauf nehmend, am Rande entlang gehen, statt den
Platz diagonal zu "uberqueren.  Um gro"se Objekte, wie zum Beispiel hohe
Mauern oder Monumente, machen sie h"aufig weite B"ogen, was zu einer
Verl"angerung der Wegstrecken f"uhrt.

Unebener Untergrund wird ebenso gemieden, wie durch Regen aufgeweichte
Trampelpfade. Sobald Pfade nicht mehr begehbar sind, entstehen parallel zu
ihnen neue Wege. Steigungen und Gef"alle im Gel"ande veranlassen die
{\fussgaenger} in 
Serpentinen zu laufen. Die dadurch entstehenden Umwege werden
in Kauf genommen, weil die Serpentinen eine geringere Steigung ausweisen.


\section{Bemessung von {\fussgaenger}anlagen}
\label{sec:bemessung}
Zur Planung von {\fussgaenger}anlagen wurden Quantit"aten definiert, die als
Bemessungsgrundlage f"ur die Dimensionierung von Fu"swegen, Treppen
etc.\ herangezogen werden.

In Analogie zur Hydrodynamik steht die Zahl der {\fussgaenger} $N$, die
w"ahrend einer Zeitperiode $T$ einen  
Querschnitt durchstr"omen in der Beziehung
\begin{equation}
  N = \rho\,b_n\,v_{h}\,T
\end{equation}
mit der {\fussgaenger}dichte $\rho$, der nutzbaren Breite des Fu"sweges $b_n$
und der Horizontalgeschwindigkeit der {\fussgaenger} $v_h$. 
Unter der Leistungsf"ahigkeit 
\begin{equation}
  L = \frac{N}{T} = \rho\,b_n\,v_h
\end{equation}
wird die maximale Anzahl von Personen, 
die pro Stunde einen Querschnitt passieren, verstanden.
In der Literatur wird h"aufig die auf die Einheitsbreite von \meter{1}
bezogenen \hi{spezifische Leistungsf"ahigkeit}
\begin{equation}
  \label{def:spez_leist}
  \hat{L} = \rho\,v_h
\end{equation}
verwendet, mittels der verschiedene Anlagentypen auf einheitliche Weise
verglichen werden k"onnen. Dabei h"angt 
die {\fussgaenger}geschwindigkeit $v_h$ empfindlich von der
vorherrschenden Dichte ab. In der Regel nimmt sie mit zunehmender
{\fussgaenger}dichte ab. So beschreibt zum Beispiel die N"aherungsrelation von
\name{Kladek}{Kladek} (nach \cite[S.\ 62]{Weidmann:1993}) die Beziehung zwischen Geschwindigkeit
und Dichte durch
\begin{equation}
  \label{def:kladek}
  v_h(\rho) = v_h^0\,\left( 1-\exp\left(-\gamma\,\left(\frac{1}{\rho} -
  \frac{1}{\rho^\mx{max}}\right) \right) \right) 
\end{equation}
mit der Geschwindigkeit auf freier Fl"ache $v_h^0$, der maximal zul"assigen
Dichte $\rho^\mx{max}$ und der Eichkonstante $\gamma$, die aus empirischen
Untersuchungen ermittelt werden kann. In \cite{Weidmann:1993} werden die Werte
$v_h^0 = \mbox{\meter{1.34}} $, $\rho^\mx{max} = \mbox{\pqmeter{5.4}}$ und
$\gamma = \mbox{\pqmeter{1.913}}$ vorgeschlagen.
Die spezifische Leistungsf"ahigkeit ergibt sich damit aus
(\ref{def:spez_leist}) und  (\ref{def:kladek}) zu
\begin{equation}
  \hat{L}(\rho) = v_h^0\,\rho\,\left(
  1-\exp\left(-\gamma\,\left(\frac{1}{\rho} - 
  \frac{1}{\rho^\mx{max}}\right) \right) \right) 
\end{equation}
beziehungsweise als Funktion der Geschwindigkeit zu
\begin{equation}
  \hat{L}(v_h) = v_h\,\left( \frac{1}{\rho^\mx{max}} - \frac{1}{\gamma}\,\ln
  \left(1-\frac{v_h}{v_h^0}\right) \right)^{-1} 
\end{equation}
Die maximale Leistung mit \pmetersec{1.225} wird demnach bei einer
Geschwindigkeit von $v_h = \mbox{\meter{0.70}}$ und einer Dichte von
\pqmeter{1.75} erzielt \cite{Weidmann:1993}. Das bedeutet, da"s die maximale
Leistungsf"ahigkeit weder bei der h"ochsten Geschwindigkeit noch bei der
gr"o"sten Dichte erreicht wird. 

Allgemein interessieren bei der Bewertung der Leistungsf"ahigkeit von
{\fussgaenger}anlagen folgende Zusammenh"ange:
\begin{itemize}
\item Geschwindigkeit in Abh"angigkeit der {\fussgaenger}dichte
\item (spezifische) Leistungsf"ahigkeit in Abh"angigkeit der {\fussgaenger}dichte
\item (spezifische) Leistungsf"ahigkeit in Abh"angigkeit der Geschwindigkeit
\end{itemize}
Die Abh"angigkeiten werden h"aufig als sogenannte
\hii{Fundamentaldiagramme}{Fundamentaldiagramm} grafisch dargestellt. Ihre
Kurvenverl"aufe gleichen  denen des motorisierten Individualverkehrs
\cite{Weidmann:1993}.
  
Zur Bemessung von {\fussgaenger}anlagen ist die Verwendung der
maximalen Leistungsf"ahigkeit ungeeignet. Eine {\fussgaenger}dichte von
\pqmeter{1.75} ist in den seltensten F"allen zumutbar. Selbst in
Extremsituationen wie bei der Evakuierung von Geb"auden
w"are vorauszusetzen, da"s sich die in Panik geratene Menschenmenge soweit
unter Kontrolle bringen l"a"st, da"s sie sich mit \metersec{0.70} fortbewegt,
was der halben mittleren Gehgeschwindigkeit entspricht.

Die Auslegung einer Anlage mu"s deshalb gro"sz"ugiger ausfallen.
Belastungsspitzen werden in Relation zur der H"aufigkeit ihres Auftretens
ber"ucksichtigt. F"ur hohe aber seltene Belastungen wird dabei eine niedrigere
Qualit"atsstufe in Kauf genommen als bei Normalbelastung.

Anfang der siebziger Jahre wurde erstmals von \name{Fruin}{Fruin, John J.} das
\hi{Level-of-Service}-Konzept (LOS) zur detaillierten Unterscheidung auf den
{\fussgaenger}verkehr adaptiert \cite{Fruin:1971}. "Ahnliche "Uberlegungen
machte \name{Oeding}{Oeding, Detlef} bereits 1963 \cite{Oeding:1963}.
Allgemein werden folgende Kriterien verwendet \cite{Weidmann:1993}:
\begin{itemize}
\item  [K1]  M"oglichkeit zur freien Geschwindigkeitswahl 
\item  [K2]  H"aufigkeit eines erzwungenen Geschwindigkeitswechsels 
\item  [K3]  Zwang zur Beachtung anderer Fu"sg"anger 
\item  [K4]  H"aufigkeit eines erzwungenen Richtungswechsels
\item  [K5]  Behinderung bei Querung eines {\fussgaenger}stromes 
\item  [K6]  Behinderung bei entgegengesetzter Bewegungsrichtung 
\item  [K7]  Behinderung beim "Uberholen 
\item  [K8]  H"aufigkeit unbeabsichtigter Ber"uhrungen 
\end{itemize}
 
Mit diesen Kriterien l"a"st sich die Benutzungsqualit"at einer
{\fussgaenger}anlage einordnen. Durch die Unterscheidung von Erf"ullung bzw.
Nichterf"ullung einzelner Kriterien bei einer bestimmten {\fussgaenger}dichte
werden verschiedene Qualit"atsstufen definiert.  Weiterhin kann
aufgrund der dargestellten Abh"angigkeiten jedem LOS eine typische
Geschwindigkeit und Leistungsf"ahigkeit zugeordnet werden. Die
Leistungsf"ahigkeit einer {\fussgaenger}anlage ist dann nicht nur in Funktion
von Geschwindigkeit und Dichte, sondern auch der Benutzungsqualit"at
bestimmbar. 
In Abschnitt \ref{sec:bewertungskriterien} werden parallel hierzu Qualit"atsma"se
eingef"uhrt, die in der Simulation von {\fussgaenger}mengen berechnet werden.

Wie die Bewertung einer {\fussgaenger}anlage, bzw. eines Teilst"uckes davon,
aussehen kann, ist in Tabelle \ref{tab:beispiel_los} aus \cite{Weidmann:1993}
am Beispiel einer Ebene gezeigt. 
\begin{table}[htbp]
  \begin{center}
    \leavevmode
    \begin{tabular}[t]{|l|l|cccccccc|l|}
      \hline
      {\bf LOS} & {\bf Dichte} & \multicolumn{8}{l|}{\bf Kriterien} & 
      {\bf Charakterisierung}\\
      & \pqmeter{} & K1 & K2 & K3 & K4 & K5 & K6 & K7 & K8 & \\
      \hline \hline
      A & 0.00--0.10 & + & + & + & + & + & + & + & + & absolut freie Bewegung\\
      B & 0.10--0.30 & + & + & = & + & + & + & + & + & freie Bewegung \\
      C & 0.30--0.45 & = & + & = & = & = & = & = & + & schwache Behinderung \\
      D & 0.45--0.60 & = & = & = & = & -- & -- & -- & + & m"a"sige Behinderung \\
      E & 0.60--0.75 & -- & -- & -- & = & -- & -- & -- & + & starke Behinderung \\
      F & 0.75--1.00 & -- & -- & -- & -- & -- & -- & -- & + & dichter Verkehr \\
      G & 1.00--1.50 & -- & -- & -- & -- & -- & -- & -- & = & m"a"siges Gedr"ange \\
      H & 1.50--2.00 & -- & -- & -- & -- & -- & -- & -- & -- & starkes Gedr"ange \\
      I & 2.00--5.40 & -- & -- & -- & -- & -- & -- & -- & -- & massives Gedr"ange \\
      \hline
    \end{tabular}
  \end{center}
  \mycaption{Level-of-Service in der Ebene}
  {Beispiel f"ur die Charakterisierung verschiedenerer
    Level-of-Service beim Gehen in der Ebene.  
    Die Beurteilung geschieht qualitativ durch + gut, = mittelm"a"sig und
    -- schlecht.}
  \label{tab:beispiel_los}
\end{table}
Auf Treppenanlagen nehmen {\fussgaenger} beim selben Grad des Wohlbefindens
h"ohere Dichten in Kauf. Gleichzeitig f"allt die Geschwindigkeit gegen h"ohere
Dichten langsamer ab als in der Ebene. Daher verschiebt sich die
Qualit"atsbeurteilung gegen h"ohere Dichten. Davon abgesehen ist die typische
Verteilung der Merkmale f"ur Treppen und ebene Verkehrsfl"achen gleich. 

\section{Modellierung}
\label{sec:modellierung}
Zur mathematischen Beschreibung und Berechnung von {\fussgaenger}anlagen
wurden verschiedene Modelle entwickelt. Viele
verkehrswissenschaftliche  Arbeiten verwenden einfache Regressionsmodelle um
Gr"o"sen 
wie {\fussgaenger}flu"s, -dichte und -geschwindigkeit anhand der empirisch
erhobenen Daten in Beziehung zu setzen \cite{Weidmann:1993}.
Da weder Verhaltensregeln noch 
Interaktionen der {\fussgaenger} einbezogen werden, lassen sich die gewonnenen
mathematischen Relationen nur f"ur bekannte Geb"audeformen
generalisieren, die Vorhersage des Verhaltens von {\fussgaengerstroeme}n in
neuen, noch in der Planung befindlichen Anlagen ist jedoch problematisch.

\subsection{Makroskopische Modelle}
\label{sec:makroskopisch}
In den Arbeiten von \name{Henderson}{L. F.} und Mitarbeitern
\cite{Henderson:1971}, 
\cite{HendersonJenkins:1974}, \cite{Henderson:1974} werden empirisch erhobene
Daten von {\fussgaenger}mengen mit gaskinetischen und hydrodynamischen
Modellen verglichen. Die Verwendung der Boltzmann-Gleichungen f"ur
gew"ohnliche Fl"ussigkeiten und Gase beinhaltet allerdings die Annahme der
Impuls- und Energieerhaltung, die f"ur {\fussgaengerstroeme} unrealistisch ist.

\name{Helbing}{Dirk} entwickelte ein fu"s"-g"anger"-spezifisches makroskopisches
Modell, das auf Boltz"-mann-artige Gleichungen aufbaut und ohne Annahme der
Impuls- und Energieerhaltung auskommt \cite{Helbing:1992}, \cite{Helbing:1993}.
Darin werden die paarweisen Wechselwirkungen zwischen {\fussgaenger}n sowie
Wunschgeschwindigkeit und Gehrichtung beschrieben. Durch die Unterteilung der
{\fussgaenger}menge in Subpopulationen mit verschiedenen
Wunschgeschwindigkeiten und Zielrichtungen, lassen sich mit dem Modell auch
unterschiedliche Zusammensetzungen des {\fussgaenger}verkehrs untersuchen. 

Makroskopische Modelle gehen von einer kontinuierlichen Dichte aus, die in
{\fussgaengerstroeme}n auch n"aherungsweise nicht gegeben ist.
{\fussgaenger} befinden sich an wohldefinierten Orten im Raum, und ihre Anzahl
pro Fl"achenelement  variiert in ganzzahligen Schritten.
Das hat starke r"aumliche Schwankungen zur Folge, die nur durch die Betrachtung
von Mittelwerten "uber ausreichend gro"se Fl"achen, Geschwindigkeitsbereiche
und Zeitintervalle verschwinden.

\subsection{Mikroskopische Modelle}
\label{sec:mikroskopisch}
Bei der mikroskopischen Modellierung wird das Verhalten einzelner Individuen
meist durch einfache Regeln beschrieben. Das Zusammenspiel 
vieler Einzelakteure f"uhrt zu makroskopischen Effekten, die zur Untersuchung
der Systeme dienen. Durch die erforderliche hohe Anzahl von Individuen sind
mikroskopische Modelle in der Regel nur durch Computersimulationen zu
behandeln. Das erkl"art auch den erst vor einigen Jahren entstandenen
Trend zu mikroskopischen Modellen und Simulation: Zum einen werden lokale,
einfache und plausible Verhaltensregeln f"ur die Individuen verwendet, zum
anderen bieten moderne Computer die notwendige Rechenleistung, um detaillierte 
Modelle berechnen zu k"onnen. 

Seit Mitte der achtziger Jahren wurden mehrere mikroskopische Modelle mit zum
Teil unterschiedlichen Zielsetzungen entwickelt.  Die Modelle unterscheiden
sich auch in der Komplexit"at des simulierten menschlichen Verhaltens und der
Darstellung und Simulation des Prozesses der Bewegung.

Eine Einteilung kann grob anhand der Darstellung
der {\fussgaenger}dynamik in den mikroskopischen Modellen geschehen.

\subsubsection{Walker-Modelle} 
Die in \cite{GibbsMarksjo:1985} und \cite{EbiharaOhtsukiIwaki:1992}
vorgeschlagenen Ans"atze lassen die {\fussgaenger} auf einem quadratischen
Raster laufen, das den zweidimensionalen physikalischen Raum darstellt. Die
Bewegung eines {\fussgaenger}s wird durch den Wechsel von einem
Rasterpunkt auf einen benachbarten realisiert.
Dabei ist die Schrittweite durch die
Maschenweite des Rasters fest vorgegeben. \name{Gibbs}{Gibbs, P. G.} und
\name{Marksj"o}{Marksj"o} \cite{GibbsMarksjo:1985} k"onnen die {\fussgaenger}
in f"unf verschieden Geschwindigkeiten $v_1\dots v_5$ mit 0.5 bis
\metersec{2.5}
auf einem \meter{0.5} Raster laufen lassen, indem sie die unterschiedlich
schnellen {\fussgaenger} in der Reihenfolge 
5, 4, 3, 5, 2, 4, 5, 3, 4, 5, 1, 4, 5, 3, 2 simulieren.

Die {\fussgaenger} 
belegen ihre umgebenden Rasterpunkte mit Besetzungzahlen, deren Werte mit der
Entfernung 
von ihrer Position abnehmen. Die Besetzungszahlen verschiedener Individuen
auf einem Rasterpunkt werden addiert. F"ur ihren n"achsten Schritt
bevorzugen die {\fussgaenger} die Nachbarposition mit der niedrigsten
Besetzungszahl. 

\name{Ebihara}{Manabu}, \name{Ohtsuki}{Akira} und \name{Iwaki}{Hideaki}
entwickelten ein Modell zur Simulation von Geb"aude"-evakuierungen bei Feuer
oder Erdbeben. Die {\fussgaenger} bewegen sich hierbei auf einem gr"o"seren
Raster, deren Knoten daf"ur aber mit mehreren Personen besetzt sein k"onnen.
Die Rasterpunkte enthalten Informationen "uber die Besetzung, den Stand der
Evakuierung sowie "uber Zustand und Richtung der n"achsten Ausg"ange, sofern
sie von dieser Position im realen Geb"aude sichtbar oder entsprechend
beschildert sind. Das Modell f"ur das Verhalten der {\fussgaenger} kann
Entscheidungen und Lernf"ahigkeit der Individuen ber"ucksichtigen
\cite{EbiharaOhtsukiIwaki:1992}.

In dem von \name{Helbing}{Dirk} entwickelten Modell 
bewegen sich die {\fussgaenger} dagegen in einem zweidimensionalen
kontinuierlichen Raum. Die simulierten {\fussgaenger} laufen dabei mit
bestimmten Schrittweiten gem"a"s ihrer Geschwindigkeit in ihrer Zielrichtung.
Treffen sie dabei auf ein Hindernis, so "andern sie ihre Richtung um
einen bestimmten Winkel. Wenn diese Kurs"anderung nicht ausreicht, oder in der
neu gew"ahlten Richtung ein anderes Hindernis angetroffen wird, werden die
Richtungs"anderungen mehrmals mit gr"o"seren Winkeln versucht. Sind diese
Ausweichman"over nicht erfolgreich, stoppen die {\fussgaenger} vor dem
Hindernis \cite{Helbing:1990}\cite{Helbing:1991}. 

In Kapitel \ref{cha:pedestrianmodell} wird ein Soziale-Kr"afte-Modell f"ur
die {\fussgaenger}dynamik vorgestellt, das ebenfalls in die Klasse der
mikroskopischen Modelle geh"ort. Es beschreibt die Positionierung 
und  Bewegung der {\fussgaenger} ebenfalls in einem kontinuierlichen
physikalischen Raum.  

\subsubsection{Warteschlangen-Modelle (queueing models)}
Ebenfalls zur Simulation von Evakuierungsma"snahmen wurden die Modelle aus
\cite{YuhaskiMacgregorsmith:1989} und \cite{Lovas:1993}\cite{Lovas:1994}
entwickelt. Die Darstellung der Position und Bewegung der {\fussgaenger}
unterscheidet sich von den oben genannten Modellen: Ein Geb"aude wird als
Netzwerk seiner R"aume dargestellt. Dabei entsprechen die R"aume den
Knoten des Netzwerkes, die  mit den im jeweiligen Raum befindlichen
Personen besetzt sind. Die Verbindungst"uren zwischen den
einzelnen R"aumen stellen die Kanten des Netzwerkes dar 
(vgl.\ Abschn.\ \ref{sec:darstellung}, Abb.\ \ref{fig:korridornetz_zwei}).

Die {\fussgaenger} k"onnen die R"aume 
unter der Ber"ucksichtigung der Leistungsf"ahigkeit des Durchgangs
und der Kapazit"at des angestrebten Raumes wechseln.
Diese Gr"o"sen m"ussen aus anderen
Modellen oder empirischen Untersuchungen gewonnen werden. Ferner k"onnen die
Effekte der Geometrie des Geb"audes in den Warteschlangen-Modellen nicht
ber"ucksichtigt werden.  

\subsubsection{Routen-Wahl}
Modelle zur Routen-Wahl der {\fussgaenger}
\cite{BorgersTimmermans:1986} \cite{TimmermansHagenBorgers:1992} besch"aftigen
sich mit der Auswahl der Strecken und Ziele, die von den {\fussgaenger}n
angesteuert werden. H"aufig besteht die Intention dieser Modelle darin,
die Akzeptanz einzelner Gesch"afte oder St"ande und die Auslastung von
Verbindungswegen in gr"o"seren Einkaufsanlagen und Innenst"adten zu ermitteln.
Diese Modelle k"onnen in einer den Bewegungsmodellen "ubergeordneten Ebene
angesiedelt werden. Dieses Thema wird in Kapiteln \ref{cha:erweiterungen}
"uber Erweiterungen des Soziale-Kr"afte-Modells und in Kapitel
\ref{cha:wegenetze} "uber Wegenetze behandelt.

\subsubsection{Vorz"uge der Mikrosimulationen}
Mikroskopische Modelle und Simulationen erlauben eine detaillierte Untersuchung
der Eigenschaften der simulierten Individuen. So k"onnen Gr"o"sen ermittelt
werden, die sich in der Beobachtung realer {\fussgaengerstroeme} nur mit sehr
hohem Aufwand bestimmen lassen. Solche Gr"o"sen sind zum Beispiel die
H"aufigkeit der Richtungs- und Geschwindigkeits"anderung. In der Simulation
lassen sich solche \hi{Me"swerte} f"ur alle Individuen auf einfache Weise
ermitteln. 

{\fussgaenger} k"onnen mit zus"atzlichen Verhaltensregeln ausgestattet werden,
die von der momentanen, lokalen Situation der einzelnen Individuen abh"angen 
(vgl.\ Kapitel \ref{cha:erweiterungen}). 

In der Simulation l"a"st sich  die Komposition des {\fussgaenger}verkehrs
detailgetreu nachempfinden. 
Die unterschiedlichen individuellen
Eigenschaften der {\fussgaenger} wie etwa Alter oder Geschlecht k"onnen ebenso
wie der Zweck des Ganges Ber"ucksichtigung finden.

\chapter[Soziale Wechselwirkungen]
{Theorien der sozialen Wechselwirkungen im Raum}
\label{cha:theorien}

In der Psychologie und in den Sozialwissenschaften wurden verschieden Ans"atze
unternommen, die Beziehung zwischen Individuen und die St"arke ihrer
gegenseitigen Beeinflussung zu beschreiben. 
Dazu wurde "ubereinstimmend angenommen, da"s die M"oglichkeit zur gegenseitigen
Beeinflussung durch Distanzen zwischen den Individuen bestimmt werden. Diese
Annahme setzt die Definition eines Raumes voraus, an der sich die Ans"atze
sehr unterscheiden. In \cite{LiuLatane} werden  Darstellungm"oglichkeiten eines
sozialen Raumes aufgef"uhrt: 
\begin{itemize}
\item als mehrdimensionales Feld aus sozialen Merkmalen wie etwa
  Alter, Geschlecht, ethnische Zugeh"origkeit, Religion oder Besch"aftigung.
  Der Abstand zwischen zwei Punkten in diesem mehrdimensionalen Raum h"angt
  davon ab, in welchen Kategorien eine "Ubereinstimmung auftritt.
  (Nach \name{Blau}{}.)
\item als Netzwerk mit paarweisen Verbindungen zwischen den
  Individuen. Die Abst"ande k"onnen durch die L"angen der Verbindungskanten
  definiert werden. (Nach \name{Moreno}{}.)
\end{itemize}

Die folgenden Abschnitte behandeln zwei Theorien, die das soziale Verhalten
r"aumlich verteilter Individuen beschreiben. In beiden
Ans"atzen kann von einem allgemeinen sozialen Raum ausgegangen werden, der sowohl
die Anordnung der Individuen im physikalischen Raum, als auch deren Meinung
und soziale Unterschiede in seiner Topologie ber"ucksichtigt. 
Eine Vereinfachung eines solchen sozialen Raumes, f"ur den bis jetzt noch
keine befriedigende Definition gefunden wurde \cite{LiuLatane},
\cite{Latane:1995}, stellt der physikalische Raum dar.  

\section{Dynamic-Social-Impact-Theory}
\label{sec:sozialereinfluss}
Die Theorie der sozialen Wirkung (Dynamic-Social-Impact-Theory) von
\name{Latan\'e}{Bibb} \cite{Latane:1981} gibt einen Ansatz zur Beschreibung
sozialer Wechselwirkungen 
unter der Ber"ucksichtigung der r"aumlichen Anordnung der Individuen. 
Der Begriff "`soziale Wechselwirkung"' gilt dabei f"ur die verschiedenen
sozialen Prozesse, wie etwa die "Uberzeugung eines Diskussionspartners oder die
Entwicklung von Gruppenidentit"aten. Die St"arke der Beeinflussung wird
abk"urzend 
"`soziale Wirkung"' genannt. Diese Wirkung ist von einer Quelle ausgehend auf
ein (Ziel-)Individuum gerichtet. 

Wenn mehrere soziale Quellen auf ein Individuum einwirken, kann der Betrag
der  Wirkung als ein Produkt aus der St"arke (strength)  $S$,
Direktheit (immediacy) $I$ und der Anzahl (number) der Quellen $N$ durch
eine Funktion 
\begin{equation}
  \hat{\i} = f(SIN) 
  \label{def:socialimpact}
\end{equation}
beschrieben werden. Die St"arke bezeichnet hierbei die Macht, die Bedeutung
oder die Intensit"at, mit der die Quellen einwirken. Die Direktheit gibt die
Beziehung zwischen dem Beeinflussenden und dem Beeinflu"sten wieder. Sie kann
zum Beispiel durch die geographische Entfernung oder die H"aufigkeit, mit der
zwei Personen Kontakt aufnehmen, bestimmt werden. \name{Latan\'e}{Bibb}
verwendet den reziproken Abstand oder dessen Quadrat bei Modellen, die auf den
physikalischen Raum aufbauen\footnote{Im Soziale-Kr"afte-Modell der
  {\fussgaenger}dynamik (Kapitel \ref{cha:pedestrianmodell}) wird die
  Einflu"sst"arke durch eine exponentiell abfallende Abstandsfunktion
  bestimmt, die auch bei verschwindendem Abstand zwischen den Individuen im
  endlichen Wertebereich bleibt.}.

Die soziale Wirkung steigt nicht proportional mit der  Anzahl der Quellen.
Diese Annahme baut auf die Gesetzm"a"sigkeiten von \name{Fechner}{}
und \name{Stevens}{S. S.}, die das Verh"altnis zwischen objektiver und
subjektiver Realit"at als nichtlinear beschreiben: Nach dem Gesetz von
\name{Stevens}{} ist die subjektive Wahrnehmungsintensit"at $\Psi$
proportional zu
einer gewissen Potenz $\beta$ des objektiven physikalischen Reizes 
\begin{equation}
  \Psi = \kappa\,\Phi^\beta
  \label{def:stevens}
\end{equation}
Parallel dazu wird die Abh"angigkeit der sozialen Wirkung von der Anzahl der
Quellen durch 
\begin{equation}
  \hat{\i} = s\,N^t
\end{equation}
mit $t<1$ beschrieben.

\section{Theorie der Sozialen Kr"afte}
\label{sec:theoriedersozialenkraefte}
Das Konzept der sozialen Kr"afte basiert auf der \hi{Feldtheorie der
  Sozialwissenschaften } \cite{Lewin:1951} und  wurde von \name{Helbing}{Dirk}
mathematisch formuliert \cite{Helbing:1994}.

Dabei  repr"asentiert ein soziales Kraftfeld die "au"seren Einfl"usse auf ein
Individuum und bestimmt sein 
Verhalten. Analog zu der Theorie der sozialen Wirkung  kann ein solches Feld
auch durch Wechselwirkungen zwischen Individuen erzeugt werden.

Im Soziale-Kr"afte-Modell wird 
der momentane Zustand eines Individuums $\alpha$ durch seine Position 
$\vec{x}_\alpha = (x_1,\dots,x_d)$  
in einem mehrdimensionalen, kontinuierlichen Raum beschrieben. Die durch den
Einflu"s einer sozialen Kraft $\vec{f}_\alpha$ bedingte "Anderung erfolgt
in Analogie zu physikalischen Systemen durch die Langevin-Gleichung
\begin{equation}
  \frac{d\vec{x}_\alpha}{dt} = \vec{f}_\alpha(\vec{x}_\alpha, t) + \vec{\cal F}
\end{equation}
In dieser Form kann das Verhalten von Individuen modelliert werden, die nach 
der Maxime  des geringsten Widerstandes, oder der wahrscheinlichsten
Reaktionen auf einen Einflu"s handeln.

Der Zufallsterm $\vec{\cal F}$ steht f"ur stochastische Abweichungen vom
regul"aren Verhalten, die in ambivalenten Situationen auftreten k"onnen.
Zudem umfa"st $\vec{\cal F}$  kleinere Einfl"usse, die in der Modellierung 
nicht im einzelnen ber"ucksichtigt werden.

Ist das soziale Kraftfeld wirbelfrei, das hei"st die partiellen Ableitungen
der Kraft nach den Komponenten von $\vec{x}$ sind vertauschbar
\begin{equation}
  \frac{\partial}{\partial x_i}f_j(\vec{x}_\alpha,t) =
  \frac{\partial}{\partial x_j}f_i(\vec{x}_\alpha,t)
  \qquad\mbox{f"ur alle $i$,
    $j=1\dots d$ und f"ur alle $t$} 
\end{equation}
liegt dem Kraftfeld ein Potential $U$ mit
\begin{equation}
  \vec{f}(\vec{x},t) = -\nabla U(\vec{x},t)
\end{equation}
zugrunde. Dieses Potential l"a"st sich nach \name{Lewin}{}'s Theorie als
soziales Feld interpretieren. Es 
stellt zum Beispiel die "offentliche Meinung, soziale 
Normen oder aktuelle Trends dar.
 
Im Gegensatz zu  physikalischen Systemen wird das Newtonsche Gesetz
"`actio = reactio"' nicht gefordert, da die Individuen durchaus
unterschiedlichen Einflu"s aufeinander aus"uben k"onnen.
Ebensowenig wird bei sozialen Wechselwirkungsprozessen eine Energie- und
Impulserhaltung nach dem Vorbild der klassischen Mechanik angenommen.

Soziale Kr"afte k"onnen sowohl Paarwechselwirkungen  beschreiben, als auch den
Einflu"s globaler Felder. In den F"allen, in denen ein Individuum
mehreren Einflu"sgr"o"sen gleichzeitig ausgesetzt ist, gilt das
Superpositionsprinzip. Die soziale Kraft setzt sich dabei aus mehreren
Kr"aften zusammen
\begin{equation}
  \vec{f}_\alpha(\vec{x}_\alpha,t) = \sum_l \vec{f}_{\alpha l}(\vec{x}_\alpha,t)
\end{equation}
wobei mit $l$ die einzelnen Quellen indiziert sind.

In der Theorie der sozialen Kr"afte lassen sich Verhaltensweisen modellieren,
die aus mehreren Regeln neu enstehen. Die Funktion $f$ aus
(\ref{def:socialimpact}) wird ebenfalls als soziale Kraft bezeichnet. Beide
Ans"atze weisen Gemeinsamkeiten auf. So gelten die Betrachtungen der Theorie
der sozialen Wirkung (Dynamic-Social-Impact-Theory) auch f"ur das 
Soziale-Kr"afte-Modell.

\chapter[{\fussgaenger}dynamik]{Soziale-Kr"afte-Modell der 
{\fussgaenger}dynamik}
\label{cha:pedestrianmodell}
\label{cha:fussgaengermodell}
\label{cha:sozialekraeftemodell}
Die einfache Aufgabe eines {\fussgaenger}s von einem Ort zum anderen zu gehen,
erfordert im Gegensatz zur Auswahl der Zielorte keine gro"sen "Uberlegungen.
Hat ein {\fussgaenger} seine Richtung bestimmt, geht er fast automatisch auf
sein Ziel zu. Automatisch in dem Sinne, da"s er Hindernissen und
entgegenkommenden Passanten ausweicht, ohne weiter dar"uber nachzudenken.
Dabei hilft ihm seine Erfahrung, optimal auf ein Ereignis zu reagieren.

Die Beobachtungen zeigen, da"s sich die Bewegungsabl"aufe einzelner
{\fussgaenger} stark gleichen. Unterschiede treten lediglich bei der
Gehgeschwindigkeit und bei den Abst"anden, die sie zu Hindernissen und anderen
{\fussgaenger}n halten, auf. Sie stehen h"aufig mit dem Alter, dem Geschlecht
und der kulturellen Herkunft in Zusammenhang.

Die Einfachheit der {\fussgaenger}bewegung erlaubt eine mathematische
Beschreibung durch die Theorie der sozialen Kr"afte.  Trotz der guten
Vorhersagbarkeit der Bewegungen eines einzelnen {\fussgaenger}s ergeben sich im
Zusammenspiel der Wechselwirkungen zwischen mehreren {\fussgaenger}n komplexe
r"aumliche und zeitliche Strukturen. Die {\fussgaenger} sind sich meistens weder
bewu"st, da"s sie Teil einer solchen Strukturierung sind, noch da"s sie
selbst dazu beitragen.

\section{{\fussgaenger}-Modell}
\label{sec:fussgaengermodell}
Im folgenden sollen die beobachteten Bewegungsmuster durch soziale Kr"afte
beschrieben werden. Dabei wird immer vom Blickpunkt eines {\fussgaenger}s
$\alpha$, auf den die sozialen Kr"afte wirken, ausgegangen. Andere
{\fussgaenger} werden mit $\beta$, umgebende Hindernisse mit $B$ und
Anziehungspunkte mit $i$ 
bezeichnet. Alle Gr"o"sen des Modells sind durch eines dieser Symbole als Index
einem bestimmten Objekt zugeordnet. Die sozialen Kr"afte $\vec{f}$ und ihre
Potentiale $U$ h"angen in der Regel vom Abstand zwischen dem Verursacher
($\beta$, $B$ oder $i$) und 
dem {\fussgaenger} $\alpha$ ab. Auch andere Gr"o"sen, wie zum Beispiel die
Geschwindigkeit $\Vb$ eines entgegenkommenden {\fussgaenger}s k"onnen die
Wechselwirkung bestimmen. Da die Abh"angigkeiten aus den Definitionen
ersichtlich sind, werden die Kraft- und Potentialfunktionen nicht zus"atzlich
gekennzeichnet.

Im Soziale-Kr"afte-Modell der {\fussgaenger}dynamik wird die Masse auf $m = 1$
gesetzt.  Die sozialen Kr"afte haben damit die Dimension der Beschleunigung
$\left[\frac{m}{\sq{s}}\right]$.

\subsection{Bewegung eines {\fussgaenger}s}
Das Verhalten eines {\fussgaenger}s "au"sert sich in der "Anderung seiner
Geschwindigkeit und damit in der Bewegung im Raum.
Die Geschwindigkeit eines {\fussgaenger}s $\Va$ wird durch das Zusammenspiel
aller auf den {\fussgaenger} wirkenden sozialen Kr"afte 
mit
\begin{equation}
  \frac{d\Va}{dt} 
  =\:\underbrace{\Fanull}_{\mx{Ziel}} \: 
  +\:\underbrace{\sum_{\beta}\vec{f}_{\alpha \beta}}
  _{\mx{{\fussgaenger}}}\:
  +\:\underbrace{\sum_B \vec{f}_{\alpha B}}_{\mx{Hindernisse}} \:
  +\:\underbrace{\sum_i \vec{f}_{\alpha i}}_{\mx{Attraktionen}} \:
  +\:\underbrace{\sum_{\alpha^\prime} {\vec{f}_{\alpha \alpha^\prime}}}
  _{\mx{Gruppen}}\: 
  +\:\underbrace{\vec{\cal F}_\alpha}_{\mx{Fluktuation}}
\label{tot_soz_kraft}
\end{equation}
ver"andert. Die einzelnen Beitr"age der Bewegungsgleichung werden in den
folgenden Abschnitten vorgestellt.

Im Gegensatz zu
physikalischen Systemen, sind {\fussgaenger} aktive Teilnehmer, die
ihre Bewegung eigenst"andig aufbringen. 

Es kann vorkommen, da"s die sozialen Kr"afte zu einer Geschwindigkeit
f"uhren, die die maximal
m"ogliche Geschwindigkeit eines {\fussgaenger}s $v_\alpha^{max}$
"uberschreitet. In diesem Fall will er schneller laufen, als er kann.
Die Bewegung eines {\fussgaenger}s ist daher durch
\begin{equation}
  \label{motivation}
  \frac{d \Ra}{dt} = \theta \,\Va 
\end{equation}
mit einer Begrenzungsfunktion
\begin{equation}
  \theta = \left\{ 
  \renewcommand{\arraystretch}{1.2}
  \begin{array}{r@{\quad:\quad}l}
    1 & \|\Va\| \le v_\alpha^{max} \\
    \frac{v_\alpha^{max}}{\|\Va\|} & \|\Va\| > v_\alpha^{max} 
  \end{array} \right.
\end{equation}
bestimmt. 

Neben den sozialen Kr"aften aus (\ref{tot_soz_kraft}), die durch das Modell
definiert werden, wird das
Verhalten eines Individuums durch einen stochastischen Term $\vec{\cal
  F}_\alpha$ beeinflu"st. Er ver"andert die Bewegungsgleichung zu jedem
Zeitpunkt um einen zuf"allig gew"ahlten Wert. Die {\einfuehrung} einer
Zufallsgr"o"se kann dabei in den folgenden drei Absichten geschehen
\cite{Gilbert:1995}:
\begin{enumerate}
\item Der Zufallsterm dient als Stellvertreter aller Einflu"sgr"o"sen, die
  wegen ihres geringen Beitrags im Modell nicht explizit definiert sind.
\item Durch kleine, zuf"allige Beitr"age werden Artefakte vermieden,
  die durch das Modell oder die Simulation entstehen k"onnen.
  Zus"atzlich l"a"st sich dadurch verhindern, da"s das System in einem
  (indifferenten) Gleichgewicht stehen  bleibt. 
\item Zugef"ugtes Rauschen testet die Robustheit eines Modells.
  Dies ist besonders beim Auftreten von Selbstorganisationseffekten   
  von Bedeutung, um sicherzustellen, da"s diese ausschlie"slich durch die
  Modellwechselwirkungen hervorgerufen werden.  
  \end{enumerate}

Das Modell der {\fussgaenger}bewegung besteht aus zahlreichen Akteuren, die
zuf"allig verteilt, mit unterschiedlichen Anfangsbedingungen in das System
eintreten. Dadurch ist f"ur st"andige, "au"sere Einfl"usse, die auch als
Rauschen gesehen werden k"onnen, gesorgt. Um instabile Gleichgewichte zu
vermeiden wird im {\fussgaenger}modell ein Fluktuationsterm 
\begin{equation}
  \vec{\cal F}_\alpha  =  \vp{\Eanull}{\vec{f}_\alpha}\,\N(0,\mu)\,\Esanull
\end{equation}
mit der Zufallsvariablen $\N(0,\mu)$ eingef"uhrt. Die Verteilungsfunktion
\begin{equation}
  P(\N(0,\mu)) = \frac{1}{\sqrt{2\pi\mu}} e^{-\sq{x}/2\mu}
\end{equation}
weist dabei die Form einer Normalverteilung auf. Der Einheitsvektor $\Esanull$
steht senkrecht zum Einheitsvektor der Zielrichtung $\Eanull$, soda"s
$\vp{\Eanull}{\Esanull} = 0$ gilt.  Der Fluktuationsterm repr"asentiert
au"serdem die zuf"alligen Abweichungen vom Regelfall. Sein Einflu"s ist
besonders stark, wenn die gesamte soziale Kraft $\vec{f}_\alpha$ gegen die
Zielrichtung des {\fussgaenger}s zeigt.

\subsection{Antriebskraft}
Ein {\fussgaenger} geht in direkter Linie und mit einer individuellen, "uber
den Weg nahezu konstanten Geschwindigkeit auf sein n"achstes Ziel zu. Dabei
         
versucht er nach einer Ablenkung oder Abbremsung wieder auf seine
Wunschgeschwindigkeit zu beschleunigen und Kurs auf sein Ziel zu nehmen.

Dieses Verhalten, den gew"unschten Kurs einzuschlagen, kann durch die
Antriebskraft 
\begin{equation}
  \Fanull = \frac{1}{\tau_\alpha}\left(\Vanull - \Va\right) 
  = \underbrace{\frac{v^0_\alpha}{\tau_\alpha}\,
    \vec{e}\,^0_{\!\alpha}}_{\mbox{I}} + 
  \underbrace{\frac{-1}{\tau_\alpha}\,\Va}_{\mbox{II}}
  \label{f_antrieb}
\end{equation}
mit der Wunschgeschwindigkeit $\Vanull$, der momentanen 
Geschwindigkeit $\Va$  und der Relaxationszeit $\tau_\alpha$ ausgedr"uckt
werden. Der Richtungseinheitsvektor 
\begin{equation}
  \Eanull(t) =  \frac{\vec{p} - \Ra}{\|\vec{p} - \Ra\|}
\end{equation}
ergibt sich aus der momentanen Position des {\fussgaenger}s $\Ra$ und dem
{\naechsten} Zielpunkt $\vec{p}$.
Man kann die beiden Teile des Kraftterms 
aber auch als konstante Beschleunigungskraft (I) und Reibungskraft (II)
interpretieren.

\begin{figure}[tb]
  \begin{center}
    \leavevmode
    \includegraphics[width=0.7\textwidth]{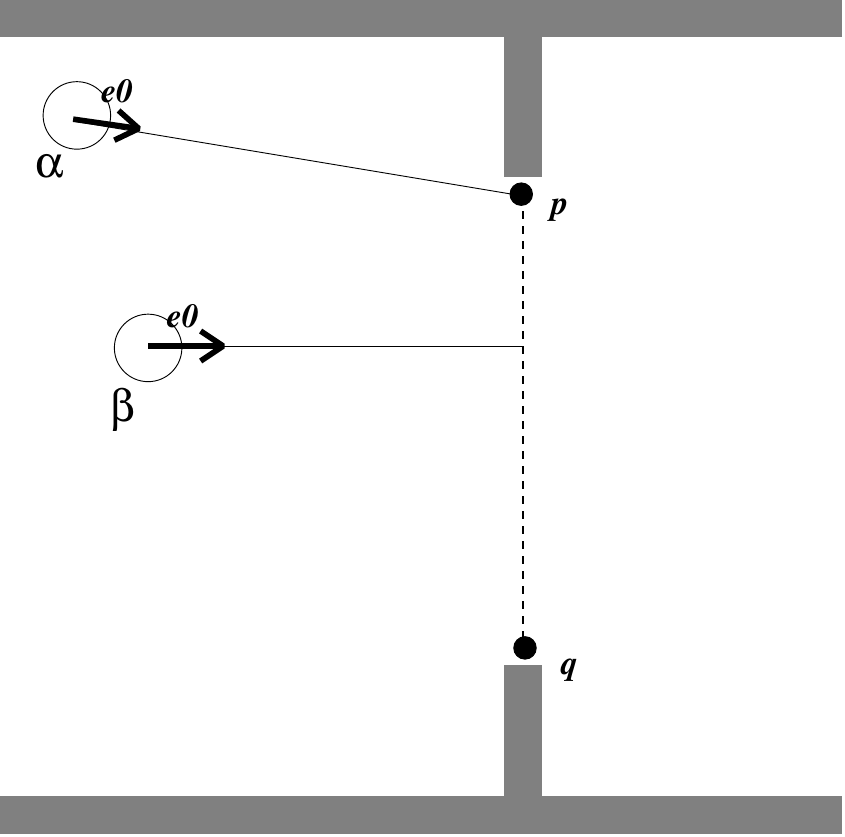}
  \end{center}
  \mycaption{Richtungsvektor auf ein (Zwischen-)Ziel}
  {Bestimmung Richtung zu einen Durchgang. Der Richtungseinheitsvektoren
    $\Eanull$, $\Ebnull$
    werden durch den k"urzesten Abstand zu der Verbindungslinie
    zwischen $\vec{p}$ und $\vec{q}$ nach (\ref{abstand_tor}) bestimmt. Der
    Richtungsvektor $\Eanull$ steht senkrecht auf der Verbindungslinie. 
    $\beta$  hat den k"urzesten Abstand zum Eckpunkt $\vec{p}$ des
    Durchgangs.}
  \label{fig:skizze_zielrichtung}
\end{figure}

Soll anstelle  eines Zielpunkts ein gr"o"serer Zielbereich beschrieben werden,
etwa das Ende eines Korridors oder ein Durchgang (vgl.\
\ref{sec:darstellung}), so kann dies durch zwei Punkte $\vec{p}$ und 
$\vec{q}$ geschehen, zwischen denen der {\fussgaenger} durchgehen mu"s.
Die Zielrichtung 
\begin{equation}
  \Eanull = \frac{\vec{s}}{\|\vec{s}\|}
\end{equation}
ergibt sich aus dem k"urzesten Abstand $\vec{s}$ zum Geradenabschnitt
zwischen den Punkten 
\begin{equation}
  \vec{s} = \left\{ 
  \renewcommand{\arraystretch}{1.2}
  \begin{array}{r@{\quad:\quad}l}
    \vec{p} - \Ra & \vp{\Ra-\vec{p}}{\vec{e}_{qp}} \le 0 \\
    \vec{p} - \Ra - \vp{\vec{e}_{qp}}{\vec{p}-\Ra}\,\vec{e}_{qp} & 0 <
    \vp{\Ra-\vec{p}}{\vec{e}_{qp}} < \|\vec{q}-\vec{p}\| \\ 
    \vec{q} - \Ra  & \|\vec{q}-\vec{p}\| \le
    \vp{\Ra-\vec{p}}{\vec{e}_{qp}} 
  \end{array} \right.
  \label{abstand_tor}
\end{equation}
mit $\vec{e}_{qp} = (\vec{q}-\vec{p})/(\|\vec{q}-\vec{p}\|)$ (vgl.\ Abb.\ \ref{fig:skizze_zielrichtung}).
F"ur {\fussgaenger}, die einen geraden Korridor von einem Ende zum anderen
durchwandern sollen, ist die Gehrichtung $\Eanull$ "uber die ganze Strecke
konstant.


\subsection{Wechselwirkung zwischen {\fussgaenger}n}
Die h"aufigste Wechselwirkung, die man in {\fussgaenger}mengen beobachtet, ist
das Abstandhalten zwischen den {\fussgaenger}n. Jeder {\fussgaenger} r"aumt seinen
Mitmenschen dabei einen bestimmten Freiraum ein. Die Gr"o"se dieses
zugestandenen Territoriums ist situationsabh"angig. Eine Rolle spielt
insbesondere die {\fussgaenger}dichte. Wird das Territorium eines anderen
verletzt, so entfernt sich der {\fussgaenger} auf direktem Wege. 

Zur Beschreibung dieses Verhaltens wird eine repulsive Wechselwirkung
entlang der Verbindungslinie der beiden {\fussgaenger} angenommen. Die St"arke
der Absto"sung h"angt von der Entfernung ab. Sie ist beim geringsten Abstand
am gr"o"sten und f"allt f"ur zunehmenden Abstand monoton gegen Null ab.
Au"serdem darf die Intensit"at der  Einfl"usse selbst beim kleinstm"oglichen
Abstand ein maximales Limit nicht "uberschreiten.

Diese Bedingungen erf"ullt zum Beispiel der Ansatz mit einer
Exponentialfunktion der Form
\begin{equation}
  U_{\alpha \beta} = p_\alpha\,
  \exp\left(-\frac{\|\Ra-\Rb\|}{\sigma_\alpha}\right)   
  \label{Uped}
\end{equation}
und der daraus resultierenden sozialen Kraft
\begin{equation}
  \vec{f}_{\alpha \beta} = -\nabla  U_{\alpha \beta} = 
  \frac{p_\alpha}{\sigma_\alpha}\,
  \exp\left(-\frac{\|\vec{r}_\alpha-\vec{r}_\beta\|}{\sigma_\alpha}\right)
  \left( \Ra - \Rb \right).
\end{equation}
Die Parameter $p_\alpha$ und $\sigma_\alpha$ bestimmen dabei die maximale
St"arke und die Reichweite der Paarwechselwirkung. Durch sie wird die
Bereitschaft eines {\fussgaenger}s $\alpha$ ausgedr"uckt, einem anderen Platz
zu machen.  

\subsubsection{Erweiterung 1}
Ist die {\fussgaenger}dichte gering, so behalten die {\fussgaenger} einen
gewissen "Uberblick "uber das Verhalten anderer Passanten und k"onnen deren
Bewegung absch"atzen. Bei Ausweichman"overn wird die Gehrichtung des anderen
eingeplant und entsprechend mehr Raum freigelassen.  Das Absto"sungspotential
wird daf"ur um den Raumbedarf der {\naechsten} paar Schritte eines
{\fussgaenger}s erweitert:
\begin{equation}
  U^\prime_{\alpha \beta} = p_\alpha\,e^{-\rp/\sigma_\alpha}
  \label{Uellipse} 
\end{equation}
mit 
\begin{equation}
  \rp = \frac{1}{2} \sqrt{
    \sq{\left(\|\Ra-\Rb\| + \|\Ra-\Rb - \Delta t \Vbnull \| \right)} 
    - \sq{\left(\| \Delta t \,\Vbnull \| \right)}}    
  \label{rellipse}
\end{equation}
Die Gleichung (\ref{rellipse}) definiert hierbei eine Ellipse mit den
beiden Brennpunkten bei $\vec{r}_\beta $ und $\vec{r}_\beta + \Delta
t \, \Vbnull$. Der zweite Brennpunkt entspricht dem Ort, den der
{\fussgaenger} $\beta$ in der Zeit $\Delta t$ erreicht haben wird (vgl.\ Abb.\
\ref{ped_pot_contour} Mitte). 

Da {\fussgaenger} die Einfl"usse der umgebenden Passanten  in
Abh"angigkeit der Richtung deren Ursprungs wahrnehmen, reagieren sie
unterschiedlich stark darauf. So wird ein {\fussgaenger} hinter ihm laufende
Passanten weniger stark beachten als solche, die vor ihm laufen, und
denen er gegebenenfalls ausweichen mu"s.   

Die Einschr"ankung des Wahrnehmungsbereiches auf einen Blickwinkel $\phi$ 
kann dieses Verhalten in der Modellierung
ber"ucksichtigen. Der Blickwinkel ist dabei auf das Ziel des {\fussgaenger}s
gerichtet. Alle Einfl"usse, deren Ursprung  au"serhalb dieses Bereichs liegt,
werden durch einen Vorfaktor  
\begin{equation}
  w  = \left\{ 
  \renewcommand{\arraystretch}{1.2}
  \begin{array}{r@{\quad:\quad}l}
    1 & \vp{-\vec{f}_\alpha}{\Eanull\/} \ge \cos \phi/2  \\[2mm]
    \omega < 1 & \vp{-\vec{f}_\alpha}{\Eanull\/}  < \cos \phi/2 
  \end{array} \right.
  \label{blickwinkel}
\end{equation}
abgeschw"acht. Auf diese Weise wird au"serdem das Dr"angeln von nachkommenden
{\fussgaenger}n verhindert. In den Simulationen haben sich die Werte $\phi =
200^\circ$  und $\omega =0.2\dots 0.5$ bew"ahrt.

\subsubsection{Erweiterung 2}
Durch die vorgestellten Potentialformen weichen die {\fussgaenger} einander sehr
abrupt aus. Daf"ur ist die 
Form des elliptischen Potentials verantwortlich. Bei
einem zentralen Zusammensto"s erfahren beide {\fussgaenger} eine Kraft entgegen
ihrer Bewegungsrichtung, aber keine senkrechte Ablenkung, die ein Ausweichen
m"oglich machen w"urde.

Dieses Problem wird durch einen neuen Ansatz behoben: Es wird
ebenfalls angenommen, da"s ein {\fussgaenger} auf vor ihm stattfindende
Ereignisse st"arker reagiert als auf solche, die hinter ihm ablaufen.

Die Wechselwirkungspotentiale der umgebenden {\fussgaenger} werden dabei
entgegen der Zielrichtung $\Eanull$\, ausgedehnt. Durch die {\einfuehrung}
eines neuen Koordinatensystems $\{\vec{x},\,\vec{y}\}$, das
in $\Eanull$-Richtung ausgerichtet ist, l"a"st sich die mathematische
Beschreibung verdeutlichen.  Der Zielrichtungsvektor $\Eanull$\, bildet den
ersten Teil der Basis des neuen Koordinatensystems. Durch das Schmidtsche
Orthonormalisierungsverfahren \cite{Fischer:1986} kann die Basis zu
\begin{equation}
  \renewcommand{\arraystretch}{1.2}
  \begin{array}{l@{\qquad}l}
    \vec{y} = \vec{e}\,^0_\alpha & \\
    \vec{x} \perp \vec{y} & \mbox{mit}\quad \vp{\vec{x}}{\vec{x}} = 1 \\
  \end{array}
  \label{neue_koord}
\end{equation}
erg"anzt werden.

Die repulsive Wirkung eines {\fussgaenger}s $\beta$ wird durch das Potential
\begin{equation}
  U^{\prime\prime}_{\alpha \beta} 
  = p\,e^{-\rpp/\sigma_\alpha}
  \label{Udehn}
\end{equation}
und die soziale Kraft
\begin{equation}
  \vec{f}^{\prime\prime}_{\alpha \beta} = - \frac{1}{\sigma\rpp}
  \,  e^{-\rpp/\sigma_\alpha}\,(\Ra-\Rb)
\end{equation}
mit
\begin{equation}
  \rpp = \sqrt{\left<(\Ra-\Rb),\vec{x}\right>^2 + 
        \gamma^2 \left<(\Ra-\Rb),\vec{y}\right>^2}
\end{equation}
und
\begin{equation}
  \gamma = \left\{ 
  \renewcommand{\arraystretch}{1.2}
  \begin{array}{r@{\quad:\quad}l}
    1 & \vp{\Ra-\Rb}{\vec{y}} \ge 0 \\    \frac{1}{1+\lambda \Va} &  \vp{\Ra-\Rb}{\vec{y}} < 0 \\
  \end{array} \right.
\end{equation}
dargestellt. 

Im Vergleich zu den Potentialfunktionen aus (\ref{Uped}) und (\ref{Uellipse})
wird bei diesem Ansatz ein {\fussgaenger} von vor ihm gehenden Passanten 
fr"uher abgelenkt. Einfl"usse von hinter ihm befindlichen Ereignissen k"onnen 
analog durch eine Stauchung des positiven Teils der $\vec{y}$-Achse mit
$\gamma^\prime > 1$ 
f"ur $\vp{\Ra-\Rb}{\vec{y}} > 0$ abgeschw"acht werden. Diese Beschreibung
ersetzt den Blickwinkel-Vorfaktor $w$ aus (\ref{blickwinkel}). 

\begin{figure}[tb]
  \begin{center}
    \leavevmode
    \includegraphics[width=0.7\textwidth,angle=-90]{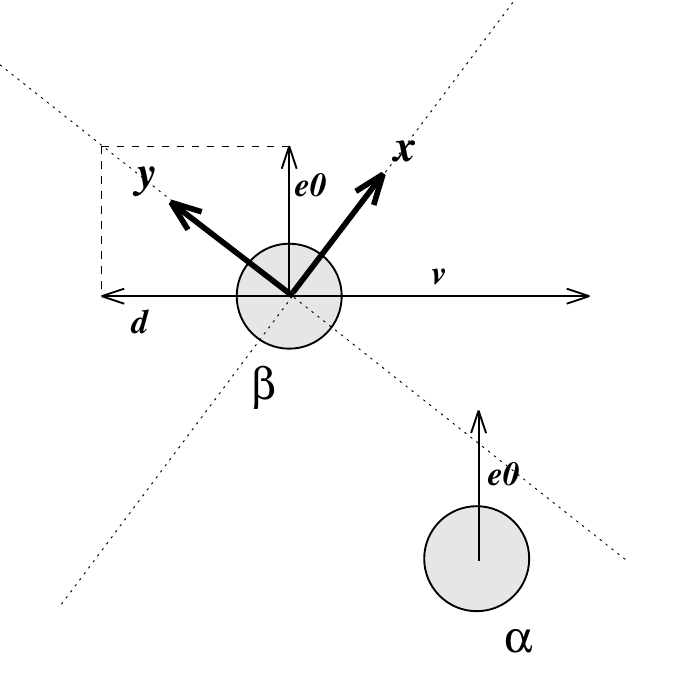}
  \end{center}
  \mycaption{Skizze des Zielrichtungs-Koordinatensystems}
  {Das Koordinatensystem des zweiten erweiterten Potentialansatzes 
    $U^{\prime\prime}_{\alpha \beta}$ aus (\ref{Udehn}) ist durch die
    momentane Geschwindigkeit des {\fussgaenger}s $\beta$ und die
    Zielrichtung von $\alpha$ definiert. Die Vektoren der Abbildung sind 
    $e0 =\Eanull$, $v = \Vb$ 
    und $d = \delta_\alpha\,(v_\beta^0 / v_\alpha^0)$.
    Die Basisvektoren des neuen Koordinatensystems sind $y = \vec{y}$ aus
    (\ref{rel_zielrichtung}) und $x = \vec{x}$ aus (\ref{neue_koord}).}
  \label{fig:skizze_koordinatensystem}
\end{figure}

Um die Bewegung der anderen {\fussgaenger} zu ber"ucksichtigen, wie das im
ersten erweiterten Ansatz (\ref{Uellipse}) bereits geschehen ist,
wird das Koordinatensystem $\{\vec{x},\,\vec{y}\}$ nach der
relativen Zielrichtung 
\begin{equation}
  \vec{y} = 
  \frac{ 
    \Eanull - \delta_\alpha \frac{v_\beta^0}{v_\alpha^0}\,\Ebnull
    } 
  {
    \| \Eanull - \delta_\alpha \frac{v_\beta^0}{v_\alpha^0}\,\Ebnull\|
    }
  \label{rel_zielrichtung}
\end{equation}
der {\fussgaenger} $\alpha$ und $\beta$ ausgerichtet. Der Parameter
$\delta_{\alpha}$ gibt an, wie stark die Gehrichtung des anderen
{\fussgaenger}s ber"ucksichtigt wird (vgl.\ Abb.\
\ref{fig:skizze_koordinatensystem}).

\begin{figure}[tb]
  \includegraphics[width=4cm]{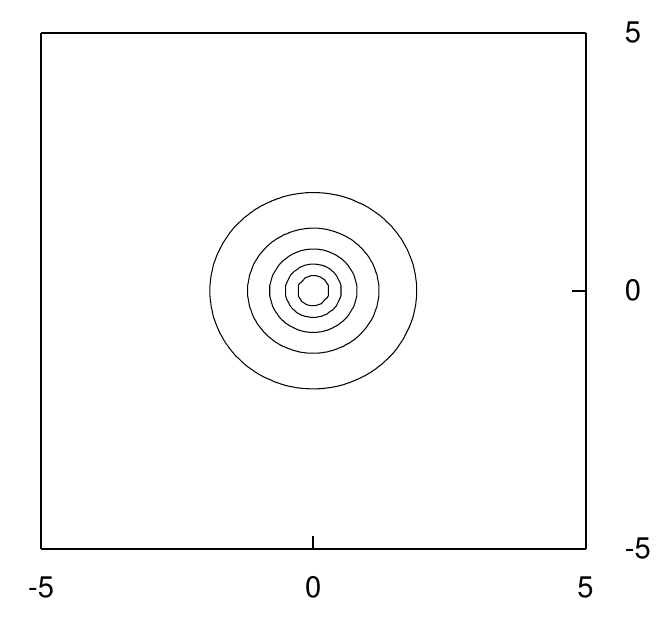}
  \hfill
  \includegraphics[width=4cm]{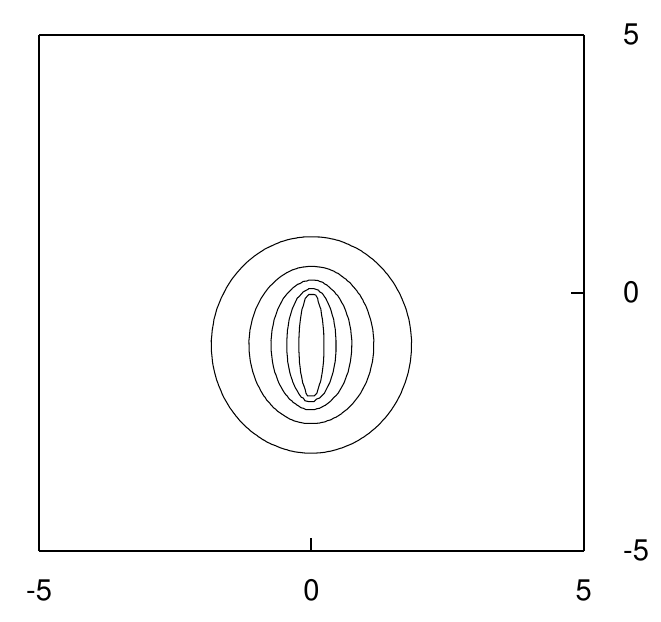}
  \hfill
  \includegraphics[width=4cm]{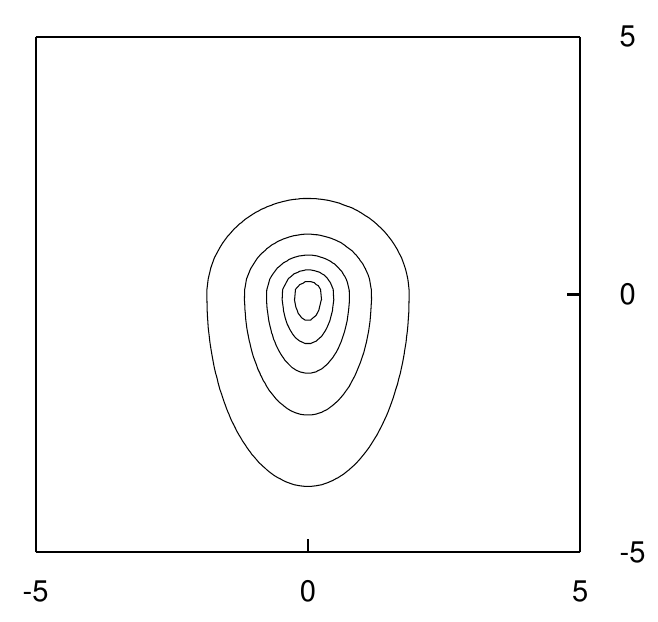}
  \mycaption{Verschiedene Potentialans"atze zur Beschreibung des Abstandsverhaltens gegen"uber {\fussgaenger}n}
  {Verschiedene
    Potentialtypen der repulsiven {\fussgaenger}-Wechselwirkung. Die
    "Aquipotentiallinien beschreiben die Form des Territoriums eines Passanten
    $\beta$ aus der Sicht des {\fussgaenger}s $\alpha$, dessen Zielrichtung
    nach oben gerichtet ist. Von links nach rechts: $U_{\alpha \beta}$ aus
    (\ref{Uped}), die erste Erweiterung $U^\prime_{\alpha \beta}$ aus
    (\ref{Uellipse}) und die zweite Erweiterung 
    $U^{\prime\prime}_{\alpha \beta}$ aus (\ref{Udehn}).} 
  \label{ped_pot_contour}
\end{figure}

Die Potentialfunktionen aus (\ref{Uped}), (\ref{Uellipse}) und
(\ref{Udehn}) sind in Abb. \ref{ped_pot_contour} zum Vergleich als
"Aquipotentiallinien dargestellt.  

\begin{figure}[t]
  \begin{center}
    \fbox{
      \includegraphics[width=10cm,angle=-90]{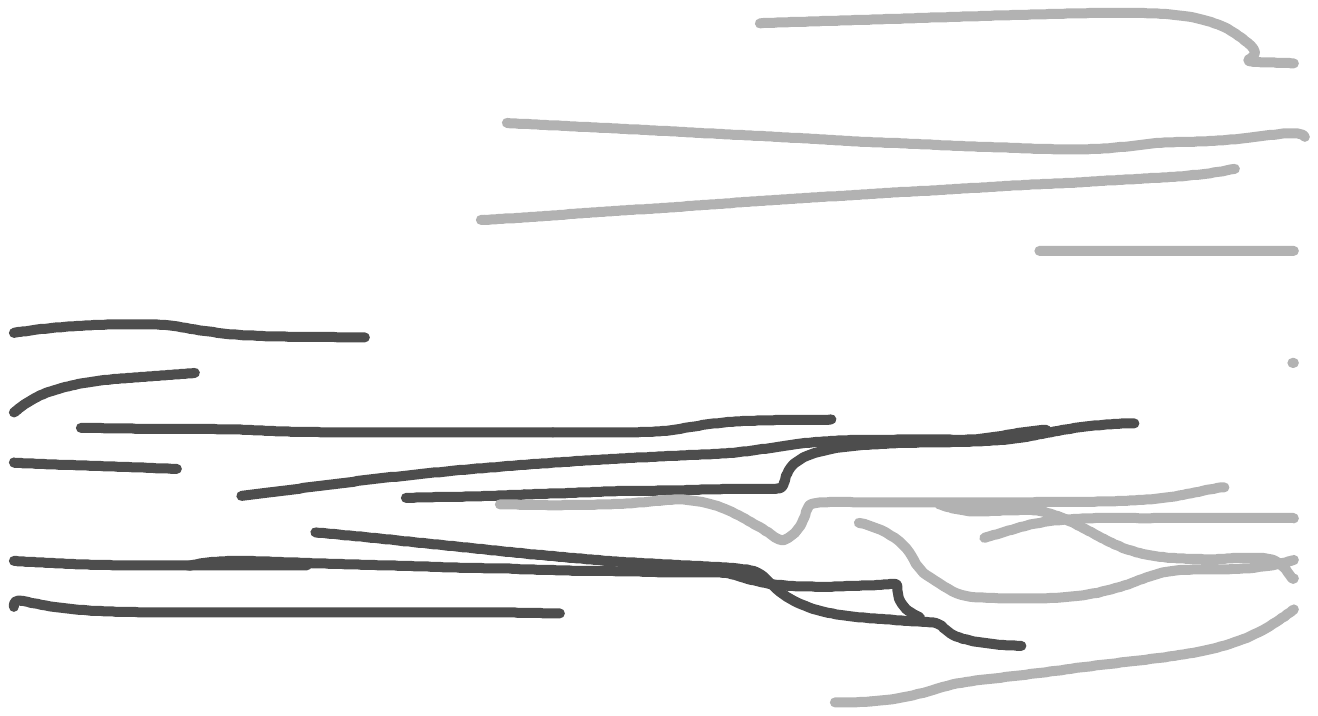}
      }
  \end{center}
  \mycaption{Trajektorien der {\fussgaenger}dynamik mit elliptischen
    Absto"sungspotential} 
  {Trajektorien zweier entgegengesetzter {\fussgaenger}str"ome. Die
    Ausweichman"over werden durch die sozialen Kr"afte des elliptischen
    Potentials zum  Abstandhalten (\ref{Uellipse}) erzeugt.}
  \label{ped_pot_eins_traj}
  \begin{center}
    \fbox{
      \includegraphics[width=10cm,angle=-90]{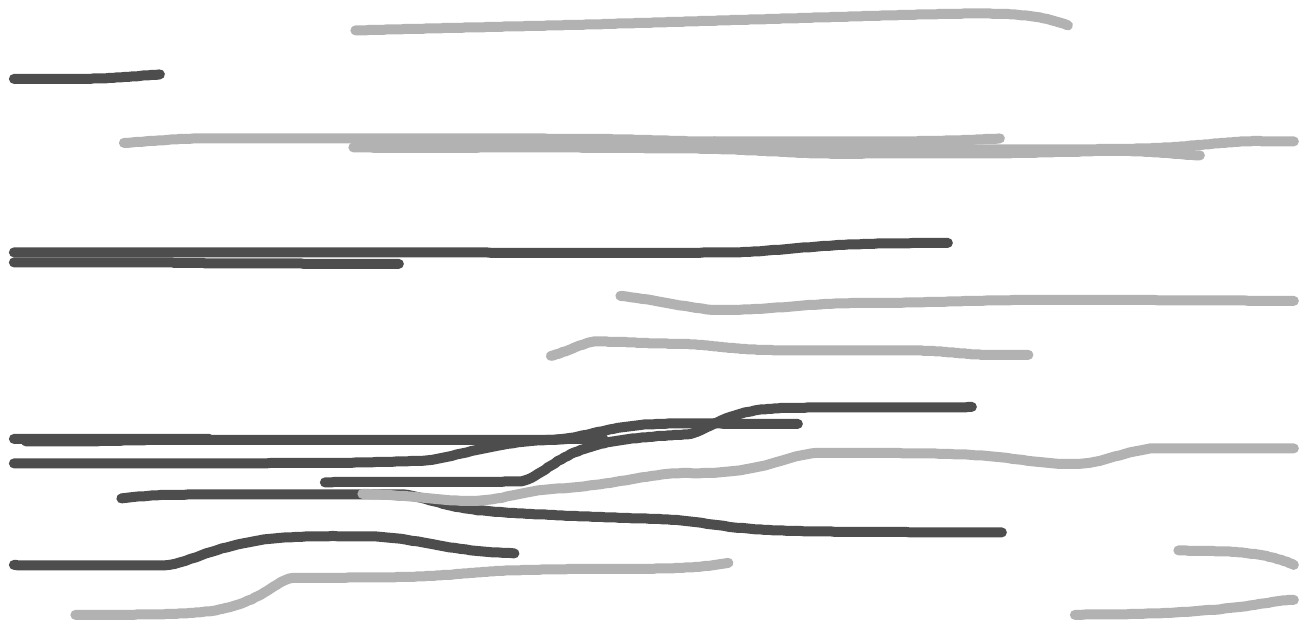}
      }
  \end{center}
  \mycaption{Trajektorien der {\fussgaenger}dynamik mit gestrecktem
    Absto"sungspotential}
  {Im Vergleich dazu erm"oglicht der Potentialansatz aus (\ref{Udehn}) 
    wesentlich glattere/weichere Ausweichman"over.}
  \label{ped_pot_zwei_traj}
\end{figure}

Mit den bis hierhin vorgestellten zwei Verhaltensregeln, dem Antrieb
ein Ziel zu erreichen und der Wahrung des Territoriums der anderen, kann
bereits ein realistischer {\fussgaenger}strom beschrieben werden.
Das Zusammenspiel der sozialen Kr"afte bei der "Anderung der
Bewegung eines {\fussgaenger}s
\begin{equation}
  \frac{d \Va}{dt} = \Fanull + \sum_{\beta \ne \alpha} \vec{f}_{\alpha \beta}
\end{equation}
sorgt daf"ur, da"s die {\fussgaenger} einander ausweichen.
Die Abbildungen \ref{ped_pot_eins_traj} und \ref{ped_pot_zwei_traj} zeigen die
Simulationen zweier {\fussgaenger}str"ome, die sich in entgegengesetzter Richtung
durchdringen. Man erkennt deutlich, da"s der Potentialansatz aus
(\ref{Udehn}) weichere Ausweichman"over hervorruft, als das elliptische
Potential aus (\ref{Uellipse}).

\subsection{Abstandsverhalten gegen"uber Hindernissen}
Neben der gegenseitigen Beeinflussung der {\fussgaenger} hat die bauliche
Struktur der {\fussgaenger}anlage eine starke Wirkung auf deren Verhalten.
Begrenzungen durch W"ande und Hindernisse, wie zum Beispiel S"aulen oder
Bestuhlung, versperren den {\fussgaenger}n den Weg und zwingen sie zum
Ausweichen.

"Ahnlich wie bei den umgebenden Passanten, halten {\fussgaenger} auch von
Hindernissen einen gewissen Abstand, der die beim Gehen verursachten
Schwankungen ber"ucksichtigt.
Die Abst"ande sind sehr stark durch die Beschaffenheit der Hindernissen und
Begrenzungen bestimmt. Die L"ange einer Wand hat dagegen nur einen geringen
Einflu"s (vgl.\ Abschn.\ \ref{sec:bewegungsverhalten}).

Der Ansatz zur Beschreibung von Hindernissen gleicht dem des
{\fussgaenger}abstandhaltens. Die zur"ucksto"sende Wirkung eines Hindernisses
wird durch das Potential 
\begin{equation}
  U_{\alpha B} = b_{\alpha B}\, e^{-{\|\vec{r}_{\alpha
        B}\|}/{\vartheta_{\alpha B}}}
\end{equation}
beschrieben. Auch hierbei f"allt die St"arke des Einflusses mit dem Abstand
exponentiell ab. Die Parameter $b_{\alpha B}$ und $\vartheta_{\alpha B}$
werden einerseits dem {\fussgaenger} $\alpha$ zugeordnet, andererseits
beschreiben sie auch die Eigenschaften des Hindernisses $B$.  Als Abstand
zwischen {\fussgaenger} und Hindernis $\vec{r}_{\alpha B}$ wird die direkte
Entfernung angenommen.

F"ur eine Wand, die als Geradenst"uck zwischen den Punkten $\vec{p}$ und
$\vec{q}$ definiert ist, ist der direkte Abstandsvektor durch
\begin{equation}
  \vec{r}_{\alpha B} = \left\{
  \renewcommand{\arraystretch}{1.2} 
  \begin{array}{r@{\quad:\quad}l}
    \vec{p} - \Ra & \vp{\Ra-\vec{p}}{\vec{e}_{qp}} \le 0 \\
    \vec{p} - \Ra - \vp{\vec{e}_{qp}}{\vec{p}-\Ra}\,\vec{e}_{qp} & 0 <
    \vp{\Ra-\vec{p}}{\vec{e}_{qp}} < \|\vec{q}-\vec{p}\| \\ 
    \vec{q} - \Ra  & \|\vec{q}-\vec{p}\| \le
    \vp{\Ra-\vec{p}}{\vec{e}_{qp}} \\ 
  \end{array} \right.
  \label{abstand_wand}
\end{equation}
gegeben (vgl.\ Abb.\ \ref{fig:skizze_zielrichtung}).
Der direkte Abstand zu einem kreisf"ormigen Hindernis mit dem Durchmesser
$d$ und Mittelpunkt bei $\vec{m}$ ist durch 
\begin{equation}
  \vec{r}_{\alpha B} = \frac{\|\Ra - \vec{m}\|+\frac{1}{2}d}{\|\Ra -
    \vec{m}\|^2}\, (\Ra - \vec{m})
\end{equation}
definiert.

\subsubsection{Mehrere Begrenzungen}
{\fussgaenger}umgebungen sind meist durch mehrere W"ande begrenzt und bergen
zus"atzlich einige Hindernisse.  Liegen die Hindernisse und Begrenzungen eng
beieinander, so stellt sich die Frage, wie die soziale Kraft zusammengesetzt
werden soll, die den gesamten repulsiven Einflu"s der gebauten Umgebung
repr"asentiert. Dies kann auf unterschiedliche Weise geschehen:
\begin{enumerate}
\item Superposition aller Einfl"usse. Das bedeutet, da"s "uber die
  Beitr"age aller Elemente der Umgebung summiert wird.
\item K"urzester Abstand. Der {\fussgaenger} ber"ucksichtigt nur das Hindernis,
  das am  {\naechsten} zu ihm liegt. 
\item  Gr"o"ste Wirkung. Der {\fussgaenger} reagiert auf das Hindernis,
  das den gr"o"sten Einflu"s auf ihn aus"ubt. Alle anderen Eindr"ucke werden
  abgeschirmt. 
\end{enumerate}
Das Problem wird am Beispiel einer Wand deutlich: Eine gerade Wand l"a"st sich
als Strecke zwischen zwei Punkten beschreiben. Die Potentialfunktion und der
Abstandsvektor $\vec{r}_{\alpha B}$ sind dann nach (\ref{abstand_wand})
wohldefiniert. Eine beliebig gekr"ummte Wand kann durch einen Polygonzug
nachgebildet werden. Dabei bestehen die M"oglichkeiten, da"s im Fall der
Superposition jedes Teilst"uck des Polygons als einzelnes Wandst"uck zu dem
Potential beitr"agt, oder da"s nur ein einzelnes auf den {\fussgaenger} wirkt.
Da eine gerade Wand als Sonderfall eines Polygons betrachtet werden kann, ist
die Beschreibung der Wechselwirkung nicht eindeutig. 

 Im folgenden wird
gezeigt, wie beim Entwurf eines Modells ein Ansatz gew"ahlt werden kann, der
die nachzubildende Situation am besten beschreibt.
\begin{figure}[tb]
  \begin{center}
    \leavevmode
    \includegraphics[width=6cm]{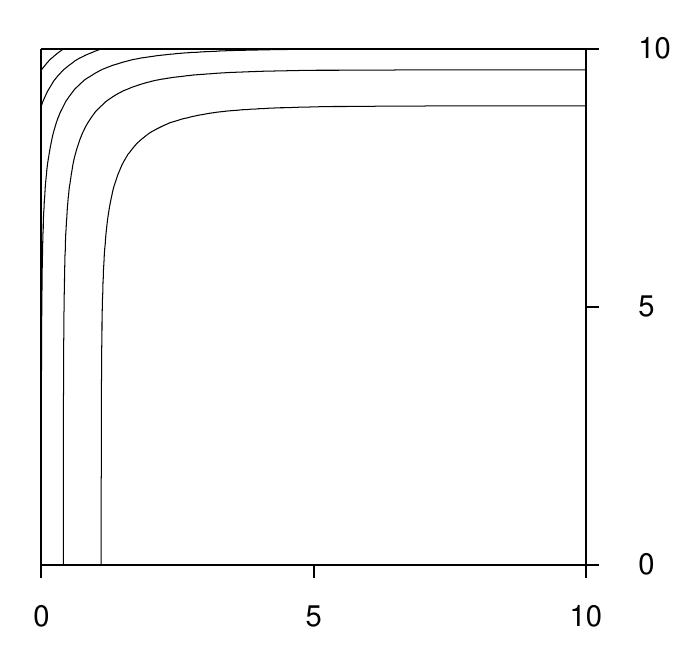} \hspace{1.5cm}
    \includegraphics[width=6cm]{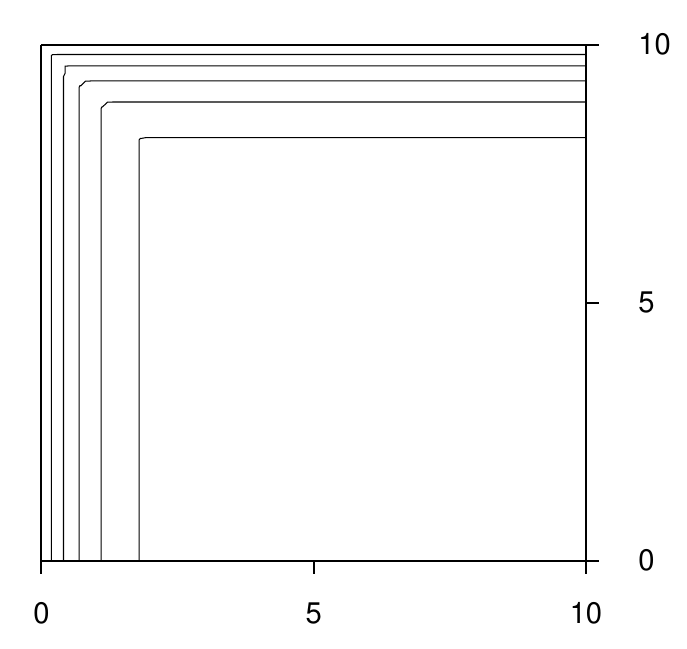} 
  \end{center}
  \mycaption{Verschiedene Potentialformen zur Beschreibung des
    Abstandsverhaltens gegen"uber Begrenzungen}
  {Vergleich zweier M"oglichkeiten zur Darstellung der
    repulsiven Wirkung senkrecht aufeinander stehender W"ande. Im ersten Ansatz
    (links) werden die Potentialbeitr"age beider W"ande zusammengez"ahlt. Im
    zweiten Ansatz (rechts) wirkt nur die Wand mit dem k"urzesten Abstand auf
    den {\fussgaenger}.}
  \label{fig:build_pot_contour}
\end{figure}

Beispielhafte Potentialfunktionen f"ur die Ans"atze 1.\ und 2.\ sind in
Abbildung \ref{fig:build_pot_contour} zum Vergleich dargestellt. Zwei
Wandst"ucke stehen senkrecht zueinander. Wird ein {\fussgaenger} nur von der
Wand, der er am {\naechsten} steht, beeinflu"st, herrscht in der Ecke die
gleiche Absto"sung wie an anderen Positionen vor der Wand. Die Fl"ache kann
von den {\fussgaenger}n vollst"andig ausgenutzt werden. Der Ansatz, bei dem
ein {\fussgaenger} ausschlie"slich das Hindernis mit der gr"o"sten Wirkung
ber"ucksichtigt, f"uhrt bei diesem Beispiel zum selben Ergebnis.

Im zweiten Fall werden die {\fussgaenger} durch die Potentialbeitr"age der
beiden Wandst"ucke gleicherma"sen beeinflu"st. In der Ecke herrscht dadurch
eine st"arkere Absto"sung. Sie wird von den {\fussgaenger}n gemieden. Dieser
Fall ist zu beobachten, wenn {\fussgaenger} einen abgewinkelten
Korridor entlanglaufen. Da sie abrupte Richtungs"anderungen vermeiden wollen,
weichen sie rechtzeitig aus.

\subsubsection{Design einer Potentialfunktion}
Der Entwurf eines Potentials h"angt stark von der Funktion und vom Typ eines
Hindernisses ab. H"aufig l"a"st sich die Reaktion eines {\fussgaenger}s auf
Hindernisse mittels plausibler Annahmen und Erfahrungswerte vorhersagen und 
eine dazu passende Beschreibung der sozialen Kr"afte finden.

Ein \meter{10} breiter Korridor ist durch zwei W"ande an den
L"angsseiten begrenzt. Die W"ande seien in $\vec{e}_x$-Richtung
ausgerichtet. Zur Beschreibung der zwei parallelen W"ande kann die
Potentialfunktion der Form
\begin{equation}
  U_x = e^{-|\vp{\vec{e}_y}{\Ra}|} + e^{-|10-\vp{\vec{e}_y}{\Ra}|}
\end{equation}
mit dem Richtungsvektor $\vec{e}_y$ senkrecht zu den W"anden gew"ahlt werden.
Auf Parameter wird der Einfachheit halber bei diesem Beispiel verzichtet.  

\begin{figure}[p]
  \begin{center}
    \leavevmode
    \includegraphics[width=12cm]{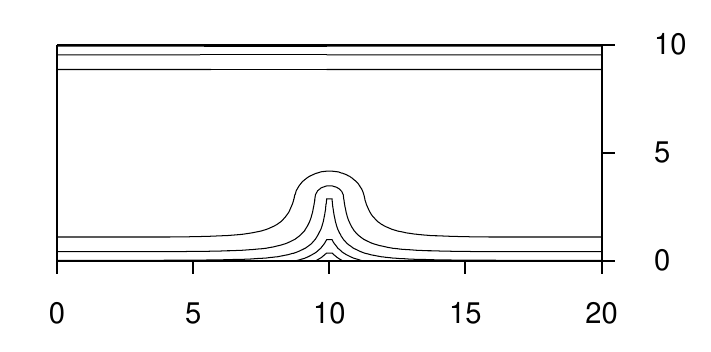} 
  \end{center}
  \mycaption{Potentialdarstellung eines Korridors mit Trennwand}
  {Potentialdarstellung 
    eines \meter{10} breiten Korridors mit einer senkrecht nach innen
    ragenden Trennwand nach (\ref{Ucorridoreins}).}
  \label{corridor_pot1}
  \begin{center}
    \leavevmode
    \includegraphics[width=12cm]{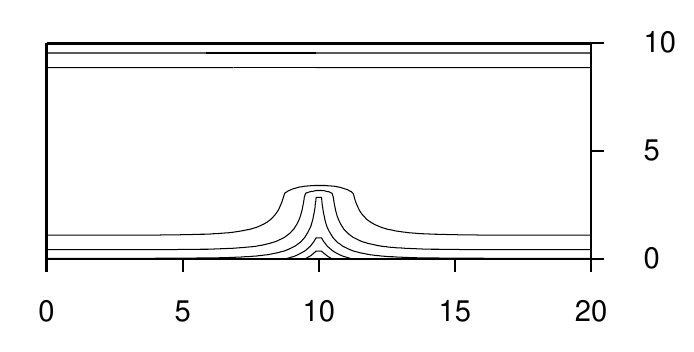} 
  \end{center}
  \mycaption{Verbesserte Potentialdarstellung eines Korridors mit Trennwand}
  {Durch die Abflachung des Potentials nach (\ref{Ucorridorzwei}) wird die
    nutzbare Fl"ache in Mitte des Korridors vergr"o"sert. Das Ausweichen des
    Hindernisses gelingt den {\fussgaenger}n weiterhin.}
  \label{corridor_pot2}
\end{figure}

Eine senkrecht in diesen Korridor ragende Trennwand von \meter{3} L"ange
l"a"st sich durch einen
Potentialbeitrag 
\begin{equation}
  U_y = \left\{
  \renewcommand{\arraystretch}{1.2} 
  \begin{array}{r@{\quad:\quad}l}
    e^{-\left|\vp{\vec{e}_x}{\Ra}-10\right|} & \vp{\vec{e}_y}{\Ra} \le 3 \\
    e^{-\|\Ra-3\vec{e}_y-10\vec{e}_x\|} & \vp{\vec{e}_y}{\Ra} > 3 \\
    \end{array} \right.
    \label{Ucorridoreins}
\end{equation}
beschreiben. In Abbildung \ref{corridor_pot1} sind die "Aquipotentiallinien der
Gesamtwechselwirkung $U_{\alpha B} = U_x + U_y$ dargestellt. Je st"arker die
repulsive Wechselwirkung der Trennwand ist, desto mehr weichen die
{\fussgaenger} aus. 
F"ur die {\fussgaenger} in der Mitte des Korridors besteht jedoch keine
Veranlassung, einen so gro"sen Abstand zum offenen Ende der Trennwand zu
halten. Daher kann das Potential an der Spitze durch
\begin{equation}
  \label{Ucorridorzwei}
  U^{\prime}_y = \left\{
  \renewcommand{\arraystretch}{1.2} 
  \begin{array}{r@{\quad:\quad}l}
    e^{-\left|\vp{\vec{e}_x}{\Ra}-10\right|} 
    & \vp{\vec{e}_y}{\Ra} \le 3 \\
    e^{-\sqrt{\sq{\left( \vp{\vec{e}_y}{\Ra} - 3 \right)} / \sq{\mu} 
        + \sq{\left( \vp{\vec{e}_x}{\Ra} - 10\right)}}} 
    & \vp{\vec{e}_y}{\Ra} > 3 \\
  \end{array} \right.
\end{equation}                                
mit $\mu > 1$ abgeflacht werden (siehe Abb. \ref{corridor_pot2}). Die
{\fussgaenger} werden dadurch vom Rand in die Mitte des Korridors geleitet und
k"onnen das Hindernis knapp passieren. 

\subsubsection{Hindernisse als Abschirmung}
Soziale Kr"afte entstehen durch die Wahrnehmung von Ereignissen, die in der
Regel durch Blickkontakt erm"oglicht wird. Versperrt ein Hindernis, wie zum
Beispiel eine Trennwand, einem {\fussgaenger} die Sicht auf andere Passanten, so
wird er von ihnen auch nicht beeinflu"st. Durch die "Uberlappung von
{\fussgaenger}- und dem Wandpotentialen kann in manchen F"allen ein
schwacher 
Einflu"s der verdeckten {\fussgaenger} auf die andere Seite dringen.  Um dies zu
verhindern, mu"s f"ur jede Wechselwirkung gepr"uft werden, ob sie durch ein
Hindernis verdeckt wird. Dazu werden aus der Verbindungslinie zwischen zwei
{\fussgaenger}n und der Verbindungslinie der beiden Enden der Wand zwei Geraden
gebildet. Wenn sich die beiden Geraden schneiden, und der Schnittpunkt
zwischen den beiden Enden der Wand liegt, findet keine Wechselwirkung statt.

\subsection{Attraktionen}
Schaufensterauslagen, Plakate, Stra"senk"unstler und vieles mehr veranlassen
die vorbeigehenden {\fussgaenger} n"aherzukommen und manchmal f"ur einen Moment
stehenzubleiben. "Ahnlich wie bei den anderen Potentialtypen geht die
St"arke der Anziehung f"ur gro"se Abst"ande gegen Null. 
Beim N"aherkommen, steigt der Einflu"s einer Attraktion an, bis er bei einem
gewissen Abstand wieder abf"allt. Der Abstandsbereich, in dem die
Wechselwirkung wieder verschwindet, schafft einen 
Aufenthaltsbereich f"ur den {\fussgaenger} um die Attraktion.
 
Durch die Lage dieses Aufenthaltsbereiches k"onnen verschiedene Typen
von Attraktionen unterschieden werden. 
Stra"senk"unstlern n"ahert man sich bis auf einen gewissen
Abstand, um alles gut beobachten zu k"onnen. Meist sorgen die Darsteller
selbst daf"ur, da"s ihre Zuschauer nicht zu nahe kommen. Ein "ahnliches
Verhalten ist bei gro"sen Objekten zu beobachten, die man nicht mehr so gut
sieht, wenn man zu dicht steht.
Eine Potentialfunktion
\begin{equation}
  U_{\alpha i} = 
  \underbrace{-e^{-\|\vec{r_i}-\Ra\| / \chi_{\alpha i}}}_{\mbox{ I}}
  \: + \: 
  \underbrace{ e^{-\|\vec{r_i}-\Ra\| / \varphi_{\alpha i}}}_{\mbox{II}}
  \label{Ukuenstler}
\end{equation}
die dieses Verhalten beschreibt, besteht aus zwei
Teilen, einem attraktiven mit langer Reichweite (I) und einen repulsiven mit
wesentlich k"urzerer Reichweite (II).

Ein anderer Typ von Attraktionen, etwa ein Schaufenster, ist durch
eine bestimmte Stelle gekennzeichnet, an der sich die {\fussgaenger}
aufhalten.  Da das zugeh"orige Potential an der Stelle $\vec{r_i}$ eine
flache Mulde aufweisen soll, wird es durch
\begin{equation}
  U^\prime_{\alpha i} = - e^{-\sq{\|\vec{r_i}-\Ra\|} / (2\chi_{\alpha i})}
  \label{Uverkaeufer}
\end{equation}
auf eine andere Weise als der erste Typ beschrieben.

\begin{figure}[tb]
  \begin{center}
    \leavevmode
    \includegraphics[width=6cm]{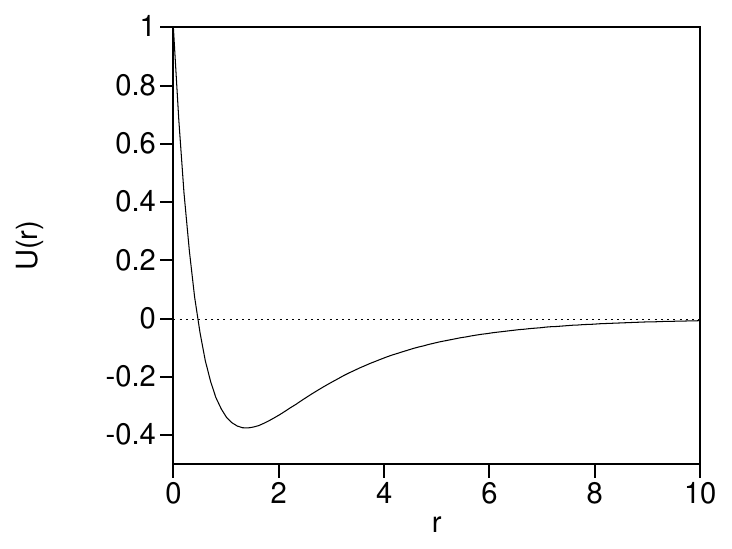} \hfill
    \includegraphics[width=6cm]{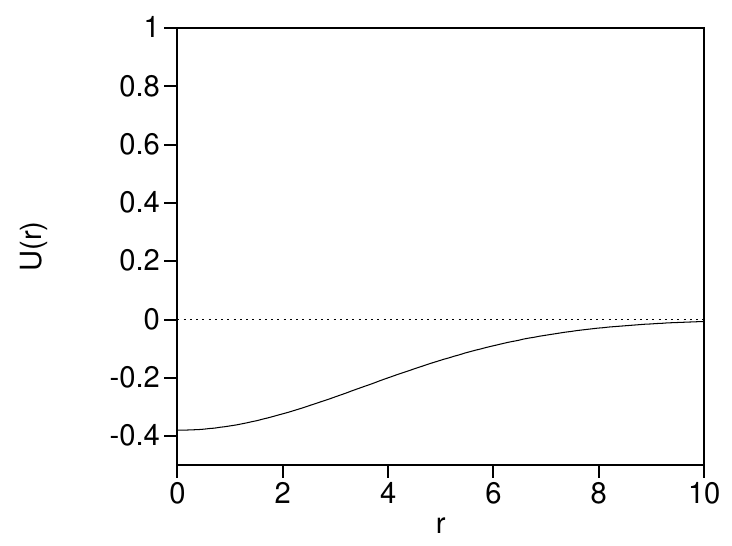}
  \end{center}
  \mycaption{Potentialdarstellung von Attraktionen} 
  {Potential f"ur
    verschiedene Typen von Attraktionen. Die Potentialfunktion aus
    (\ref{Ukuenstler}) h"alt die Passanten durch den inneren repulsiven Teil
    auf eine gewisse Distanz, w"ahrend das Potential aus (\ref{Uverkaeufer})
    rein anziehend ist.}
  \label{fig:pot_attraktionen}
\end{figure}
Attraktionen sind nicht allein auf Anziehungspunkte
beschr"ankt. Auch andere r"aumliche Ausdehnungen von Attraktionen sind
realisierbar. Zum Beispiel kann die Schaufensterfront eines Kaufhauses 
durch ein von zwei Punkten begrenztes Geradenst"uck beschrieben werden.

Das Interesse $a_{\alpha i}$ eines {\fussgaenger}s an einer Darstellung oder
Auslage $i$ nimmt meist mit der Zeit ab. 
Die zeitliche "Anderung des Interesses wird durch 
\begin{equation}
  \frac{d a_{\alpha i}}{dt} = - \frac{a_{\alpha i}^0}{T^i}\,
  e^{-\|\vec{r}^{min}_{i}-\Ra\| / \chi_{\alpha i}}
\end{equation}
mit $a_{\alpha i}(t_0) = a_{\alpha i}^0$ beschrieben.
Das Anfangsinteresse $a_{\alpha i}^0$ an einer Attraktion ist dabei
individuell verschieden. F"ur manche {\fussgaenger} kann eine Attraktion
"uberhaupt nicht interessant sein. 
H"alt sich ein {\fussgaenger} im Aufenthaltsbereich einer Attraktion auf, f"allt
sein Interesse ann"ahernd linear mit der Zeit ab. Kommt ein {\fussgaenger}
dann ein weiteres Mal in die N"ahe einer Attraktion, bleibt er von ihr
unbeeinflu"st, sofern das Interesse an dieser Attraktion bereits auf Null
abgefallen ist.

\subsection{Wechselwirkung in Gruppen}
{\fussgaenger} treten h"aufig in kleineren Gruppen auf. Besonders in
Einkaufsbereichen und Freizeitparks ist der {\fussgaenger}verkehr durch einen
gro"sen Anteil an Zweier- und Dreier Gruppen gepr"agt. Die Gr"o"sen der Gruppen
folgen dabei ungef"ahr einer Poissonverteilung
\cite{ColemanHopkinsJames:1961}. Gruppen mit vier, f"unf oder noch mehr
Mitgliedern treten sehr viel seltener auf.

Mitglieder einer Gruppe versuchen in der Regel nebeneinander
herzugehen. 
Eine Gruppe kann sich  f"ur kurze Momente aufl"osen, falls das zum
Beispiel bei Ausweichman"overn notwendig wird. Sobald das Hindernis
"uberwunden ist, streben sie wieder zueinander.

Die Beschreibung des Gruppenverhaltens wird in diesem Modell auf das Bestreben
der einzelnen Partner nebeneinander zu gehen beschr"ankt. Dies kann durch eine
zeitlich und "uber die Entfernung konstante Anziehungskraft erreicht werden.
Die anziehende Gruppenwechselwirkung ist mit der Kraft zum Zielpunkt zu kommen
vergleichbar. 

Das Potential f"ur das Zusammenhalten der Gruppe
\begin{equation}
  \label{Ugruppe}
  U_{\alpha\alpha^\prime} = g_\alpha \|\vec{r}_{\alpha}-
  \vec{r}_{\alpha^\prime}\|
\end{equation}
verl"auft dabei proportional zum 
Abstand $\vec{r}_{\alpha}- \vec{r}_{\alpha^\prime}$ 
zweier Partner $\alpha$ und $\alpha^\prime$.
Der Betrag der sozialen Kraft
\begin{equation}
  f_{\alpha\alpha^\prime} = -\nabla U_{\alpha\alpha^\prime} = - g_\alpha
  \frac{\vec{r}_{\alpha}- \vec{r}_{\alpha^\prime}}{\|\vec{r}_{\alpha}-
    \vec{r}_{\alpha^\prime}\|}
\end{equation}
ist "uber den gesamten Raum konstant.  Da das Potential selbst f"ur sehr
kleine Abst"ande nicht vollst"andig verschwindet, schw"acht es die repulsive
Kraft des anderen {\fussgaenger}s etwas ab. Damit halten die Mitglieder einer
Gruppe geringere Abst"ande zueinander als zu den anderen Passanten.

\section[Bestimmung der Parameter]
{Untersuchung der Kraftterme, Bestimmung der
  Potentialparameter} 
\label{sec:potentialparameter}
Das Soziale-Kr"afte-Modell verwendet zahlreiche Parameter zur Anpassung an die
realen {\fussgaengerstroeme}. Einige davon lassen 
sich durch stark vereinfachende Annahmen aus empirischen Daten
ermitteln. Andere dagegen k"onnen erst durch die Simulation des Modells und den
Vergleich der Ergebnisse mit beobachteten Str"omen gefunden werden.

\subsubsection{Wunschgeschwindigkeit und Antriebkraft}
Durch empirische Beobachtungen (Abschn.\ \ref{sec:bewegungsverhalten})
wurde f"ur geringe {\fussgaenger}dichten (LOS A mit weniger als
\pqmeter{0.1}, \cite{Weidmann:1993}) eine Normalverteilung der
{\fussgaenger}geschwindigkeit festgestellt. Sie hat die Form
\begin{equation}
  P(v) = \frac{1}{\sqrt{2\pi\sigma}}\, 
  e^{-\frac{ \left( v - \sq{\ave{v}}\right)} {2\sigma}}.
\end{equation}
mit dem Mittelwert von $\ave{v} = $\metersec{1.34} mit der 
Standardabweichung $\sigma = $\metersec{0.26}.
Diese Werte k"onnen als Wunschgeschwindigkeit $\vanull$ "ubernommen werden.

Aus der Definition der Antriebskraft (\ref{f_antrieb}) ergibt sich eine
exponentielle Beschleunigung. Ein {\fussgaenger}, der zur Zeit $t_0$ zum
Stillstand gekommen ist, erreicht nach einer Zeit $t$ die Geschwindigkeit 
\begin{equation}
  v(t_0+t) = \vanull\,\left(1-e^{-t/\tau_\alpha}\right)
\end{equation}
Der Parameter $\tau_\alpha$ gibt dabei die Dauer bis zum Erreichen von 63\%
der Wunschgeschwindigkeit an. In der Literatur wurden keine Angaben
"uber die Beschleunigungszeiten f"ur {\fussgaenger} gefunden. Die Simulationen
produzieren f"ur $\tau_\alpha = 0.2\dots 0.5$ realistische Ergebnisse. 

Die Gr"o"se $\tau_\alpha$ macht sich im Zusammenspiel mit anderen Wechselwirkungen
bemerkbar. In der Simulation ergaben Werte $\tau<0.5$ ein aggressives
Verhalten der {\fussgaenger} mit geringer Bereitschaft anderen Passanten
auszuweichen. 
Bei Werten von $\tau>1.0$ lie"sen sie sich dagegen weit von ihrer Bahn
abdr"angen. Bei unterschiedlicher Wunschgeschwindigkeit aber gleicher
Relaxationszeit weichen die schnellen {\fussgaenger} wesentlich seltener aus
als die langsamen. Da dies der Erfahrung widerspricht, in der Literatur aber
keine Angaben "uber das verschieden offensive Verhalten einzelner {\fussgaenger}
zu finden waren, wird im Modell nicht $\tau_\alpha$  sondern der Quotient
$\vanull/\tau_\alpha$ zur Charakterisierung eines {\fussgaenger}s herangezogen.

\subsubsection{Ausweichverhalten}
Die Parameter der Abstandspotentiale 
sind so zu w"ahlen, da"s Zusammenst"o"se zwischen den
{\fussgaenger}n  vermieden werden.

Angenommen, zwei aufeinander zulaufende {\fussgaenger} k"onnen aufgrund
seitlicher Begrenzungen einander nicht
ausweichen, dann stellt sich bei einem minimalen Abstand $r^{min}$ zwischen der
Antriebskraft und der repulsiven Wechselwirkung der {\fussgaenger} das 
Gleichgewicht 
\begin{eqnarray}
  \Fanull & = & \vec{f}_{\alpha\beta} \nonumber \\
  \Leftrightarrow \qquad \frac{v_\alpha^0}{\tau_\alpha}  & = &
  \frac{p_\alpha}{\sigma_\alpha}\,e^{-\frac{r^{min}}{\sigma_\alpha}} 
\end{eqnarray}
ein. F"ur $r^{min} = \sigma_\alpha$ stehen die Parameter in der Abh"angigkeit
\begin{equation}
  \frac{p_\alpha}{\sigma_\alpha} = \frac{v_\alpha^0\, e}{\tau_\alpha}
  \approx 2.71628\,\frac{v_\alpha^0}{\tau_\alpha} \quad\mbox{und}\quad
  p_\alpha \approx 7\sigma_\alpha 
\end{equation}
mit $\vanull = 1.34$ und $\tau_\alpha = 0.5$.
In den Simulationen haben sich die Parameter $p_\alpha = 2.1$ und
$\sigma_\alpha = 0.3$ bew"ahrt. 

Auf diese Weise lassen sich auch die Parameter f"ur die
Abstandspotentiale von Begrenzungen, Hindernissen und 
Anziehungspunkten absch"atzen.

\subsubsection{Gruppenzusammenhalt}
Zwei Individuen  einer Gruppe erfahren die konkurrierenden Wechselwirkungen des
Abstandhaltens gegen"uber {\fussgaenger}n und des Zusammenhaltens einer
Gruppe. Im Gleichgewichtsabstand $r$ heben sich die Potentiale aus (\ref{Uped})
und (\ref{Ugruppe}) gegenseitig auf:
\begin{equation}
  p_\alpha\,e^{-r/\sigma_\alpha} = g_\alpha\, r
\end{equation}
Mit den oben genannten Werten und einem Gleichgewichtsabstand zwischen den
Gruppenmitgliedern von \meter{0.8} ergibt sich der Parameter der
Gruppenattraktion zu $g_\alpha = 0.18$.

\section[Ph"anomene der Selbstorganisation]{Ph"anomene der Selbstorganisation
  in {\fussgaenger}str"omen}
\label{sec:phaenomenederselbstorganisation}
Im folgenden werden Simulationen des Soziale-Kr"afte-Modells der
{\fussgaenger}dynamik aus Abschnitt \ref{sec:fussgaengermodell}
vorgestellt. Im Vordergrund steht dabei die Emergenz kollektiven
Verhaltens. Obwohl die simulierten {\fussgaenger} selbst"andig handeln und
kein \hi{Imitations- oder Kooperationsverhalten} im Modell enthalten ist, kommt
es durch die \hi{Selbstorganisation} der individuellen Dynamik der
{\fussgaenger} zu \hi{r"aumlichen und zeitlichen Strukturen}
im {\fussgaenger}strom.

\subsection{Bahnbildung}
\begin{figure}[p]
  \begin{center}
    \leavevmode
    \includegraphics[width=0.8\textwidth]{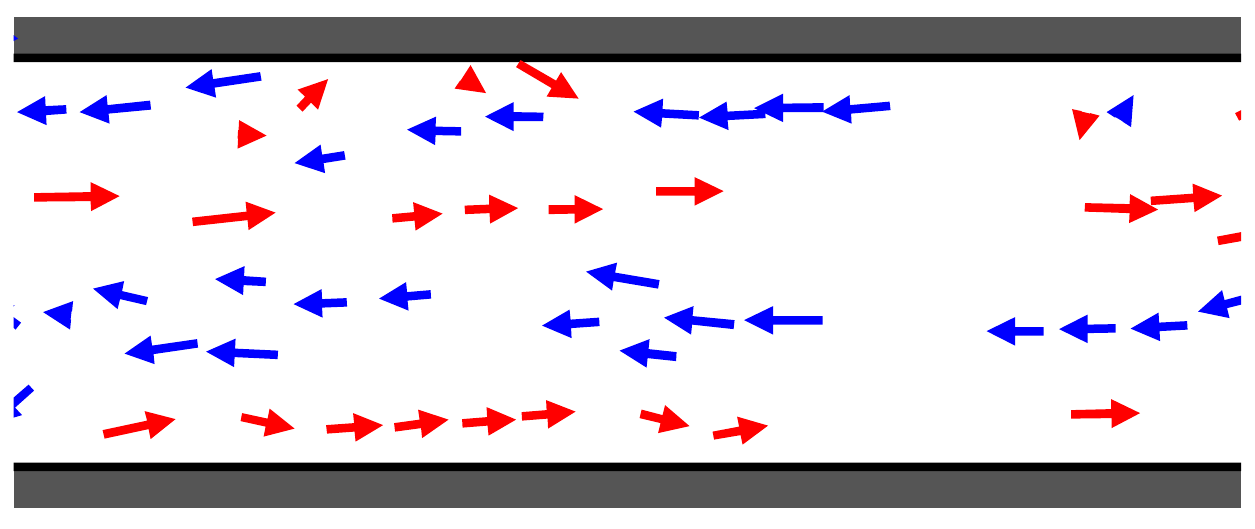}
  \end{center}
  \mycaption{Bahnbildung in {\fussgaengerstroeme}n 1 \farbe}
  {Ausbildung von Spuren. {\fussgaenger} einer Bewegungsrichtung bilden
    gemeinsam Spuren durch den Strom. Dieses kollektive Verhalten ensteht
    allein aus den Wechselwirkungen zwischen den {\fussgaenger}n und der
    Wirkung der Korridorw"ande.}
  \label{fig:bahnbildung1}
  \begin{center}
    \leavevmode
    \includegraphics[width=0.8\textwidth]{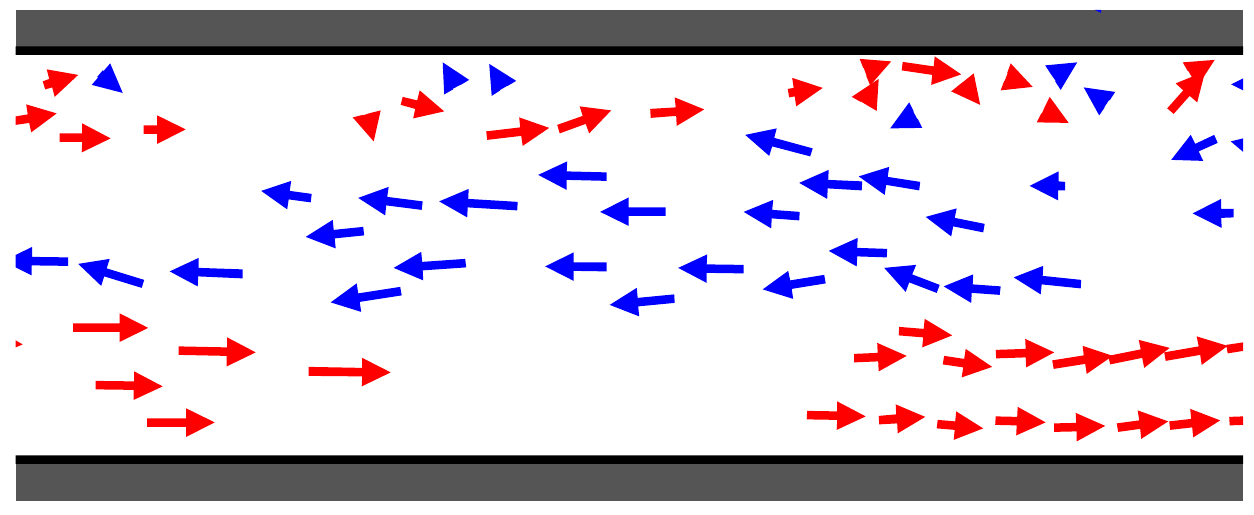}
  \end{center}
  \mycaption{Bahnbildung in {\fussgaengerstroeme}n 2 \farbe}
  {Teilweise breiten
    sich die Bahnen einer Gehrichtung soweit aus, da"s fast alle
    {\fussgaenger} in Bezug auf ihre Gehrichtung rechts laufen. Eine
    Situation, in der fast alle auf der linken Seite gehen, kann mit der
    gleichen Wahrscheinlichkeit eintreten. Im rechten oberen Teil des
    Korridors sind Turbulenzen zu sehen: Die aufeinandertreffenden
    {\fussgaenger} k"onnen nicht ausweichen, weil sie auf der einen Seite von
    der Wand und auf der anderen von einem sehr stabilen {\fussgaenger}strom
    eingeschlossen sind.}
  \label{fig:bahnbildung2}
\end{figure}
In einem Korridor, in dem {\fussgaenger}mengen von den Enden zu der jeweils
anderen  Seite des Korridors laufen, bilden sich aus {\fussgaenger}n mit
derselben Gehrichtung Spuren durch den Strom
(Abb.\ \ref{fig:bahnbildung1}). Obwohl diese Spuren keine festen Laufbahnen
sind und durch entgegenkommende oder "uberholende {\fussgaenger} immer
wieder gest"ort werden, stellen sie dennoch stabile Strukturen dar.

Zum Teil k"onnen sich die Spuren soweit verbreitern, da"s fast alle
{\fussgaenger} bezogen auf ihre Gehrichtung auf derselben Seite laufen.
Die urspr"ungliche Modellsymmetrie der Ausweichrichtung  wird hierbei
zugunsten einer Seite allein aufgrund der Wechselwirkungen zwischen den
{\fussgaenger}n gebrochen (Abb.\ \ref{fig:bahnbildung2}).

Die Simulationen zeigen, da"s eine unterschiedlich starke Tendenz nach links
oder nach rechts auszuweichen zur Spurbildung nicht notwendig ist. Gleichwohl
ist gelegentlich eine Pr"aferenz zur rechten Seite zu beobachten. Dies h"angt
allerdings nicht mit der Regelung des Stra"senverkehrs
zusammen. So hat \name{Older}{S.~J.} auch bei {\fussgaenger}n in London die
Rechtstendenz beobachtet, obwohl dort links gefahren wird \cite{Older:1968}.    
  
\subsection{{\fussgaenger}kreuzung}
\begin{figure}[p]
  \begin{center}
    \leavevmode
    \includegraphics[width=0.8\textwidth,clip]{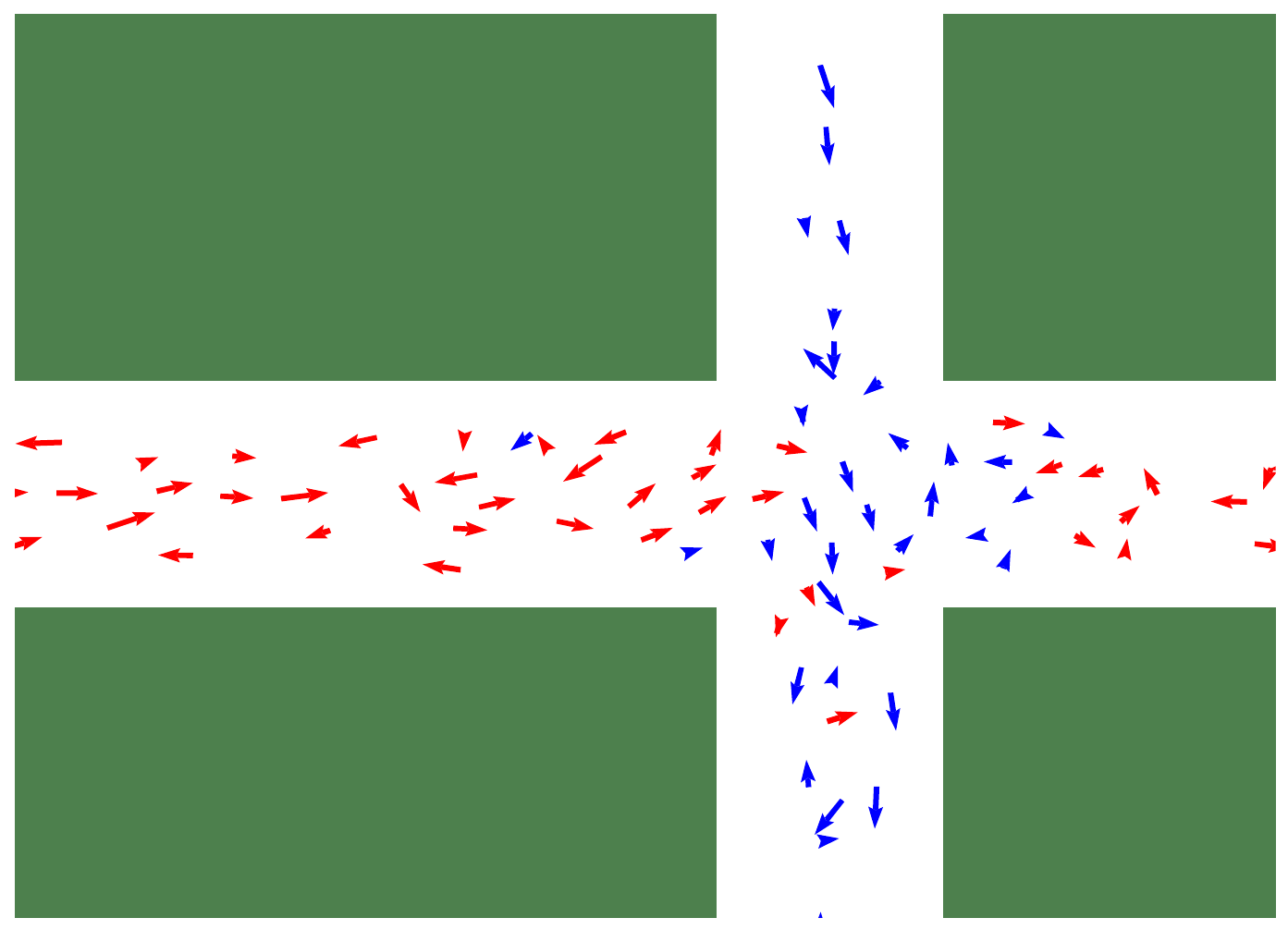}
  \end{center}
  \mycaption{Oszillation in senkrecht zueinander stehenden Str"omen \farbe}
  {Eine
    Kreuzung mit {\fussgaengerstroeme}n in vier Richtungen: Von oben nach
    unten, von rechts nach links und jeweils in die Gegenrichtung. Die
    Kreuzung wird abwechselnd von den vertikal und den horizontal laufenden
    {\fussgaenger}n bev"olkert.}
  \label{fig:crossblue}
  \begin{center}
    \leavevmode
    \includegraphics[width=0.8\textwidth,clip]{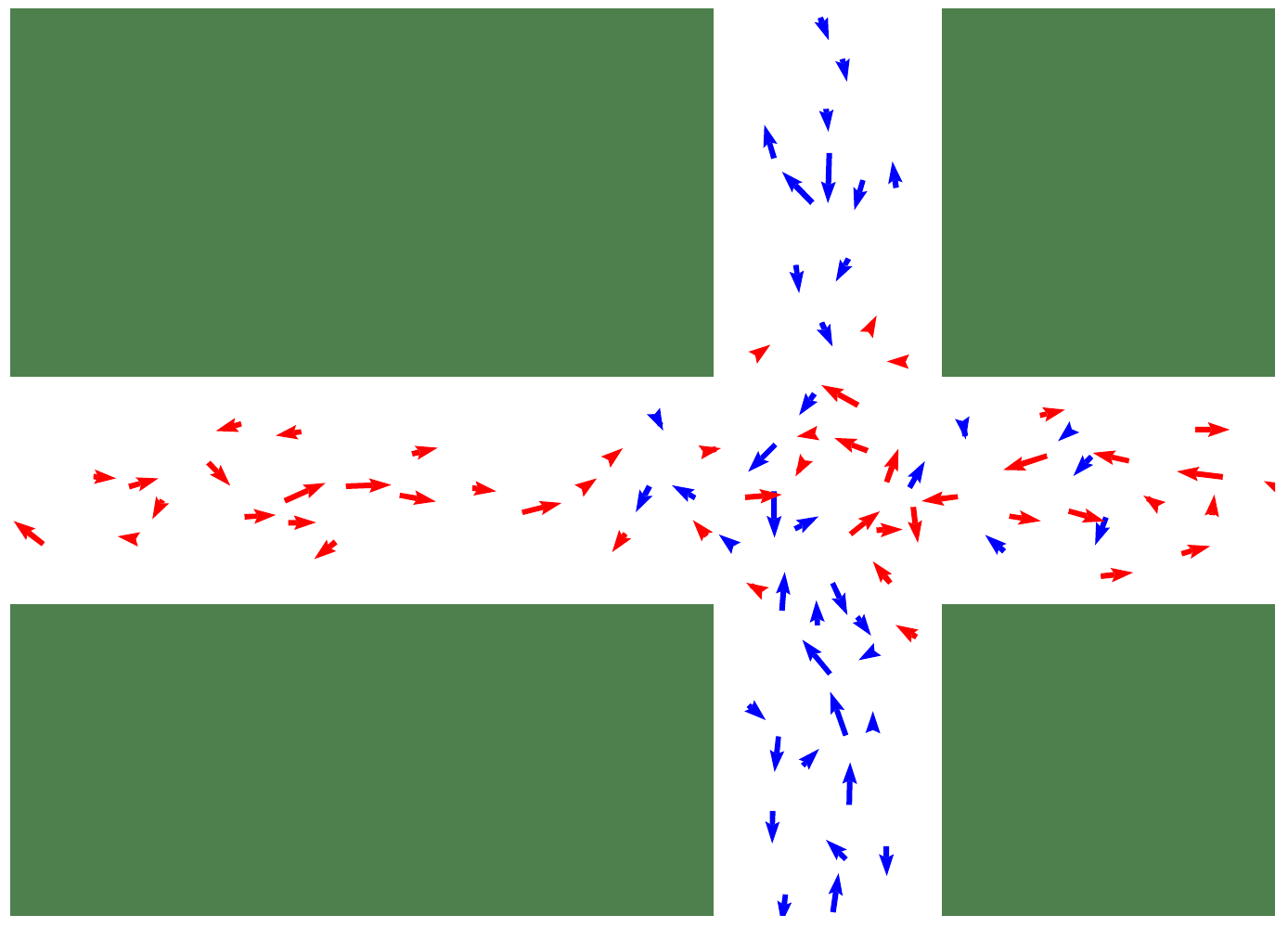}
  \end{center}
  \mycaption{Selbstorganisierter Kreisverkehr \farbe}
  {Durch die Wechselwirkungen
    zwischen den {\fussgaenger}n ensteht ein Kreisverkehr. Diese Struktur ist
    jedoch nur von kurzer Lebensdauer. Sie kann jedoch durch ein Hindernis im
    Zentrum der Kreuzung stabilisiert werden (vgl.\ Abschn.\
    \ref{sec:leistungsmasse}).}
  \label{fig:crossmixed}
\end{figure}
Auf einer {\fussgaenger}kreuzung treten die selbstorganisierten Muster weniger
deutlich hervor. Manchmal kann eine Oszillation zwischen den horizontalen und
den vertikalen {\fussgaengerstroeme}n beobachtet werden (Abb.\
\ref{fig:crossmixed}). Als zweite Struktur bilden sich Wirbel in Form eines
Kreisverkehrs aus. Diese fl"uchtige Struktur kann durch ein Hindernis im
Zentrum der Kreuzung stabilisiert werden (vgl.\ Abschn.\
\ref{sec:leistungsmasse}).

\subsection{Oszillation der Durchgangsrichtung}
\label{sec:oszillation_durchgang}
\subsubsection{Einfacher Durchgang}
\label{sec:einfach_tuer} 
\begin{figure}[p]
  \begin{center}
    \leavevmode
    \includegraphics[width=0.8\textwidth]{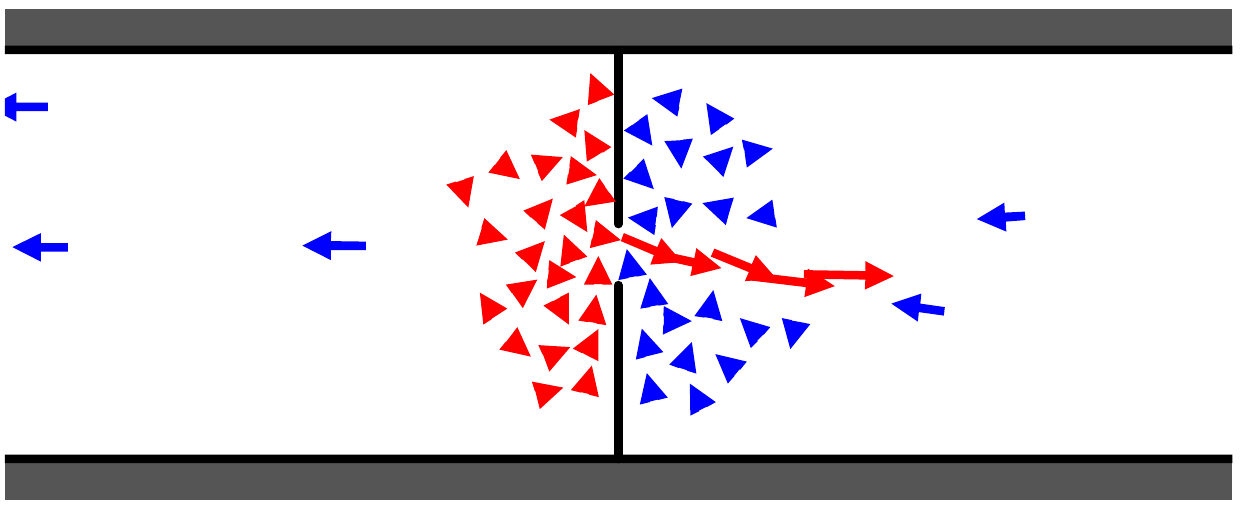} \\[1cm]
    \includegraphics[width=0.8\textwidth]{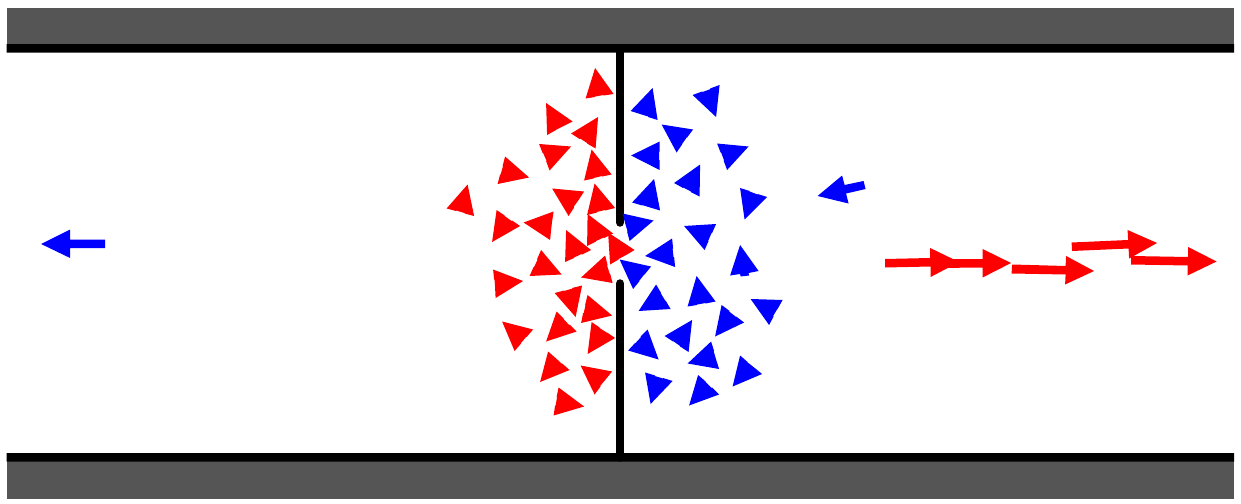} \\[1cm]
    \includegraphics[width=0.8\textwidth]{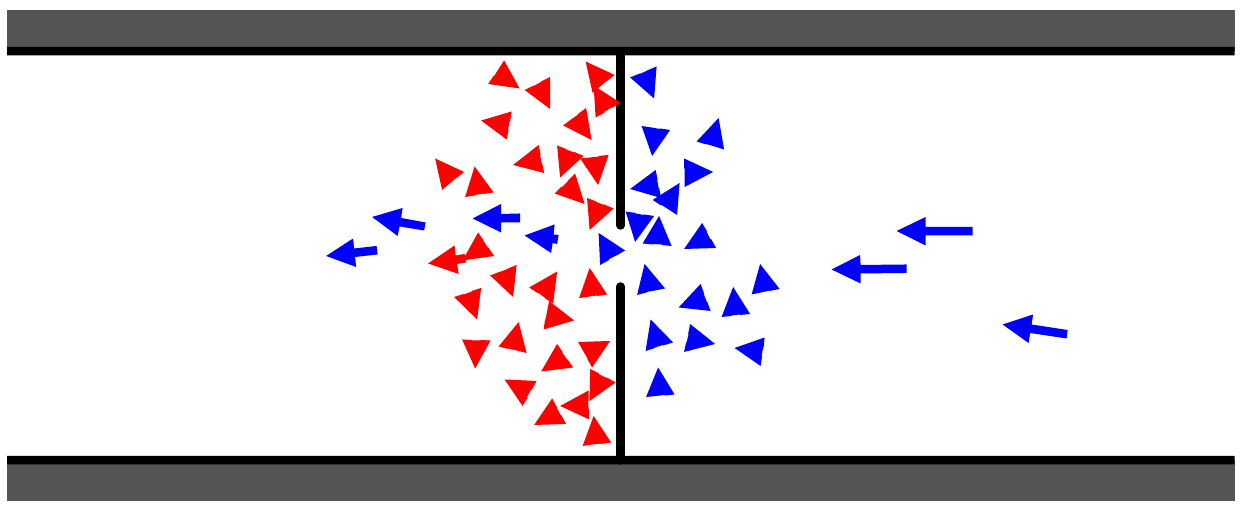}
  \end{center}
  \mycaption{Geteilter Korridor mit einem schmalem Durchgang \farbe}
  {Durch die schmale "Offnung des
    Durchgangs gelangen jedesmal nur ein paar {\fussgaenger} (oben), bis der
    Flu"s durch den Druck der Gegenseite gestoppt wird. Nach einer Periode des
    Gleichgewichts (Mitte) kann die andere Seite zum Zuge kommen (unten). Die
    Proze"s   wiederholt sich, solange neue {\fussgaenger} nachkommen.}
  \label{fig:tuer}
\end{figure}
Wird ein Korridor durch eine Wand mit einem schmalen Durchgang getrennt, 
deren "Offnung gerade f"ur einen {\fussgaenger} breit genug ist (Abb.\
\ref{fig:tuer}), kann in einer Zeitperiode nur ein Schwall von {\fussgaenger}n
einer Seite die Barriere "uberwinden. Durch den Druck der Gegenseite wird ihr
Flu"s unterbrochen. 
Dadurch stellt sich eine Oszillation der Gehrichtung durch die "Offnung ein.
Dieses zeitliche Muster wird durch nachfolgende {\fussgaenger} und durch
Fluktuationen aufrechterhalten. Andernfalls stellt sich ein
Gleichgewicht auf beiden Seiten des Korridors ein, und der {\fussgaenger}strom
kommt zum erliegen.

In einer solchen Situation "andern die {\fussgaenger} ihr Verhalten. Sie
werden aggressiver und dr"angeln st"arker, oder sie treffen Absprachen oder
suchen sich einen anderen Weg. 

\subsubsection{Doppelter Durchgang}
\label{sec:doppeltuer}
Zur Betrachtung eines Szenarios mit zwei Durchg"angen wurde das
{\fussgaenger}modell um ein Entscheidungsmodell erweitert, das in Abschnitt 
\ref{sec:entscheidungsmodell} vorgestellt wird.
Ein Ergebnis wird an dieser Stelle bereits
vorweggenommen, weil man mittels eines sehr simplen Entscheidungsverhaltens
bereits kollektive Strukturen erzeugen kann. Die {\fussgaenger} w"ahlen den
Durchgang, dem sie am n"achsten sind. Die Verkehrsdichte, die vor den
Durchg"angen herrscht, wird in der Entscheidungsfindung nicht ber"ucksichtigt.

\begin{figure}[tb]
  \begin{center} 
    \leavevmode
    \includegraphics[width=0.8\textwidth]{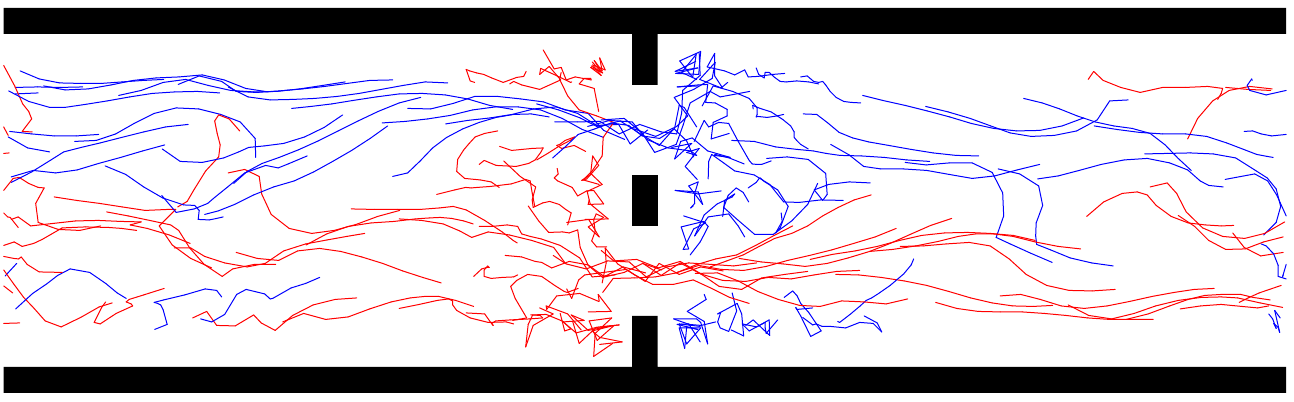}
  \end{center}
  \mycaption{Geteilter Korridor mit zwei Durchg"angen \farbe}
  {Die {\fussgaenger} der beiden Bewegungsrichtungen haben die zwei
    Durchg"ange untereinander aufgeteilt. Die Voraussetzung daf"ur ist, das sie
    sehr geduldig sind und viel Abstand voneinander halten. Wenn von jeder
    Subpopulation gen"ugend {\fussgaenger} nachkommen, ist diese Struktur sehr
    stabil. Diejenigen, die den \glqq falschen \grqq Durchgang gew"ahlt haben,
    bleiben in den Ecken gefangen.}
  \label{fig:doppeltuer}
\end{figure}

Unter der Voraussetzung, da"s die {\fussgaenger} sehr geduldig sind und
verh"altnism"a"sig viel Abstand voneinander halten, kann sich eine Situation
einstellen, in der jede Gruppe einen Durchgang besetzt h"alt (Abb.
\ref{fig:doppeltuer}).
Ein anderes Muster weist Oszillationen der Bewegungsrichtung "ahnlich der
Situation mit einem Durchgang auf. Jeweils eine Subpopulation besetzt dabei
gleichzeitig beide Durchg"ange. 

Die gleichen Effekte k"onnen auch mit einem Salzl"osung-Wasser-Oszillator
produziert werden, der experimentell und in der Theorie in
\cite{YoshikawaOyamaShojiNakata:1991} untersucht wurde. 
Dabei wird in einen mit Wasser gef"ullten
Beh"alter ein Plastikbecher gesetzt, der mit einer Salzl"osung gef"ullt ist.
Versieht man den Becher am Boden mit einer kleinen "Offnung, kann Salzl"osung
aus dem Becher in den Wasserbeh"alter str"omen, und umgekehrt Wasser in den
Becher eindringen. Bei geeignet gew"ahlten Werten f"ur die
L"osungskonzentration, Einf"ullmenge und Durchmesser der "Offnung stellt sich
eine Oszillation der Durchflu"srichtung ein.

Versieht man den Becher mit zwei "Offnungen, so k"onnen zwei
unterschiedliche Effekte auftreten: Liegen die "Offnungen eng zusammen, weisen
die beiden Durchflu"srichtungen dieselbe Orientierung auf und oszillieren wie
beim Experiment mit einer "Offnung. Werden die "Offnungen dagegen weit
auseinander gesetzt, so ist ein zyklischer Flu"s zu beobachten. Durch eine
"Offnung entweicht die Salzl"osung, durch die andere str"omt Wasser
ein. Hierbei treten keine Oszillationen der Flu"srichtungen auf.

\section[Bewertungskriterien]
{Bewertungskriterien von {\fussgaengerstroeme}n}
\label{sec:bewertungskriterien}
In den vorangehenden Abschnitten wurden verschiedene Ph"anomene von
{\fussgaengerstroeme}n beschrieben.

Neben der Beobachtung des  simulierten {\fussgaenger}verkehrs und der
Untersuchung der im vorangehenden Abschnitt beschriebenen Ph"anomene
interessieren auch quantitative Ergebnisse der Simulation, mit denen sich die
Eigenschaften eines Geb"audes bewerten lassen. Schwachstellen, an denen es
zu kritischen Situationen im {\fussgaenger}verkehr kommt, k"onnen damit
bereits im Planungsstadium aufgedeckt und beseitigt werden.
In Analogie zu den in Abschnitt
\ref{sec:bemessung} eingef"uhrten Kriterien K1 bis K8 der
\hi{Level-of-Service}-Betrachtungen werden im folgenden
die Bewertungsma"se $Y^1$ der Effizienz, $Y^2$ der erzwungenen
Geschwindigkeitswechsel, $Y^3$ des Wohlbefindens, $Y^4$ des Zusammenbleibens
von Gruppen und $Y^5$ des Grades der Segregation verschiedener Subpopulationen
definiert. 

Die Ma"se werden f"ur ein bestimmtes Segment der {\fussgaenger}anlage
durch die durchwandernden {\fussgaenger} ermittelt. Die momentanen Me"swerte
werden bei jedem Zeitschritt berechnet und "uber die Zeitdauer $T$
integriert, die
ein {\fussgaenger} braucht, um die Anlage zu durchwandern.
Der Index $\alpha$ kennzeichnet die individuellen Gr"o"sen eines
{\fussgaenger}s $\alpha$. 
  
Die Bewertungsma"se geben die Eigenschaften eines (Teil-)St"uckes der
{\fussgaenger}anlage bei einer bestimmten Verkehrsbelastung und
Zusammensetzung der {\fussgaenger}menge wieder. Daher m"ussen w"ahrend der
Simulation die Bedingung des {\fussgaenger}aufkommens konstant gehalten werden.
 
\subsection{Effizienz}
\label{subsect_eff} 
Die Effizienz einer {\fussgaenger}anlage macht eine Aussage dar"uber, wie
schnell die {\fussgaenger} in Bezug auf ihre Wunschgeschwindigkeit durch die
Anlage (oder durch ein Teilst"uck davon) kommen. Dazu wird die in Zielrichtung
liegende Komponente  der momentanen
Geschwindigkeit eines {\fussgaenger}s $\Va$ 
in Relation zu seiner Wunschgeschwindigkeit gesetzt und "uber den
Zeitraum $T$ gemittelt.
\begin{equation}
  Y^1_\alpha =
  \frac{1}{T}\int\limits_{t_0}^{t_0+T}\!dt\,\frac{\vp{\Eanull}{\Va}}{\vanull} 
  = \underbrace{\frac{1}{\vanull T}}_{\mx{I}}
  \underbrace{\int\limits_{t_0}^{t_0+T}\!dt\,
    \vp{\Eanull}{\Va}}_{\mx{II}}
  \label{def:effizienz}
\end{equation}
In einem Korridor ist die Zielrichtung $\Eanull$ zeitlich konstant.
Term II aus  (\ref{def:effizienz}) entspricht dann der L"ange des Korridors 
\begin{equation}
  \int\limits_{t_0}^{t_0+T}\!dt\,\vp{\Eanull}{\Va} = l_{\mx{Korridor}}
\end{equation}
und die Effizienz dr"uckt das Verh"altnis zwischen der Wegl"ange zum Ziel
und der Strecke, die der {\fussgaenger} in der Zeit $T$ auf freier Fl"ache ohne
Hindernisse zur"uckgelegt h"atte. 
Daraus l"a"st sich eine \hi{effektive L"ange} des Wegest"uckes als 
\begin{equation}
  \lanull = \frac{l_\mx{Korridor}}{Y^1_\alpha} = \vanull\,T
\end{equation}
definieren. 

Die {\fussgaenger} k"onnen im Gedr"ange auch einmal
schneller als ihre Wunschgeschwindigkeit laufen. Das f"uhrt zu einer Effizienz
$Y^1_\alpha$ gr"o"ser 1.
Wenn die Gehgeschwindigkeit laut (\ref{motivation}) nach oben begrenzt ist, 
wird $\Va$ durch $\theta \Va$ ersetzt:
\begin{equation}
  Y^1_\alpha =
  \frac{1}{T}\int\limits_{t_0}^{t_0+T}\!dt\,
  \frac{\vp{\Eanull}{\theta\Va}}{\vanull} 
\end{equation}
F"ur $\vamax = \vanull$ gilt dann stets $Y^1_\alpha \le 1$.

\subsection[Geschwindigkeitswechsel]{Geschwindigkeitswechsel, Varianz der
  Geschwindigkeit} 
In Anlehnung an das Kriterium K2 aus Abschnitt \ref{sec:bemessung}, der
H"aufigkeit eines erzwungenen Geschwindigkeitswechsels, l"a"st sich in der
Simulation die Varianz der Geschwindigkeit in Zielrichtung  als
Beurteilungsgr"o"se definieren:
\begin{eqnarray}
  Y^2_\alpha & = & 
  \frac{1}{T}\int\limits_{t_0}^{t_0+T}\!dt\,
  \sq{\left(\frac{\vp{\Eanull}{\Va}}{\vanull}\right)} 
  - \sq{\left( \frac{1}{T}\int\limits_{t_0}^{t_0+T}\!dt\, 
    \frac{\vp{\Eanull}{\Va}}{\vanull} \right) }  \nonumber \\
  & = &   \frac{1}{T}\int\limits_{t_0}^{t_0+T}\!dt\,
  \sq{\left(\frac{\vp{\Eanull}{\Va}}{\vanull}\right)} 
  - \sq{\left(Y^1_\alpha\right)}
  \label{def:geschw_varianz}
\end{eqnarray}

Die Betrachtung der Geschwindigkeit in Zielrichtung $\vp{\Eanull}{\Va}$ 
bedeutet, da"s die  Varianz sowohl die Effekte von Abbrems-, als auch von
Ausweichman"overn der {\fussgaenger} enth"alt. Mit $Y^2_\alpha$ wird daher
auch das Kriterium K4, die H"aufigkeit eines erzwungenen Richtungswechsels, in
die Bewertung aufgenommen.

\subsection[Wohlbefinden]{Wohlbefinden, Varianz der Beeinflussung}
\label{subsect_stress} 
Die St"arke der Reaktionen eines {\fussgaenger}s auf seine Umgebung erlaubt
eine Bestimmung seiner Situation.
Je mehr Einfl"ussen er ausgesetzt ist,
desto unwohler f"uhlt er sich, und umso mehr mu"s
er sich auf seine Umgebung konzentrieren. Eine Gr"o"se, die diesen Sachverhalt
quantisiert, hat auch eine praktische Bedeutung: Mu"s sich ein {\fussgaenger}
sehr stark auf die umgebenden Passanten konzentrieren, schenkt er anderen
Einfl"ussen, wie zum Beispiel Warenangeboten im Kaufhaus, weniger
Aufmerksamkeit. 

Eine passende Gr"o"se wird durch den Betrag der sozialen Kr"afte, die zu einem
Zeitpunkt $t$ auf den {\fussgaenger} $\alpha$ wirken, in der Form
\begin{equation}
  \frac{1}{N_{\bf F}}\sum_{l\in{\bf F}}
  \sq{\|\vec{f}_{\!\alpha}^{\,l}\|}\qquad\mbox{mit der Anzahl der Kraftterme
    $N_f$} 
\end{equation}
geboten. Die Differenz dieses Ausdrucks zur ausge"ubten sozialen Kraft 
$\vec{f}_{\!\alpha}^{\,tot} = \sum_{l\in {\bf F}}\vec{f}_{\!\alpha}^{\,l}$, 
ergibt die Varianz der gesamten sozialen Kraft 
\begin{equation}
   \sq{\sigma}(\vec{f}_{\!\alpha}^{\,tot}) =
  \frac{1}{N_{\bf F}}\sum_{l\in {\bf F}}
  \sq{\|\vec{f}_{\!\alpha}^{\,l}\|}\,- 
  \,\frac{1}{\sq{N_{\bf F}}} \sq{\|\vec{f}_{\!\alpha}^{\,tot}\|} 
\end{equation}
die als momentane Belastung definiert wird.
Die Menge der Kr"afte
${\bf F}$, die in diesem Ma"s ber"ucksichtigt werden, kann dabei auf bestimmte
Wechselwirkungstypen eingeschr"ankt werden. Zum Beispiel nur die
{\fussgaenger}einfl"usse:
\begin{equation}
  \vec{f}_{\!\alpha}^{\,l} = \vec{f}_{\!\alpha\beta}
\end{equation}
Das entspricht dem Kriterium K3 aus Abschnitt \ref{sec:bemessung}, dem
Zwang zur Beachtung anderer {\fussgaenger}. 

Die mittlere Belastung eines {\fussgaenger}s auf der im Zeitraum $T$
zur"uckgelegten Strecke ergibt sich durch
\begin{equation}
  Y^3_{\alpha} = \frac{1}{T}\int\limits_{t_0}^{t_0+T}\!dt\,
  \frac{1}{N_{\beta}}\sum_{\beta} 
  \sq{\|\vec{f}_{\!\alpha\beta}\|}\,- 
  \,\frac{1}{\sq{N_{\beta}}} 
  \sq{\left\| \sum_{\beta} \vec{f}_{\!\alpha\beta} \right\|}
  \label{def:beeinfl_varianz}
\end{equation}
Je geringer diese Belastung f"ur den {\fussgaenger} ausf"allt, desto h"oher
ist die Qualit"at des Weges.

\subsection{Gruppenabstand}
Ein weiteres Qualit"atskriterium ist das Ma"s des Zusammenbleibens einer
Gruppe im {\fussgaenger}strom. Bei den \hi{Level-of-Service}-Betrachtungen
(vgl. Abschnitt \ref{sec:bemessung}) fand dies keine Ber"ucksichtigung.
Dieses Kriterium ist besonders im \hii{Freizeit-}{Freizeitverkehr} und
\hi{Einkaufsverkehr} f"ur das Zusammenbleiben von Familien relevant.

Die zu $\alpha$ geh"orenden Gruppenmitglieder $\{\alpha_k\}$ wollen so dicht wie
m"oglich zusammenbleiben. Durch das Ausweichen von Hindernissen werden sie
jedoch zeitweise auseinander gedr"angt. Ein Ma"s zur Bestimmung, wie dicht eine
Gruppe "uber eine Wegstrecke zusammenbleiben kann, ist der mittlere Abstand
zwischen den einzelnen Gruppenmitgliedern
\begin{equation}
  Y^4_\alpha = \frac{1}{T}\int\limits_{t_0}^{t_0+T}\!dt\,
  \frac{1}{(n_\alpha-1)}\,\sum_{\alpha^\prime} \|\vec{r}_{\alpha} -
  \vec{r}_{\alpha^\prime}\| 
  \label{def:gruppenabstand}
\end{equation}
mit der Gruppengr"o"se $n_\alpha$.



F"ur gr"o"sere Gruppen ist der \hi{mittlere Gruppenabstand} gr"o"ser,
weil nicht mehr jeder neben jedem gehen kann.  Die Aussagef"ahigkeit dieses
Ma"ses besteht jedoch nur unter der Voraussetzung, da"s die Attraktion der
einzelnen Gruppenmitglieder stark genug ist, und da"s sie ungef"ahr die
gleiche Wunschgeschwindigkeit haben. Ist diese Voraussetzung nicht erf"ullt,
wird die Gruppe selbst auf freien Strecken allein durch die unterschiedlichen
Geschwindigkeiten auseinandergezogen.

\subsection[Gemittelte Ma"se]{Gemittelte Ma"se eines (Teil-)Systems}
\label{sec:mittelung}
Die Bewertungsgr"o"sen $Y_\alpha$, die in den vorhergehenden
Abschnitten 
eingef"uhrt wurden, beschreiben die individuelle Situation eines
{\fussgaenger}s. Zur Bewertung von {\fussgaenger}anlagen lassen sich daraus
Me"swerte f"ur ein bestimmtes (Teil-)System bestimmen, indem w"ahrend der
gesamten \hi{Simulationsdauer} die \hi{individuellen Me"swerte} "uber alle
$N$ {\fussgaenger} gemittelt werden, die diesen Wegeabschnitt
durchwandert haben.  Der \hi{mittlere Me"swert} eines Abschnittes ist dann
durch 
\begin{equation}
  \ave{Y} = \frac{1}{N} \sum_{\alpha} Y_\alpha
\end{equation}
mit der Varianz
\begin{equation}
  \sq{\sigma}\left(Y\right) = 
  \frac{1}{N} \sum_{\alpha} \sq{Y_\alpha} - 
  \sq{\left(\frac{1}{N} \sum_{\alpha} Y_\alpha \right)} 
\end{equation}
gegeben. F"ur detailliertere Aussagen "uber die {\fussgaengerstroeme} kann
auch "uber verschieden Subpopulationen, die sich zum Beispiel durch ihre 
Gehrichtung oder Geschwindigkeit unterscheiden, getrennt gemittelt werden.

Durch 
\begin{equation}
  \ave{Y^\prime} = \frac{1}{N_1+\dots+N_n}\left(N_1 \ave{Y_1} + \dots +
  N_n \ave{Y_n} \right)  
\end{equation}
und
\begin{eqnarray}
  \sq{\sigma}\left(Y^\prime\right) & = &\frac{1}{N_1+\dots+N_n} 
  \bigg(
    N_1 \left(\sq{\sigma}\left(Y_1\right) +  \sq{\ave{Y_1}}\right)
    + \dots +
    N_n \left(\sq{\sigma}\left(Y_n\right) +  \sq{\ave{Y_n}}\right)
  \bigg) \nonumber \\
  &- &   \frac{1}{\sq{\left(N_1+\dots+N_n\right)}}\, 
  \sq{\bigg( N_1 \ave{Y_1} + \dots +  N_n \ave{Y_n} \bigg)}
\end{eqnarray}
geschieht die Zusammensetzung von Me"swerten einzelner Wegeabschnitte oder
Subpopulationen.

\subsection{Vorhersage der Me"swerte}
\label{sec:vorhersage}
Die Simulation eines {\fussgaenger}stromes startet aus einem Anfangszustand
mit leerem System. Nach und nach werden die {\fussgaenger} aktiviert und
treten in das System ein. Nach einer gewissen Zeit bleibt die Anzahl der
aktiven {\fussgaenger} konstant, da f"ur jeden, der das System verl"a"st, ein
neuer eintritt. Es dauert jedoch sehr lange, bis die
Effekte des Anfangszustandes abgeklungen sind und das System einen
\hi{station"aren Zustand} erreicht.   

Die Ermittlung der oben genannten Bewertungsma"se ben"otigt daher eine sehr
lange Simulationsdauer und somit einen sehr hohen
Rechenaufwand. Typischerweise konvergiert die Entwicklung eines Ma"ses $Y$
von einem willk"urlichen Wert startend monoton auf den endg"ultigen Wert. Der
zeitliche Verlauf hat dabei ungef"ahr die Form einer Exponentialfunktion $y(t)
= a_0 + a_1 \exp(-t/a_2)$ mit den Konstanten $a_0$, $a_1$ und $a_2$.

An die bis zu einem Zeitschritt $t_i$ gesammelten  
mittleren Me"swerte $\ave{Y}_i$ mit den Fehlern $\sigma_i =
\sqrt{\sq{\sigma}(Y)}$ kann eine entsprechende Funktion angepa"st werden,
um eine Vorhersage des endg"ultigen Resultates zu treffen.

Ein geeignetes Verfahren ist die \hi{Methode des Minimalen Quadratischen
Fehlers} \cite[Kap. 15]{NRC}, bei
der die Gr"o"se
\begin{equation}
  \label{chi_sqare_def}
  \sq{\chi} \equiv \sum_{i=1}^N\sq{\left(\frac{\ave{Y}_i-y(t_i,
    a_0...a_{M-1})}{\sigma_i}\right)} 
\end{equation}  
durch Anpassung der Parameter $a_0...a_{M-1}$ minimiert wird.
Die Bedingung f"ur ein Minimum lautet:
\begin{equation}
  \label{chi_sqare_deriv}
  0 = \sum_{i=1}^N\left(\frac{\ave{Y}_i-y(t_i)}{\sq{\sigma_i}}\right)
  \left(\frac{\partial y(t_i,...a_k...)}{\partial a_k}\right)
  \qquad k=0,...,M-1
\end{equation}

Handelt es sich bei der anzupassenden Funktion um eine Linearkombination von
Basisfunktionen 
\begin{equation}
  y(t) = \sum_{k=0}^{M-1}a_k X_k(t) 
\end{equation}
so wird aus (\ref{chi_sqare_deriv}) das \hi{lineare Gleichungssystem}
\begin{equation}
  0 = \sum_{i=1}^N 
  \left.
    \frac{1}{\sq{\sigma_i}}
    \left( \ave{Y}_i - \sum_{j=0}^{M-1} a_j X_j(t_i) \right) X_k(t_i)
    \right. \qquad k=0,...,M-1.
\end{equation}
Durch vertauschen der  Reihenfolge der Summationen l"a"st sich das
Gleichungssystem  als Matrizengleichung 
\begin{equation}
  \sum_{j=0}^{M-1} \alpha_{kj} a_j = \beta_k
\end{equation}
mit der $M \times M$ Matrix
\begin{equation}
  \alpha_{kj} = \sum_{i=1}^N\frac{X_j(t_i)\,X_k(t_i)}{\sq{\sigma_i}}
\end{equation}
und dem Vektor der L"ange $M$
\begin{equation}
  \beta_k = \sum_{i=1}^N\frac{\ave{Y}_i\,X_k(t_i)}{\sq{\sigma_i}}
\end{equation}
schreiben. Die Inversion der Matrix $ \left(\alpha\right)_{M\times M}$,
die nach dem
\hi{Gau"sschen Eliminationsverfahren}
\cite[S. 61 ff.]{Klingenberg:1984}\cite[Kap. 2]{NRC} 
durchgef"uhrt werden kann, liefert direkt die Parameter 
der anzupassenden Funktion:
\begin{equation}
  a_j = \sum_{k=0}^{M-1}\left(\alpha\right)_{jk}^{-1}\,\beta_k
\end{equation}
mit der Varianz der Parameter
\begin{equation}
  \sq{\sigma}(a_j) = \left(\alpha\right)_{jj}^{-1}
\end{equation}   

Die Wahl der Anpassungsfunktion als Linearkombination verschiedener
Basisfunktionen hat den Vorteil, da"s die Berechnung nach dem obigen Verfahren
auf das Ergebnis des vorangegangenen Zeitschritts aufbauen kann. Die einzelnen
Mittelwerte und Varianzen der $N$ Simulationsschritte m"ussen dabei nicht
gespeichert werden. 

Die vorgeschlagene Exponentialfunktion l"a"st sich daher nicht
verwenden. Aber unter der Voraussetzung, da"s die Vorhersage erst nach einer
gewissen Anlaufsperiode $t_s>0$ ge"-startet wird, kann als anzupassende Funktion
\begin{equation}
  y(t) = a_0 + \sum_{j=1}^{M-1}a_j\,\frac{1}{t^j}
\end{equation}
eine Linearkombination aus Hyperbeln verwendet werden. Dabei reicht in der
Regel ein Parametersatz der L"ange $M=5$ aus, um die Entwicklung der Ma"se zu
extrapolieren. Im Grenzwert $t\rightarrow\infty$ konvergiert die
Anpassungsfunktion gegen $a_0$.

\subsection{Leistungsf"ahigkeit (Flu"s)}
\label{sec:flowmeasure}
\label{sec:leistungsfaehigkeit}
Zur Planung von  {\fussgaenger}anlagen wird oft die (spezifische)
Leistungsf"ahigkeit einer Anlage (\ref{def:spez_leist}) als
Bemessungsgrundlage herangezogen (vgl.\ Abschn.\ \ref{sec:bemessung}). 

Sie l"a"st sich auch in der Simulation bestimmen, indem an den Toren, die die
einzelnen Teilst"ucke einer Anlage verbinden (vgl.\
Abschn.\ \ref{sec:darstellung}), die Anzahl der hinein $N_l^{+}$ und hinaus
laufenden {\fussgaenger} $N_l^{-}$ ermittelt wird.

Daf"ur wird f"ur jede Durchgangsrichtung der Flu"s 
\begin{equation}
  \Phi_l^{\pm} = \frac{N_k^\pm}{T} = \rho_l^{\pm}\,v^{\pm}\,b_l
\end{equation}
und die Flu"sdichte 
\begin{equation}
  \hat{\Phi}_l^{\pm} = \frac{N_k^\pm}{Tb_l} = \rho_l^{\pm}\,v^{\pm}
\end{equation}
mit der Simulationszeit $T$, der Breite des Tores $b_l$, der partiellen Dichte
$\rho_l^{\pm}$ jeder Durchgangsrichtung und deren Geschwindigkeit $v^{\pm}$
definiert.

Die Leistungsf"ahigkeit eines Teilst"uckes einer {\fussgaenger}anlage, die
nach (\ref{def:spez_leist}) der Anzahl der in einem bestimmten
Zeitintervall die Verkehrsfl"ache durchwandernden {\fussgaenger} entspricht,
ergibt sich aus den Zufl"ussen und Abfl"ussen des {\fussgaenger}stroms
durch alle Tore der Anlage nach
\begin{equation}
  \label{def:simulationleistungsfaehigkeit}
  L = \frac{1}{2}\sum_l b_l\left(\hat{\Phi}_l^{+}+\hat{\Phi}_l^{-}\right)
\end{equation}

\subsection{Grad der (Selbst-)Organisation, Vielf"altigkeitsma"s}
In den {\fussgaenger}str"omen treten h"aufig regelm"a"sige Muster
auf, die bei der Beobachtung von {\fussgaenger}anlagen und
Simulationsergebnissen sofort auffallen
(vgl. \ref{sec:phaenomenederselbstorganisation}). 

Selbstorganisationsph"anomene werden allerdings erst erkennbar, wenn die
{\fussgaenger} nach bestimmten Merkmalen gekennzeichnet sind. Stellt man die
{\fussgaenger} je nach Zielrichtung durch unterschiedlich farbige Symbole dar,
so lassen sich die entgegengesetzten Laufrichtungen und die Ausbildung von
Bahnen in einem Korridor gut erkennen. W"urde die Aufteilung der {\fussgaenger}
jedoch nach anderen Kriterien, zum Beispiel nach ihrer Geschwindigkeit
erfolgen, w"urde dieses Ph"anomen nicht sichtbar werden, obwohl es 
weiterhin auftritt. 

Um die Organisation eines Systems nach bestimmten Merkmalen auch quantitativ
erfassen zu k"onnen, soll nun ein Ma"s f"ur die Ordnung eingef"uhrt werden.
Eine M"oglichkeit besteht darin, den Zustand des Systems durch die Entropie 
zu beschreiben. Dabei wird ein Volumen in gleichgro"se Einheiten
aufgeteilt und die anteilige Anzahl der Objekte einer Subpopulation ermittelt,
die zum Zeitpunkt $t$ in der Einheit um $\vec{x}_i$ anzutreffen sind.
Die Entropie ist als
\begin{equation}
  \label{def:negentropy}
  H(A,t) = -\sum_{\bf C}\sum_{x_i\in A} P({\bf C}, \vec{x}_i,t)\,\ln P({\bf C},
  \vec{x}_i,t) 
\end{equation}
definiert. ${\bf C}$ bezeichnet die einzelnen Subpopulationen wie etwa die
{\fussgaenger} einer Zielrichtung. $P({\bf C}, \vec{x}_i,t)$ ist der Anteil der
{\fussgaenger}, die zur Subpopulation ${\bf C}$ geh"oren und sich zum
Zeitpunkt $t$ in dem Segment um $\vec{x}_i$ aufhalten
\cite{Haken:1983}. 

In der mikroskopischen Beschreibung von {\fussgaengerstroeme}n 
ist die Definition nach (\ref{def:negentropy})
problematisch, da die Fl"acheneinheiten in einer Anlage nicht ausreichend
gro"s gew"ahlt 
werden k"onnen, um eine gr"o"sere Menge von {\fussgaenger}n aufzunehmen.
Daher wird eine Gr"o"se eingef"uhrt, die die selben Eigenschaften der Entropie
aufweist, aber durch Paarwechselwirkungen zwischen den {\fussgaenger}n
definiert ist. Da dieser Ausdruck formal der Entropie entspricht, wird er im
folgenden auch mit Entropie bezeichnet.
 
In Anlehnung an die Definition der Entropie wird der anteilige Einflu"s $\Pac$
jeder Subpopulation ${\bf C}$ auf einen {\fussgaenger} $\alpha$ bestimmt.
Der Einflu"s wird dabei in Relation zu der Summe der einwirkenden Potential 
aller Subpopulation gesetzt
\begin{equation}
  \label{def:pac}
  \Pac = 
  \frac{ \displaystyle \sum_{\mbox{\scriptsize $\beta\in \bf C$}} U_{\alpha
      \beta} }
  { \displaystyle \sum_{\mbox{\scriptsize $\bf C^\prime$}}  
    \sum_{\mbox{\scriptsize $\beta\in \bf C^\prime$} } U_{\alpha \beta} }
\end{equation}
und die Ordnung f"ur einen {\fussgaenger} durch
\begin{equation}
  \label{def:entropyalpha}
  S_\alpha = -k\,\sum_{\bf C} \Pac\, \ln \Pac
\end{equation}
definiert. Die Konstante $k>0$ dient zur Normierung des Ausdrucks. 

Das Ordnungsma"s ist wegen des negativen Vorzeichens in
(\ref{def:entropyalpha}) eigentlich ein Unordnugsma"s, denn die Entropie ist
am geringsten, wenn ein {\fussgaenger} nur von einer einzigen Subpopulation
umgeben ist und von ihr beeinflu"st wird. Das Maximum der 
Entropie wird erreicht, wenn alle Subpopulationen einen gleichstarken
Einflu"s auf die Person aus"uben.  

Im Fall zweier Subpopulationen ${\bf C}$ und ${\bf C^\prime}$ mit den
anteiligen Einflu"sst"arken (Proportionen)
\begin{eqnarray}
  \Pac & = & P \qquad \mbox{mit}\quad 0\le P\le 1  \nonumber \\
  \Pacprime & = & 1-\Pac \\
\end{eqnarray}
ist die momentane Entropie zum Zeitpunkt $t$ durch
\begin{equation}
  S_\alpha =  -k\,\left(P\, \ln P\,+\,(1-P)\, \ln(1-P)\right)
\end{equation}
bestimmt. Wie gefordert wird sie f"ur $P^0 = 1/2$ maximal. Dies kann durch die
notwendige (\ref{bed:entropy:n}) und hinreichende (\ref{bed:entropy:h})
Bedingung gezeigt werden: 
\begin{eqnarray}
  0  & = & 
   \frac{d}{dP} S_{\alpha} \nonumber \\
   \Leftrightarrow \: 0  & = &  
    -k\, \frac{d}{dP} \big( 
    P\, \ln P\,+\,(1-P)\, \ln(1-P) \big)
        \nonumber \\
 \Rightarrow \:0 & = & \left. \ln P + \frac{P}{P} - \frac{1}{1-P} - 
    \ln (1-P) + \frac{P}{1-P}  \right|_{P = P^0}  \nonumber \\
    & = &  \ln P - \ln (1-P) \bigg|_{P = P^0} \nonumber \\
    \Rightarrow \: P^0 & = &\frac{1}{2} 
    \label{bed:entropy:n}
  \end{eqnarray}
  \begin{equation}
    \label{bed:entropy:h}   
  \frac{d}{dP}\left.\left(-k\,\ln\frac{P}{1-P}\right)\right|_{P=P^0}
  =\; \left.\frac{-k}{P} + \frac{-k}{1-P}\right|_{P=P^0} =\; -4k < 0
\end{equation}
F"ur Systeme mit $n$ Subpopulationen gilt allgemein
$\Pac^0 = 1/n$ f"ur alle Populationen ${\bf C}$. Bei maximaler Entropie
ist die {\fussgaenger}menge vollst"andig durchmischt. In diesem Fall hat die
Entropie nach (\ref{def:entropyalpha}) den Wert
\begin{equation}
  S_\alpha = -k\,\sum_{\bf C} \Pac\, \ln \Pac 
  = -k\,\sum_{\bf C} \frac{1}{n}\, \ln \frac{1}{n} = k\,\ln n
\end{equation}
Um den Wert der maximalen Entropie von der Anzahl $n$ zu l"osen, wird die
Entropie mit $k = 1/\ln n$ normiert.  Wenn die Menge vollst"andig
getrennt ist, und der {\fussgaenger} $\alpha$ nur von einer einzigen
Subpopulation beeinflu"st wird, ist die Entropie gleich Null.

In einem System mit insgesamt $N$ {\fussgaenger}n und $n$ Subpopulationen ist
die \hi{momentane Entropie} als
\begin{equation}
  \label{def:momententropy}
  S = -\frac{1}{\ln(Nn)} \sum_\alpha \sum_{\bf C}
  \frac{\Pac}{\sum_{\alpha^\prime} \Pacp} \,\ln 
  \frac{\Pac}{\sum_{\alpha^\prime}\Pacp}  
\end{equation}
definiert. Hierbei werden die anteiligen Einfl"usse auf einen
{\fussgaenger} in Relation zu den Einfl"ussen auf alle Personen gesetzt.

Au"serdem kann auch die \hi{Entropie "uber eine Zeitperiode} $T$ in der Form
\begin{equation}
  \label{def:zeitentropy}
  S_T = -\frac{1}{\ln(NnT)} \int\limits_{t_0}^{t_0+T}\!\!dt\,
  \sum_\alpha \sum_{\bf C} \left.
  \frac{\Pac}{\int\limits_{t_0}^{t_0 + T}\!\!dt^\prime\,
    \sum_{\alpha^\prime}\Pacp}
  \,\ln 
  \frac{\Pac}{\int\limits_{t_0}^{t_0 + T}\!\!dt^\prime\,
    \sum_{\alpha^\prime}\Pacp}
\right.
\end{equation}
definiert werden. Dadurch werden neben der r"aumlichen Verteilung der
{\fussgaenger} auch zeitliche "Anderungen des Systems in das Ordnungsma"s
aufgenommen.  



Die drei Gr"o"sen $S_\alpha$, $S$ und $S_T$ sind jeweils eigenst"andige
Ma"se und treffen unterschiedliche Aussagen "uber den Zustand des Systems.
Die momentane Entropie des Systems mit $N$ {\fussgaenger}n unterscheidet sich
vom Mittelwert der \hi{individuellen Entropie} "uber alle {\fussgaenger}
$1/N\,\sum_\alpha S_\alpha$ darin, da"s sie auch die Unterschiede in der
{\fussgaenger}dichte anzeigt. 
Zur Erl"auterung seien zwei {\fussgaenger} $\alpha$ und
$\alpha^\prime$ angenommen, die jeweils gleichm"a"sig von allen Populationen
beeinflu"st werden. {\fussgaenger} $\alpha$ soll dabei aber einen st"arkeren
Einflu"s erfahren, weil die ihn umgebenden Personen dichter stehen. Die
momentane Entropie des Systems $S$ zeigt in diesem Fall die unterschiedliche
Dichte als Ordnung ($S \rightarrow 0$) an, w"ahrend der Mittelwert einen
Wert f"ur gleichm"a"sige Verteilung annimmt ($S < 1/N\,\sum_\alpha
S_\alpha$).

In gleicher Weise unterscheidet sich auch die gesamte Entropie des Systems
w"ahrend einer Zeitperiode $S_T$ von der Zeitmittelung der momentanen Entropie
$S$: W"ahrend erstere zeitliche Schwankungen mit einem niedrigen Wert f"ur
Ordnung anzeigt, kann der zeitliche Mittelwert der momentanen Entropie h"oher
liegen.

An der Entropie ist zu erkennen, in welchem Ma"se sich
das System gleichm"a"sig im Raum und im Zeitverlauf verh"alt. Der Begriff
Ordnung ist daher als Verschiedenheit von Teilen eines Systems zu
interpretieren.

Als ein weiteres Bewertungsma"s f"ur den {\fussgaenger}verkehr l"a"st sich aus
der Definition des \hi{anteiligen Einflu"s}es (\ref{def:pac}) und der 
\hi{individuellen Entropie} (\ref{def:entropyalpha}) mit 
\begin{equation}
  \label{def:pact}
  \PPac = 
  \frac{\displaystyle \int\limits_{t_0}^{t_0 + T}\!\!dt\,
    \sum_{\mbox{\scriptsize $\beta\in \bf C$}} U_{\alpha\beta} }
  { \displaystyle \sum_{\mbox{\scriptsize $\bf C^\prime$}}   
    \int\limits_{t_0}^{t_0 + T}\!\!dt\,
    \sum_{\mbox{\scriptsize $\beta^\prime \in \bf C^\prime$} } 
    U_{\alpha\beta^\prime} }
\end{equation}
und 
\begin{equation}
  S_\alpha^\prime = - \frac{1}{nT} \sum_{\bf C} \PPac \ln \PPac
\end{equation}
der Grad der Segregation 
\begin{equation}
  \label{def:segregation}
  Y^5_\alpha = 1 - S_\alpha^\prime = 1 + \frac{1}{nT} 
  \sum_{\bf C} \PPac \ln \PPac
\end{equation}
ableiten. Dabei ist $t_0$ bis $t_0+T$ gerade die Zeit, in der der
{\fussgaenger} $\alpha$ das (Teil-)System durchwandert. Damit
wird zum Ausdruck gebracht, in welchem Ma"se ein 
{\fussgaenger} auf seinem Weg 
dem Einflu"s
unterschiedlicher Subpopulationen ausgesetzt war. In einer  gleichm"a"sig
vermischten Menschenmenge ist $Y^5_\alpha =0$, bei einem hohen Grad an
Segregation geht $Y^5_\alpha \rightarrow1$. Analog zu den anderen
\hi{Bewertungsma"s}en kann der \hi{Grad der Segregation} "uber alle
{\fussgaenger}, die eine Anlage durchwandern, gemittelt werden (vgl.\ 
\ref{sec:mittelung}). Zudem lassen sich Vorhersagen "uber die zeitliche
Entwicklung treffen (vgl.\ \ref{sec:vorhersage}).

Dieses Ma"s ist "uberall dort von Interesse, wo Menschen
unterschiedlicher kultureller oder politischer Zugeh"origkeit auf
"offentlichen Pl"atzen zusammenkommen. Bereits gespannte Situationen k"onnen
durch aufkommendes Gedr"ange leicht eskalieren. Zum Beispiel bei den Fans
zweier Fu"sballmannschaften im Stadion. Zur Vermeidung von Konflikten ist es
hilfreich, den einzelnen Gruppen durch geeignete (Bau-)Ma"snahmen die
M"oglichkeit zu r"aumlicher Separation zu geben und das Vermischen der Mengen
zu verhindern.

\section{Leistungsma"se}
\label{sec:leistungsmasse}

Die im vorigen Abschnitt eingef"uhrten Bewertungskriterien wurden f"ur zwei
einfache Systeme ermittelt, einen Korridor und eine Kreuzung. 
Dabei wird deutlich, da"s neben der Verkehrsdichte auch 
die geometrische Form eines Geb"audes das Verhalten der {\fussgaengerstroeme} bestimmt.

\subsection{Leistungsma"se eines Korridors}
Zur Untersuchung der {\fussgaengerstroeme} in Korridoren wurde die Simulation
f"ur die Korridorbreiten 4, 6 und \meter{8} und die Populationsgr"o"sen der
beiden entgegenlaufenden {\fussgaenger}gruppen von 102, 204 und 408 Individuen
durchgef"uhrt. Erreicht ein {\fussgaenger} sein Ziel, wird er durch
ein neues Mitglied aus seiner Subpopulation ersetzt, das mit vom Vorg"anger
unabh"angigen Eigenschaften startet. Dadurch bleibt die Anzahl der aktiven
{\fussgaenger} jeder Subpopulation konstant.

\begin{figure}[p]
  \includegraphics[width=0.49\textwidth]{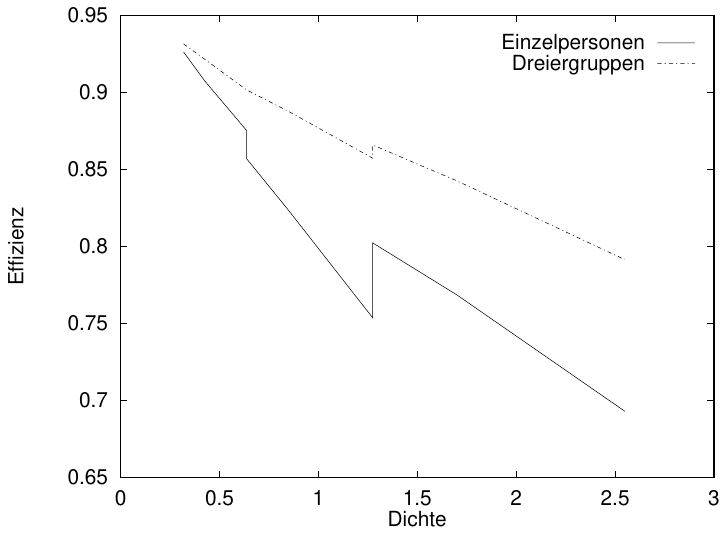} \hfill
  \includegraphics[width=0.49\textwidth]{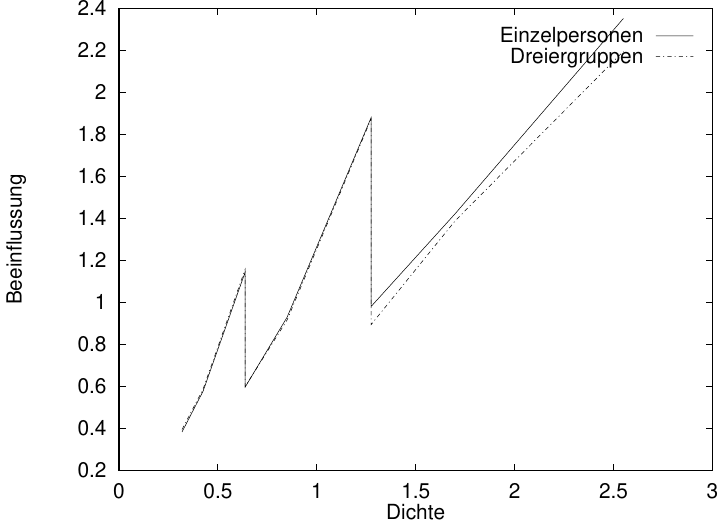} \\[0.5cm]
  \includegraphics[width=0.49\textwidth]{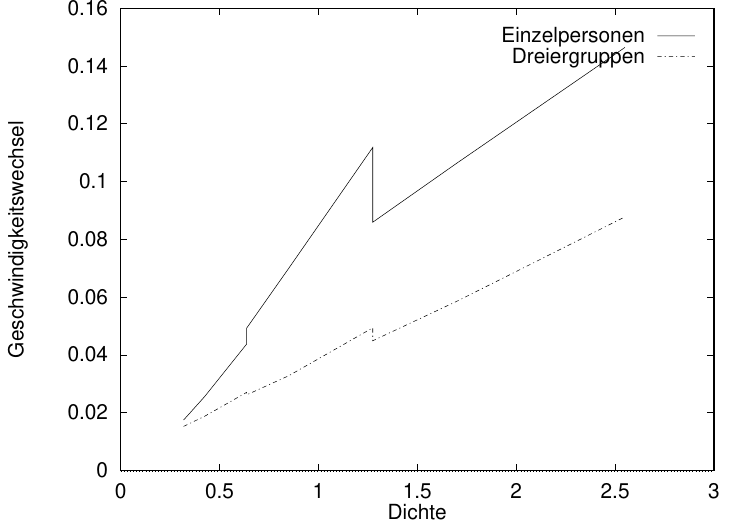} \hfill
  \includegraphics[width=0.49\textwidth]{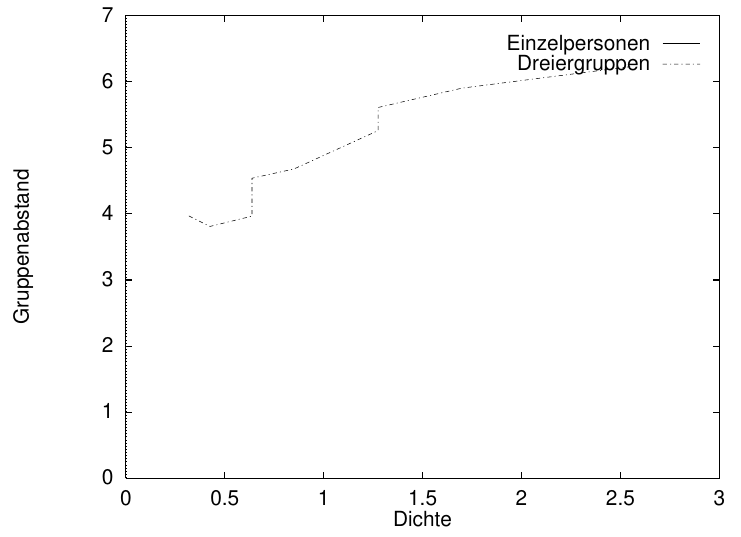}
  \mycaption{Leistungsma"se eine Korridors} {Leistungsma"se eines Korridors.
    Die einzelnen Datenpunkte sind "uber die {\fussgaenger}dichte aufgetragen.
    Sie stehen in der Reihenfolge aus den Tabellen \ref{tab:leistung_korridor}
    und \ref{tab:leistung_korridor_dreier}.  Die Effizienz der Anlage (links
    oben) f"allt mit zunehmender Dichte nahezu linear ab. Erstaunlicherweise
    f"allt die Effizienz f"ur Dreigruppen flacher ab.  Trotz der zus"atzlichen
    Gruppenwechselwirkung k"onnen die Gruppen ihren Strom bei gro"sen Dichten
    effizienter gestalten. Die Varianz der Geschwindigkeit (links unten)
    verh"alt sich genau spiegelbildlich zur Effizienz. Der Vergleich zwischen
    den Messungen im 4 und \meter{8} breiten Korridor bei gleicher
    {\fussgaenger}dichte zeigt, da"s die Effizienz und die Varianz der
    Geschwindigkeit nicht nur durch die Dichte, sondern auch durch die
    Korridorform bestimmt sind.  Noch deutlicher wird die Abh"angigkeit von
    der Geb"audeform im der Varianz der Beeinflussung (rechts oben). Die
    Verengung des Korridors l"a"st diese Ma"s wesentlich st"arker ansteigen,
    als die zunehmende Dichte. Der Gruppenabstand (rechts unten) nimmt
    erwartungsweise bei hohen Dichten zu.}
  \label{fig:leistung_korridor}
\end{figure}

Im ersten Satz der Simulationen (Tabelle
\ref{tab:leistung_korridor}) liefen die {\fussgaenger} einzeln, im
zweiten Satz in Dreiergruppen  (Tabelle \ref{tab:leistung_korridor_dreier}).
Die Unterschiede zwischen einem {\fussgaenger}strom aus Einzelpersonen und
einem aus Dreiergruppen sind in den Abbildungen \ref{fig:leistung_korridor}
erkennbar. Die Simulationsergebnisse werden im Bildtext
\ref{fig:leistung_korridor} diskutiert. 

\begin{table}[tp]
  \begin{center}
    \leavevmode
    \begin{tabular}{|l|l|l|l|l|l|l|l|}
      \hline 
      $\rho$ & $w$ & $N$ & $N_\mx{tot}$  & Effizienz & Varianz &
      Beeinflussung & Gruppe \\ \hline \hline
      0.318750 & 8 & 102 &  2779 & 9.263520e-01 & 1.750491e-02 & 3.846491e-01 & 0.000000 \\
      0.425000 & 6 & 102 &  2718 & 9.076417e-01 & 2.563989e-02 & 5.782057e-01 & 0.000000 \\ 
      0.637500 & 4 & 102 &  2598 & 8.753088e-01 & 4.383204e-02 & 1.144845e+00 & 0.000000 \\ 
      0.637500 & 8 & 204 &  5027 & 8.570126e-01 & 4.934287e-02 & 5.975415e-01 & 0.000000 \\
      0.850000 & 6 & 204 &  4658 & 8.234650e-01 & 7.001586e-02 & 9.307962e-01 & 0.000000 \\
      1.275000 & 4 & 204 &  3828 & 7.537521e-01 & 1.118642e-01 & 1.881226e+00 & 0.000000 \\
      1.275000 & 8 & 408 &  8700 & 8.024332e-01 & 8.599004e-02 & 9.819080e-01 & 0.000000 \\ 
      1.700000 & 6 & 408 &  7417 & 7.687076e-01 & 1.064169e-01 & 1.423571e+00 & 0.000000 \\
      2.550000 & 4 & 408 &  6063 & 6.931106e-01 & 1.464716e-01 & 2.351771e+00 & 0.000000 \\ 
      \hline
    \end{tabular}
  \end{center}
  \mycaption{Leistungsma"se eines Korridors f"ur Einzelpersonen}
  {Leistungsma"se eines Korridors f"ur {\fussgaengerstroeme} aus
    Einzelpersonen. Die Simulationen wurden f"ur verschiedene Korridorbreiten
    $w$ und Populationsgr"o"sen $N$ durchgef"uhrt. Mit $\rho$ ist die mittlere
    Dichte und mit $N_\mx{tot}$ die Gesamtzahl aller "`gemessenen"'
    {\fussgaenger} angegeben. Aufgetragen sind die Leistungsma"se Effizienz
    $Y^1$ aus (\ref{def:effizienz}), Varianz der Geschwindigkeit $Y^2$ aus
    (\ref{def:geschw_varianz}) und Varianz der Beeinflussung $Y^3$ aus
    (\ref{def:beeinfl_varianz}). Der Gruppenabstand $Y^4$ aus
    (\ref{def:gruppenabstand}) ist f"ur Einzelpersonen gleich Null.}
  \label{tab:leistung_korridor}
  \begin{center}
    \begin{tabular}{|l|l|l|l|l|l|l|l|}
      \hline 
      $\rho$ & $w$ & $N$ & $N_\mx{tot}$ & Effizienz & Varianz &
      Beeinflussung & Gruppe \\ \hline \hline
      0.318750 & 8 & 102 &  1107 &  9.315365e-01 &  1.533524e-02 & 3.957259e-01 & 3.970409\\
      0.425000 & 6 & 102 &  1092 &  9.216949e-01 &  1.880761e-02 & 5.883629e-01 & 3.811176 \\
      0.637500 & 4 & 102 &  1075 &  9.016998e-01 &  2.712228e-02 & 1.158885e+00 & 3.966465  \\
      0.637500 & 8 & 204 &  2128 &  9.016516e-01 &  2.624432e-02 & 6.006539e-01 & 4.540360  \\
      0.850000 & 6 & 204 &  2090 &  8.877308e-01 &  3.285849e-02 & 9.173967e-01 & 4.681740  \\
      1.275000 & 4 & 204 &  2036 &  8.573940e-01 &  4.942715e-02 & 1.874801e+00 & 5.262294  \\
      1.275000 & 8 & 408 &  4045 &  8.659953e-01 &  4.494565e-02 & 8.955614e-01 & 5.612268  \\
      1.700000 & 6 & 408 &  3848 &  8.427222e-01 &  5.863792e-02 & 1.389480e+00 & 5.904872  \\
      2.550000 & 4 & 408 &  3312 &  7.915510e-01 &  8.780707e-02 & 2.195521e+00 & 6.230577  \\
      \hline
    \end{tabular}
  \end{center}
  \mycaption{Leistungsma"se eines Korridors f"ur Dreiergruppen}
  {Leistungsma"se eines Korridors f"ur {\fussgaengerstroeme} aus
    Dreiergruppen.  Die aufgetragenen Gr"o"sen entsprechen denen aus Tabelle
    \ref{tab:leistung_korridor}.}
  \label{tab:leistung_korridor_dreier}
\end{table}

\subsection{Effizienz einer Kreuzung}
\begin{figure}[p]
  \begin{center}
    \leavevmode
    \includegraphics[width=0.8\textwidth,clip]{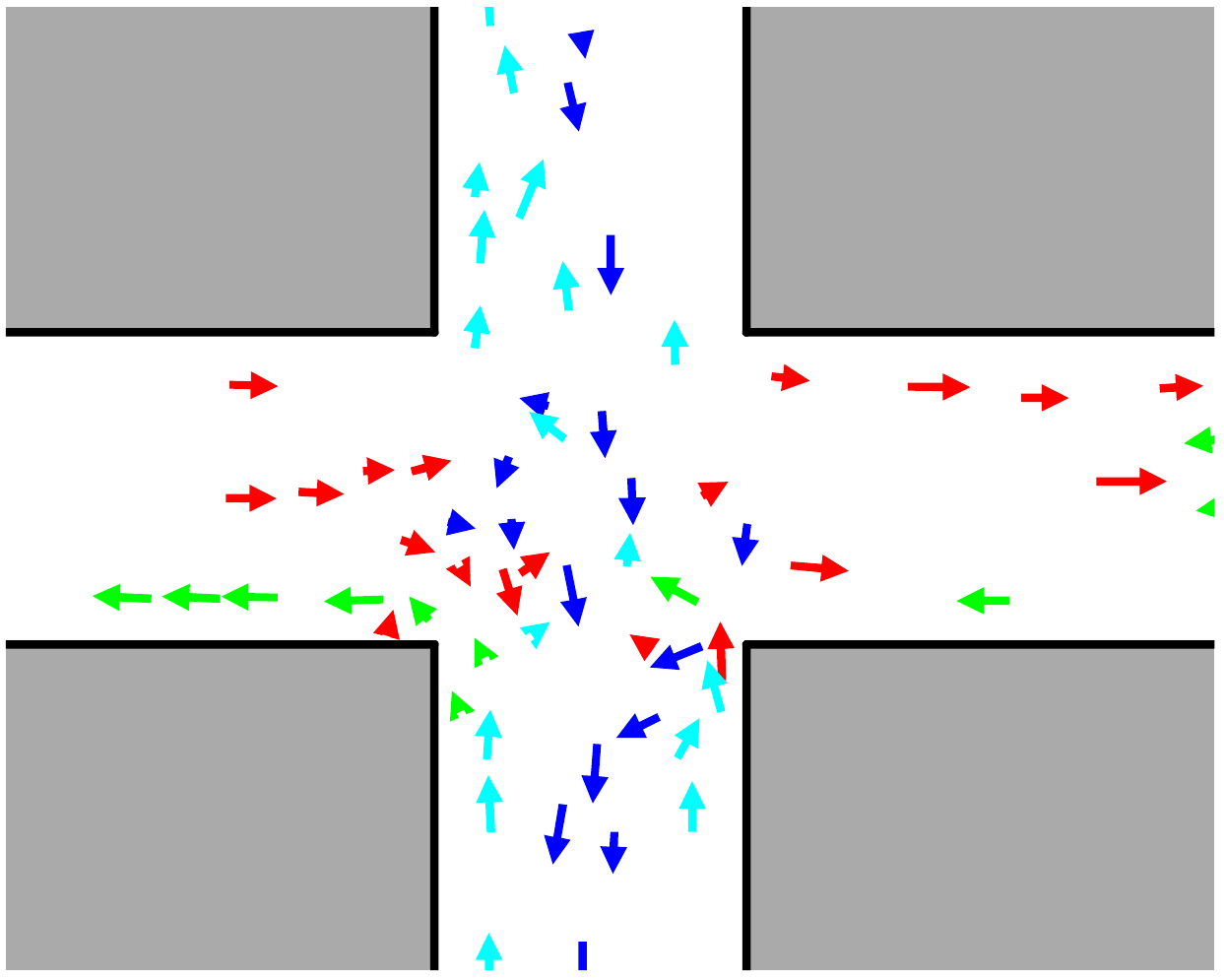}
    \\[0.5cm]
    \includegraphics[width=0.8\textwidth,clip]{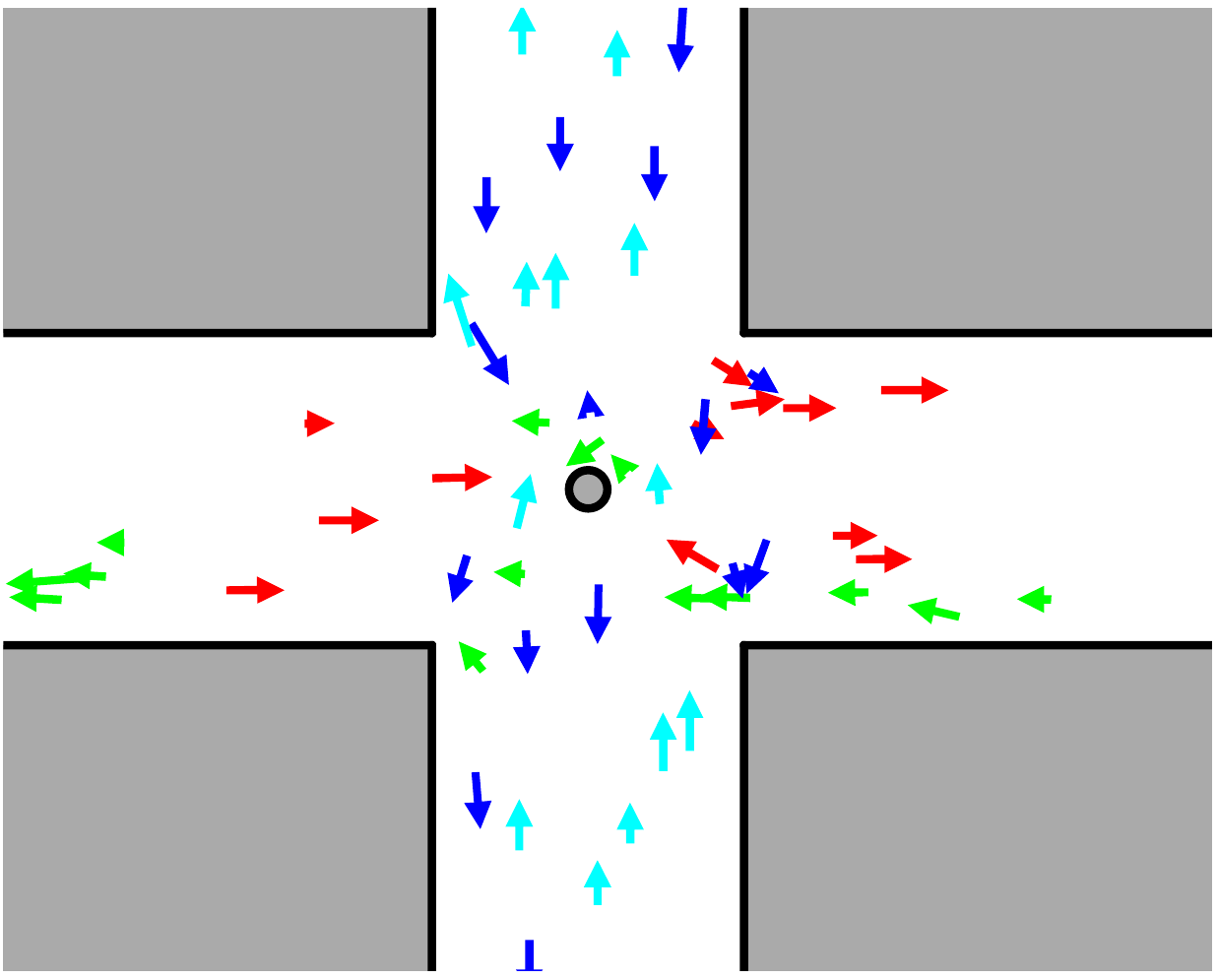}
  \end{center}
  \mycaption{Kreuzung mit Hindernis \farbe}
  {Die Leistungseigenschaften des {\fussgaenger}verkehrs auf Kreuzungen
    (oben) k"onnen durch ein zus"atzliches Hindernis (unten) erheblich
    verbessert werden.}
  \label{fig:kreuzung_mit_baum}
\end{figure}
{\fussgaengerstroeme} werden gerade auf Kreuzungen durch h"aufige
Ausweichman"over empfindlich gest"ort. Eine Methode zur Verbesserung der
Leistungsf"ahigkeit von Kreuzungen besteht darin, ein Hindernis in die Mitte
der Kreuzung zu setzen. Dadurch kann die Anzahl von Ausweichman"overn soweit
reduziert werden, da"s der durch die verkleinerte Verkehrsfl"ache
entstandene Nachteil nicht ins Gewicht f"allt.

In einer Simulation des {\fussgaenger}verkehrs auf Kreuzungen wurde dieser
Effekt deutlich:
F"ur die beiden in Abbildung \ref{fig:kreuzung_mit_baum} dargestellten
Anlagen wurden die Simulation mit gleicher Anzahl der {\fussgaenger}
sowie gleichen Parametern und Anfangsbedingungen durchgef"uhrt.

\begin{figure}[t]
  \begin{center}
    \leavevmode
    \includegraphics[width=0.9\textwidth,clip]{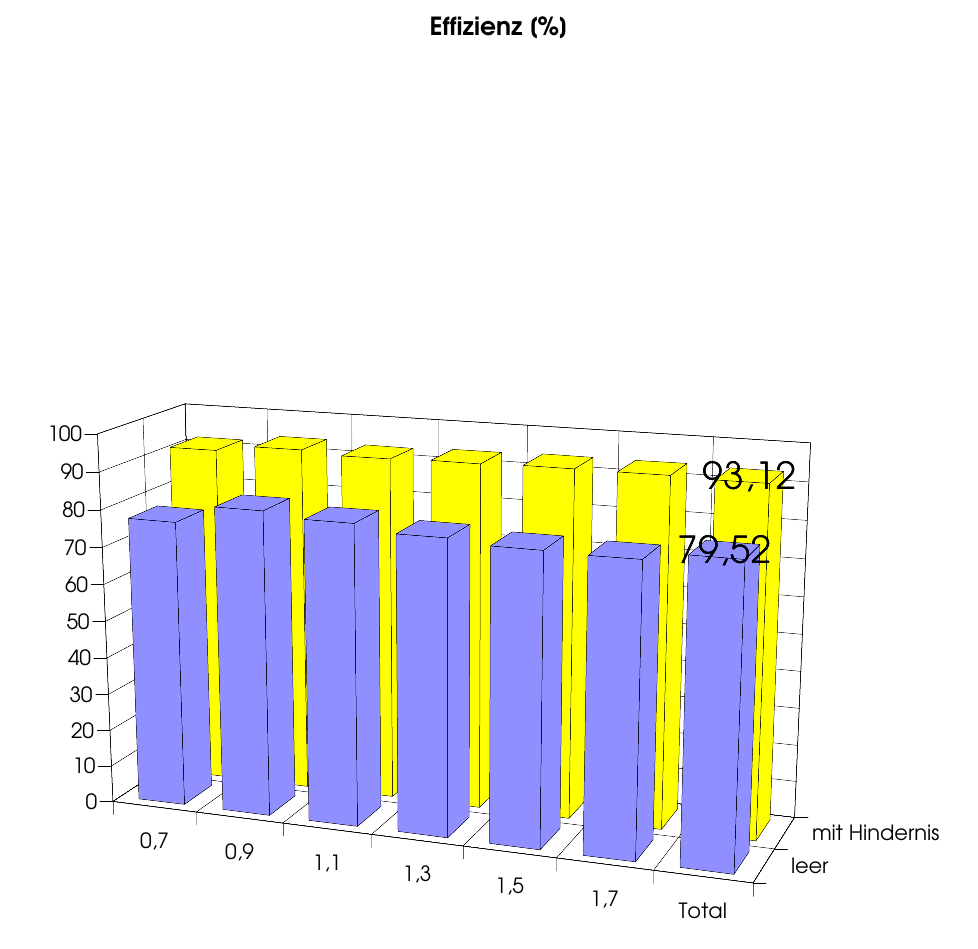}
  \end{center}
  \mycaption{Effizienzma"s einer Kreuzung \farbe}
  {Effizienzma"s $Y^1$ f"ur eine Kreuzung. Die Simulation wurde mit den
    gleichen Parametern f"ur die leere Kreuzung und die Kreuzung mit dem
    Hindernis durchgef"uhrt. Das Hindernis reduziert dabei die Zahl der
    Ausweichman"over von {\fussgaenger}n. Dadurch kann die Effizienz des
    Systems signifikant gesteigert werden.}
  \label{fig:effizienzcross}
\end{figure}

Abbildung \ref{fig:effizienzcross} zeigt die Ergebnisse der Simulationen einer
leeren Kreuzung und einer Kreuzung mit Hindernis. Neben der gesamten Effizienz
(Total) wurde das Leistungsma"s auch f"ur f"unf verschiedene
Geschwindigkeitsklassen ermittelt. Die Steigerung, die durch das Hindernis
erzielt wird, tritt klar hervor. Dagegen sind die "Anderungen in
der Verteilung "uber die einzelnen Geschwindigkeitsklassen weniger stark ausgepr"agt. 

\chapter[Formoptimierung]{Formoptimierung durch Evolution"are Programme}
\label{cha:formoptimierung}
Die Simulationen von {\fussgaenger}mengen zeigen, da"s die Str"ome sehr
empfindlich von der geometrischen Form des Geb"audes abh"angen. Durch die
Wechselwirkung zwischen den {\fussgaenger}n und ihren Reaktionen auf die
bauliche Umgebung k"onnen bereits kleinere gezielte Ver"anderungen am
Grundri"s eines Geb"audes den {\fussgaenger}verkehr signifikant
verbessern. Wie bereits am Beispiel einer Kreuzung in Abschnitt
\ref{sec:leistungsmasse} gezeigt wurde, 
l"a"st sich in manchen F"allen sogar durch verkleinern der begehbaren Fl"ache
eine Effizienzsteigerung erreichen. 

Die Optimierung des Grundrisses eine Anlage
l"a"st sich dabei mit Hilfe der Modellierung evolution"arer Prozesse
durchf"uhren.
Diese Methode wird seit den siebziger Jahren auf vielf"altige Weise zur
technischen Optimierung eingesetzt (vgl. \cite{Rechenberg:1973}). 

Dabei haben sich ausgehend von der Grundidee der Evolution zahlreiche
Methoden entwickelt, die in ihrer Ausf"uhrung zum Teil stark variieren.
Die historische Unterscheidung zwischen \hi{Genetischen Algorithmen} und
\hi{Evolutionsstrategien} ist durch ihre fast zeitgleiche Entstehung und die
gegenseitige Unkenntnis begr"undet \cite{Michalewicz:1994}.
Ihr Bestehen setzt sich bis in
die heutige Literatur fort. Der gro"se Fundus dieser Ans"atze und
ihrer Weiterentwicklungen bietet zahlreiche Komponenten, aus denen sich das
zum eigenen Problem passende Evolution"are Programm zusammensetzen l"a"st.

In den Abschnitten \ref{sec:genetischealgorithmen} und
\ref{sec:evolutionsstrategie} wird auf die urspr"unglichen Ans"atze beider
Methoden eingegangen. In Abschnitt \ref{sec:implementierung} wird dann die
Implementierung eines evolution"aren Programms vorgestellt, mit dem
verschiedene Methoden anhand von Beispielfunktionen untersucht wurden.

\section{Idee des Evolutionsprinzips}
\label{sec:ideederevolution}
In der Natur mu"s sich eine Population von Individuen im allt"aglichen
Leben behaupten. Nur die erfolgreichsten Individuen k"onnen
"uberleben und sich fortpflanzen. Durch diesen Selektionsproze"s werden die
in den Genen kodierten Eigenschaften der st"arksten Individuen
an die nachfolgende Generation weitergegeben.
Sofern sich die Anforderungen der Umwelt nicht "andern, enstehen dadurch immer
erfolgreichere Generationen. 

Das Kernst"uck ist die Reproduktion einer neuen Generation. Die
Erbinformation wird in leichter Variation an die Nachkommen
weitergegeben (\hi{Mutation}). Gibt es mehrere Vorfahren bezeichnet man den
Proze"s als \hi{mehrgliedrige Evolution}. Die neue Erbinformation kann dann
zus"atzlich durch die Kombination der Gene zweier Individuen erzeugt werden
(\hi{Crossover}).  Auf diese Weise profitieren die Nachkommen von mehreren
erfolgreichen Eigenschaften ihrer Eltern. 

In \hi{mathematischen und technischen Anwendungen} sind es  die \hi{potentiellen
  L"osungen} eines Problems, die dem \hi{Evolutionsprinzip} unterzogen
werden. Sie entsprechen bestimmten Parameters"atzen,
Startwerten oder Ger"ateeinstellungen. In der Terminologie der Evolution"aren
Methoden bezeichnet man diese L"osungen als \hi{Individuen}, die aber nicht
mit den ebenfalls als Individuen bezeichneten {\fussgaenger}n in Zusammenhang
stehen.

Die numerische Repr"asentation der Individuen erfolgt in Form von Genen, die 
aus {\Binaer}zahlen oder reellwertigen Vektoren eines mehrdimensionalen Raumes
bestehen k"onnen. Zus"atzlich werden Regeln zur Ver"anderung der Gene
definiert. 

Das Verfahren der evolution"aren Optimierung l"auft in den folgenden
Schritten ab: aus zuf"allig gew"ahlten Werten wird eine Ursprungspopulation
von potentiellen L"osungen des Problems erzeugt. Anhand des Erfolges eines
jeden Individuums werden 
durch Reproduktion und zuf"allige Variation die neuen Individuen der
nachfolgenden Generation geschaffen. Die Bewertung und Reproduktion werden
solange wiederholt, bis das Optimum gefunden wurde. Dazu bedarf es auch der
Definition geeigneter Abbruchkriterien.

Die folgenden f"unf Komponenten machen den Erfolg einer Optimierung mit
evolution"aren Methoden aus \cite[S.~17, 18]{Michalewicz:1994}: 
\begin{enumerate}
\item eine numerische Kodierung (Repr"asentation) der potentiellen L"osungen
  des Problems, 
\item eine Methode zur Erzeugung einer Anfangspopulation potentieller
  L"osungen,
\item eine M"oglichkeit die Fitness einer L"osung zu ermitteln
  (Bestimmungsfunktion),
\item Operatoren zur Ver"anderung und Kombination von Erbmaterial,
\item verschiedene Parameter, wie Populationsgr"o"se und H"aufigkeit der
  Anwendung von Mutationsoperatoren.
\end{enumerate}
Allein aufgrund dieser zahlreichen M"oglichkeiten zur Realisierung und
Durchf"uhrung von evolution"aren Optimierungsmethoden gibt es zur Anwendung 
kein
allgemeing"ultiges Rezept. Das zeigt auch die Optimierung der
Beispielfunktionen aus Abschnitt \ref{sec:beispiele}. Vielmehr gilt, je besser 
man das zu optimierende System experimentell oder theoretisch untersucht hat,
desto h"oher sind die Aussichten auf eine erfolgreiche Optimierung. 

\section[Evolution"are Optimierung von {\fussgaenger}anlagen]
{Evolution"are Optimierung \\ von {\fussgaenger}anlagen}
\label{evolutionvonfussgaengeranlagen}
\begin{figure}[t]
  \begin{center}
    \leavevmode
    \includegraphics[width=0.8\textwidth]{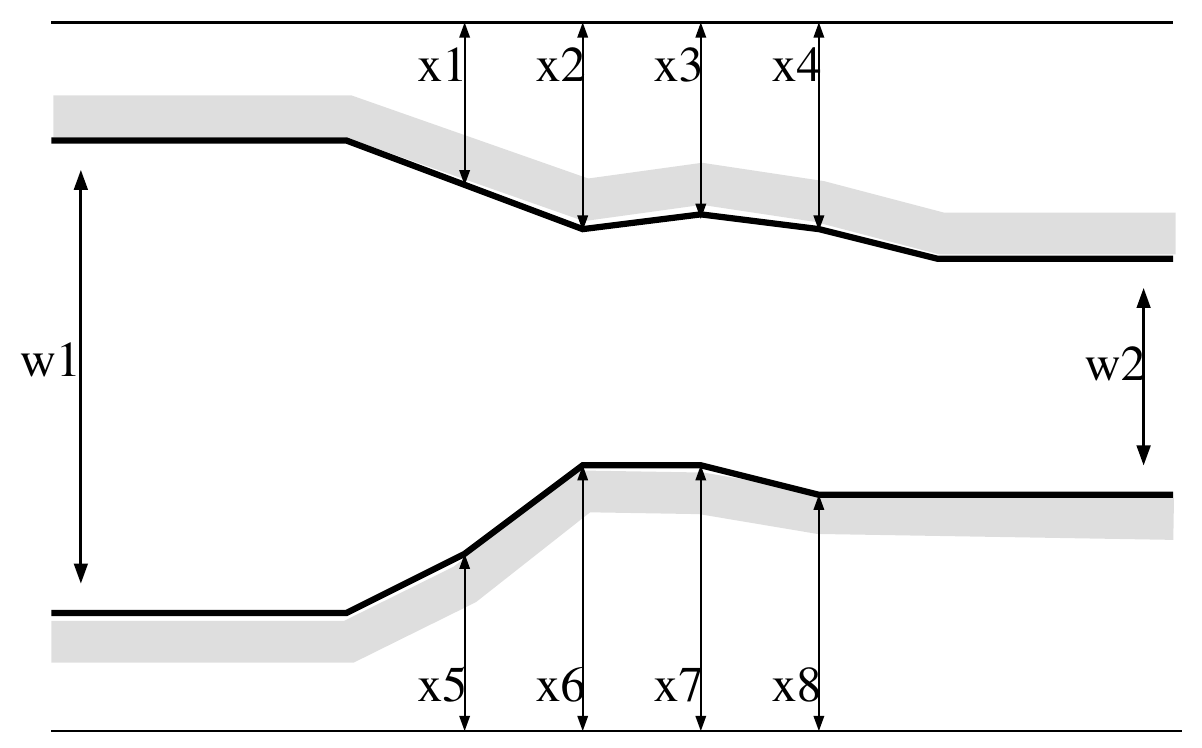} 
  \end{center}
  \mycaption{Variationsm"oglichkeiten eines Korridors}
  {Variationsm"oglichkeiten eines Korridors mit unterschiedlich
    breiten Enden $w1$\/ und $w2$.\/ Die Ausrichtung der einzelnen Wandst"ucke
    $x_1\dots x_8$\/ kann im Evolutionsproze"s optimiert werden.} 
  \label{fig:evo_korridor}
\end{figure}
\begin{figure}[t]
  \begin{center}
    \leavevmode
    \includegraphics[width=0.8\textwidth]{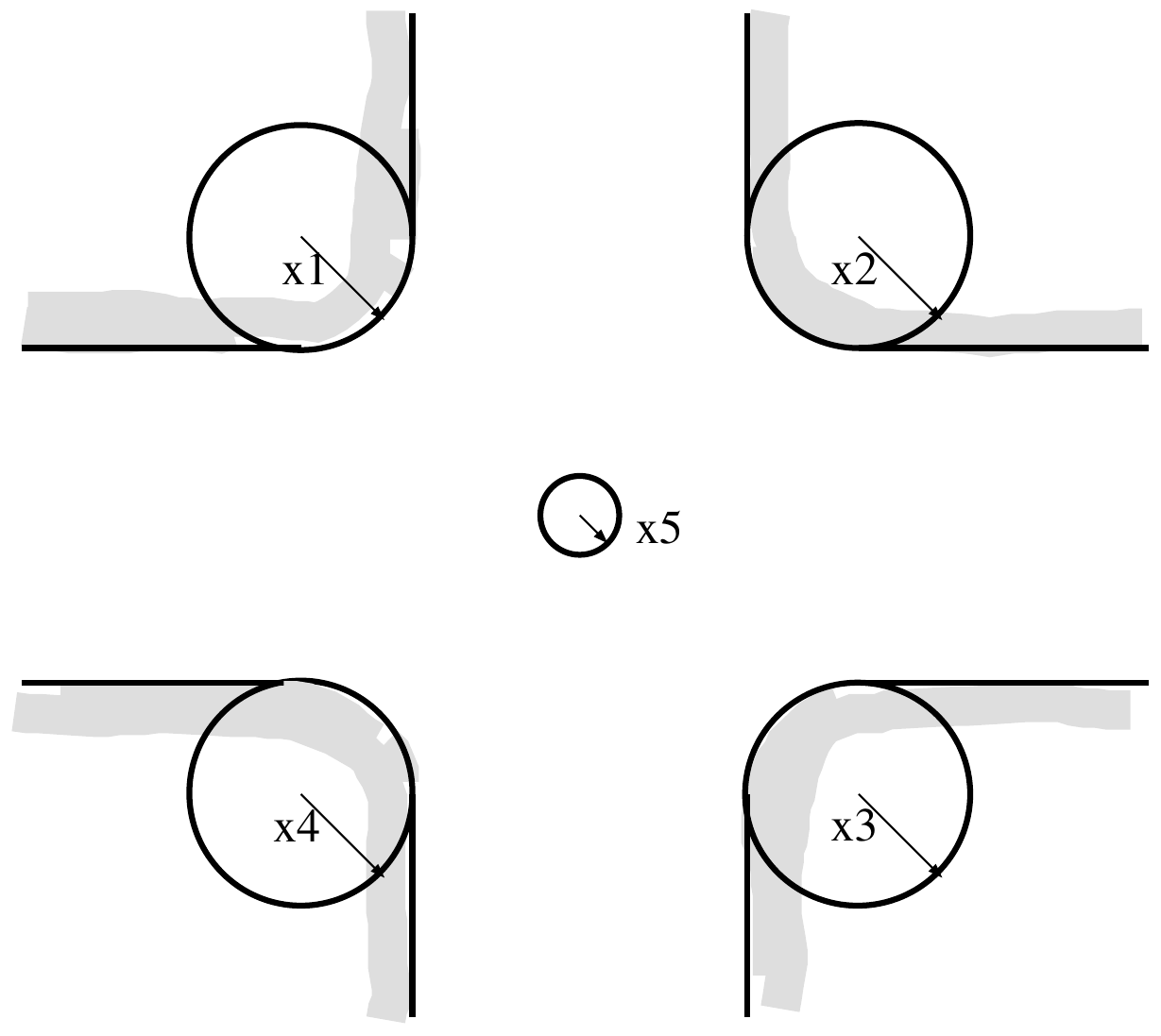}    
  \end{center}
  \mycaption{Variationsm"oglichkeiten einer Kreuzung}
  {Variationsm"oglichkeiten einer Kreuzung. Durch die Variation der
    Rundungen $x1\dots x4$\/ an den Ecken und dem Hindernis in der Mitte $x5$\/
    kann der {\fussgaenger}verkehr optimiert werden.}
  \label{fig:evo_cross}
\end{figure}
Mittels der Evolution"aren Optimierung lassen sich auch die Grundrisse von
{\fussgaenger}anlagen in Bezug auf die verschieden Bewertungkriterien
verbessern, die in Abschnitt \ref{sec:bewertungskriterien} eingef"uhrt wurden.
H"aufig steht dabei die Steigerung der Effizienz des Verkehrs im
Vordergrund.  Die Fitness eines Individuums wird dann durch das Effizienzma"s 
$Y1$ aus  (\ref{def:effizienz}) bestimmt. Zur Gestaltung von Verkaufsfl"achen
kann dagegen das Wohlbefinden der Kunden und deren
Aufmerksamkeit gegen"uber den angebotenen Waren als Optimierungskriterium
dienen. 

Die Abbildungen \ref{fig:evo_korridor} und \ref{fig:evo_cross} zeigen
Beispiele f"ur die Darstellung des Grundrisses einer {\fussgaenger}anlage.
Die W"ande eines Korridores lassen sich in verschiebbare Segmente aufteilen,
durch die der Querschnitt des Korridors variabel ist. Die Individuen des
Optimierungsprozesses setzen sich aus den Abst"anden $x1\dots x8$ zusammen.
Kreuzungen k"onnen durch unterschiedliche Radien $x1\dots x4$ an den Ecken
und der Gr"o"se eines Hindernisses $x5$ modifiziert werden.

Im Evolutionsproze"s wird zu jedem Individuum f"ur
eine bestimmte Dauer der {\fussgaenger}strom im zugeh"origen Grundri"s
simuliert. Es ist zu bemerken, da"s die  
daraus ermittelten Bewertungskriterien ausschlie"slich f"ur eine
bestimmte Zusammensetzung des {\fussgaenger}aufkommens gelten. Diese bleibt
f"ur den gesamten Optimierungsproze"s unver"andert.

Verschiedene {\fussgaenger}mengen (vgl. Abschn. \ref{sec:bewegungsverhalten})
in der Simulation k"onnen daher auch zu unterschiedlichen Ergebnissen der
Optimierung f"uhren.  Zur erfolgreichen Anwendung des
Evolutionsverfahrens auf Verkehrsfl"achen ist daher auch die Kenntnis "uber
den Zweck des Geb"audes und des erwarteten {\fussgaenger}aufkommens notwendig.  

\section{Genetische Algorithmen}
\label{sec:genetischealgorithmen}
Die Genetischen Algorithmen lehnen sich sehr stark an ihr biologisches
Vorbild. In Analogie zur \hi{DNS} (Desoxyribonukleins"aure), die aus vielen,
aneinandergereihten \hi{Aminos"auren} besteht, 
werden die Gene als vielstellige {\Binaer}zahlen dargestellt.
Die Interpretation dieser
Darstellung kann f"ur verschiedene Probleme sehr unterschiedlich
sein. Beispiele sind die Kodierung von unterschiedlichen Spielstrategien
\cite{Probst:1995}, der Reihenfolge von Punkten, die nach dem
Traveling-Salesman-Problem nacheinander besucht werden
sollen \cite{Michalewicz:1994} oder die numerische
Darstellung eines Wertes. 

\subsubsection{Kodierung}
Ein Wert $x$ aus dem Intervall $[-1,1]$ kann zum Beispiel durch
eine zehnstellige {\Binaer}zahl in der Form 
\begin{equation}
  x = 2\frac{s}{2^{10}}-1 = \frac{s}{2^9} - 1;
\end{equation}
repr"asentiert werden. Das Gen $s=(0101000111)$ steht dann f"ur $x = 327/512
-1 \approx -0.3613$. Bei mehrdimensionalen Problemen werden entsprechend viele
Bin"arzahlen aneinandergeh"angt.

\subsubsection{Ausgangspopulation}

Am Anfang des Evolutionsprozesses wird eine Population der Gr"o"se $\mu$ mit
zuf"allig bestimmten Individuen erzeugt. Die Erzeugung der ersten Generation
hat bereits Einflu"s auf die Geschwindigkeit des Evolutionsprozesses. Je besser
das zu optimierende Problem bekannt ist, desto dichter sollten die Individuen 
im Bereich um das vermutete Optimum gesetzt werden.   

\subsubsection{Bewertungsfunktion und Auswahl}

F"ur jedes Individuum wird die \hi{Fitness} ermittelt, das hei"st die
Qualit"at der L"osung des Problems. Dies kann durch eine
\hi{Bewertungsfunktion} $f:\{x\}\rightarrow {\rm I\!R}$, ein Experiment oder
eine Simulation geschehen. Der Fitnesswert bestimmt die
Wahrscheinlichkeit mit der ein Individuum in der {\naechsten} Epoche
reproduziert wird. Die Wahrscheinlichkeit l"a"st sich durch 
\begin{equation}
  p_k = \frac{f(x_k)^\tau}{\displaystyle \sum_{k^\prime=1}^\mu
    f(x_{k^\prime})^\tau} 
\end{equation}
mit dem Exponenten $\tau > 0$ definieren. 
Eine neue Generation ensteht aus der $\mu$-fachen Auswahl der Individuen der
Vorfahrengeneration anhand deren Reproduktionswahrscheinlichkeit $p_i$.
Dabei k"onnen besonders erfolgreiche Individuen auch mehrfach reproduziert
werden. Je gr"o"ser der Exponent $\tau$ gew"ahlt 
wird, desto st"arker setzen sich die erfolgreichen L"osungen durch, und umso
schneller kann der Evolutionsproze"s ablaufen. Kleinere Werte f"ur $\tau$
lassen auch weniger erfolgreiche Individuen zur Reproduktion zu und
verhindern damit, da"s der Proze"s in einem lokalen Optimum stecken bleibt.

\subsubsection{Genetische Operatoren}
In jeder Generation k"onnen die Gene  mit einer gewissen Wahrscheinlichkeit
$p$ ver"andert werden. Die Ver"anderung kann dabei auf zwei Arten geschehen:
\begin{itemize}
\item Mutation. Eine zuf"allig ausgew"ahlte Stelle der Bin"arzahl wird
  ver"andert. Hatte sie vorher den Wert 1, dann wird sie auf 0 gesetzt und
  umgekehrt.
  Bei einer numerischen Interpretation der Gene f"uhrt das Umdrehen einer
  h"oherwertigen Stelle zu einer gr"o"seren Betrags"anderung.
\item Crossover. Zwei Gene werden an einer zuf"allig bestimmten Stelle
  geteilt. Dann wird das Anfangsst"uck des ersten Gens mit dem Endst"uck des
  zweiten zusammengesetzt. Mit den beiden "ubrigen St"ucken wird genauso
  verfahren.  Es entstehen dadurch zwei neue, unterschiedliche
  Gene, die Eigenschaften von beiden Vorfahren "ubernommen haben. 
\end{itemize}
Je nach Kodierung fallen manchmal Gene der neuen Generation aus dem g"ultigen
Wertebereich heraus. L"a"st sich das durch eine geeignete Definition des
Mutations- und Crossoveroperators nicht verhindern, so kann eine
Instanz eingef"uhrt werden, die die ung"ultigen Gene
repariert\index{Genreperatur}.
In Bereichen der numerischen Optimierung werden h"aufig auch Strafterme
\index{Strafterm} in die Bewertungsfunktion einf"ugt. Diese sorgen daf"ur,
da"s ung"ultige Gene keinen Erfolg haben.
F"ur sehr (zeit-)aufwendige Bewertungfunktionen sind Strafterme jedoch nicht
geeignet. 

\section{Evolutionsstrategie}
\label{sec:evolutionsstrategie}
Der Ansatz der Evolutionsstrategie stammt aus dem Bereich der numerischen
Optimierung. Das erste Verfahren war ein Experiment von
\name{Rechenberg}{Ingo} und \name{Schwefel}{}, bei dem mit einer Gelenkplatte
im Windkanal das str"omungs"|g"unstigste Profil gesucht wurde.  Die Variablen
des Problems waren die Winkel, in dem die angrenzenden Platten zueinander
standen. Zuf"alliges Ver"andern einer Gr"o"se und das Verwerfen aller
L"osungen, die schlechter als ihre Vorfahren waren f"uhrt nach zahlreichen
Schritten zur optimalen L"osung \cite{Rechenberg:1973}. 

Anfangs bestanden Evolutionsstrategien aus Populationen mit genau einem
Individuum. Der einzige genetische Operator war die Mutation. Anders als bei
den Genetischen Algorithmen wird ein Individuum als Vektorenpaar
$(x, \sigma) \in {\rm I\!R^n \times I\!R^n}$ dargestellt.
Der Vektor $x$ gibt dabei die Position im L"osungsraum an, $\sigma$ 
ist ein Vektor aus Standardabweichungen, mit denen die
Mutation durch
\begin{equation}
  x^\prime = x + \N(0,\sigma)
\end{equation}
realisiert werden. Dabei ist $\N(0,\sigma)$ ein Vektor aus unabh"angigen,
normalverteilten Zufallszahlen mit dem Mittelwert $0$ und der Standardabweichung
$\sigma$. Dieser Ansatz entspricht der Beobachtung aus der Biologie,
da"s kleinere "Anderungen wesentlich h"aufiger als gro"se auftreten.
Ist das neugeschaffene Individuum erfolgreicher, das hei"st n"aher am Optimum,
ersetzt es seinen Vorfahren. Im anderen Fall wird es verworfen.      

\subsubsection{Konvergenz und Schrittweitensteuerung}

Der Beweis, da"s dieses Verfahren f"ur regul"are Optimierungsprobleme
konvergiert,  wurde in \cite{BaeckHoffmeisterSchwefel:1991} erbracht.
Allerdings l"a"st sich dabei nichts "uber die Konvergenzgeschwindigkeit
aussagen. Um den Evolutionsproze"s zu beschleunigen, kann man eine
Schrittweitensteuerung einf"uhren: Die Standardabweichung $\sigma$ pa"st sich 
in Abh"angigkeit des Erfolges der letzten Evolutionsschritte an.
\name{Rechenberg}{Ingo} f"uhrt dazu die Erfolgswahrscheinlichkeit
\begin{equation}
  W = \frac{\mx{Zahl der erfolgreichen Mutationsschritte}}
  {\mx{Gesamtzahl der Mutationsschritte}}
\end{equation}
und sogenannte $\frac{1}{5}$-Erfolgsregel ein \cite{Rechenberg:1973} \cite[S.~
128--132]{Schwefel:1977}: 
F"ur jeweils $k$ Evolutionsschritte wird die
mittlere Erfolgswahrscheinlichkeit $W$ bestimmt und der
Standardabweichungsvektor gem"a"s
\begin{equation}
  \sigma^\prime = \left\{
  \begin{array}{r@{\quad:\quad}l}
    c_d\, \sigma & W < 1/5 \\
    \sigma & W = 1/5 \\
    c_i\, \sigma & W > 1/5 \\ 
  \end{array}
\right.
\end{equation}
mit $c_d < 1$ und $c_i > 1$ ver"andert. In \cite{Schwefel:1977}
werden die Werte $c_d = 0.85$, $c_i = 1/0.85$ und $k = 10$ vorgeschlagen.

\subsubsection{Mehrgliedrige Evolutionsstrategien}
Bei einem Anfangsindividuum und einem Nachkommen spricht man von
einer \hi{zweigliedrigen Evolutionsstrategie}.
Gr"o"sere Populationen werden dann \hi{mehrgliedrige Evolutionsstrategie}
genannt. 
Allgemein werden die Populationsgr"o"se mit $\mu$ und die Zahl der Nachkommen
mit $\lambda$ angegeben. Die Auswahl kann dann aus $\mupluslam$ Individuen
getroffen werden. Neben dieser $\mupluslam$-Methode kann man die Auswahl
auch auf die $\lambda$ Nachkommen beschr"anken,
$(\mucommalam)$-Evolutionsstrategie. Erfolgreiche Individuen weisen dadurch
eine l"angere Lebensdauer auf. Produziert eine Generation keine Individuen,
die erfolgreicher sind als ihre Vorfahren, bleibt die bis dahin erreichte
Qualit"at der Evolution bei
der $\mupluslam$-Strategie erhalten, w"ahrend sich bei $\mucommalam$ das
Ergebnis wieder verschlechtern kann.

\subsubsection{Lernende Evolutionsstrategien}
Bei $\mupluslam$ und $\mucommalam$ Strategien kann man die Regelung der
Schrittweiten statt durch deterministische Algorithmen, wie etwa der
$\frac{1}{5}$-Erfolgsregel, auch durch das Einbeziehen der
Standardabweichungsvektoren $\sigma$ in den Evolutionsproze"s erreichen
\cite{Michalewicz:1994}.

Die genetischen Operatoren werden dabei auf beide Teile des Vektorpaares
$(x,\sigma)$ angewendet. Das Verfahren arbeitet in mehreren Stufen:
Zuerst wird aus zwei Individuen 
\begin{eqnarray}
  (x^1,\sigma^1) & = & ((x^1_1,\dots,x^1_n),(\sigma^1_1,\dots,\sigma^1_n))
  \nonumber \\
  (x^2,\sigma^2) & = & ((x^2_1,\dots,x^2_n),(\sigma^2_1,\dots,\sigma^2_n))
\end{eqnarray}
durch Vermischen ein neues
\begin{equation}
  (x^\prime,\sigma^\prime) = ((x^{q_1}_1,\dots,x^{q_n}_n),
    (\sigma^{q_1}_1,\dots,\sigma^{q_n}_n)) 
    \label{evo_crossover}
  \end{equation}
  mit zuf"allig gew"ahltem $q_i = 1,2$ f"ur alle $i=1\dots n$ erzeugt.
  Alternativ k"onnen die Nachkommen auch durch eine arithmetische Mittelung
  der  Eltern 
  \begin{equation}
    (x^\prime,\sigma^\prime) = (( (x^1_1+x^2_1)/2, \dots, (x^1_n+x^2_n)/2 ),
    ((\sigma^1_1+\sigma^2_1)/2, \dots, (\sigma^1_n+\sigma^2_n)/2 ) )
  \end{equation}
  entstehen. Auf das Produkt dieser Crossover-Operation wird dann die Mutation 
  $(x^\prime,\sigma^\prime)$ mit 
  \begin{eqnarray}
    \sigma^{\prime\prime} & = & \sigma^\prime\, e^{N(0,\Delta\sigma)} 
    \nonumber \\ 
    \mbox{und}\qquad x^{\prime\prime} & = & x + N(0,\sigma^{\prime\prime})
    \label{evo_mutation}
  \end{eqnarray}
  angewendet. $\Delta\sigma$ ist dabei ein Steuerungs"|parameter des Verfahrens.

\section[Untersuchung von GA und ES]{Untersuchung von Genetische Algorithmen und Evolution"are Strategien}
\label{sec:anderemethoden}
\label{sec:Untersuchung}
Sowohl Genetische Algorithmen, als auch Evolution"are Strategien verwenden
vorhandene L"osungen und erzeugen daraus neue potentielle 
L"osungen, die sich gegeneinander und gegen ihren Vorfahren behaupten m"ussen. 
W"ahrend bei klassischen Genetischen Algorithmen die Individuen durch
{\Binaer}zahlen dargestellt werden, existieren auch Ans"atze, die mit
reellwertigen 
Vektoren arbeiten und dadurch f"ur numerische Optimierungen ebenso geeignet
sind, wie die Evolution"aren Strategien. Die bedeutendsten Unterschiede sind
in Tabelle \ref{ga_es_unterschiede} zusammengefa"st.
\begin{table}[htbp]
  \begin{center}
    \begin{tabularx}{0.99\textwidth}{|l|X|X|}
      \hline {\bf Kriterium} & {\bf Genetische Algorithmen} & {\bf Evolution"aren Strategien}
      \\ \hline \hline
      Kodierung & Darstellung der Individuen durch {\Binaer}zahlen 
      & Darstellung durch reellwertige Vektoren \\ \hline
      Generation & F"ur den n"achste Generation werden $\mu$
      Individuen ausgew"ahlt. Dabei haben starke Individuen eine gute
      Chance mehrfach ausgew"ahlt zu werden. Andererseits bleibt auch f"ur
      die Schw"achsten eine gewisse Wahrscheinlichkeit ausgew"ahlt zu
      werden 
      & Es wird eine tempor"are
      Generation mit $\lambda$ bzw. $\mu+\lambda$ Individuen erzeugt. Die
      Population wird durch Entfernen der Schw"achsten wieder auf $\mu$
      Individuen reduziert. \\ \hline
      Auswahl & Die Individuen werden zuerst ausgew"ahlt
      und dann den Ver"anderungs"|operatoren unterzogen 
      & Durch Crossover- und Mutations"|operatoren
      wird zuerst eine tempor"are Generation der Gr"o"se $\lambda$ geschaffen,
      und dann werden die St"arksten ausgew"ahlt. \\ \hline
      Steuerung & Das Optimierungsverfahren wird durch die
      Mutations"|wahrscheinlichkeit $p_m$ und die
      Crossoverwahrscheinlichkeit $p_c$ gesteuert. Sie bleiben w"ahrend
      des ganzen Prozesses konstant.
      & Die
      Schrittweite des Mutationsoperators richtet sich nach dem
      $\sigma$-Vektor, der ebenfalls dem Evolutionsproze"s unterliegen kann
      \\ \hline
      Ung"ultige Individuen & Wird durch die Reproduktion ein Individuum
      erzeugt, das die Randbedingungen des Problems nicht erf"ullt, kann
      es entweder repariert werden, oder durch Strafterme in der
      Bewertungsfunktion zum Ausscheiden gebracht werden
      & Ung"ultige Individuen der tempor"aren
      Generation werden verworfen. Erzeugen die Reproduktions"|operatoren zu
      viele ung"ultige Individuen, m"ussen die Steuerungsparameter
      angepa"st werden. \\ \hline
    \end{tabularx}
  \end{center}
  \mycaption{Unterschiede zwischen GA und ES}
  {Unterschiede zwischen den klassischen Ans"atzen
    von Genetischen Algorithmen und
    Evolution"aren Strategien}
  \label{ga_es_unterschiede}
\end{table}

Beide Verfahren wurden in den letzten zwei Jahrzehnten weiterentwickelt, und
"ubernahmen auch Ideen aus dem jeweilig anderen Ansatz. Jede Anwendung der
Evolution"aren Optimierung erfordert eine besondere Anpassung an die
Problemstellung. Bei den heutigen Entwicklungen verwischt die Grenze
zwischen Genetischen Algorithmen und Evolution"aren Strategien immer mehr.

F"ur die Optimierung der {\fussgaenger}anlagen und die Strategieoptimierung
in den Modellerweiterungen (Kap.\ \ref{cha:erweiterungen}) wurde ein
Evolutionsprogramm implementiert, das auf den Ans"atzen der vorangehenden
Abschnitte aufbaut. 
Im folgenden werden damit die Eigenschaften der einzelnen Methoden an zwei
Beispielfunktionen untersucht.

\subsection[Implementierung der Evolution"aren Optimierung]{Implementierung der
  Evolution"aren Optimierung}
\label{sec:implementierung}
Die im vorigen Abschnitt behandelten Darstellungsform der
{\fussgaenger}anlagen legt auch bei der Implementierung der Genetischen
Algorithmen die Verwendung  von n-dimensio"-nalen, reellwertigen 
Vektoren $x \in {\rm I\!R^n}$ nahe. Der G"ultigkeitsbereich jeder 
Komponente $x_i$ ist dabei
durch ein eigenes Intervall $u_i \le x_i \le l_i$ festgelegt.

Die Individuen einer Anfangspopulation k"onnen auf zwei Arten erzeugt werden:
\begin{enumerate}
\item Gleichm"a"sige Verteilung "uber den gesamten Wertebereich. Dabei wird
  jeder Komponente 
  \begin{equation}
    x_i = u_i + (l_i-u_i)\,\Z 
  \end{equation}
  eine Zufallszahl aus dem Wertebereich zugeordnet.
\item Randwertverteilung. Eine Komponente 
  \begin{equation}
    x_i = \left\{\begin{array}{r@{\quad:\quad}l}
      u_i & \Z \le 0.5 \\
      l_i & \Z > 0.5 
    \end{array}\right.
  \end{equation}
  nimmt entweder den kleinsten oder den gr"o"sten Wert des Intervalls an.
\end{enumerate}
Die Zufallszahl $\Z$ hat dabei eine uniforme Verteilung "uber das Intervall
$[0\dots1]$.
Die Erzeugung Anfangspopulationen der Beispielrechnugnen verwendet beide Arten
zu jeweils 50\%.

Die Reproduktion einer neuen Generation von Individuen beinhaltet die Auswahl
der Eltern nach ihrer Fitness und die Erzeugung der Nachkommen. Hierf"ur
wurden vier unterschiedliche Verfahren implementiert, die im folgenden mit
Standardversion und Plus-Version des Genetischen Algorithmus und
$\mucommalam$- und $\mupluslam$-Evolutionsstrategie bezeichnet werden.

Die Auswahl eines Individuums $k$ geschieht nach einer bestimmten
Wahrscheinlichkeit  
\begin{equation}
  p_k = \frac{f(x_k)^\tau}{\displaystyle \sum_{k^\prime=1}^\mu f(x_{k^\prime})^\tau} 
  \label{def:evowahrsch}
\end{equation}
in Abh"angigkeit von seiner  Fitness $f(x_k)$ und dem Exponenten $\tau$.
Dabei kann ein Individuum auch mehrfach ausgew"ahlt werden.
In der Plus-Version des Genetischen Algorithmus wird die Auswahl aus zwei
aufeinander folgenden Generationen getroffen. Analog dazu "uberleben in der 
$\mupluslam$-Evolutionsstrategie die $\mu$ besten Individuen aus der
Gesamtheit der Vorfahrenpopulation und der Zwischengeneration.  

Der Crossoveroperator in der Implementation der Genetisch Algorithmen bestimmt
einen zuf"alligen Index $i_c$, an dem zwei Gene $x$ und $y$ geteilt
werden. Die neuen Individuen 
\begin{eqnarray}
  \label{genet_crossover}
  x^\prime  & = & \left(x_1,\dots,x_{i_c},y_{i_c+1},\dots,y_n\right) \nonumber \\
  y^\prime  & = & \left(y_1,\dots,y_{i_c},x_{i_c+1},\dots,x_n\right)  
\end{eqnarray}
setzen sich aus den vertauschten Teilst"ucken zusammen.
Die Wahrscheinlichkeit $p_c$, mit welcher der Crossover-Operator auf zwei Gene
angewendet wird, bleibt w"ahrend des Evolutionsprozesses konstant.

Danach wird mit einer Wahrscheinlichkeit $p_m$ eine Mutation an den Genen
durchgef"uhrt. Dazu bestimmt der Mutationsoperator einen zuf"alligen Index
$i_m$, an dem die Komponente einen Zufallswert zugewiesen bekommt:
\begin{equation}
  \label{genet_mutation}
  x^{\prime\prime} =  \left(x^\prime_1,\dots, x^\prime_{i_m-1},
  u_{i_m} + (l_{i_m} - u_{i_m})\,\Z, x^\prime_{i_m+1}, \dots, x^\prime_n\right)
\end{equation}
Auch $p_m$ bleibt w"ahrend des Evolutionsprozesses unver"andert.

In der Implementation der Evolutionsstrategien besteht das Individuum aus
einem Vektorenpaar $(x,\sigma)$. Die Reproduktion  erfolgt gem"a"s der lernenden
Strategien aus Abschnitt \ref{sec:evolutionsstrategie} mit dem
Crossover-Operator nach (\ref{evo_crossover}) und dem Mutationsoperator nach
(\ref{evo_mutation}). Im Gegensatz zu den Genetischen Algorithmen wenden die
Evolutionsstrategien die Operatoren auf alle Individuen an.

\subsection[Beispiele zur Evolution"aren Optimierung]
{Beispiele zur Evolution"aren Optimierung}
\label{sec:bespielfuerevolution}
\label{sec:beispiele}
\begin{figure}[tb]
  \begin{center}
    \leavevmode
    \includegraphics[width=0.8\textwidth]{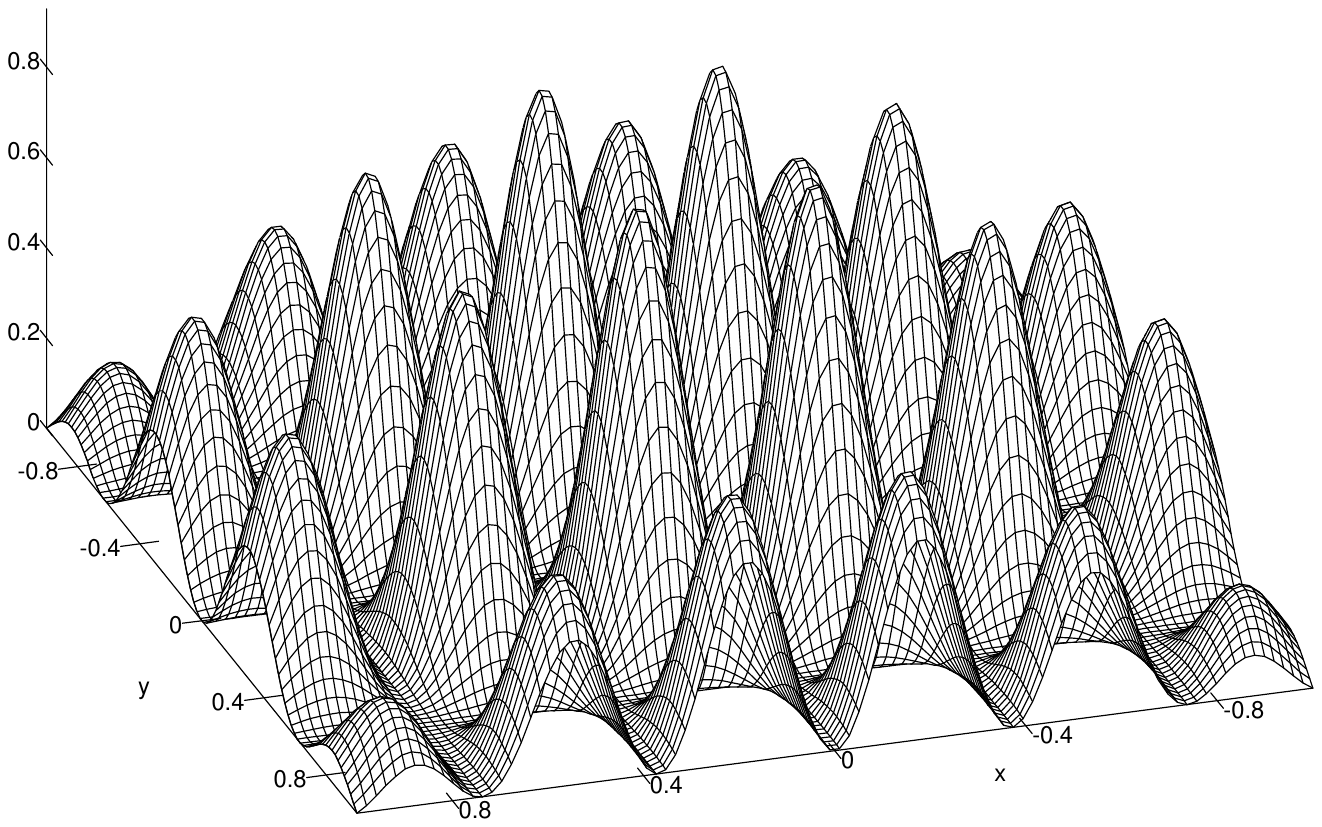}
  \end{center}
  \mycaption{Bewertungsfunktion zur Untersuchung der Evolutionsmethoden}
  {Bewertungsfunktion aus
    (\ref{beispiel_g}), deren Maximum
    gefunden werden soll. Die Darstellung gibt einen zweidimensionalen Schnitt
    mit ${\rm x}=x_1$ und ${\rm y}=x_2$ f"ur $x_3 = x_4 = x_5 = 0$. wieder}
  \label{fig:beispiel_g}
\end{figure}

Anhand der beiden Testfunktionen 
\begin{equation}
  \label{beispiel_f}
  f(x) = \frac{1}{5}(5 - \sq{x_1} - \sq{x_2} - \sq{x_3} - \sq{x_4} - \sq{x_5}) 
\end{equation}
und 
\begin{eqnarray}
  \label{beispiel_g}
  g(x_1,x_2,x_3,x_4,x_5) & 
  = & \frac{1}{5}\left(5 - \sum_{i=1}^5 \sq{x_i}\right) 
  \cdot \frac{1}{2}\left(\cos\left(17x_1\right)+1\right)  \\ \nonumber 
  & \cdot & \frac{1}{2}\left(\cos\left(12x_2\right)+1\right)
  \cdot\frac{1}{2}\left(\cos\left(23x_3\right)+1\right) \\ \nonumber 
  & \cdot & \frac{1}{2}\left(\cos\left(6x_4\right)+1\right) 
  \cdot \frac{1}{2}\left(\cos\left(15x_5\right)+1\right)
\end{eqnarray}
wurde die Implementierung der Evolution"aren Optimierung untersucht. Beide
Funktionen haben ihr Maximum an der Stelle $x_i = 0\mbox{, f"ur }i=1\dots5$,
mit dem Maximalwert 1. Die Funktion $g$ birgt dar"uberhinaus noch mehrere
Nebenmaxima, die die Optimierung erschweren (vgl.\ Abb.\ \ref{fig:beispiel_g}).

Die Optimierung wurde f"ur die unterschiedlichen Versionen mit
verschiedenen Parametern durchgef"uhrt
und die Zahl der insgesamt getesteten Individuen bestimmt, die bis zum
Erreichen von 95\% der maximalen Fitness ben"otigt wurden. 
Sp"atestens nach zwanzig Generationen wurde der Proze"s abgebrochen.

Zur Untersuchung der Genetischen Algorithmen wurden die
Operatorwahrscheinlichkeiten $p_c$ und $p_m$ in $0.1$ Schritten von $0$ bis $1$
variiert. Die anderen Parameter bekamen die Werte $\mu = 2,4,10,10$ und
$\tau=0.5,1,2$ zugewiesen.

Bei der Optimierung der parabolischen Funktion $f$ ben"otigten
sowohl die Standardversion, als auch die Plus-Version 
mit vielen Parameterkombinationen nicht mehr als zwanzig Individuen.
Die Plus-Version war etwas erfolgreicher, weil sie alte L"osungen, die besser
sind als die neuen, beibehielt. Diesen Vorteil konnte sie gegen die
Standardversion, die dagegen unempfindlicher f"ur  lokale Optima ist,
bei der einfachen Testfunktion $f$ mit nur einem Maximum ausspielen.

Bei der Optimierung von $g$ 
brauchte die Standardversion mindestens 3400 Individuen.
Die Plusversion konnte im besten Fall bereits mit
zweihundert Individuen einen Erfolg aufweisen. 
Auf\/f"allig war dabei, da"s die Wahrscheinlichkeit zur Anwendung der
Crossover-Operation $p_c$ f"ur alle Parameterkombinationen, die die 95\%
Grenze erreichten 
verschwindet.
Damit wurde der Vorzug der Evolution"are Programme gegen"uber anderen
numerischen Methoden, Teilst"ucke von erfolgreichen L"osungen weiter zu verwenden und unter den
L"osungen auszutauschen,  nicht eingesetzt.

Zur Untersuchung der Evolutionsstrategien $\mucommalam$ und $\mupluslam$
wurde versucht, die beiden Testfunktionen $f$ und $g$ mit den folgenden
Parametern zu optimieren:
$\mu  = 1, 5, 10, 20, 100$, $\lambda = 1, 5, 10, 20, 100$, $\tau = 0,
1, 2$ und $\Delta\sigma = 0.1, 0.5, 1, 2$. 
Die Erfolge glichen denen der Genetischen Algorithmen.
Bei der Optimierung der Funktion $g$ ben"otigte die $\mucommalam$-Strategie
eine gro"se Nachkommengeneration $\lambda$, weil h"aufiger ung"ultige
Individuen produziert wurden. Auch hier
zeigte sich, da"s die $\mupluslam$-Strategie der $\mucommalam$ bei diesem
Problem "uberlegen war.

Insgesamt ergaben die Optimierungsversuche der beiden Beispielfunktionen
jedoch keine aussagekr"aftigen Empfehlungen f"ur bestimmte
Parameterkombinationen. Beide Richtungen der Evolution"aren
Optimierung, die Genetischen Algorithmen und die Evolutionsstrategien,
arbeiteten gleicherma"sen erfolgreich. 

Da der zeitaufwendigste Teil des Optimierungsprozesses von
{\fussgaenger}anlagen die Bestimmung der 
Fitness durch die Simulation des Verkehrs ist, sollte ein 
Evolutionsverfahren gew"ahlt werden, das mit relativ wenigen Individuen
auskommt. Diesbez"uglich scheinen die Plus-Version des Genetischen Algorithmus
und die $\mupluslam$-Evolutionsstrategie gegen"uber den anderen
beiden Implementationen erfolgversprechender.

\chapter{Erweiterungen des Soziale-Kr"afte-Modells}
\label{cha:erweiterungen}
Das Verhalten von {\fussgaenger}n kann in verschiedenen Ebenen
unterschiedlicher Komplexit"at betrachtet werden. Die Gehbewegung, durch die die
{\fussgaenger} zu ihrem Ziel gelangen, legt dabei der Grundstock zu weiteren
Modellierungsebenen. 

"Uber die Ebene des Soziale-Kr"afte-Modells der {\fussgaenger}dynamik l"a"st
sich etwa das Entscheidungsmodell der Zielwahl, das im folgenden
Abschnitt vorgestellt wird, ansiedeln. 

Die Ebenen stehen in gegenseitiger Abh"angigkeit ihrer Modellzust"ande: So
wird die Entscheidung f"ur eine Zielrichtung durch die Gehbewegung
ausgel"ost. Das Ergebnis der Entscheidungsfindung gibt wiederum die neue
Zielrichtung vor.

Mikroskopische Modelle, wie das der {\fussgaenger}, k"onnen mit nahezu
beliebig vielen komplexen Verhaltensregeln ausgebaut werden. Auf diese
sogenannten Multi-Agent-Modelle soll hier aber nicht weiter eingegangen
werden.

\section{Entscheidungsmodell}
\label{sec:entscheidungsmodell}
Wie bereits in Abschnitt \ref{sec:doppeltuer} erw"ahnt wurde, l"a"st sich das
Soziale-Kr"afte-Modell mit einem Entscheidungsmodell erweitern, das
den {\fussgaenger}n die Wahl zwischen  alternativen Durchg"angen erlaubt.

Der Entscheidungs-Findungs-Proze"s kann dabei durch den Zeitpunkt oder -raum
spezifiziert werden:
\begin{itemize}
\item Spontane Entscheidungen werden zum Zeitpunkt des Auftretens der
  Alternativen getroffen, etwa an einem Eingangsbreich mit mehreren T"uren.
  Rufen sie keine Folgeentscheidungen hervor, wird die Auswahl einer
  Alternative allein anhand der momentanen Situation getroffen. Dies ist
  im Beispiel des doppelten Durchgangs der Fall: die {\fussgaenger}
  entscheiden sich f"ur 
  einen Durchgang, um in den n"achsten Raum zu gelangen. Der Fortsetzung
  ihres Weges 
  ist dabei unabh"angig von der vorher getroffenen Entscheidung.
  Im {\fussgaenger}modell wird die Entscheidungsfindung nach dem unten
  beschriebenen Multi-Nomial-Logit-Ansatz  realisiert.   
\item Steht die (ann"ahernd) vollst"andige Information "uber das System
  zu Beginn des Weges zur Verf"ugung, kann die Route bereits im voraus
  bestimmt werden. 
  In Kapitel \ref{cha:wegenetze} wird hierzu eine Methode vorgestellt, um
  die Streckenbelastungen eines Wegesystems festzustellen, die sich durch
  die Routenwahl der {\fussgaenger} ergeben.
\item Regelm"a"sig gelaufene Routen werden in der Regel ebenfalls zu Beginn
  des Weges ausgew"ahlt. Der Verlauf der Entscheidungsfindung setzt sich 
  dabei "uber die mehrmalige Benutzung der Wegstrecke fort. Die Routenwahl
  basiert auf den  Erfahrungswerte fr"uherer Entscheidungen. 
  Entscheidungsprozesse dieser Art lassen sich durch
  evolution"are Verfahren (vgl.\ \ref{sec:evolution_verhalten},
  \ref{cha:formoptimierung}) und andere lernf"ahige Systeme\footnote{Einen
    ausf"uhrlichen "Uberblick "uber lernf"ahige Systeme gibt 
    \name{Starke}{Jens} in \cite{Starke:1994}} realisieren. 
\end{itemize}

Entscheidungskriterien sind dabei:
\begin{itemize}
\item L"ange des Weges und zu erwartende Anstrengung 
\item {\fussgaenger}aufkommen
\item Beschaffenheit des Weges 
\end{itemize}
Diese Beurteilungskriterien werden in einer \hi{subjektiven L"ange} 
zusammengefa"st (vgl.\ Abschn.\ \ref{subsect_eff}). 
Eine andere Gr"o"se stellt die \hi{erwartete Zeitdauer} dar, die ein
{\fussgaenger} f"ur das Zur"ucklegen der Strecke einplant.

In der Simulation treffen die Individuen ihre Entscheidungen mit gewissen
Wahrscheinlichkeiten, die von der aktuellen Situation bestimmt werden. Das
Entscheidungsverhalten kann durch das 
Multi-Nomial-Logit-Modell \cite{DomencichMcFadden:1975} beschrieben werden. Dabei ist die
Wahrscheinlichkeit, mit der ein {\fussgaenger} $\alpha$ seine Entscheidung
"andert  durch
\begin{equation}
  \label{multinomial}
  p_{j\leftarrow i} = \frac{e^{(U_j-U_i - S_{j\leftarrow i}\delta_{ij})/\xi}}
  {\displaystyle \sum_{j^\prime} e^{(U_j-U_i - S_{j\leftarrow
        i}\delta_{ij})/\xi}}
  = \frac{e^{(U_j - S_{j\leftarrow i}\delta_{ij})/\xi}}
  {\displaystyle \sum_{j^\prime} e^{(U_j - S_{j\leftarrow
        i}\delta_{ij})/\xi}}
\end{equation}
mit den Nutzenfunktionen $U_j$ und $U_i$ der Alternativen $j$ und $i$ sowie dem 
beim Wechsel enstehenden Verlust $S_{j\leftarrow i}$ definiert. Der Parameter
$\xi$ gibt die Bereitschaft der Individuen an, auch schlechtere Alternativen
zu w"ahlen. F"ur das Kronecker-Symbol $\delta_{ij}$ gilt $\delta_{ij} = 1$ 
f"ur $i=j$ und  $\delta_{ij} = 0$ in allen anderen F"allen. 

Im Beispiel der zwei Durchg"ange aus \ref{sec:doppeltuer} wird der Nutzen durch
die effektive Zeit $T_{\alpha j}$ definiert, die zum Passieren einer der
beiden T"uren 
ben"otigt wird. Jeder Wechsel verursacht eine
Verz"ogerungszeit $\Delta T$, die im Modell f"ur alle Alternativen gleich ist.
Mit der Nutzenfunktion $U_j = - T_{\alpha j}$ und dem Verlust $S_{j\leftarrow i} = \Delta T$, f"ur alle $i, j$,  
ist die Wahrscheinlichkeit f"ur einen  Wechsel von Durchgang $i$ nach $j$
durch
\begin{equation}
  p_{j\leftarrow i} = \frac{e^{-(T_{\alpha j} + \Delta T\delta_{ij})/\xi_\alpha}}
  {\displaystyle \sum_{j^\prime} e^{-(T_{\alpha j^\prime} + \Delta T\delta_{ij^\prime})/\xi_\alpha}}
\end{equation}
definiert. Die effektive Zeit ergibt sich aus den Zeiten, die zum
Erreichen des Durchgangs und zum Passieren erwartet werden.
\begin{equation}
  T_{\alpha j} = \underbrace{\frac{l_{\alpha j} - \Delta
      l_j}{\vanull}}_{\mx{Zeit zum Erreichen}} 
    + \underbrace{\frac{\Delta l_j}{v_j}}_{\mx{Zeit zum Passieren}}
\end{equation}
mit der momentanen Entfernung $l_{\alpha j}$ des {\fussgaenger}s $\alpha$ zum
Durchgang $j$. Der an dem Durchgang 
herrschende {\fussgaenger}verkehr wird mit der Durchgangsgeschwindigkeit $v_j$
beschrieben. Sie ergibt sich  aus der mittleren Geschwindigkeit der 
{\fussgaenger} $\ave{v}_j$, die zu diesem Zeitpunkt im Umkreis $\Delta l_j$ um
den Durchgang  anzutreffen sind.

Falls der Bereich um den Durchgang frei ist, wird f"ur $v_j$ die
Wunschgeschwindigkeit des {\fussgaenger}s $\vanull$ angenommen. Im anderen
Fall gilt
\begin{equation}
  \label{def:v_tuer}
  v_j = \left\{
  \renewcommand{\arraystretch}{1.2}
  \begin{array}{r@{\quad:\quad}l}
    v_\mx{min} & \ave{v}_j \le 0 \\
    \ave{v}_j & 0 < \ave{v}_j \le \vanull \\
    \vanull &  \ave{v}_j > \vanull 
  \end{array}\right.
\end{equation}
mit
\begin{equation}
  \ave{v}_j = \frac{1}{N_j}\sum_{\alpha^\prime \in \{\alpha^\prime |
    l_{\alpha^\prime j} \le \Delta l_j \}} \Vaprime\,\vec{e}_{\alpha j}
\end{equation}
Der Einheitsvektor $\vec{e}_{\alpha j}$ gibt dabei die Durchgangsrichtung
von $\alpha$ an. Die Durchgangsgeschwindigkeit $v_j$ ist nach oben durch die
Wunschgeschwindigkeit begrenzt. Um das Auftreten von unendlich langen
Duchgangszeiten zu vermeiden, wird die Untergrenze $v_\mx{min}$
eingef"uhrt. Je h"oher der Wert f"ur  $v_\mx{min}$ liegt, desto
wahrscheinlicher ist auch die Entscheidung f"ur bev"olkerte Durchg"ange.

Die Parameter $v_\mx{min}$, $\xi_\alpha$, und $\Delta T$ stellen  die
Entscheidungsstrategie eines {\fussgaenger}s dar. Da es sich hierbei nicht um
me"sbare  Gr"o"sen handelt, ist ihre Bestimmung aus empirischen Daten 
problematisch.

Unter der Annahme, da"s {\fussgaenger} aufgrund ihrer Erfahrungen eine optimale
Strategie entwickeln, k"onnen die Parameter einem Evolutionsproze"s unterzogen
werden, der im n"achsten Abschnitt beschrieben wird.

\section{Evolution der Verhaltensstrategie}
\label{sec:evolution_verhalten}

{\fussgaenger} sind in der Lage, ihre Bewegung und Ausweichman"over durch
st"andige Neubewertung und wiederholtes Ausprobieren zu verbessern.
Daher kann man annehmen, da"s sie durch ihre Erfahrungen ein optimales
Verhalten im {\fussgaenger}verkehr entwickeln.
Diese F"ahigkeiten  lassen sich in der Modellierung
nachempfinden, indem die Modellparameter einem evolution"aren
Optimierungsproze"s unterzogen werden.

Ein Teil der Modellparameter, die das Bewegungs- und Entscheidungsverhalten der
{\fussgaenger} repr"asentieren, l"a"st sich im Evolutionsproze"s 
optimieren. Andere Parameter bleiben dagegen als
Eigenschaften "uber die gesamte Simulationsdauer unver"andert.

Zu den ver"anderbaren Parametern geh"oren die Potentialparameter aus dem
Soziale-Kr"af"-te-Modell (Abschn.\ \ref{sec:fussgaengermodell}) und die
Strategieparameter des Entscheidungsmodells (Abschn.\
\ref{sec:entscheidungsmodell}), wobei die in Abschnitt
\ref{sec:potentialparameter} gefundenen Relationen f"ur die modifizierten
Parameter eingehalten werden m"ussen. 
Als Eigenschaft der {\fussgaenger} kann zum Beispiel deren Wunschgeschwindigkeit
$\vanull$ angenommen werden. Die Gehrichtung, das Bestimmungsziel oder der
Zweck des Ganges stellen weitere M"oglichkeiten dar.

Zur Optimierung der Parameter eignet sich
das in Kapitel \ref{cha:formoptimierung} vorgestellte
Evolutionsprinzip in leichter Ab"anderung.
Der Satz der zu optimierenden Parameter stellt eine potentielle L"o"-sung $x_k$
im Evolutionsverfahren dar. 
Im Unterschied zu den "ublichen Ans"atzen wird jeder L"osung neben der Fitness
auch eine Eigenschaft $v_k$ zugeordnet. 

Dieses Evolutionsverfahren kennt keine Einteilung in Generationen. Neue
potentielle L"osungen, die sogenannten Nachkommen, werden jeweils beim Start eines
{\fussgaenger}s erzeugt. Die Bewertung geschieht anhand der individuellen
Leistungsma"se $Y_\alpha^1\dots Y_\alpha^5$ aus Abschnitt
\ref{sec:bewertungskriterien}, nachdem der {\fussgaenger} 
sein Ziel erreicht hat. Ferner werden den bewerteten L"osungen die
Eigenschaften der {\fussgaenger} zugewiesen. Der Parametersatz steht dann zur
Reproduktion neuer potentieller L"osungen zur Verf"ugung.

In der Simulation wird eine bestimmte Anzahl von {\fussgaenger}n erzeugt, die
"uber die Simulationsdauer konstant bleibt. 
Nachdem ein {\fussgaenger} das System durchlaufen hat, startet er von neuem aus
der Anfangsposition. Seine Eigenschaften bleiben dabei unver"andert.

Aus den gesammelten L"osungen entsteht bei jedem  Start eines {\fussgaenger}s
$\alpha$ ein neuer Parametersatz. Dabei werden analog zur
Reproduktionswahrscheinlichkeit aus (\ref{def:evowahrsch}) L"osungen
mit der Wahrscheinlichkeit 
\begin{equation}
  \label{def:evowahrsch_strat}
  p_{\alpha k} = \frac{c_{\alpha k}\, f_k^\tau}{\displaystyle
    \sum_{k^\prime=1}^\mu c_{\alpha k^\prime}\, f_{k^\prime}^\tau}
\end{equation}
zur Reproduktion ausgew"ahlt. Die Definition ber"ucksichtigt dabei nicht nur
die Fitness $f_k$, 
sondern auch wie gut die potentielle L"osung zu den vorgegebenen Eigenschaften
des neuen {\fussgaenger}s pa"st.  
Die "Ubereinstimmung der Eigenschaften kann durch
\begin{equation}
  c_{\alpha k} = \exp\left(-\frac{\sqrt{\sq{(v_k^0 - \vanull)}}}{\omega} \right)
\end{equation}
ausgedr"uckt werden. F"ur gleiche Wunschgeschwindigkeiten ist die
"Ubereinstimmung maximal. Mit $\omega$ wird der Toleranzbereich bestimmt.
Der neue Parametersatz ensteht dann durch Anwendung des Crossover- und des
Mutationsoperators gem"a"s Abschnitt \ref{sec:implementierung}.

Die Eigenschaften der {\fussgaenger} bewirken eine wahrscheinlichere
Reproduktion von L"o"-sungen mit gleichen Eigenschaften. Dies ist in der Natur
mit Gruppen aus Individuen einer Art vergleichbar, die in regional
unterschiedlichen Lebensr"aumen angesiedelt sind.
Da die Fortpflanzung unter den Individuen eines Lebensraums wahrscheinlicher
ist als die unter Individuen verschiedener Regionen, k"onnen sich in den
Lebensr"aumen unterschiedliche Subspezies  entwickeln.
Im Modell der Evolution von {\fussgaenger}strategien w"urde das bedeuten, da"s
sich f"ur langsame und schnelle {\fussgaenger} unterschiedliche
Verhaltensstrategien  ausbilden.
Sind die  Eigenschaften durch diskrete Gr"o"sen gegeben, etwa durch die
Bewegungsrichtung,  so lassen sich nur L"osungen mit exakt gleichen
Eigenschaften miteinander kombinieren.

\chapter{Wegenetze}
\label{cha:wegenetze}
Betrachtet man gr"o"sere {\fussgaenger}anlagen, so liegen die Zielpunkte der
{\fussgaenger} meist nicht in Sichtweite ihrer Startpositionen. Es werden
\hi{Zwischenziele} ausgew"ahlt, die zun"achst angesteuert werden.
Mit der Anlage vertraute {\fussgaenger} haben eine genaue
Vorstellung von den Verbindungswegen  und deren Beschaffenheit.
Die Wegewahl nach diesen kognitiven Karten \cite{DownsStea:1982} erfolgt
dabei nicht nur nach Kriterien wie Streckenl"ange oder Anstrengung, sondern
auch nach anderen subjektiven Empfindungen, die h"aufig auch durch das
{\fussgaenger}aufkommen in den einzelnen Streckenabschnitten bestimmt werden
(vgl. Abschnitte \ref{sec:bemessung} und \ref{sec:bewertungskriterien}).

Das {\fussgaenger}aufkommen in den einzelnen Abschnitten h"angt von der
Produktionsrate der Eintrittspunkte ab, ferner von der Attraktivit"at der
Zielknoten sowie von den Wegen, die die {\fussgaenger} benutzen.  Um das
{\fussgaenger}aufkommen auf den einzelnen Strecken im Wegesystem zu ermitteln,
kann ein Suchalgorithmus f"ur alle Start-Ziel-Knotenpaare diejenigen Wege
bestimmen, welche die {\fussgaenger} in bezug auf ihre pers"onlichen
Bewertungskriterien w"ahlen w"urden.  Die Streckenabschnitte werden dabei in
Abh"angigkeit ihres Vorkommens in den ausgew"ahlten Wegen mit
Belastungsfrequenzen besetzt.

Auf diese Weise lassen sich bereits aus dem {\fussgaenger}aufkommen, den
Bed"urfnissen der {\fussgaenger} und den Eigenschaften einzelner Teilst"ucke
einer {\fussgaenger}anlage Problemstellen aufdecken, die durch "uberlastete
Strecken oder Durchg"ange enstehen. Auch besonders stark frequentierte und
deshalb f"ur Verkaufsfl"achen interessante Stellen werden dadurch sichtbar. 

Durch die Belastung der Teilstrecken einer Anlage "andern sich auch deren
Leistungsmerkmale  und damit die Wahrscheinlichkeit f"ur die Benutzung der
einzelnen  Wege. F"ur die Kanten des Wegenetzes kann eine
\hi{subjektive L"ange} eingef"uhrt werden, die das Verhalten der
Streckencharakteristika bez"uglich des {\fussgaenger}aufkommens
ber"ucksichtigt (vgl.\ Abschn.\ \ref{subsect_eff}).
 F"ur die {\fussgaenger} dauert es meist l"anger, sich durchs
Gedr"ange zu bewegen, als den Umweg durch eine Seitenstra"se zu nehmen.

In Abh"angigkeit der Streckenbelastungen des Netzwerks lassen sich mit den
subjektiven L"angen die (subjektiv) k"urzesten Wege und die
Streckenbelastungen wiederholt berechnen, bis sich gegebenenfalls  
eine Gleichgewicht einstellt.

Zur Simulation der {\fussgaengerstroeme} in gro"sen Anlagen kann das System in
Teilst"ucke aufgeteilt werden, deren charakteristische Eigenschaften in
Abh"angigkeit von {\fussgaenger}menge und -zusammensetzung durch die
Mikrosimulation des Soziale-Kr"afte-Modells
(Kap.\ \ref{cha:sozialekraeftemodell}) bestimmt werden. 

\section{Darstellung einer {\fussgaenger}anlage als Netzwerk}
\label{sec:darstellung}
\begin{figure}[p]
  \begin{center}
    \leavevmode
    \includegraphics[width=12cm]{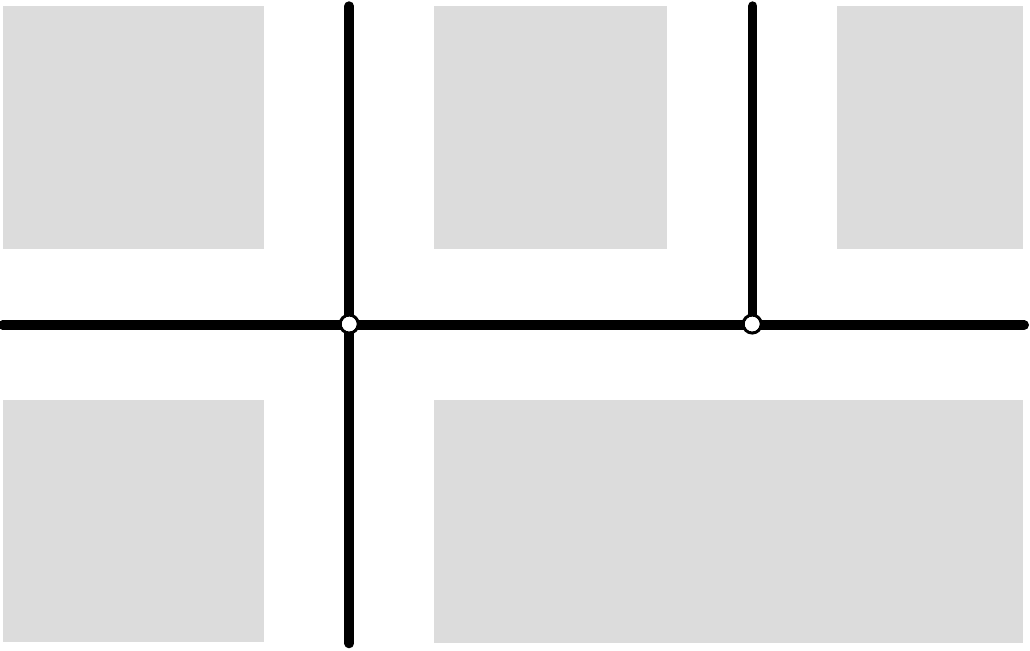}
  \end{center}
  \mycaption{Netzwerk aus mehreren Korridoren}
  {Einfache Netzwerkdarstellung
    einer {\fussgaenger}anlage. Die einzelnen Kreuzungen werden als Knoten
    (wei"se Punkte), die Verbindungskorridore als Kanten (dicke Balken)
    dargestellt.}
  \label{fig:korridornetz_eins}
\end{figure}

In der Betrachtung  einer {\fussgaenger}anlage als Netzwerk wird die
Beschaffenheit des Ge"-b"au"-des auf eine Orte-Verbindungswege-Beziehung
reduziert. Die Darstellung erfolgt durch einen Graphen $G(V,E)$ mit $V$ Knoten
und $E$ Kanten. F"ur die Interpretation der Knoten und Kanten werden hier drei
M"oglichkeiten vorgeschlagen:

Die einfache Darstellung einer {\fussgaenger}anlage (Abbildung
\ref{fig:korridornetz_eins}) behandelt die Kreuzungen als (Netzwerk-)Knoten und die
Stra"sen, bzw.\ Korridore, als (Netzwerk-)Kanten.  Sie vernachl"assigt dabei 
den zweidimensionalen Charakter des {\fussgaenger}verkehrs, sowie die Form der
Kreuzungen. Daher kann diese Darstellung nur f"ur Systeme mit langen
Verbindungswegen zwischen den einzelnen Punkten angewendet werden.   

\begin{figure}[p]
  \begin{center}
    \leavevmode
    \includegraphics[width=12cm]{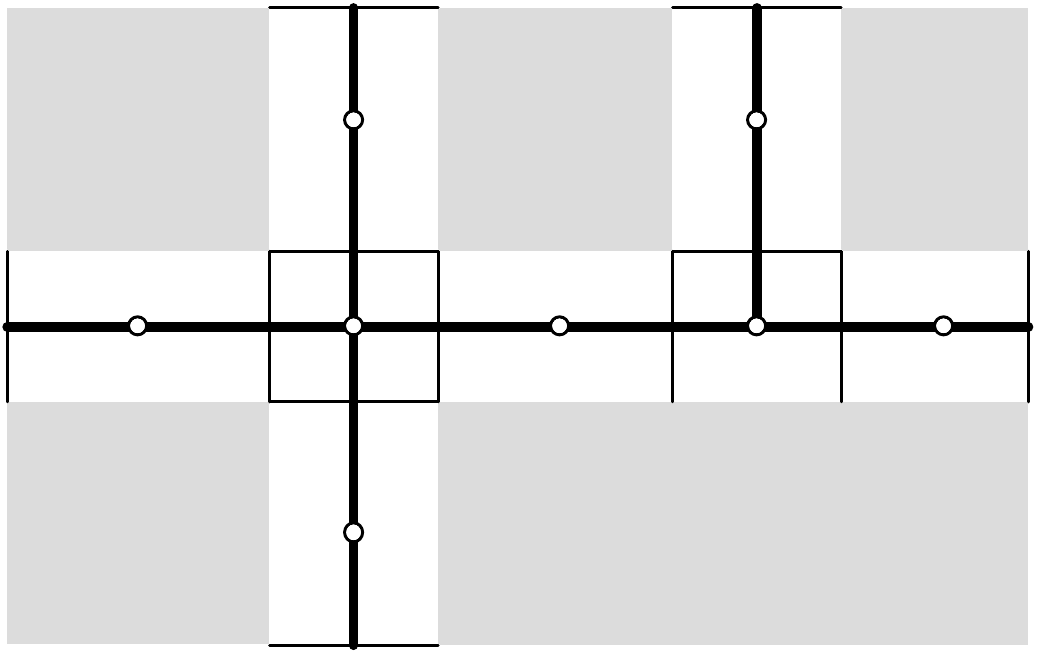}
  \end{center}
  \mycaption{Netzwerk aus R"aumen} {Die {\fussgaenger}anlage wird durch Tore
    (d"unne Linien) in einzelnen Segmente aufgeteilt, in denen sich die
    {\fussgaenger} aufhalten. Die Segmente bilden die Knoten und     
    die Schnittstellen (Tore) die Verbindungskanten des Netzwerks.
    Diese Darstellung wird in  Warteschlangen-Modellen verwendet
    (vgl. Abschn. \ref{sec:mikroskopisch}).}
  \label{fig:korridornetz_zwei}
\end{figure}

\begin{figure}[t]
  \begin{center}
    \leavevmode
    \includegraphics[width=12cm]{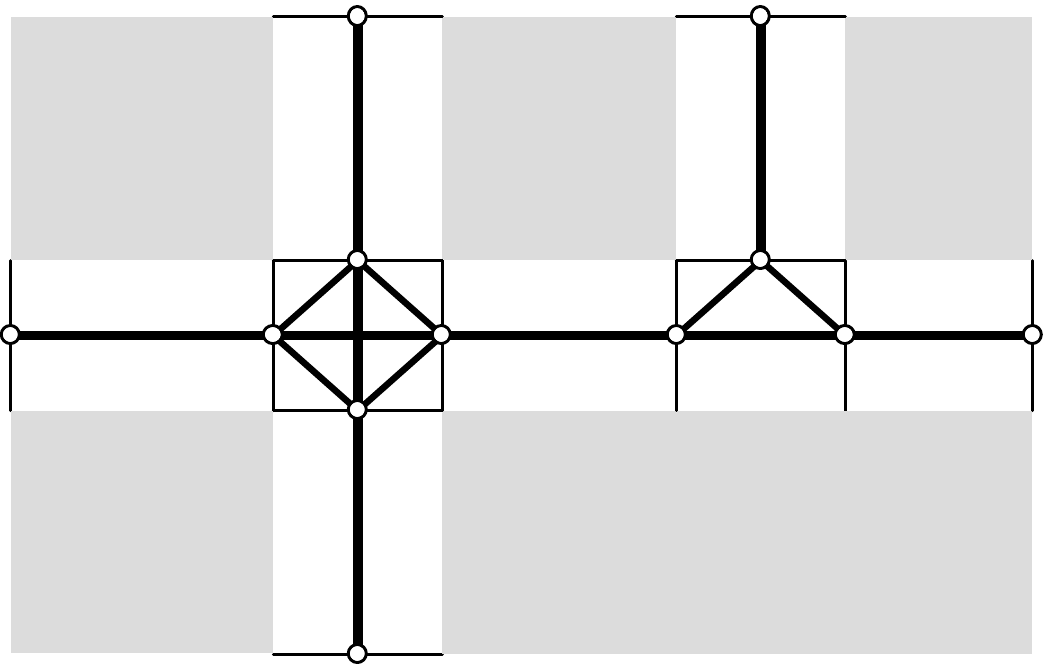}
  \end{center}
  \mycaption{Netzwerkdarstellung des Modells f"ur {\fussgaengerstroeme}}
  {Komplement"ar zur Darstellung in Warteschlangen-Modellen
    (Abb. \ref{fig:korridornetz_zwei}) werden im Modell f"ur {\fussgaengerstroeme}
    die Schnittstellen zwischen den Segmenten als Orientierungspunkte und
    damit auch als Knoten des Netzwerks behandelt. Durch ein Raumsegment
    k"onnen mehrere Verbindungskanten f"uhren.}
  \label{fig:korridornetz_drei}
\end{figure}

In der Darstellung aus Abbildung \ref{fig:korridornetz_zwei} werden die Korridore
durch Tore (vgl. \ref{sec:structureprog}) in einzelne Segmente
abgetrennt. Diese Segmente, die eine bestimmte Anzahl von {\fussgaenger}n
aufnehmen k"onnen, werden als Knoten des Netzwerks behandelt. Die
Tore, die den "Ubergang zwischen den Segmenten erm"oglichen, bilden die Kanten.
Diese Darstellung findet zum Beispiel bei den Warteschlangenmodellen
Verwendung (vgl. Abschn. \ref{sec:mikroskopisch}).
x
Eine dritte M"oglichkeit zur Darstellung (Abb. \ref{fig:korridornetz_drei})
besteht in darin, die Tore als 
Netzwerkknoten zu betrachten. Sie dienen auch als (Zwischen)-Ziele zur
Orientierung der {\fussgaenger}. Alle an ein Segment grenzenden Tore sind
miteinander verbunden, da gerade das St"uck der eingeschlossenen
Verkehrsfl"ache den Zugang zu allen diesen Toren gew"ahrt. Die Fl"ache einer
Kreuzung ist dadurch von mehreren Netzwerkkanten durchzogen.

Alle Kanten eines Segmentes sind gleicherma"sen vom {\fussgaenger}aufkommen
betroffen und werden daher bei der Bestimmung der Kantenbelastung als Einheit
behandelt. 

Die in den folgenden Abschnitten behandelten Untersuchungen beziehen sich
sowohl auf die Betrachtung der Wegesysteme als einfache Graphen, als auch 
auf die Betrachtung in der zuletzt vorgestellten Weise.  

\section{Streckenbelastung in Wegenetzen} \label{sec:streckenbelastung}
Die Belastung der Strecken in einem Wegenetz ergibt sich aus der Anzahl der
{\fussgaenger}, die von einem bestimmten Punkt zu einem anderen laufen, und
der Route, die sie dazu ausw"ahlen. In der Regel benutzen {\fussgaenger} den
k"urzesten Weg zu ihrem Ziel, sind aber bei ung"unstiger Beschaffenheit der
Strecken oder bei hohem {\fussgaenger}aufkommen auch zu Umwegen bereit. Der
k"urzeste Weg bezieht sich daher auf die \hi{subjektive L"ange}.

Die Belastungsh"aufigkeit ist definiert als die H"aufigkeit des
Vorkommens einer Kante in allen k"urzesten Wegen $(a,b)$ multipliziert mit der
Anzahl $N_{a,b}$ der {\fussgaenger}, die von $a$ nach $b$ laufen. Die von der
Gesamtzahl der Wanderungen unabh"angige Gr"o"se $F_i$ wird als {\hi Frequenz der
  Belastung} 
\begin{equation}
  \label{def:belastungsfrequenz}
  F_i = \frac{ \displaystyle 
    \sum_{a=1}^{V}\sum_{b=1}^{V} N_{(a,b)}\,\delta_{i\in (a,b)}
    }
  { \displaystyle \sum_{a=1}^{V}\sum_{b=1}^{V} N_{(a,b)}
    }
\end{equation}
mit 
\begin{equation}
  \delta_{i\in (a,b)} = \left\{ 
  \renewcommand{\arraystretch}{1.2}
  \begin{array}{r@{\quad:\quad}l}
    1 & \mbox{Kante $i$ ist ein Teil des k"urzesten Weges ist} \\
    0 & \mbox{sonst}
    \end{array}
  \right.
\end{equation}
bezeichnet.

\subsection{Bestimmung der k"urzesten Wege}
Eine sehr elegante Methode zur Bestimmung des k"urzesten Pfades von 
Knoten $a$ nach $b$ baut auf einer Methode zur Bestimmung 
aller paarweisen Verbindungen eines Graphen auf, die von \name{Warshall}{S.}
eingef"uhrt wurde. Dabei wird folgender Sachverhalt ausgenutzt: 
\begin{quote}
  \glqq Wenn es einen Weg von Knoten $a$ nach $b$ und von $b$
  nach $c$ gibt, dann gibt es auch einen Weg von $a$ nach $c$.\grqq
\end{quote}
Dies l"a"st sich sogar noch etwas strenger fassen, was die
Berechnung aller paarweisen Verbindungen eines Graphen in einem Durchlauf
erlaubt. Dazu werden die Knoten in eine Reihe gesetzt und
indiziert. Es gilt:
\begin{quote}
  \glqq 
  Wenn es einen Weg von Knoten $a$ nach $b$ gibt, auf dem nur Knoten mit einem
  Index  kleiner $b$ benutzt werden, und einen Weg von $b$ nach $c$, dann gibt
  es auch einen Weg von $a$ nach $c$, auf dem nur Knoten mit dem Index kleiner  
  $b+1$ angelaufen werden.\grqq 
\end{quote}
Diese f"ur \hi{topologische Graphen} aufgestellte Beobachtung l"a"st sich auch
auf \hi{metrische, gerichtete Graphen} anwenden. 
\begin{quote}
  \glqq 
  Der k"urzeste Weg von einem Knoten $a$ zu einem anderen Knoten $c$, auf dem
  nur Knoten mit einem Index kleiner $b+1$ benutzt werden, ist entweder 
  der k"urzeste Weg von $a$ nach $c$ unter ausschlie"slicher
  Verwendung von Knoten mit einem Index $b$, oder, falls  dieser k"urzer
  ist, der k"urzeste Weg von $a$ nach $b$ plus der Distanz von $b$ nach
  $c$.\grqq  
\end{quote}

\begin{figure}[tb]
  \begin{center}
    \leavevmode
    \begin{minipage}[t]{0.9\textwidth}
\begin{verbatim}
for (a=1; a<=V; a++) 
  for (b=1; b<=V; b++)  
    if (M[a][b] < Infinity) 
      for (c=1; c<=V; c++) 
        if (M[a][b] + M[b][c] < M[a][c]) M[a][c] = M[a][b] + M[b][c];
\end{verbatim}
    \end{minipage}
  \end{center}
  \mycaption{Warshall-Floyd-Algorithmus}
  {Warshall-Floyd-Algorithmus zur Bestimmung des jeweils k"urzesten
    Weges f"ur jedes Knotenpaar des Graphen besteht lediglich aus drei
    ineinandergeschachtelten Wiederholungsschleifen ({\tt for}).
    Die beiden "au"seren Schleifen "uber $a$ und
    $b$ durchlaufen alle Knoten. Wenn es einen Weg von $a$ nach $b$ gibt, d.h.
    {\tt M[a][b]} $< \infty $, wird f"ur alle Wege zwischen $b$ und $c$
    gepr"uft, ob sie Teil eines k"urzeren Weges von $a$ nach $c$ sind.}
  \label{fig:floydalgorithmus}
\end{figure}
Daraus l"a"st sich  ein Algorithmus zur Bestimmung aller paarweise
k"urzesten Verbindungen in einem Netzwerk ableiten, der allgemein
\name{Floyd}{R. W.} zugeschrieben wird. Der Unterschied 
besteht lediglich in der Vergleichsabfrage: \glqq Gibt es eine Verbindung?\grqq
\/bei \name{Warshall}{} und \glqq Ist die Verbindung k"urzer?\grqq\/ bei
\name{Floyd}{} (s. Abb. \ref{fig:floydalgorithmus}). Der im folgenden mit
Warshall-Floyd bezeichnete Algorithmus l"ost das Problem in $O(V^3)$
Schritten, das hei"st die drei ineinander geschachtelten Schleifen werden 
jeweils h"ochstens $V$ mal durchlaufen.

Zus"atzlich zu der Abstandsmatrix $\left(M_{ab}\right)_{V\times V}$, die die
Streckenl"ange des Weges von $a$ nach $b$ beinhaltet, wird die Wegematrix
$\left(R_{ab}\right)_{V\times V}$ 
eingef"uhrt. Auf einem Weg von $a$ nach $b$ ist der n"achste anzulaufende Knoten
$R_{ab}$. Aus der Transitivit"at der Bedingungen folgt: Der k"urzeste Weg von
$a$ nach $b$ geht "uber $R_{ab}$ und von $R_{ab}$ "uber $R_{R{ab}b}$ und so
weiter. Auf diese Weise sind die k"urzesten Wege f"ur alle Knotenpaare 
in $\left(R_{ab}\right)_{V\times V}$ und deren L"ange 
in $\left(M_{ab}\right)_{V\times V}$ gespeichert.

Die Methode des Warshall-Floyd-Algorithmus 
garantiert, da"s es keinen k"urzeren Weg
als die gefundene L"osung gibt. Wenn ein Graph mehrere Kanten mit der gleichen
L"ange enth"alt, ist die L"osung jedoch nicht eindeutig.

Im Vergleich zu andere Algorithmen von \name{Dijkstra}{E.}, \name{Prim}{R.}
oder \name{Kruskal}{Joseph.~B.~JR.} zeichnet sich dieser Algorithmus durch die
einfache Implementierung und die Effizienz bei dichten Graphen,
d.h.\ $E>V$, aus \cite{Sedgewick:1992}\cite{Kruskal:1956}. 

\subsection{Verteilung der Benutzungsh"aufigkeit}
\label{sec:pareto}
Die Verteilung der \hi{Benutzungsh"aufigkeit}en der Kanten macht eine Aussage
"uber die Qualit"at eines Wegenetzes bez"uglich der Belastung der
Verbindungswege. 

Bei den folgenden Berechnungen wird von einer gleichm"a"sigen Verteilung der
Start- und Zielknoten ausgegangen und von jedem der $V$ Knoten eines
Netzwerkes jeweils ein {\fussgaenger} zu den anderen $V-1$ Knoten geschickt.
Damit wird f"ur alle Knotenpaare $N_{(a,b)} =1$ angenommen.

\begin{figure}[ptb]
  \begin{center}
    \leavevmode 
    \includegraphics[width=0.6\textwidth]{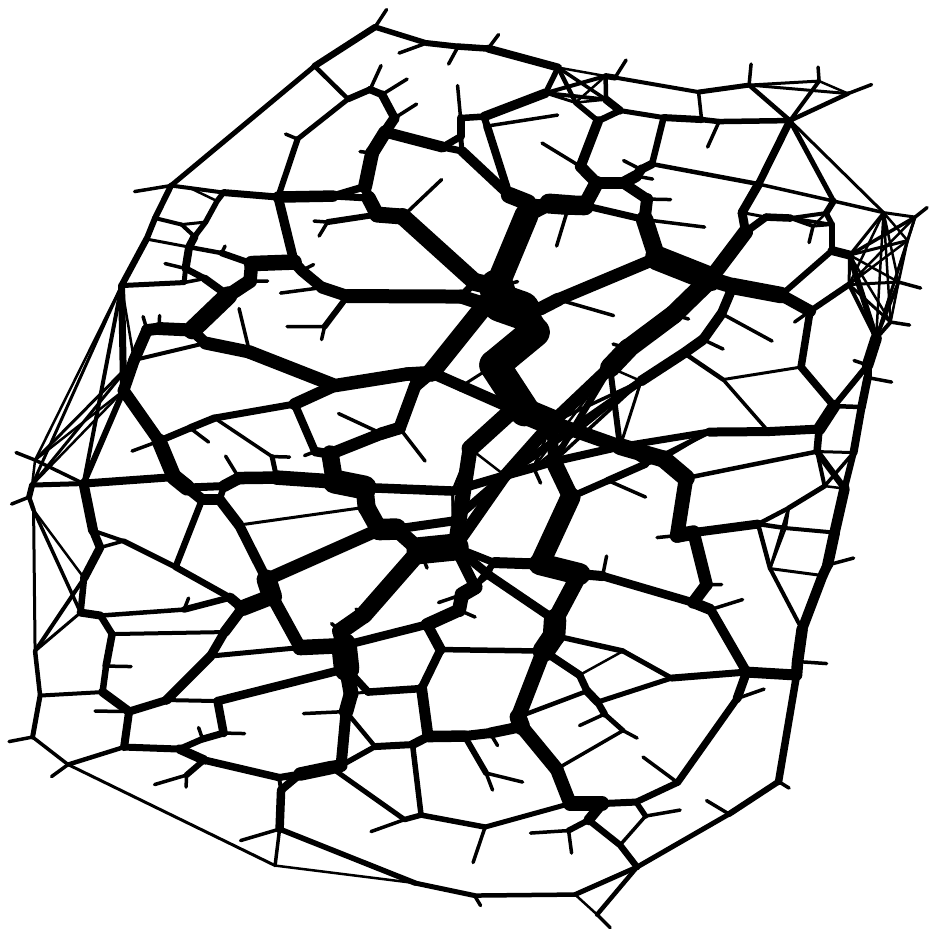}
  \end{center}
  \mycaption{Wegesystem von Martina Franca 1}
  {Das Netzwerk stellt die italienische Stadt Martina Franca dar. Die
    H"aufigkeiten, mit der die Kanten auf den paarweise k"urzesten Wegen liegen,
    sind durch die Liniendicke wiedergegeben.}
  \label{fig:martina}
  \begin{center}
    \leavevmode
    \includegraphics[width=0.6\textwidth]{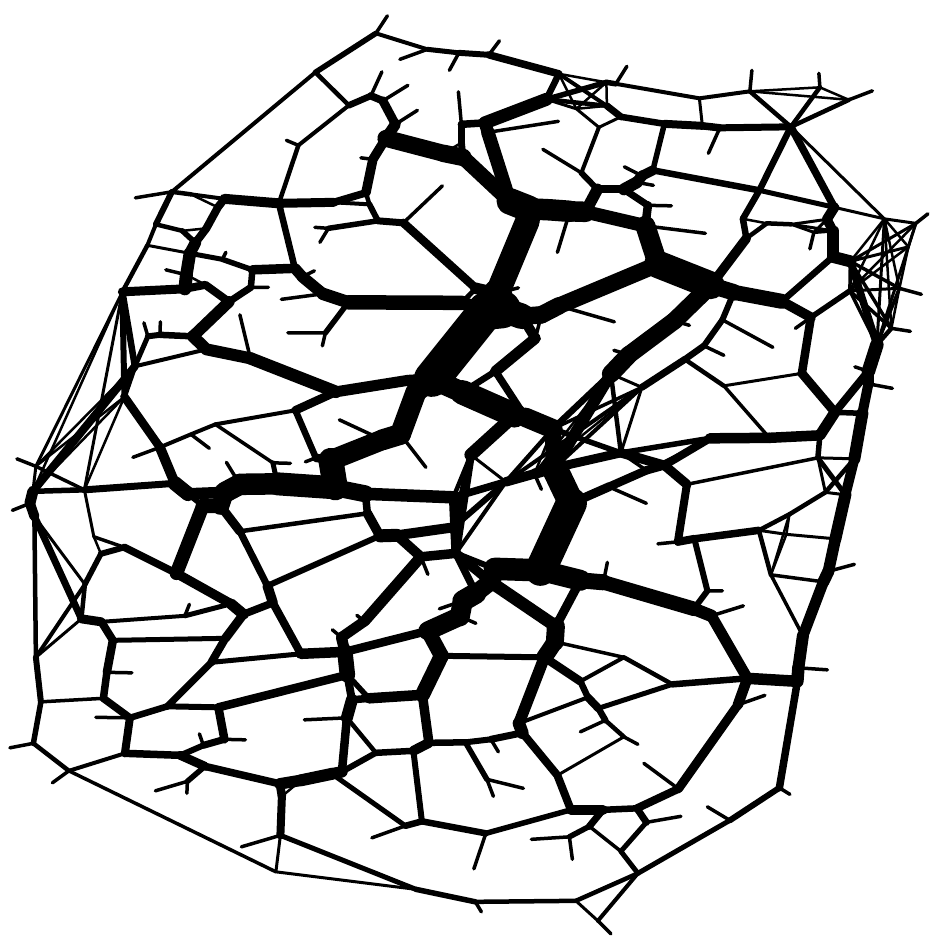}
  \end{center}
  \mycaption{Wegesystem von Martina Franca 2} {Benutzungsh"aufigkeit nach 20
    Iterationen des Random-Warshall-Floyd-Algorithmus. Die 
    Kantenl"angen wurden in 15\% Klassen eingeteilt. Dadurch werden
    Streckenf"uhrungen mit kleinem L"angenunterschied gleichstark belastet
    (vgl.\ Abschn.\ \ref{sec:randomwarshall}, S.\ \pageref{fig:martinavert}).}
  \label{fig:martinarandom}
\end{figure}

In Analogie zur \hi{Pareto-Verteilung}, die von
\name{Pareto}{Vilfredo Marquis} zum Vergleich der Relationen zwischen
Einkommensbezieher und Einkommen eingef"uhrt wurde \cite{Gabler:1993}, 
werden die Kanten eines Netzwerkes nach der H"aufigkeit ihrer Benutzung
sortiert und die Ordnungszahl $z_i$ der Kanten "uber ihre
Benutzungsh"aufigkeit $F_i$ aufgetragen.
Die Verteilung der \hi{Benutzungsh"aufigkeit} in (nat"urlichen) Netzwerken nimmt
dabei h"aufig die Form einer exponentiellen Verteilungsfunktion
\begin{equation}
  \label{def:expoverteil}
  z_i = k\,\exp(-F_i/\alpha)\qquad i=1\dots E
\end{equation}
mit der Anzahl der Kanten $E$ an. Je steiler die Verteilung zu hohen
Frequenzen abf"allt, desto
gleicherm"a"siger ist die Belastung der einzelnen Strecken im Netz verteilt.

Die \hi{Pareto-Verteilung} 
\begin{equation}
  \label{def:pareto}
  z_i = k^\prime\,F_i\/^{-\alpha^\prime}\qquad i=1\dots E
\end{equation}
gibt die Benutzungsh"aufigkeit schlechter als die Exponentialverteilung
wieder (vgl.\ Abb.\ \ref{fig:martinavert}). Durch eine Transformation der
Benutzungsh"aufigkeit $\hat{F}_i = \exp(F_i)$ l"a"st sich
(\ref{def:expoverteil}) jedoch in eine  \hi{Pareto-Verteilung}
(\ref{def:pareto}) "uberf"uhren.

Die Parameter der Verteilungsfunktion $\alpha$ und $k$, bzw.\ $\alpha^\prime$
und $k^\prime$ lassen sich nach der \hi{Methode der Linearen Regression}
\cite[Kap. 15]{NRC}, (vgl. \ref{sec:vorhersage}), bestimmen: Dazu wird auf
beide Seiten von (\ref{def:expoverteil}), bzw.\ (\ref{def:pareto}), der
nat"urliche Logarithmus angewendet.  Durch die Einf"uhrung der neuen Variablen
$\tilde{z}_i = \ln z_i$ und $\tilde{k} = \ln k$ wird die
\hi{Exponentialverteilung} zu der Geradengleichung
\begin{equation}
  \tilde{z}_i = \tilde{k} - \alpha\,F_i
\end{equation}
Zur Bestimmung der Koeffizienten der Paretoverteilung mu"s zus"atzlich
$\tilde{F}_i = \ln F_i$ eingef"uhrt werden:
\begin{equation}
  \tilde{z}_i = \tilde{k} - \alpha\,\tilde{F}_i
\end{equation}

Aus den notwendigen Bedingungen zur Minimierung des \hi{quadratischen Fehler}s
\begin{equation}
  \sq{\chi}(\alpha,\tilde{k}) = \sum_{i=1}^{E} \sq{\left(
    \tilde{z}_i -\tilde{k} +\alpha\,\tilde{F}_i\right)} 
\end{equation}
\begin{eqnarray}
  0 & \stackrel{!}{=} & \frac{\partial\sq{\chi}}{\partial\tilde{k}} = 
  -2 \sum_{i=1}^{E} \left(\tilde{z}_i -\tilde{k} +\alpha\,\tilde{F}_i\right)
  \nonumber \\
  0 & \stackrel{!}{=} & \frac{\partial\sq{\chi}}{\partial\alpha} = 
  2 \sum_{i=1}^{E} \tilde{F}_i\,\left(\tilde{z}_i -\tilde{k}
  +\alpha\,\tilde{F}_i\right) 
\end{eqnarray}
werden die Parameter $\alpha$ und $\tilde{k}$ durch 
\newcommand{\Sx}{{\sum_{i=1}^{E} F_i}}
\newcommand{\Sxx}{\sum_{i=1}^{E}\sq{F_i}}
\newcommand{\Sy}{\sum_{i=1}^{E}\tilde{z}_i}
\newcommand{\Sxy}{\sum_{i=1}^{E} F_i\tilde{z}_i}
\begin{eqnarray}
  \tilde{k} & = & \frac{ 
    \left(\Sxx\right) \left(\Sy\right) 
    - \left(\Sx\right) \left(\Sxy\right) }
  { 
    E \Sxx - \sq{\left(\Sx\right)}} \nonumber \\ 
  \alpha & = &  \frac{ 
    E \Sxy - \left(\Sx\right) \left(\Sy\right) }
  { 
    E \sq{\left(\Sx\right)} - \Sxx} 
\end{eqnarray}
bestimmt. F"ur $\alpha^\prime$ und $\tilde{k}^\prime = \ln k^\prime$ ist $F_i$
durch $\tilde{F}_i$ zu ersetzen.

\begin{figure}[tb]
  \begin{center}
    \leavevmode
    \includegraphics[width=0.8\textwidth]{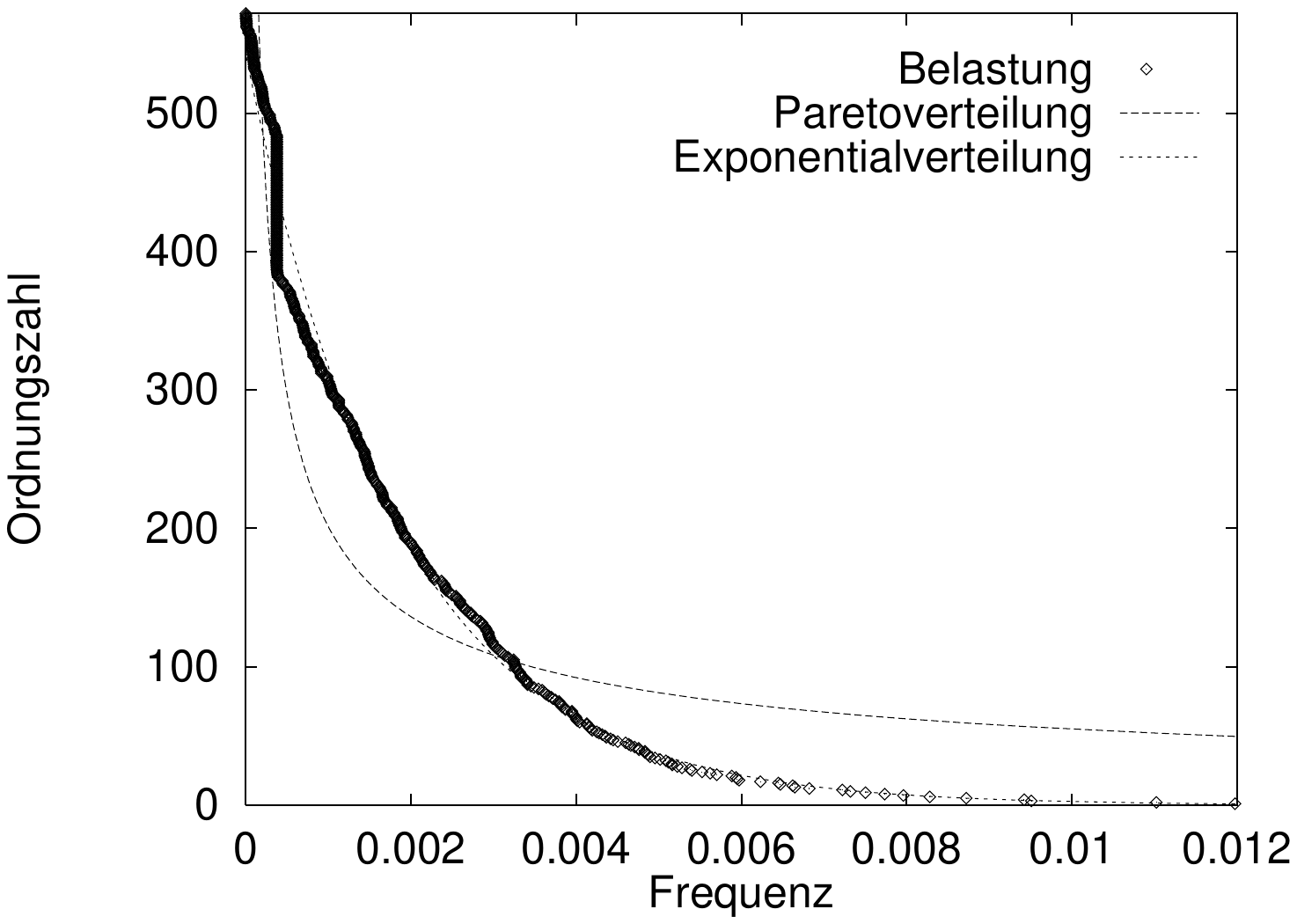} 
  \end{center}
  \mycaption{Verteilung der Streckenbelastung von Martina Franca}
  {Verteilung der
    Streckenbelastung von Martina Franca. Die 
    Ordnungszahl der Kanten $z_i$ ist "uber die Belastungsfrequenz $F_i$ 
    aufgetragen. Die Symbole stehen f"ur die durch den
    einfachen Warshall-Floyd-Algorithmus ermittelten Frequenzen. Die
    angepa"sten Verteilungfunktionen sind durch Linien dargestellt. W"ahrend
    die Exponentialfunktion gut mit der Verteilung "ubereinstimmt, gibt die
    Paretoverteilung die Beziehung zwischen Benutzungsfrequenz und
    Ordnungszahl nur unzureichend wieder.}
  \label{fig:qpareteo}
\end{figure}

\subsection{Random-Warshall-Floyd-Algorithmus}
\label{sec:randomwarshall}
\begin{figure}[ptb]
  \begin{center}
    \leavevmode
    \rule{0pt}{0pt}
    \includegraphics[width=0.9\textwidth,height=0.9\textwidth]{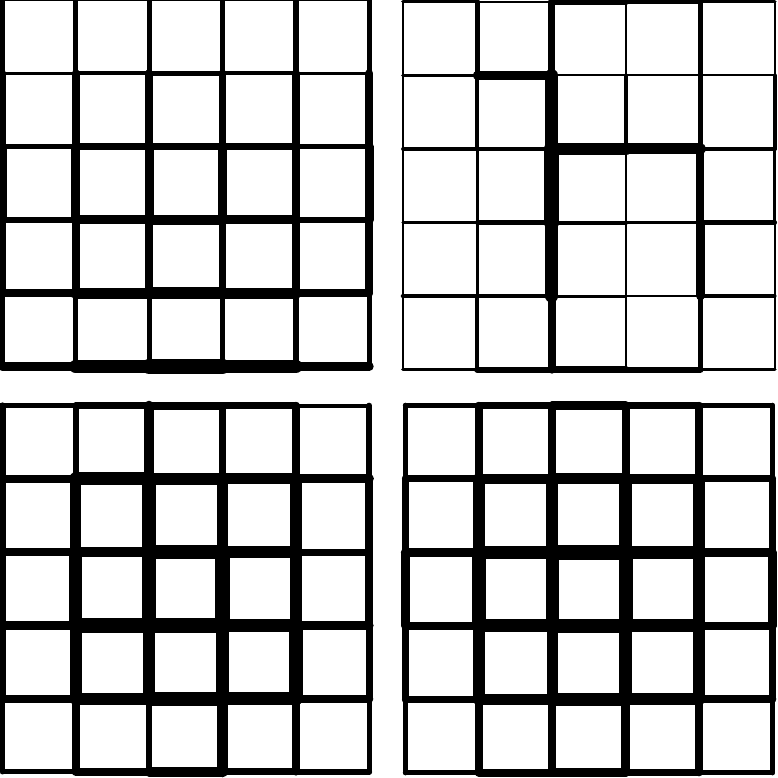}
  \end{center}
  \mycaption{Quadratisches Netzwerk} {Quadratisches Netzwerk. Die oberen
    Abbildungen zeigen die Kantenbelastungen nach einem Lauf des einfachen
    Warshall-Floyd-Algorithmus. Bei der rechts abgebildeten Berechnung wurde
    die Reihenfolge, in der die Knoten auf einen Verbindungsweg hin gepr"uft
    werden, durch einen Zufallszahlengenrator erzeugt. Die unteren Abbildungen
    zeigen die Ergebnisse nach 20 (links) und 100 (rechts) L"aufen des
    Random-Warshall-Floyd-Algorithmus.}
  \label{fig:quadrat}
\end{figure}

Ein gleichm"a"siges Netzwerk, wie zum
Beispiel das in den Abbildungen
\ref{fig:quadrat} dargestellte quadratische Raster, 
verdeutlicht die Schwachstelle des Warshall-Floyd-Algorithmus:
Gibt es mehrere gleichlange Wege
von den Knoten $a$ nach $b$, wird jeweils diejenige Verbindung genommen, die
nach der Indizierung der Knoten als erste auftritt. W"ahrend in einem
Transportwegesystem gleichwertige Verbindungen auch zu gleichen Teilen
belastet werden, belegt der Algorithmus von den gleichlangen
Strecken nur die erste mit voller Belastung, die anderen werden nicht
belegt.

Dies ist in der Abb. \ref{fig:quadrat} links oben deutlich zu erkennen: Die
unteren horizontalen Kanten sind extrem stark belastet. Die Symmetrie, die das
Wegesystem in der horizontalen und vertikalen Richtung und den beiden
Diagonalen aufweist, ist auch in der Verteilung der Belastungsh"aufigkeit zu
erwarten (vgl. Abb.\ref{fig:quadrat} rechts unten). 

Auf der Suche nach k"urzeren Verbindungswegen bearbeitet der Algorithmus die
Knoten des Graphen in einer bestimmten Reihenfolge. Wird diese Reihenfolge
willk"urlich ge"andert, so entsteht eine neue Aufteilung der Belastung auf die
Kanten (vgl. Abb. \ref{fig:quadrat} rechts oben).

\begin{figure}[tb]
  \begin{center}
    \leavevmode
    \includegraphics[width=0.45\textwidth]{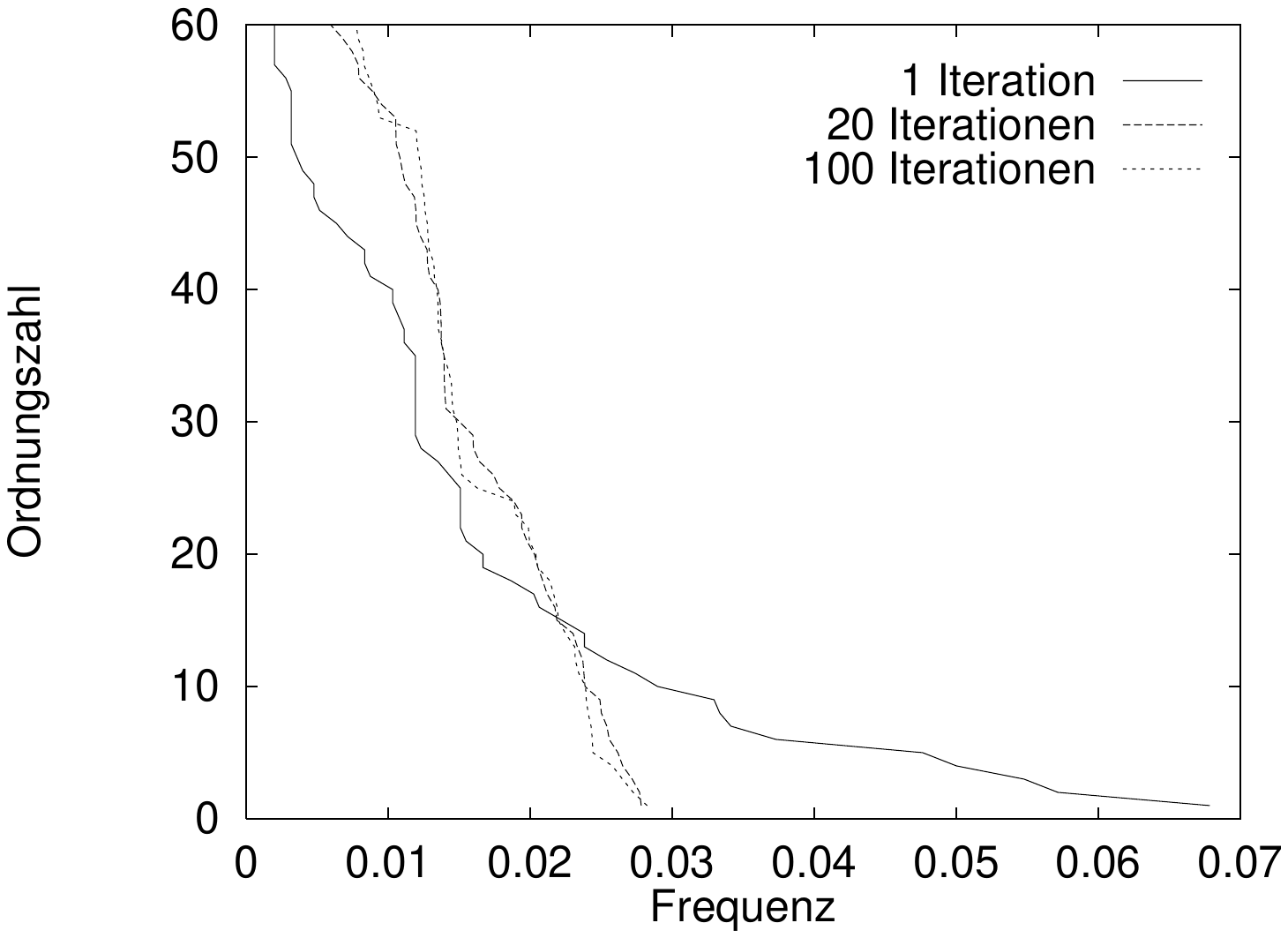} \hspace{0.5cm}
    \includegraphics[width=0.45\textwidth]{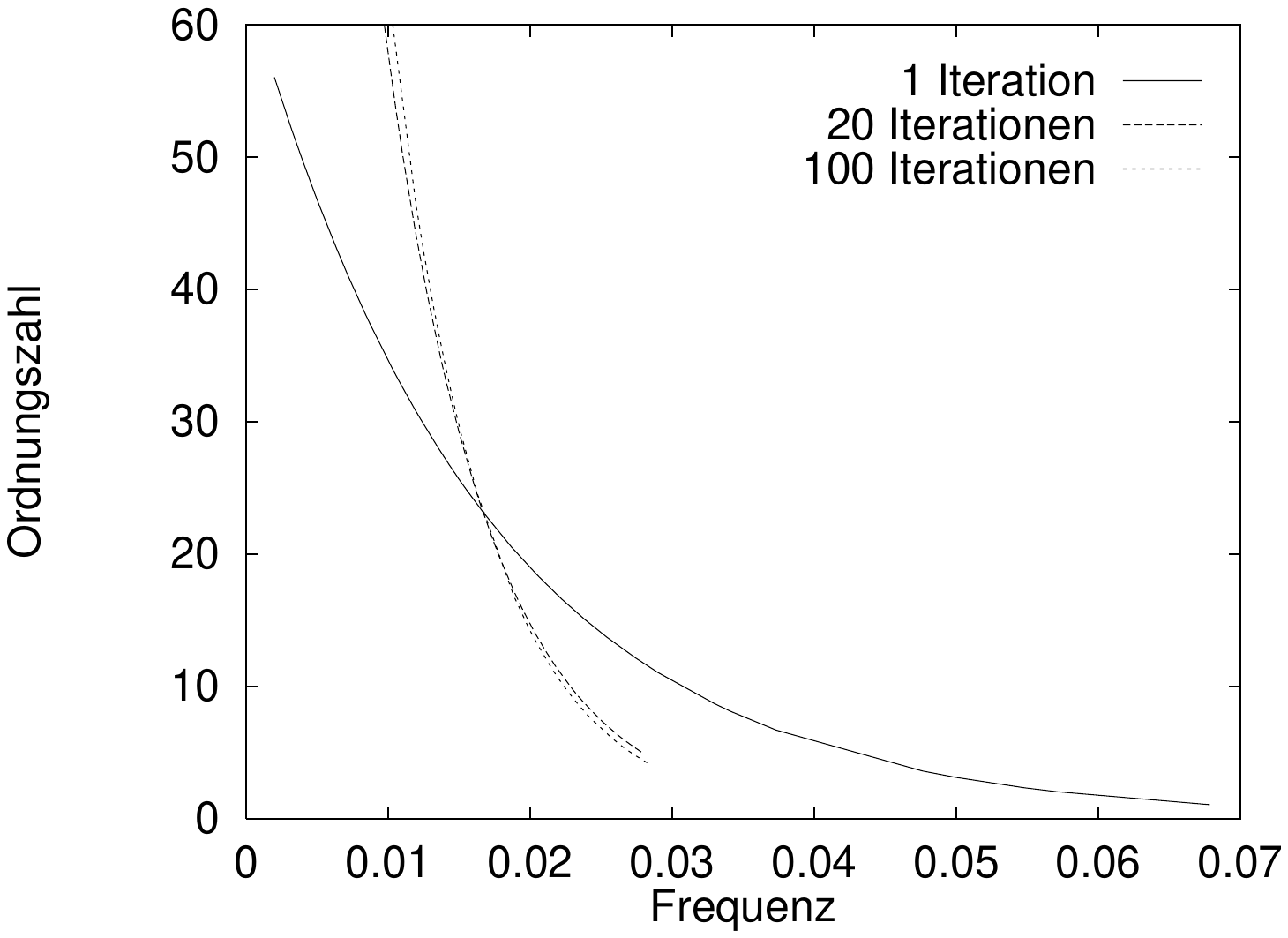}
  \end{center}
  \mycaption{Belastungsverteilung eines quadratischen Rasters}
  {Belastungsverteilung (links) und exponentielle Verteilungsfunktion
    (rechts) des quadratischen Rasters f"ur den einfachen
    Warshall-Floyd-Algorithmus und den Random-Warshall-Floyd-Algorithmus 
    mit 20 und 100 L"aufen. Bereits nach 20 Iterationen "andert sich an der
    Verteilungsfunktion kaum etwas. Die Verteilung f"allt nach dem
    Random-Warshall-Floyd-Algorithmus insgesamt breiter und gleichm"a"siger
    aus. Sie nimmt eher einen sigmoiden Verlauf als einen exponentielle Form
    an.}
  \label{fig:belastverteilungquadrat}
\end{figure}

Durch einen Trick l"a"st sich dieser Artefakt korrigieren:
\begin{quote}
  \glqq Der Warshall-Floyd-Algorithmus wird mehrmals hintereinander
  mit zuf"allig indizierten Knoten angewendet. Die Benutzungsh"aufigkeiten
  werden dabei f"ur jede Kante "uber alle Iterationen summiert.\grqq
\end{quote}
Dieser Random-Warshall-Floyd-Algorithmus  sorgt daf"ur, da"s
jede der gleichlangen Stre"cken mit einer gewissen  Wahrscheinlichkeit als
erste in der Suchfolge steht. Je mehr Wiederholungen mit
unterschiedlichen Indizierungen durchgef"uhrt werden, desto gleichm"a"siger
wird die Streckenbelastung verteilt. Der Fortschritt jeder Iteration $t$ des
Algorithmus kann durch die Fehlerfunktion 
\begin{equation}
  \epsilon_t = \sum_{i=1}^{E} \frac{\Delta F_i^{t-1}}{F_i^{t-1}} - 
  \sum_{i=1}^{E} \frac{\Delta F_i^{t}}{F_i^{t}}
\end{equation}
die auch als Abbruchkriterium des Random-Warshall-Floyd-Algorithmus dienen
kann, verfolgt werden. Abbildung \ref{fig:quadrat} zeigt die Belastung nach 20
(links) und 100 (rechts) Iterationen. W"ahrend die Unterschiede in der
Kantendicke noch deutlich erkennbar sind, "andert sich die
Exponentialverteilungsfunktion der Kantenbelastung nach 20 Iterationen kaum
(vgl. Abb. \ref{fig:belastverteilungquadrat}).

In nat"urlichen Netzwerken wie etwa dem Wegesystem von Martina
Franca kommen Kanten mit exakt gleicher L"ange sehr selten vor.
\label{pag:martinavert}
Da aber das
subjektive Empfinden der {\fussgaenger} f"ur Distanzen Toleranzen
von bis zu 15\% zul"a"st \cite{Humpert:1994}, werden auch in solchen
Wegesystemen ann"ahernd gleich 
lange Strecken "ahnlich h"aufig genutzt. Zur Bestimmung der Belastung in
diesen Netzen werden die Kanten in Klassen bestimmter L"ange aufgeteilt. Die
Kanten einer Klasse werden als gleichlang betrachtet und damit bei der
wiederholten Anwendung des Algorithmus auch ann"ahernd gleichm"a"sig belastet.
Der Vorgang sollte dabei mit verschiedenen Klassengrenzen durchgef"uhrt
werden, um einen Einflu"s der L"angeneinteilung auf das Ergebnis zu unterbinden.

\begin{figure}[tb]
  \begin{center}
    \leavevmode
    \includegraphics[width=0.8\textwidth]{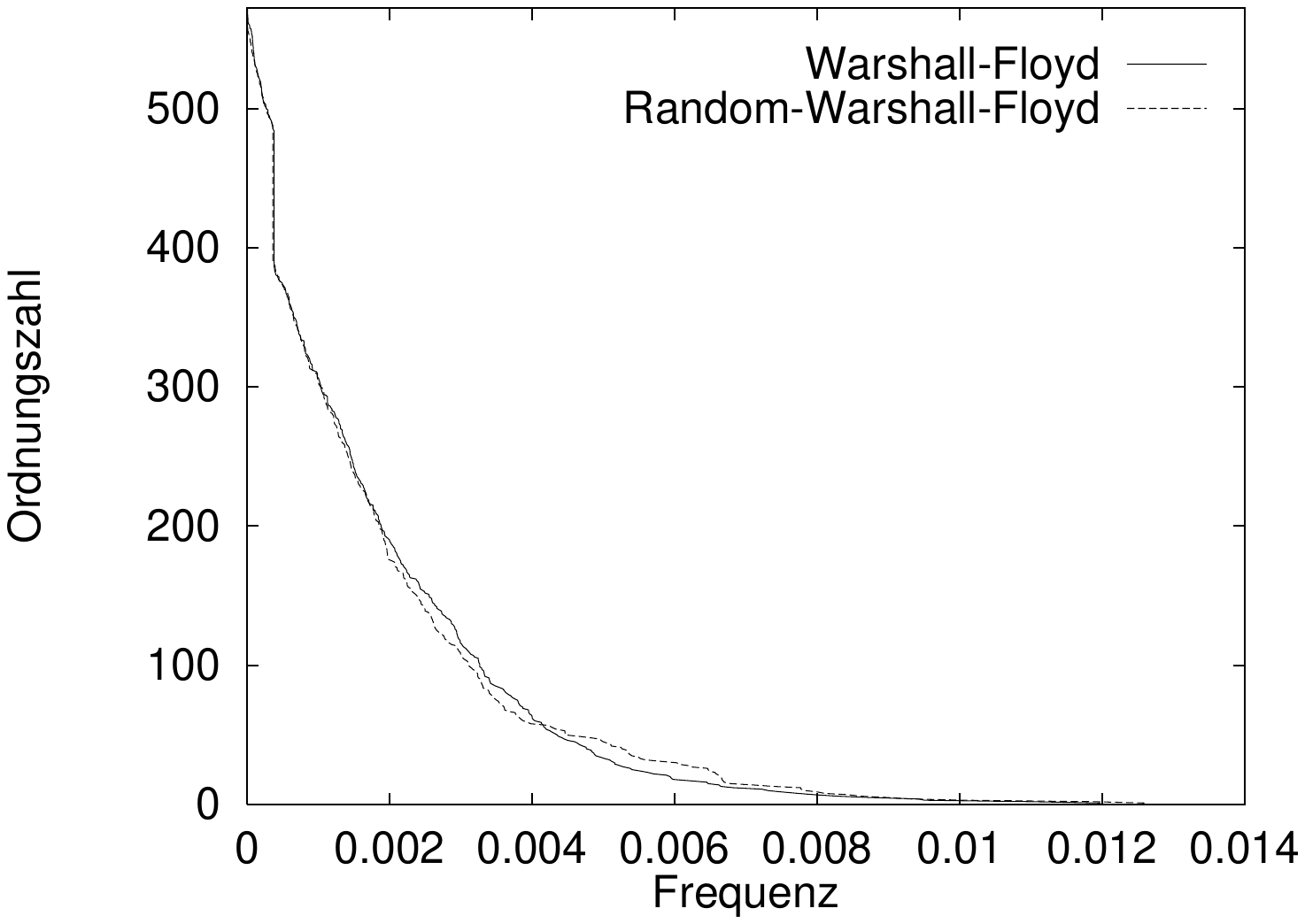}
  \end{center}
  \mycaption{Vergleich der Belastungsverteilungen aus unterschiedlichen
  Verfahren f"ur Martina Franca}
  {Belastungsverteilung von Martina Franca. Im Vergleich
    die Berechnung durch den einfachen Warshall-Floyd-Algorithmus mit den
    Koeffizienten der Exponetialverteilung $\alpha
    = 538.05$ und $k = 542.90$ (durchgezogene Linie) und des
    Random-Warshall-Floyd-Algorithmus (gestrichelte Linie) mit $\alpha =
    508.23$ und $k = 515.33$. Die Unterschied macht sich gerade an der Spitze
    der Rangliste bemerkbar. Die Verteilung nach dem
    Random-Warshall-Floyd-Algorithmus f"allt 
    f"ur die h"aufig benutzten Kanten gleichm"a"siger aus.}
  \label{fig:martinavert}
\end{figure}

Die Einteilung der Kanten des Wegesystems von Martina Franca in 15\% Klassen
ergibt nach 20 Iterationen des Random-Warshall-Floyd-Algorithmus eine
gleichm"a"sigere Verteilung der Belastung auf die h"aufig benutzen Stra"sen.
Die unterschiedliche Streckenbelastung wird im Vergleich der Abbildungen
\ref{fig:martina} und \ref{fig:martinarandom} deutlich. Die
Belastungsverteilungen f"ur beide F"alle sind in Abbildung
\ref{fig:martinavert} dargestellt.

Weitere Methoden zur Behandlung "ahnlich langer Strecken in Netzwerken wurden
in \cite{Hilliges:1995} und \cite{Nagel:1995} vorgeschlagen. Dabei werden
neben dem k"urzesten Weg, der nach \name{Warshall}{}-\name{Floyd}{},
\name{Deijkstra}{}-\name{Prim}{} oder \name{Kruskal}{} ermittelt
werden kann, auch die anderen vom Startknoten ausgehenden Kanten betrachtet.
Bei der Bestimmung der Wegl"ange wird dabei angenommen, da"s die
{\fussgaenger} "uber diese Kanten und dann weiter auf dem k"urzesten Wege zum
Ziel laufen. Die Belastungen werden den L"angenverh"altnissen entsprechend auf
die einzelnen Kanten portioniert.

\chapter{Entwicklung von Trampelpfaden}
\label{cha:entwicklungvontrampelpfaden}
\label{cha:trailformation}
Im Unterschied zu den 
im vorhergehenden Kapitel behandelten Wegenetzen
sind in der Natur h"aufig Wegesysteme anzutreffen, die einen Kompromi"s
zwischen der Gesamtl"ange eines Netzwerkes und dem (Material-)Aufwand f"ur die
Wegstrecken schlie"sen. Man spricht hierbei von \hi{Minimalen Umwegen}, da die
Wege zwischen jeweils zwei Knoten l"anger sein k"onnen als ihre direkten
Verbindungen.  Die Summe aller Strecken ist jedoch k"urzer als die
Gesamtl"ange eines \hi{vollst"andigen Graphen}, bei dem es von jedem Knoten
zu jedem anderen einen direkten Weg gibt \cite{Otto:1991}.

Viele Transportsysteme in der Natur weisen diese Form auf. Dazu geh"oren zum
Beispiel die elektrische Entladung in Blitzen, Versorgungssysteme in Pflanzen
und Verkehrswegenetze von Tieren und Menschen.  Die Ursache der Entstehung ist
bei allen Systemen gleich: F"ur das zu transportierende Teilchen oder das
Individuum, das zu seinem Zielknoten laufen will, ist es einfacher, d.h.\ 
weniger anstrengend, bereits existierende Spuren zu benutzen, anstatt eigene
Pfade zu produzieren. Aus diesem Unterschied ergibt sich die Bereitschaft zu
Umwegen.

Minimale Umwege Systeme sind auch in {\fussgaenger}anlagen von gro"sem
Interesse, da einerseits Verbindungswege in ausreichendem Umfang zur
Verf"ugung gestellt werden sollen, andererseits nicht die ganze Fl"ache der
Anlage durch Fu"swege zerschnitten werden darf. Im Gegensatz zu einem Ger"ust
eines Netzwerkes, das die kleinste m"ogliche Gesamtl"ange der Verbindungswege
aufweist, erf"ullt ein System minimaler Umwege besser die Bed"urfnisse des
{\fussgaenger}verkehrs.
      
\begin{figure}[tb]
  \begin{center}
    \leavevmode
    \includegraphics[width=0.9\textwidth]{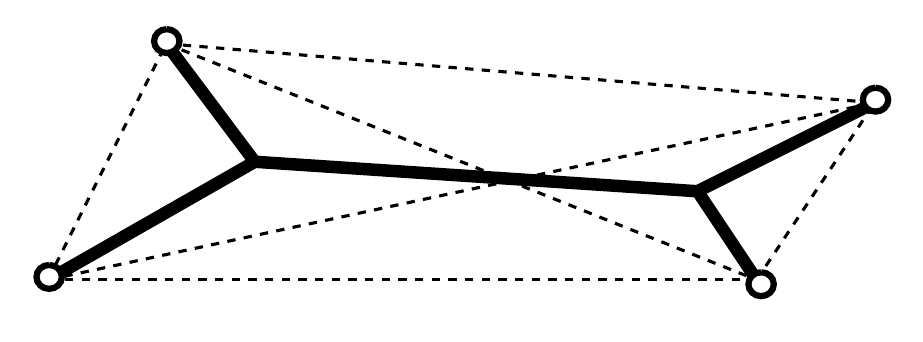}
  \end{center}
  \mycaption{Schematische Darstellung eines Minimale-Umwege-Systems}
  {Schematische Darstellung eines Minimale-Umwege-Systems mit vier Knoten
    (dicken Linien). In nat"urlichen Systemen 
    weisen die Gabelungen kurvige Verl"aufe auf. Die gestrichelten Linien
    zeigen das Direktwegesystem zum Vergleich.}
  \label{fig:bundle}
\end{figure}

W"ahrend das Ausd"unnen eines vollst"andigen Graphen zu einem Ger"ust durch
Suchalgorithmen \cite[S. 423ff]{Sedgewick:1992} bewerkstelligt werden kann, erweist
sich die Konstruktion von Systemen mit \hi{minimalen Umwegen} als
komplizierter: sie enstehen durch die \hi{B"undelung von Verbindungen}, die
nahe beieinander laufen. Im Gegensatz zur Konstruktion eines \hi{Ger"ustes}
m"ussen hierbei zus"atzliche Knoten in das Netzwerk eingef"ugt werden.  Die
\hi{Gesamtl"ange} des Systems wird durch gemeinsam genutzte Verbindungen
reduziert. Bei einzelnen Verbindungen zwischen zwei Knoten werden dagegen
l"angere Wege in Kauf genommen. In Abbildung \ref{fig:bundle} ist ein
\hi{minimales Umwegesytem} f"ur vier Knoten schematisch dargestellt.

Die Konstruktion solcher Wegesysteme kann mit einem sehr arbeitsaufwendigen,
mechanischen Fadenmodell durchgef"uhrt werden \cite{Kolodziejczyk:1991}: Dabei
werden die Knoten eines Wegesystems im Modell mit F"aden aus Baumwolle oder
nat"urlicher Seide verbunden, deren L"ange gr"o"ser als der Abstand zwischen
den Befestigungspunkten ist. Die "Uberl"ange der F"aden entspricht dabei der
maximal erlaubten L"ange eines Umwegs. Werden die F"aden befeuchtet, b"undeln
sie sich und kleben aneinander. Die dadurch enstehende Struktur bleibt selbst
nach dem Trocknen fest.

In der Simulation k"onnen Wegesysteme, die den {\fussgaenger}bed"urfnissen
entsprechen, von diesen selbst erzeugt werden. Wie bereits in Abschnitt
\ref{sec:orientierung} erw"ahnt wurde, ist es auch f"ur {\fussgaenger}
einfacher, existierende Pfade zu benutzen, als neue anzulegen. Unter den
folgenden Annahmen kann dieses Verhalten in das Soziale-Kr"afte-Modell der
{\fussgaenger}bewegung implementiert werden:
\begin{itemize}
\item Die {\fussgaenger} bevorzugen auf bereits existierenden Pfaden zu
  laufen. Auf Pfaden m"ussen sie sich nicht st"andig neu orientieren. 
\item Spuren, die h"aufig genutzt werden, werden damit breiter, auff"alliger
  und damit noch attraktiver f"ur andere {\fussgaenger}.
\item Selten benutzte Pfade verschwinden wieder.
\end{itemize}

Im Modell wird der Untergrund, auf dem die {\fussgaenger}
laufen, durch ein zeitabh"angiges Potential $\Utr(\vec{r},t)$ und die
neuproduzierten Fu"sspuren 
\begin{equation}
  \label{def:markers}
  Q_\beta(\vec{r},t) = -q\,\exp\left(-\|\vec{r}_\beta(t)-\vec{r}\|/\gamma\right)   
\end{equation}
dargestellt. Die Parameter $q > 0$ und $\gamma > 0$ geben die Form der des
"`Fu"sabdruckes"' an, der zur Zeit $t$ vom {\fussgaenger} $\beta$ gesetzt wird.  
Die Spuren entsprechen dabei
negativen Werten des Potentials. Die Dynamik des Spurpotentials ist dabei durch
\begin{equation}
  \label{dyn:trail}
  \frac{d\Utr}{dt} = -\frac{1}{T}\,\Utr + \sum_\beta Q_\beta
\end{equation}
mit der Zerfallskonstante $T$ gegeben. Anfangs ist das Spurpotential "uberall
Null. Bei jedem Zeitschritt werden dann die Fu"sspuren zu dem Potential
addiert. Gleichzeitig zerfallen die bereits existierenden Spuren um den Faktor
$dt/T \equiv \Delta t/T$. Die Spuren auf dem Grund erzeugen eine attraktive soziale
Kraft der Form 
\begin{equation}
  \Ftr = - \nabla \Utr
\end{equation}
die zu den anderen Termen in (\ref{tot_soz_kraft}) eingef"ugt wird.

\begin{figure}[tb]
  \begin{center}
    \leavevmode
    \includegraphics[width=0.9\textwidth]{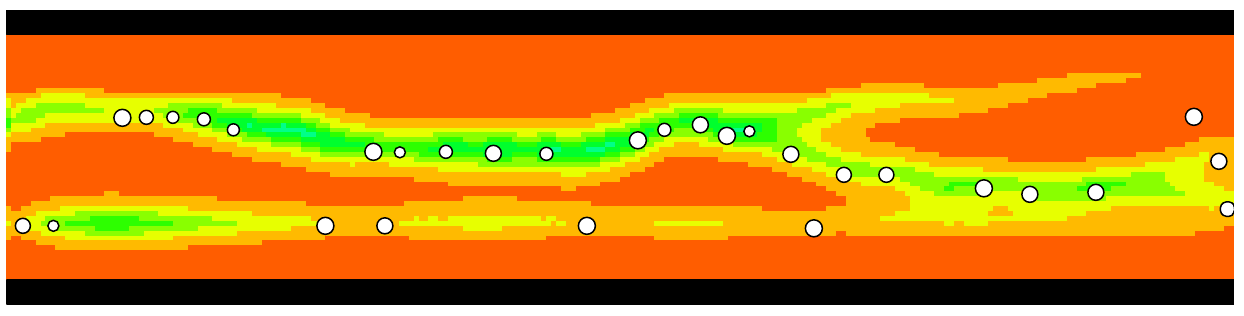} 
  \end{center}
  \mycaption{Spurbildung eines {\fussgaenger}stroms in einem Korridor \farbe}
  {Die {\fussgaenger} starten am rechten Ende
    des Korridors und laufen geradeaus zum anderen Ende. Zus"atzlich zum
    repulsiven Einflu"s der anderen {\fussgaenger} und der W"ande werden sie
    von den existierenden Pfaden angezogen. Das Simulationsergebnis h"angt
    sehr empfindlich vom {\fussgaenger}aufkommen, der Anziehungskraft der
    Pfade und ihrer Zerfallszeit ab.}
  \label{fig:trailkorridor}
\end{figure}

Dadurch erfahren die {\fussgaenger} zus"atzlich zu den Kr"aften der anderen
{\fussgaenger} und ihrer Umgebung eine Anziehung auf bereits existierende
Pfade. In Abbildung \ref{fig:trailkorridor} ist die B"undelung der Trampelpfade
eines {\fussgaenger}stromes in einem Korridor dargestellt. Obwohl die
{\fussgaenger} die ganze Breite des Korridors nutzen k"onnten, entstehen
durch die Wechselwirkung mit den Fu"sspuren auf dem Untergrund schmale
Trampelpfade. 

Nach einem "ahnlichen Prinzip arbeitet das in
\cite{SchweitzerLaoFamily} vorgestellte active-walker-Modell: In Anlehnung an
das Prinzip der Chemotaxis\footnote{Eine Orientierungsbewegung von Tieren und
  Pflanzen, die durch chemische Reizungen ausgel"ost wird. Einige Ameisenarten
  orientieren sich an vorher gesetzten lokalen Duftmarken.} der Ameisen werden
dort die Bewegungem von Teilchen 
(active walker) simuliert. Diese bewegen sich auf einem vorgegebenen
Raster und bilden Spuren, indem sie an ihren momentanen Positionen
Markierungen setzen. Der Wechsel zu einem Nachbarknoten geschieht dabei mit
einer gewissen durch die lokalen Markierungspunkte bestimmten
Wahrscheinlichkeit. Die Spuren verschwinden wieder, wenn sie nicht
regelm"a"sig durch neue Markierungen aufgefrischt werden.

\begin{figure}[tp]
  \begin{center}
    \leavevmode
    \includegraphics[width=0.45\textwidth,clip]{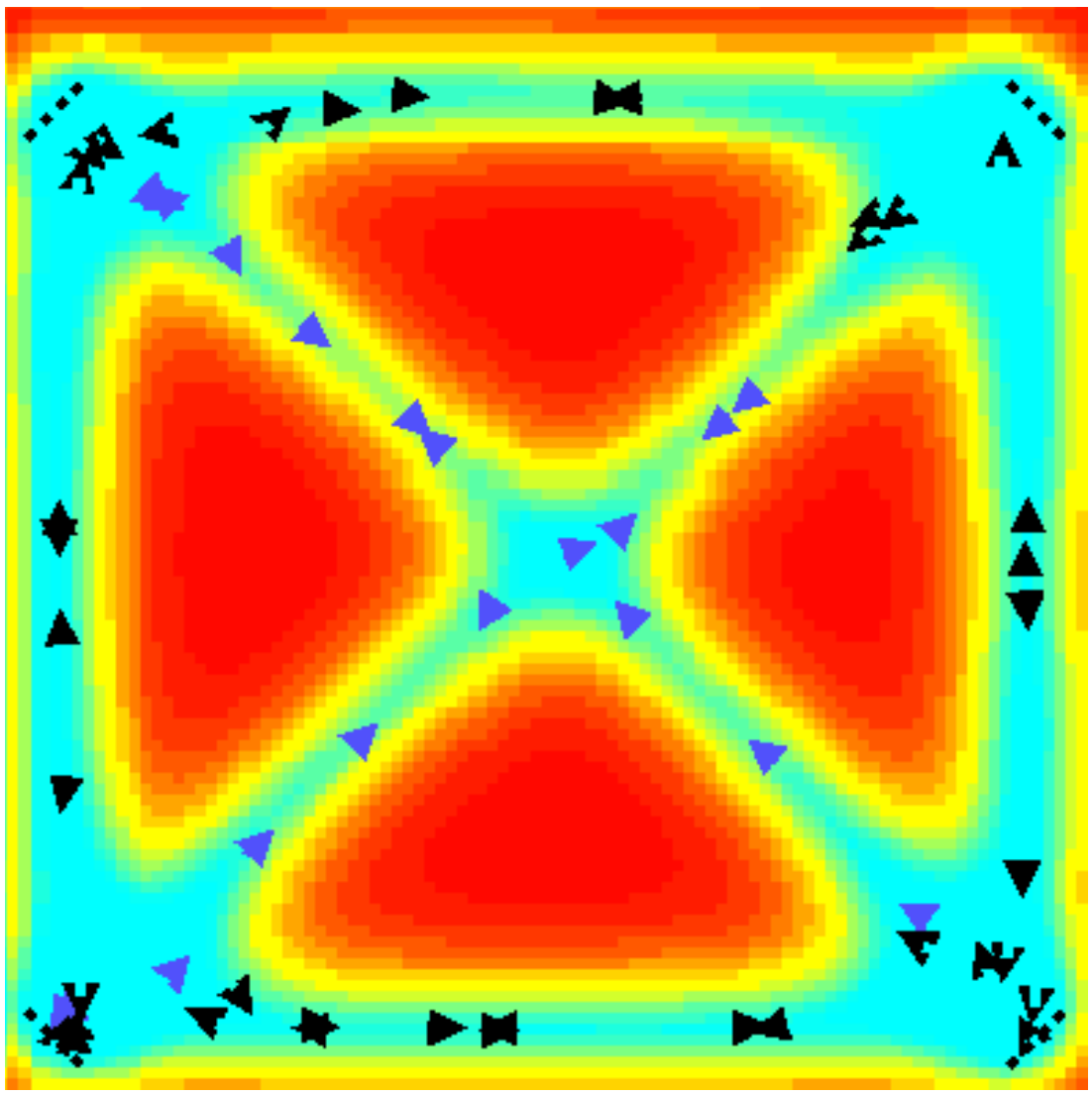} \hspace{0.5cm}
    \includegraphics[width=0.45\textwidth,clip]{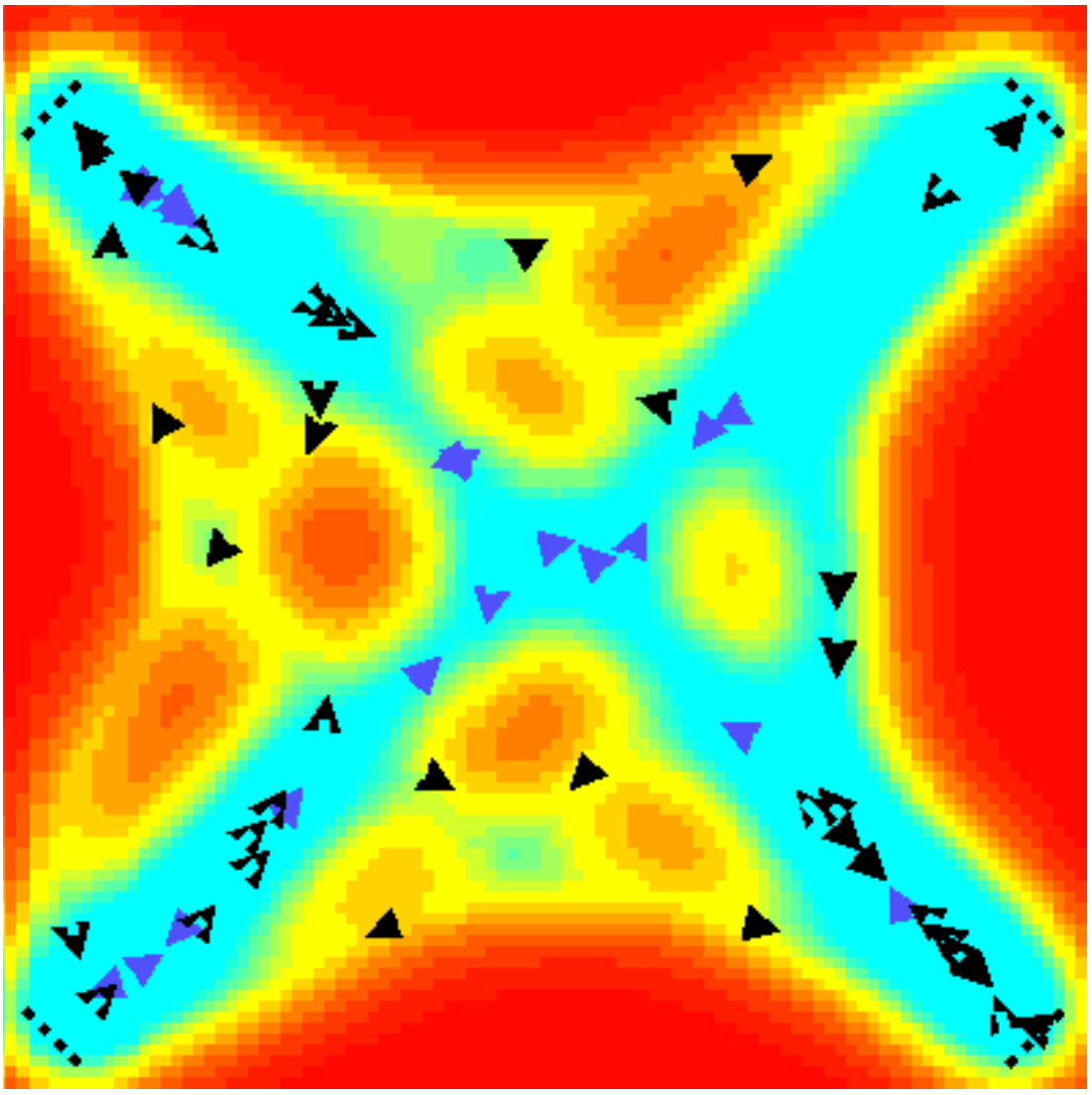} \\[0.5cm]
    \includegraphics[width=0.45\textwidth,clip]{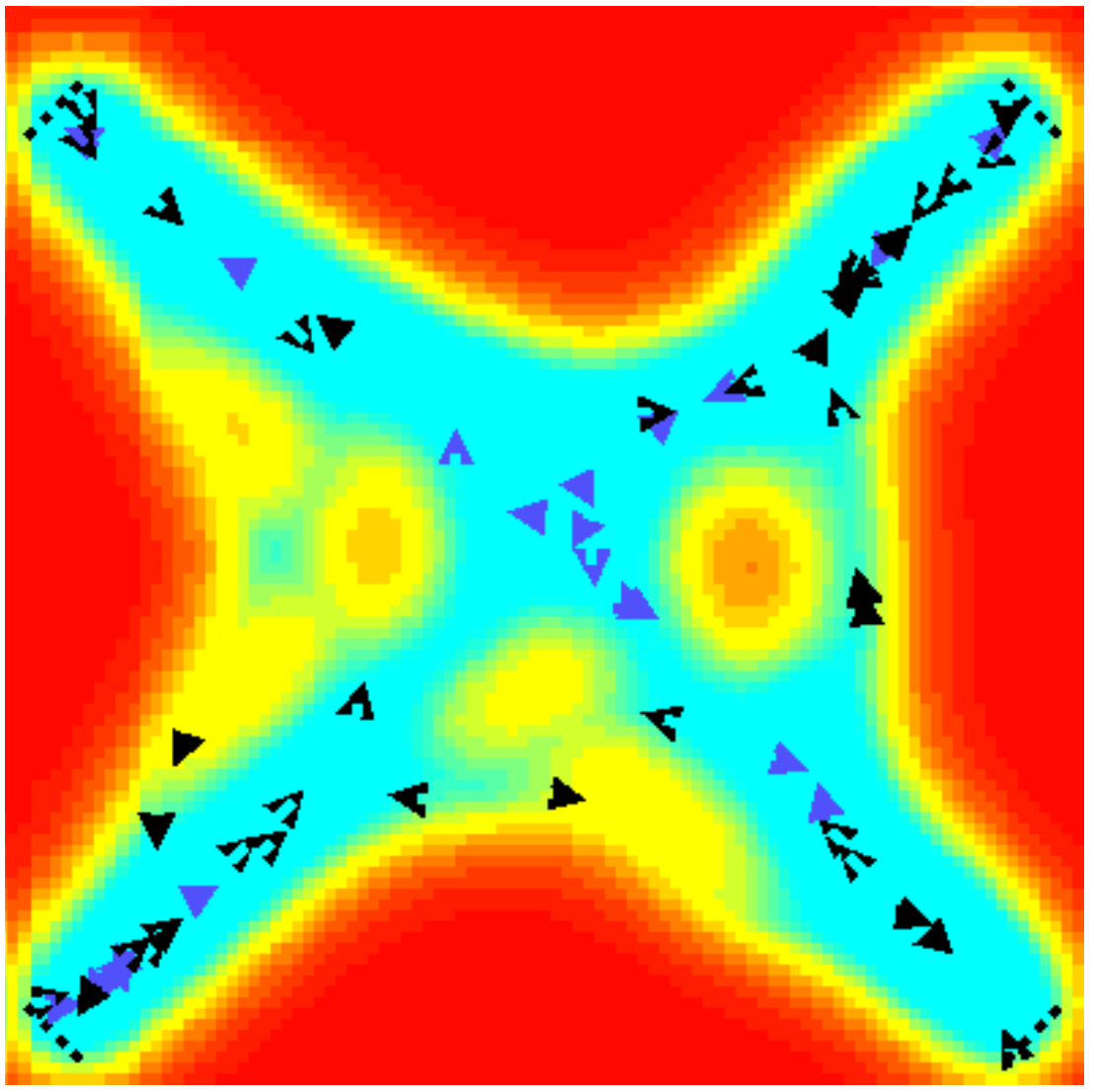} \hspace{0.5cm}
    \includegraphics[width=0.45\textwidth,clip]{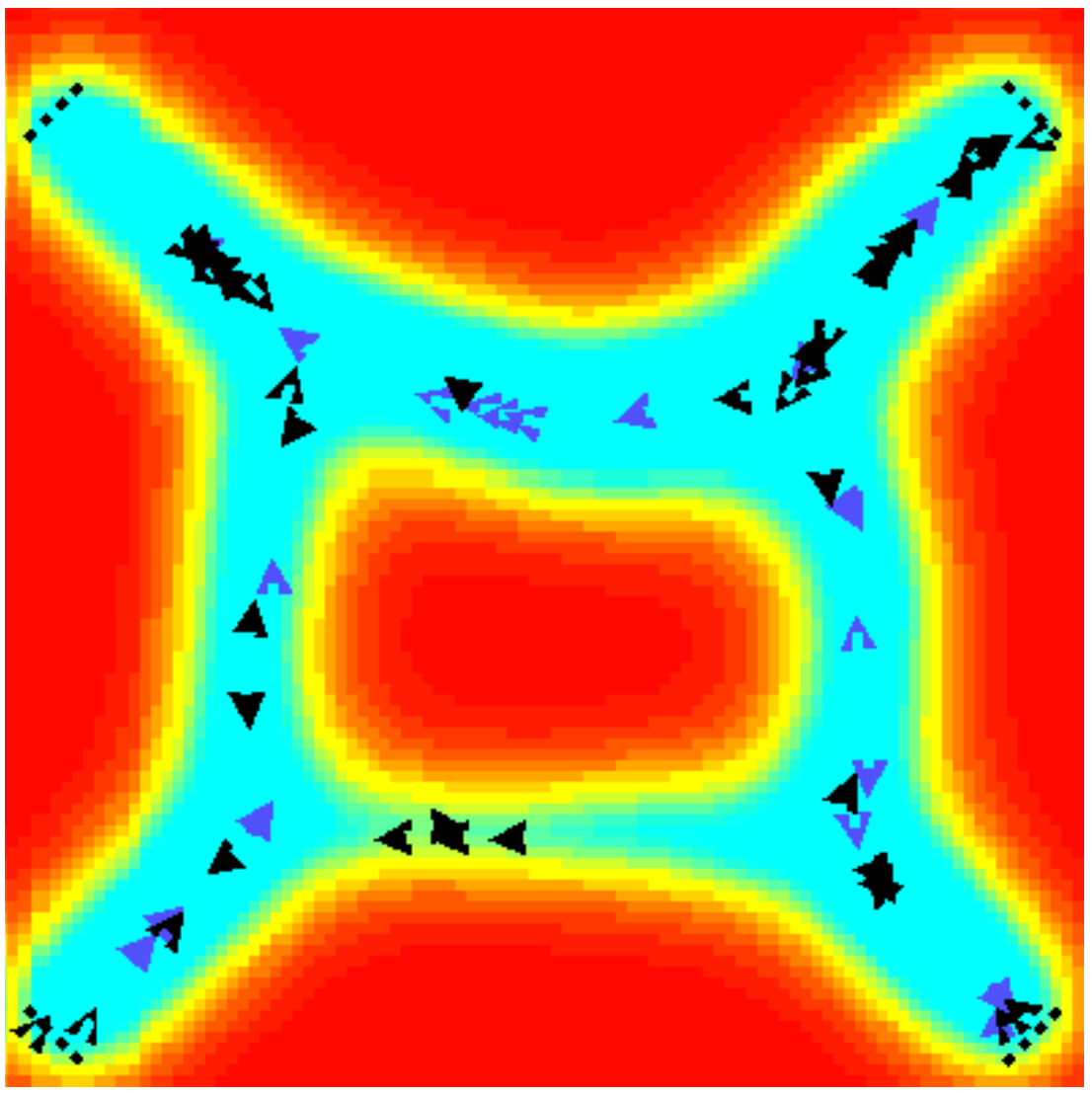} 
  \end{center}
  \mycaption{Entstehung eines Trampelpfadesystems mit vier Knoten \farbe}
  {Simulation der Entstehung eines Wegesystems
    mit vier Knoten durch die {\fussgaenger}dynamik. Von jedem Knoten laufen
    {\fussgaenger} zu allen anderen Knoten. Um den Effekt der Wechselwirkung
    mit dem Untergrund zu untersuchen, finden in diesem Beispiel keine
    Wechselwirkungen zwischen den {\fussgaenger}n statt. Der Proze"s startet
    mit einem vorgegebenen, vollst"andig verkn"upften Netzwerk (links oben).
    Mit der Zeit verbreitern sich die Wege, und die diagonalen Verbindungen
    gewinnen zunehmend an Fl"ache und Bedeutung (rechts oben). In diesem
    Verlauf sind die vier Inseln irgendwann fast v"ollig verschwunden (links
    unten). Es ensteht ein System mit minimalen Umwegen (rechts unten), das
    ausschlie"slich dreiarmige Knoten enth"alt.}
  \label{fig:trailquada}
\end{figure}

Die resultierenden Wegesysteme weisen die typischen Merkmale von \hi{minimalen
  Umwegen} auf, die B"undelung einzelner Pfade und den kurvigen Verlauf. Die
in Abbildung \ref{fig:trailquada} dargestellte Simulation eines Systems mit
vier Knoten zeigt, wie die vierarmige Kreuzung durch dreiarmige Knoten ersetzt
wird, die in nat"urlichen Transportsystemen auch am h"aufigsten angetroffen
werden \cite{Schaur:1994}.

Das Ergebnis einer Simulation h"angt sehr empfindlich von den Parametern ab.
Diese k"onnen von den Planern in gewissen Bereichen frei gew"ahlt werden: Je
gr"o"ser die Anziehungskraft der Pfade ausf"allt, desto mehr Umwege laufen die
{\fussgaenger}, und umso weniger Fl"ache der Anlage wird f"ur Fu"swege
verbraucht. Durch die Wechselwirkungen zwischen den {\fussgaenger}n
verbreitern sich die Wege bei hoher Fu"sg"angerdichte. Die daraus ebenfalls
resultierende h"aufigere Nutzung sorgt f"ur eine besonders starke Ausbildung
diese Wege.  Je mehr {\fussgaenger} von einem bestimmten Knoten zu einem
anderen laufen, desto geradliniger wird der Verbindungsweg.

In das Modell zur Spurbildung lassen sich auch durch Steigung im
Gel"ande oder durch das "Uberwinden von Hindernissen hervorgerufene Einfl"usse
einbeziehen. Die Eigenschaften des Untergrunds m"ussen dabei nicht
realistisch sein, sie orientieren sich vielmehr an den Zielsetzungen der
Planung einer {\fussgaenger}anlage.

Daher ist dieses Modell nicht nur geeignet, das Verhalten von {\fussgaenger}n 
im Gel"ande zu beschreiben, es kann vielmehr unter Ber"ucksichtigung der 
Bed"urfnisse und Eigenschaften des {\fussgaenger}verkehrs einen wesentlichen
Beitrag zur Konstruktion bedarfsgerechter Wegesysteme leisten.

Die Modellierung und Simulation der
Spurbildung auf der Basis eines Soziale-Kr"afte-Modells wird in
\cite{Keltsch:1996} ausf"uhrlich behandelt.  

\chapter{Simulationsprogramm}
\label{cha:simulationsprogramm}
\label{cha:simulatioprogram}
Aufgrund der zwei Aspekte der Modellierung von {\fussgaengerstroeme}n wurde
eine Si"-mu"-la"-ti"-ons-Software entwickelt, die zum einen als Bestandteil von
CAD-Programmen f"ur Architekten und St"adteplanern
dienen kann, zum anderen aber auch ein eigenst"andiges Anwendungsprogramm
darstellt, mit dem sich Soziale-Kr"afte-Modelle entwerfen und simulieren
lassen.

Im Hinblick auf die Verbreitung der Theorie und m"oglicher Anwender des
Simulators in unterschiedlichen (weniger computer-begeisterten) Disziplinen
wurde die Software mit einer leicht zu bedienenden, grafischen
Benutzeroberfl"ache ausgestattet.
Damit bietet sie dem Benutzer M"oglichkeiten zur Beobachtung und
Steuerung des Ablaufs der Simulation. Die Simulationsergebnisse lassen sich als
Animation auf Video aufzeichnen oder als grafische Darstellungen zu Papier
bringen.

Die Modellspezifikation geschieht in einer eigens daf"ur entwickelten
Beschreibungssprache.

Das Simulationsprogramm wurde unter dem Betriebssystem UNIX und
X-Window/OSF-Motif entwickelt. Es l"a"st sich daher auf einer Vielzahl von
Computern installieren.

\begin{figure}[tb]
  \begin{center}
    \leavevmode
    \includegraphics[width=0.9\textwidth]{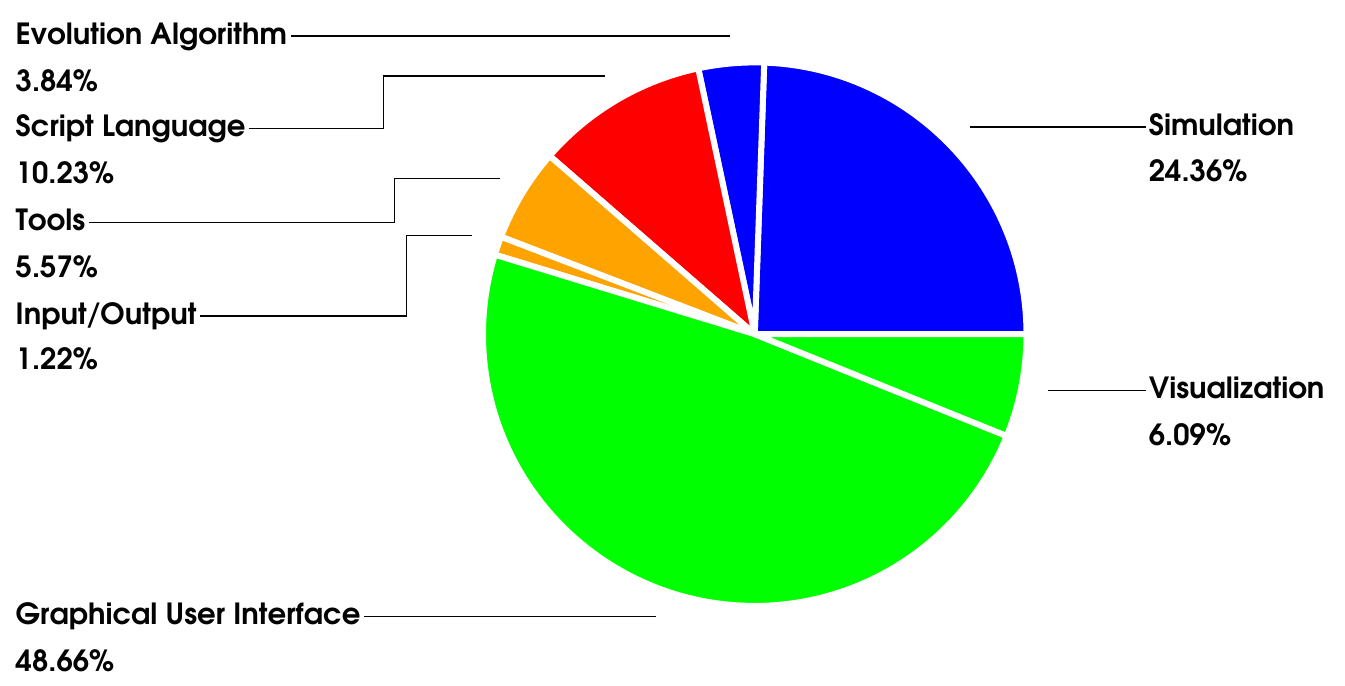}
  \end{center}
  \mycaption{Verteilung des Programmkodes auf die Funktionsbereiche des
  Simulators \farbe}
  {Das Simulationsprogramm bietet dem Benutzer (wie jede moderne Software)
    Funktionen zur interaktiven Modell-Spezifikation und zur Beobachtung und
    Steuerung des Ablaufs der Simulation. Zu jedem Zeitpunkt k"onnen 
    aus den Ergebnissen grafische Darstellungen erzeugt werden. Diese
    Funktionen sind in den Bereichen Grafische-Benutzerschnittstelle (Graphical
    User Interface) und Visualisierung (Visualization) zusammengefa"st, die
    insgesamt "uber 50\% der mehr als $17\,000$ Zeilen Programmcode in C++ und
    Motif/UIL einnehmen. Im Vergleich dazu ben"otigen die Implementierung der
    numerischen Berechnungen (Simulation) und der Evolution"aren Optimierung
    (Evolution Algorithm) nur ein Viertel der Programmzeilen. Der Rest
    verteilt sich auf den Sprachinterpreter (Script Language), verschiedene
    Werkzeuge (Tools) und die Ein- und Ausgaberoutinen (Input/Output).} 
  \label{fig:pichart}
\end{figure}

\section{Struktur des {\fussgaenger}modells}
\label{sec:struktur}
\label{sec:structureprog}
Das Modell der {\fussgaenger}dynamik besteht aus verschiedenen Komponenten.
Dazu geh"o"-ren die "`{\fussgaenger}"' und "`Korridore"', in denen sie
sich aufhalten. Die Begrenzungen und geometrische Form der Korridore sind
durch "`W"ande"' und "`Hindernisse"' bestimmt. Alle diese Komponenten
beschreiben ein "`Stockwerk"', von dem es auch mehrere in dem  Geb"aude
geben kann. Die Verbindungen zwischen den Korridoren werden als "`Tore"'
bezeichnet. Diese Tore steuern den "Ubergang der {\fussgaenger} von einem
Korridor zum anderen und geben ihnen die  Zielrichtung vor.

Zwischen den Komponenten k"onnen Verbindungen aufgebaut werden.
Objekte, die von anderen beeinflu"st werden, oder selber einen Einflu"s
auf andere aus"uben, sind miteinander verkettet.
 Eine Verbindung bedeutet,
da"s eine Komponente die andere kennt, sie enth"alt, oder Einflu"s auf diese
aus"ubt. Die Art der Wechselwirkungen ist dabei durch die Typen der
Komponenten bestimmt.

Ein Beispiel: Ein Korridor ist mit den {\fussgaenger}n verbunden, die sich in
ihm aufhalten. Zus"atzlich besteht ein Korridor aus Hindernissen, zu denen
ebenfalls Verbindungen existieren. Zwischen {\fussgaenger}n und Hindernissen
besteht eine indirekte Verbindung "uber den Korridor, mit dem sowohl
{\fussgaenger} als auch Hindernisse verbunden sind. Ein {\fussgaenger}
erf"ahrt die Wirkung der im Korridor wirkenden sozialen Kraft, die 
sich wiederum aus der  Summe der Kr"afte aller im Korridor befindlichen
{\fussgaenger} und Hindernisse zusammensetzt.

Die Eigenschaften einesObjekts, etwa die Wunschgeschwindigkeit der
{\fussgaenger} oder die St"arke des Absto"sungseffektes einer Wand, werden
durch sogenannte Simulations"-parameter definiert. Den Parametern werden in der
Modellspezifikation durch die Beschreibungssprache Werte zugewiesen.

\section{Objekte und Klassen}
\label{sec:objekteundklassen}
Das Konzept der Objekte erfordert eine spezielle Betrachtungsweise
der Aufgabe: Ein Problem wird in unabh"angige Objekte 
zerlegt. Gleichartige Objekte werden in Klassen zusammengefa"st, die die
Eigenschaften ihrer Mitglieder beschreiben. 

Gerade die Zusammensetzung einer Klasse aus mehreren Teilen stiftet h"aufig
Verwirrung. Ein kleines (biologisches) Beispiel soll den Unterschied von
einer \hi{Ist-Ein-} zu einer \hi{Besteht-Aus-Beziehung} zwischen zwei
Klassen verdeutlichen:
\begin{eqnarray}
  & & \mbox{Ein Hund {\em ist ein} S"augetier.}
  \label{hundist} \\
  & & \mbox{Ein Hund {\em besteht aus} Beinen, Rumpf, Kopf und einer
    Wirbels"aule.} 
  \label{hundhat}
\end{eqnarray}
Der erste Satz (\ref{hundist}) sagt aus, da"s Hunde unter den Oberbegriff
S"augetier fallen. 
Diese Ist-Ein-Beziehung dr"uckt die Spezialisierung einer Klasse aus. Eine
neue Klasse "ubernimmt alle Eigenschaften seines Partners und f"ugt weitere
hinzu. Man nennt diesen Vorgang Vererbung oder Ableitung. Die
Ist-Ein-Beziehung gilt nur in einer Richtung, so da"s von einer
Klassenhierarchie gesprochen werden kann. In einigen Sprachen ist auch die
\hi{mehrfache Vererbung} erlaubt. So kann ein Hund zum Beispiel von den
S"augetieren und den Haus"-tieren abgeleitet werden. Diese Hierarchie ist
jedoch nicht auf eine einfache Vererbung zur"uckzuf"uhren, da nicht alle
S"augetiere Haustiere sind und umgekehrt.

Der zweite Satz (\ref{hundhat}) dr"uckt eine Besteht-Aus-Beziehung zwischen
Hunden und 
K"opfen aus. Sie erkl"art die Zusammensetzung eines Hundes. Er hat
einen Kopf, ist aber keiner. Eine Ist-Ein-Beziehung ist nicht m"oglich, da
die Eigenschaften von K"opfen nicht auf (ganze) Hunde "ubertragbar sind. 
Verallgemeinert bedeutet eine Besteht-Aus-Beziehung die Verbindung zwischen
zwei Objekten, deren Zustands"anderungen von einander abh"angen.

Die Klasse der Wirbeltiere hat als besonderes
Merkmal die Besteht-Aus-Beziehung zu der Klasse Wirbels"aule, die auch an alle
nachfolgenden Klassen wie der Klasse der S"augetiere und der Hunde
weitergegeben wird. Die Besteht-Aus-Beziehung zur Klasse der Wirbeltiere
sollte daher nicht in der Definition der Hunde, sondern m"oglichst weit oben
in der Hierarchie definiert werden. 

\section[Objektorientierte Modellspezifikation]{Objektorientierte
  Modellspezifikation, objektorientierte Modell"-implementierung}


Prinzipiell mu"s bei Simulationen zwischen der Modellspezifizierung und der
Modellimplementation unterschieden werden \cite{Schmidt:1995}.
 Ein Simulationsprogramm kann
objektorientiert geschrieben sein, auch wenn die Modellspezifikation keine
Objekte kennt. Genauso erfordert ein objektorientiertes Modell keine
objektorientierte Implementation des Simulationsprogramms.
Trotzdem gibt es Zusammenh"ange, die in der Betrachtungsweise des
Modells begr"undet sind.  

Zur Implementierung des in dieser Arbeit vorgestellten Simulationsprogramms der
{\fussgaenger}dynamik  wurde die Programmiersprache C++
verwendet und ihre Unterst"utz"-ungs"-m"oglichkeiten zur objektorientierten
Programmierung reichlich ausgenutzt.  Teilweise werden Teile der
Programmstruktur sowohl f"ur die eigentliche Simulation, als auch zur
Steuerung des Programms, zur animierten grafischen Darstellung auf dem
Bildschirm und zur Erstellung der Grafiken auf Papier verwendet.
Viele Eigenschaften objektorientierter Programmiersprachen sind nicht sinnvoll
auf die objektorientierte Modellspezifikation "uber"-tragbar und werden daher
nicht weiter erl"autert. Dazu geh"oren zum Beispiel die Funktionsschablonen
(templates), Verkapselung der Objekte gegen unerlaubten Zugriff auf die
Eigenschaften von au"sen und die Beschr"ankung des Kommunikationsmechanismus
auf Botschaftenaustausch \cite{Schmidt:1995}\cite{Stroustrup:1993}.

Objektorientierte Modellspezifikation sollte 
die folgenden Eigenschaften aufweisen\cite{Schmidt:1995}:
\begin{itemize}
\item Komponentenweiser Aufbau: Ein reales System besteht aus einzelnen
  Komponenten, die auch f"ur sich allein ablauff"ahig sein m"ussen. Die
  Komponenten bilden damit wieder eigenst"andige Modelle, die auf einer
  h"oheren Hierarchieebene zu umfangreicheren Modelle zusammengesetzt werden
  k"onnen.

  Die Reihenfolge, in der Komponenten spezifiziert werden, darf keine
  Rolle spielen. Die Forderung der Reihenfolgeunabh"angigkeit setzt sich auch in
  der Durchf"uhrung der Simulation der Objekte fort und stellt dadurch hohe
  Anforderungen an die Modell"-implementation (vgl.\ Abschn.\
  \ref{sec:gleichzeitigkeit}).
\item Zustandsorientierte Betrachtungsweise: Eine Komponente befindet sich zu
  jeder Zeit in einem Zustand, der durch die momentanen Werte seiner Zustandsvariablen
  bestimmt wird. Das zeitliche Verhalten ist durch eine Dynamikbeschreibung
  (Aktualisierungsfunktion) festgelegt, die vom eigenen Zustand sowie von den
  Zust"anden anderer Komponenten abh"angt.
\end{itemize}

\subsection{Elemente des Simulators}
In dem Simulationsprogramm erfolgt der Aufbau der Objekte nach den beiden oben
genannten Punkten. Dabei stammen alle Simulationsobjekte von der Klasse {\em
  SimulObject} ab, die die Struktur vorgibt. 

Jedes Objekt besteht aus einem Satz von Parametern, Zustandsvariablen,
Verkettungslisten und Methoden. Die Parameter entsprechen den Gr"o"sen der
Soziale-Kr"afte-Definitionen des Modells 
(vgl.\  Kap.\ \ref{cha:pedestrianmodell}). 
Die Zustandsvariablen der {\fussgaenger}, zum Beispiel, sind ihre momentane
Position und Geschwindigkeit.

Anhand der Verkettungslisten werden die Abh"angigkeiten zwischen den Objekten
definiert. Die meisten Verkettungen entsprechen dabei bestimmten sozialen
Kr"aften, die auf die Objekte wirken (vgl.\ Abschn.\ \ref{sec:struktur}).

Die Methoden stellen Unterprogramme f"ur verschiedene Aufgaben dar, etwa
zur Aktualisierung der Zustandsvariablen. Die Funktion der f"unf
Basismethoden wird im n"achsten Abschnitt erl"autert.

Neben den Objektdefinitionen enth"alt der Simulator eine Komponente zur
Analyse der Spezifikationsdatei. Diese wird zu Beginn der Simulation
eingelesen und Modellspezifikation  durchgef"uhrt.

Ferner gibt es einen Verwalter, der die Aktualisierung der Objekte durch eine
Priorit"atswarteschlange regelt (vgl.\ Abschn.\ \ref{sec:gleichzeitigkeit}).   

\subsection{Methoden der Simulationsobjekte}
\label{sec:methodendersimulationsobjekte}
Alle Objekte, die der Simulator zur Verf"ugung stellt, stammen von der Klasse
{\em SimulObject} ab, in der die folgenden Methoden definiert sind:
\begin{enumerate}
\item {\em Create}. Bei der Erzeugung eines Objektes wird der ben"otigte
  Speicherplatz angefordert und dem Objekt eine Speicheradresse
  zugeordnet. Damit kann das Objekt bereits mit anderen verkettet werden. Die
  Zustandsvariablen haben aber noch keine definierten Werte.  
\item {\em Initialize}. Die Parameter werden aufgel"ost. Dabei wird anhand des 
  Bezeichners eines Parameters im Geltungsbereich des Objektes nach einer
  Wertzuweisung gesucht.  Besteht die
  Wertzuweisung aus einem arithmetischen Ausdruck, etwa einer Zufallsfunktion,
  so wird der Ausdruck zu diesem Zeitpunkt berechnet.  
\item {\em Reference}. Die Methode zur Einrichtung einer Verkettung mit einem
  anderen Objekt pr"uft als erstes, ob eine Verkettung zu der Klasse des
  Partners definert ist. Ist das nicht der Fall, so wird der
  Verkettungsauftrag an die Vorfahrenklasse weitergegeben. 
\item {\em Update}. Die Methode erledigt die Aktualisierung eines
  Objekts. Dabei ber"ucksichtigt die Aktualisierungsfunktion die inneren
  Zust"ande sowie die Grundzust"ande von Objekten, zu denen eine Verkettung
  besteht.  
\item {\em Clock}. Die Methode besorgt die Takt- und Zeitfortschaltung, in der
  die aktuellen Zustandsvariablen von den Grundzustandsvariablen "ubernommen
  werden. 
\end{enumerate}
In der Vererbungshierarchie stellen alle Klassen diese f"unf Methoden zur
Verf"ugung. Dabei behandeln die Methoden einer Klasse auch nur die Instanzen
(Parameter und Zustandsvariablen),
die in der Klassendefinition neu hinzugekommen sind.
Vererbte Instanzen werden von den Methoden der Klassen behandelt, in denen sie
definiert wurden.
 
Die Reihenfolge der Aufrufe der Methoden {\em Create}, {\em Initialize} und 
{\em Update} beginnt bei der Basisklasse {\em  SimulObject} und geht
die Vererbungshierarchie abw"arts.  

Bei der Methode {\em Reference} verh"alt es sich genau umgekehrt. Hier wird
zuerst in der untersten (letzten) Klassendefinition ermittelt, ob f"ur eine
Verkettung mit dem Partner eine entsprechende Wechselwirkung definiert
ist. Wenn nicht, wird der Versuch zu einer Verkettung an die Methode der
Vorg"angerdefinition weitergegeben. 

Die einzelnen Teile der Methode {\em Clock} sind
reihenfolgenunabh"angig, da sie sich nur auf die Zust"ande beziehen, die in
der Klassendefinition eingef"uhrt werden.   

Die Programmiersprache C++ regelt die Erzeugung der Teile eines Objektes in
der richtigen Reihenfolge ({\em  Create}) eigenst"andig durch eine 
sogenannte {\em Constructor}-Methode, die f"ur jedes C++-Objekt definiert wird. 
Die Bearbeitungsreihenfolge f"ur die anderen Methoden dagegen mu"s explizit
implementiert werden. 

\section{Das Problem der Gleichzeitigkeit}
\label{sec:gleichzeitigkeit}
Ein Problem mit der Reihenfolge der Aktualisierung von Objekten ergibt sich,
sobald die Zustands"anderung eines Objekts von anderen abh"angt.  So erh"alt
man f"ur zwei Komponenten $a$ und $b$ mit den Zust"anden $a=1$, $b=1$ und den
Aktualisierungsregeln $a\leftarrow 2$, $b\leftarrow a+1$ das 
Ergebnis $a=2$, $b=3$, wenn $a$ vor $b$ berechnet wird. Bei vertauschter
Reihenfolge erh"alt man dagegen $a=2$, $b=2$.

\subsubsection{Bestimmung der Abh"angigkeiten}
\name{Rasmussen}{Steen} und \name{Barrett}{Christopher L.} schlagen in ihrer
Theorie der Simulationen \cite{RasmussenBarrett:1995} einen universalen
Simulator vor, der die Abh"angigkeiten zwischen den einzelnen Objekten
untersucht: Dabei werden den $n$ Objekten $S_i$ des Systems Z"ahler $q_i$
zugewiesen. Der Universalsimulator geht die einzelnen Objekte $S_i$ der Reihe
nach durch und testet, ob ihre Aktualisierung  von anderen Objekten abh"angt.
Existieren andere Objekte, von denen das Ergebnis bestimmt wird, die aber noch
nicht aktualisiert worden sind, ist diese Abh"angigkeit gegeben.

Bei Unabh"angigkeit wird die Aktualisierung von $S_i$ durchgef"uhrt
und der Z"ahler $q_i$ um Eins erh"oht. Im anderen Fall bleibt das Objekt
unver"andert. 
Danach geht der Simulator zum n"achsten Objekt $S_{i+1}$ "uber. Der Vorgang
wird solange wiederholt, bis alle Objekte aktualisiert worden sind.

Am Beispiel zweier Objekte $S_1$ und $S_2$ kann dieses Verfahren erl"autert
werden. Dabei h"angt $S_1$ von  $S_2$ ab. Der Simulator stellt am Objekt
$S_1$ eine Abh"angigkeit fest und geht zum n"achsten Objekt "uber. $S_2$ ist
unabh"angig und kann aktualisiert werden. Der Z"ahler $q_2$ wird um Eins
erh"oht. In der n"achsten Runde startet der Simulator wieder bei Objekt $S_1$,
das jetzt aktualisiert werden kann. 
F"ur die Z"ahler der Objekte wird auf diese Weise die Abh"angigkeit
\begin{eqnarray}
  q_1 & = & 1\, q_2 \nonumber \\
  q_2 & = & 0\, q_1
  \label{bespielzaehler}
\end{eqnarray}
festgestellt.
  
Die Anzahl der Schritte, die der Simulator zur Feststellung der
Abh"angigkeiten ben"otigt, ist durch 
\begin{equation}
  N = \sum_{i=0}^{n-1}(n-1)(n-i) = n(n+1)(n-1) = n^{\bf 3}-n
\end{equation}
nach oben begrenzt.

Der Universalsimulator berechnet in diesem Proze"s auch die \hi{Jacobi-Matrix
  der Aktualisierungsabh"angigkeiten} 
\begin{equation}
  \label{def:jacobian}
  Dq = \left(\frac{\partial q_i}{\partial q_j}\right)_{n\times n}
  = \left(\frac{\Delta q_i}{\Delta q_j}\right)_{n\times n}
\end{equation}
wobei die Ableitung $\partial q_i/\partial q_j$ angibt, wieviele
Aktualisierungsschritte von $S_j$ notwendig sind, um einmal $S_i$
aktualisieren  zu k"onnen. Werte gr"o"ser Null geben die Abh"angigkeit zweier
Objekte an. F"ur den Sonderfall $i=j$ gilt $\partial q_i/\partial q_j = 1$.
Ein System ist simulierbar, das hei"st reihenfolgenunabh"angig, wenn die
Jacobi-Matrix $Dq$ gleich der 
Einheitsmatrix ist.  Dies folgt direkt aus der Definition.

Das Beispiel ergibt die Jacobi-Matrix
\begin{equation}
  \label{beispiel_jacobi}
  Dq  = \left(
  \begin{array}{r@{\quad}l}
    \displaystyle \frac{\partial q_1}{\partial q_1}  
    &  \displaystyle \frac{\partial q_1}{\partial q_2} \\
    \displaystyle \frac{\partial q_2}{\partial q_1} 
    &  \displaystyle \frac{\partial q_2}{\partial q_2} 
  \end{array}
\right)
= \left(
\begin{array}{r@{\quad}l} 1  & 1 \\ 0 & 1 \end{array} \right) 
\end{equation}

Enth"alt $Dq$ Untermatrizen auf ihrer Diagonalen, so ist das System nicht
simulierbar. Aber durch die Zusammenfassung der Objekte in den Untermatrizen
zu einem "ubergeordneten Objekt kann das System simulierbar gemacht werden.

Im Fall einer oberen Dreiecksmatrix erreicht man die Simulierbarkeit durch
Umsortierung der Reihenfolge, in der die Objekte aktualisiert werden (vgl.\
Gau"ssches Eliminationsverfahren \cite{Klingenberg:1984})
\cite{RasmussenBarrett:1995}.

\subsubsection{Zeitfortschaltung}
Eine M"oglichkeit die {\Reihenfolgenunabhaengigkeit} zu erreichen, besteht
im Verfahren der \hi{Zeitfortschaltung}. Dabei haben die Objekte
neben ihren Zustandsvariablen $z_i$ auch noch Variablen f"ur den Grundzustand
$z_i^0$. 
Jeder Zeitschritt wird in zwei Stufen durchgef"uhrt. Zuerst werden f"ur alle
Objekte die neuen Zust"ande 
\begin{equation}
  \label{aktualisierung}
  z_i = u_i(z_1^0,\dots,z_n^0) 
\end{equation}
nach der Aktualisierungsfunktion $u_i$ in 
Abh"angigkeit der Grundzust"ande berechnet.
Danach "ubernehmen die Grundzust"ande den aktuellen Wert 
\begin{equation}
  \label{zeitfortschaltung}
  z_i^0 = z_i  
\end{equation}
Das Ergebnis ist dadurch eindeutig definiert. 
Bei diesem Verfahren wird allerdings die doppelte Anzahl an
Zustandsvariablen ben"otigt. 

\subsubsection{Takt- und Zeitfortschaltung}
Eine Erweiterung dieses Verfahrens stellt die \hi{Takt- und Zeitfortschaltung}
von \name{Eschenbacher}{Peter} dar. Die Methode garantiert die
{\Reifo} und behandelt dar"uber hinaus auch
Folgeereignisse bei Ereigniskaskaden \cite{Schmidt:1995}. Ein Beispiel f"ur
Folgeergebnisse ist die mittlere Geschwindigkeit der
{\fussgaenger} zur Zeit $t$. Sie kann erst bestimmt 
werden, wenn alle {\fussgaenger}objekte aktualisiert worden sind.

Bei der \hi{Takt- und Zeitfortschaltung} besteht ein Zeitschritt aus mehreren
Takten. Zu einem bestimmten Takt $k$ zur Zeit $t$ k"onnen alle Objekte
aktualisiert werden, die ausschlie"slich von in den Takten $k_0$ bis $ k$
behandelten Objekten abh"angen. Analog zu (\ref{zeitfortschaltung}) werden
die Zust"ande "ubernommen und ein neuer Takt $k+1$ geschaltet. Daraufhin
k"onnen weitere Objekte bearbeitet werden.
Nachdem alle Objekte aktualisiert worden sind, geht das Modell zum n"achsten
Zeitschritt $t+1$ und Takt $k_0$ "uber.

\subsubsection{Variable Zeitschritte}
Im Modell der {\fussgaenger}dynamik ist es im Sinne der
Rechenzeit unwirtschaftlich, alle Objekte
im selben Zeittakt zu aktualisieren. 
Die {\fussgaenger}objekte werden wegen ihrer kurzreichweitigen
Wechselwirkungen in sehr kleinen Zeitschritten von $\Delta t = 0.05$ Sekunden
berechnet. Ein Fu"sspurenobjekt dagegen "andert sich langsamer und sollte
daher auch seltener aktualisiert werden.  
Daher wird jedem Objekt eine individuelle Zeitschrittweite $\Delta t_i$ zugeordnet. 

Die Zeitschrittweite von {\fussgaenger}n kann sich auch dynamisch an deren
Geschwindigkeit und Situation anpassen. {\fussgaenger}, die in einem freien
Bereich der Anlage laufen und sich au"serhalb der Wechselwirkungsreichweite
von anderen Objekten befinden, m"ussen seltener aktualisiert 
werden, als diejenigen im Gedr"ange.

F"ur die \hi{variablen Zeitschritte} l"a"st sich das Verfahren der Takt- und
Zeitfortschaltung erweitern. Nach der "Ubernahme ihres aktuellen Zustands in
den Grundzustand bestimmen die Objekte den Zeitpunkt $t_i^\prime$, zu dem sie
sich wieder aktualisieren wollen. Anhand dieses Zeitpunkts werden sie dann in
eine Warteschlange eingeordnet. 

Das Simulationsprogramm bearbeitet die Objekte in der Reihenfolge, in der
sie in der Warteschlange stehen. Dabei kann auch das Verfahren der
Taktfortschaltung einbezogen werden, um Folgeereignisse zu behandeln. Sind alle
Objekte, die zu dem bestimmten Zeitpunkt $t$ aktualisiert werden sollten,
bearbeitet, schaltet die Systemzeit direkt auf den Zeitpunkt des n"achsten
Objekts aus der Warteschlange. 

Einige Systemobjekte der grafischen Benutzerschnittstelle werden durch
"au"sere Ereignisse gesteuert, zum Beispiel durch Eingaben des Benutzers
oder durch die Systemaufforderung an ein Bildschirmfensterobjekt seine
Darstellung zu rekonstruieren. 

Beim Eintreffen asynchroner Ereignisse werden diese Objekte
mit der aktuellen Systemzeit $t$ in die Warteschlange aufgenommen.
Direkt nachdem alle zur Zeit $t$ relevanten Objekte aktualisiert worden sind,
k"onnen sie dann  auf die "au"seren Ereignisse reagieren.
Die Einreihung in die Warteschlange hat
gegen"uber der unmittelbaren Bearbeitung der ereignisgesteuerten Objekte den
Vorteil, da"s der Aktualisierungsproze"s  zum Zeitpunkt $t$ wohldefiniert
bleibt.

Bedeutend f"ur die Ausf"uhrungsgeschwindigkeit des Simulationsprogramms ist
die Implementation der Warteschlange. Diese wird haupts"achlich durch
die Geschwindigkeit bestimmt, mit der die Objekte nach ihrem Zeitpunkt
$t_i^\prime$ in die
Warteschlange einsortiert werden. Da die effizientesten Suchalgorithmen auf
Baumstrukturen arbeiten, ergibt sich auch die Implementation der
Warteschlange als Baum. 

Die Effizienz eines Suchalgorithmus l"a"st sich durch die mittlere Anzahl der
Vergleichsoperationen $Z$, die beim Einsortieren eines neuen 
Elements ben"otigt werden, absch"atzen. F"ur eine sortierte Liste liegt die
Anzahl bei $Z = 1/2\,n$. Bei einem gleichm"a"sig gef"ullten {\Binaer}baum sind
dagegen mit 
\begin{equation}
  \label{anzahlbinaerbaum}
  Z \cong \frac{1}{n}\sum_{i=1}^{\ln n/\ln 2-1} i\,2^i < 
  \frac{\ln n}{n\,\ln 2}\sum_{i=1}^{\ln n/\ln 2-1} 2^i < 
  \frac{\ln n}{n\,\ln 2} 2^{\ln n/\ln 2} = \frac{\ln n}{\ln 2}
\end{equation}
wesentlich weniger Operationen notwendig. In der selben Gr"o"senordnung liegt
auch die mittlere ben"otigte Anzahl von Baumstrukturen mit mehr als 
zwei (nach unten gerichteten) Verbindungen.
Bereits bei 1000 Objekten ben"otigt 
die Implementation als Liste ungef"ahr f"unfzig mal mehr Vergleichsoperationen
als eine Implementation als Baum. 

Wenn neue Objekte ausschlie"slich am Ende der Warteschlange eingef"ugt werden,
degeneriert die Struktur eines Baumes jedoch zu der einer sortierten
Liste, und der Geschwindigkeitsvorteil ist verloren.
Zu dieser Situation kann es leicht kommen, wenn alle Objekte eines Modells
die gleiche Schrittweite $\Delta t_i = \Delta t$ haben.

Abhilfe schaffen hierbei speziell entwickelte Algorithmen f"ur sogenannte
ausgeglichene B"aume, zum Beispiel 2-3-4-B"aume oder B-B"aume von
\name{Bayer}{R.} und \name{McCreight}{E.} \cite[Kap.~15,18]{Sedgewick:1992}.
Sie sorgen daf"ur, da"s der Baum bei jedem Einsortierungsvorgang wieder
ausgeglichen wird. Das Prinzip ist vergleichbar mit einem Mobile, das durch
Anh"angen einer 
zus"atzlichen Figur an einen der F"aden ins Ungleichgewicht kommt. In
"ahnlicher Weise,  wie das Mobile durch einen neuen Aufh"angungspunkt wieder
ins Gleichgewicht gebracht werden kann, w"ahlen die Ausgleichalgorithmen einen
Knoten des Baumes als neue Wurzel aus. 
Die Verwendung von ausgeglichenen B"aumen lohnt sich trotz ihres komplizierten
Mechanismus gerade bei einer gro"sen Anzahl von Objekten.

\section{Rechenzeit und Rechengenauigkeit} 
Die Paarwechselwirkungen des Soziale-Kr"afte-Modells verursachen lange
Rechenzeiten, da sich die Zahl der ben"otigten Berechnungen f"ur jeden
Zeitschritt proportional zur Anzahl der {\fussgaenger}objekte verh"alt.
Selbst wenn der Beitrag der Wechselwirkungen mit sehr weit entfernten
{\fussgaenger}n vernachl"assigt wird, bleibt der Aufwand der Berechnung des
Abstands zwischen den Personen. 

Die Zahl der Berechnungen l"a"st sich  durch die Aufteilung der
Verkehrsfl"ache in kleinere Segmente vermindern. Dadurch ist eine grobe
Lokalisierung der {\fussgaenger} m"oglich. Die Aufteilung in Segmente erfolgt
durch die Korridorobjekte, die eine Liste mit den in ihnen
befindlichen {\fussgaenger}n verwalten. Bei der Berechnung der
Wechselwirkungen m"ussen dann nur {\fussgaenger} aus dem selben oder
angrenzenden Segmenten ber"ucksichtigt werden. Den "Ubergang der
{\fussgaenger} zwischen den einzelnen Korridoren besorgen dabei die
Torobjekte.

Die kleinen Zeitschritte $\Delta t$ von 0.05 Sekunden, die wegen der steilen
Verl"aufe der Kraftterme notwendig sind, tragen ebenfalls zu dem hohen
Rechenaufwand pro Zeitschritt bei. In dieser Zeitperiode bewegt sich ein
{\fussgaenger} bei einer Geschwindigkeit von \metersec{1.3} gerade um 6.5 cm. 
Au"serdem ber"ucksichtigt das Modell bereits die Bewegung der anderen
Passanten und deren weitere Schritte. 

Daher kann in einer N"aherung die Umgebung eines {\fussgaenger} f"ur mehrere
Zeitschritte als gleichbleibend angesehen werden. Der Genauigkeitverlust ist
dabei im Verh"altnis zu anderen St"orungen vernachl"assigbar.

F"ur die Simulation bedeutet das, da"s die Zeitschritte zwischen den
Berechnungen der {\fussgaenger}wechselwirkungen wesentlich gr"o"ser gew"ahlt
werden k"onnen. Die {\fussgaenger}objekte aktualisieren sich mit den
Zeitschritten $\Delta T$, die  Bewegungsgleichungen k"onnen (innerhalb des
Objektes) weiterhin mit $\Delta t \ll \Delta T$ integriert werden (vgl.\
Abschn.\ \ref{sec:gleichzeitigkeit}).  

\section{Beschreibung der Modellsprache}\label{sec:syntaxbeschreibung}
\label{sec:syntax}
Die Beschreibungssprache wurde entwickelt, um die Vielzahl der einzelnen
Steuerungparameter der Simulation bestimmen zu k"onnen. Ihre Struktur
lehnt sich dabei an Programmiersprachen wie C oder Pascal an. 
Zur Erkl"arung der Beschreibungssprache werden ein paar Begriffe eingef"uhrt:
\begin{description}
\item[Bezeichner] Parameter und Objekte werden durch sogenannte Bezeichner
  aufgerufen.  Ein Bezeichner besteht aus einer
  Kombination von Buchstaben und Ziffern, wobei das erste Zeichen ein
  Buchstabe sein mu"s. Diese Einschr"ankung ist in vielen Programmiersprachen
  zu finden, weil dadurch auf einfache Weise die Bezeichner von numerischen
  Werten unterschieden werden k"onnen.
\item[Konstanten] Konstanten k"onnen numerische Werte oder Texte
  darstellen. Numerische Konstanten beginnen mit einer Ziffer oder einem Plus-
  oder Minuszeichen, gefolgt von Ziffern und maximal einem Dezimalpunkt.
  Mehrdimensionale Vektorenkonstanten werden durch eine Reihe von numerischen
  Konstanten dargestellt, die mit senkrechten Strichen getrennt sind.
  Textkonstanten sind in Anf"uhrungszeichen eingeschlossen.
\item[Ausdr"ucke] Auf der rechten Seite einer Wertzuweisung k"onnen auch
  numerische Ausdr"ucke oder Textoperationen stehen. Sie setzen sich aus
  einer Kombination von Bezeichnern, Konstanten sowie vordefinierten Operatoren
  und Funktionen zusammen. Die Ausdr"ucke werden erst bei der Aufl"osung
  durch die Objekte berechnet.
\item[Geltungsbereich] Die Geltungsbereiche werden durch geschweifte Klammern
  eingefa"st. Innerhalb eines Geltungsbereiches sind alle Bezeichner bekannt
  und k"onnen verwendet werden. Au"serdem lassen sich darin weitere
  Geltungsbereiche anlegen. Dabei werden die einzelnen Bereiche als
  Baumstruktur angeordnet (vgl.\ Abb.\ \ref{fig:geltungsbereich}). Das
  bedeutet, da"s ein Bezeichner auch in allen inneren Geltungbereichen zu
  finden ist. 
  \begin{figure}[tb]
    \begin{center}
      \leavevmode
      \includegraphics[width=0.6\textwidth,clip]{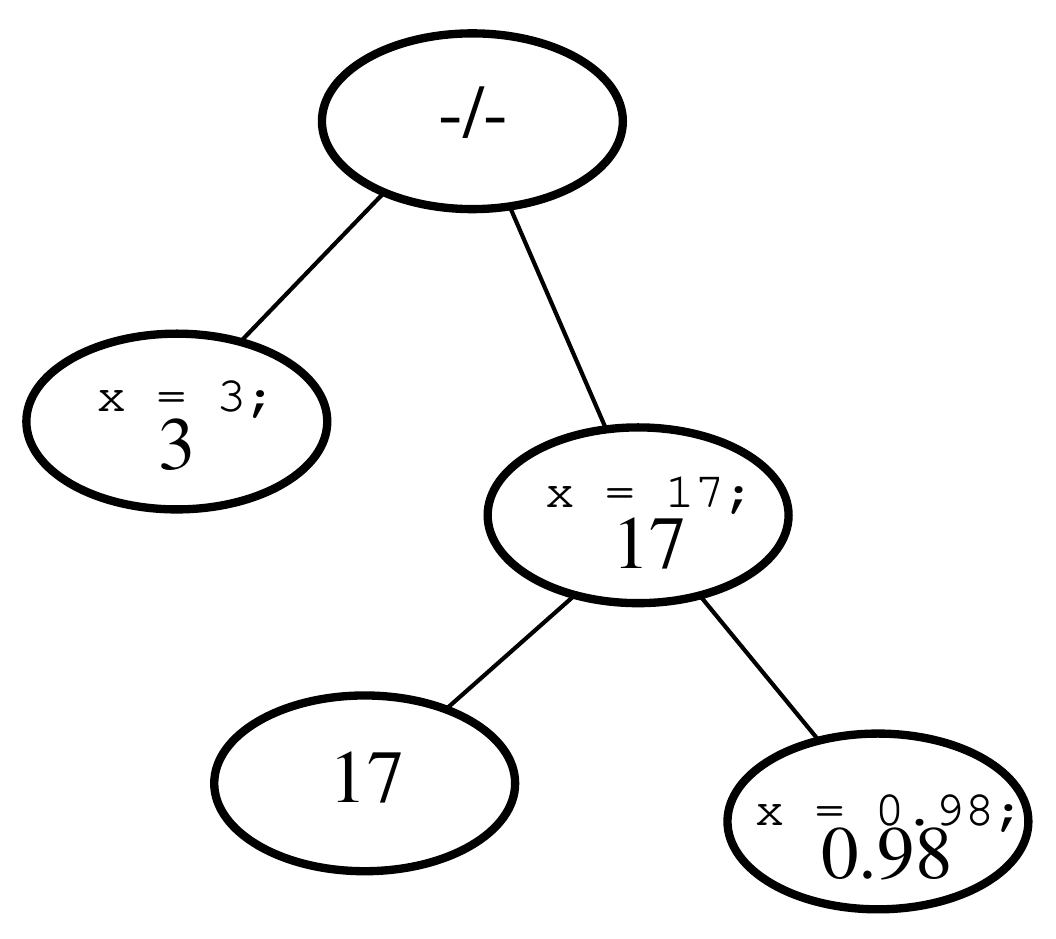}
    \end{center}
    \mycaption{Baumstruktur geschachtelter Geltungsbereiche}
    {Baumstruktur geschachtelter Geltungsbereiche, (als Ellipsen
      dargestellt). Ein Geltungsbereich umfa"st alle Bereiche, zu denen er im
      Baum eine abw"arts gerichtete Verbindung hat. Die Aufl"osung eines
      Bezeichners geschieht in entgegengesetzter Richtung: Um den Wert eines
      Bezeichners zu ermitteln, wird der Baum nach oben durchwandert, bis eine
      Zuweisung gefunden wird.  Die gro"sgedruckten Zahlen geben den mit {\tt
        x} verbundenen Wert im jeweiligen Geltungsbereich an.}
    \label{fig:geltungsbereich}
  \end{figure}
\item[Klassenname] Einige Bezeichner stehen f"ur einen festen Klassentyp. Sie
  werden Klassenname genannt und d"urfen nicht frei verwendet werden. Jede
  Objektklasse des Modells ist durch einen Klassennamen vertreten. Dazu
  geh"oren zum Beispiel: \\ {\tt pedestrian}, {\tt street} und {\tt floor}.
\item[Verkettung] Zwei Objekte k"onnen miteinander verkettet werden. In der
  Regel treten sie dadurch in gegenseitige Wechselwirkung. Die Bedeutung einer
  Verkettung wird in Abschnitt \ref{sec:methodendersimulationsobjekte}
  n"aher erl"autert.
\end{description}

Die Syntax der Beschreibungssprache unterscheidet zwischen drei Arten von
Anweisungen. Jede Anweisung wird durch ein Semikolon {\tt ;}
abgeschlossen. 
\begin{description}
\item[Wertzuweisung] Die einfachste Anweisung ist die Wertzuweisung. Sie wird
  durch ein Gleichheitszeichen {\tt =} ausgedr"uckt. Dabei geht die Zuweisung
  stets von rechts nach links. Wertzuweisungen haben die Form:
  \begin{quote}
    {\em Bezeichner1} {\tt = } {\em Konstante}{\tt;} \\
    {\em Bezeichner1} {\tt = } {\em Bezeichner2}{\tt;}
  \end{quote}
  Der Bezeichner  Nr.~1 wird eingef"uhrt und mit dem Wert der
  Konstanten, beziehungsweise dem Wert des zweiten Bezeichners, verbunden.
\item[Definition] Durch eine Anweisung der Form:
  \begin{quote}
    {\em Klassenname} {\em Bezeichner} {\tt ;}\\
    {\em Klassenname} {\em Bezeichner} {\tt \{ ... \} }
  \end{quote}
  wird ein Objekt aus der mit Klassenname bezeichneten Klasse erzeugt und mit
  dem angegebenen Bezeichner verbunden. In der zweiten Version wird
  zus"atzlich ein neuer Geltungsbereich f"ur das Objekt erzeugt.  
\item[Deklaration] Die Anweisung Deklaration erzeugt eine Verkettung zwischen
  zwei Objekten. Sie hat die Form:
  \begin{quote}
    {\em Bezeichner}{\tt;} 
  \end{quote}
  Das erste Objekt ist durch den Geltungsbereich bestimmt, in dem die
  Deklaration auftritt. Das zweite ist das mit dem Bezeichner verbundene
  Objekt. 
\end{description}
Ferner l"a"st sich die Definition eines Objektes durch zwei zus"atzliche
Anweisungen erweitern:
\begin{description}
\item[Multiplikation] Mehrere gleiche Objekte k"onnen aufeinmal erzeugt werden.
  Die Multiplikation hat die Form:
  \begin{quote}
    {\em Klassename \em Bezeichner \tt [ \em Zahl \tt ]}
  \end{quote}
  
\item["Ubernahme] Ein Objekt kann bei seiner Definition den gesamten
  Geltungsbereich und die Verkettungen eines anderen Objekts
  "ubernehmen. Zus"atzlich kann das Objekt eigene Wertzuweisungen und
  Deklarationen aufnehmen, die jedoch nicht auf das Objekt zur"uckwirken, von
  dem der Geltungsbereich "ubernommen wurde.
\end{description}
Das zu simulierende Modell kann durch diese Anweisungen aufgebaut
werden. Dabei  ist die
Reihenfolge zu beachten, in der die Anweisungen stehen: Treten in einem
Geltungsbereich zwei Wertzuweisungen f"ur denselben Bezeichner auf, so
verwirft das Objekt die erste Zuweisung und ber"ucksichtigt die zweite.
Wird zwischen den Anweisungen ein
weiteres Objekt definiert, kennt dieses nur die erste Zuweisung.
Weiterhin mu"s ein Objekt definiert werden, bevor es in einer Deklaration
mit einem anderen verkettet wird.

Die Beschreibungssprache bietet zahlreiche M"oglichkeiten zur Definition der
Struktur und der Parameterwerte einer Simulation. Die syntaktische Richtigkeit
bedeutet aber noch kein sinnvoller Simulationsaufbau. Abbildung
\ref{fig:modellspezifikation} zeigt ein Beispiel der Modellspezifikation f"ur
die Simulation des {\fussgaenger}stroms durch eine schmale "Offung aus
Abschnitt \ref{sec:einfach_tuer}. 
\begin{figure}[p]
    \leavevmode
    \begin{tabbing}
      \rule{0.5cm}{0cm}\=\rule{0.5cm}{0cm}\= \kill
      \tt   \# Simulation eines schmalen Durchgangs\\
      \tt    dt = 0.05; display\_rate = 1; 
      \\
      \tt    street GangTuer \{ \\
      \tt    \> \tt beta = 10.0; \\
      \tt    \> \tt polygon \{ -1.0|10.0; 51.0|10.0; color = 1; sig = 0.2;\} \\
      \tt    \> \tt polygon \{ 51.0|0.0; -1.0|0.0; color = 1; sig = 0.2;\} \\
      \\
      \tt    \> \tt  beta = 2.0;  sig = 0.2;  avoid = 0; color = 1; \\
      \tt    \> \tt  polygon \{ 25.2|0.0; 25.2|4.25; 24.8|4.25; 24.8|0.0;\} \\
      \tt    \> \tt  polygon \{ 25.2|10.0; 25.2|5.75; 24.8|5.75; 24.8|10.0;\} \\
      \}
      \\
      \tt    floor F \{ \\
      \tt    \> \tt  GangTuer; \\
      \tt    \> \tt  xmin = -1.0|-1.0; xmax = 52.0|11.0; \tt    dx = 0.1|0.1; \\ 
      \tt    \> \tt  vmin = 0.7; vmax = 2.3; vtau = 2.68; epsilon = 0.001; \\
      \\
      \tt    \> \tt  gate LINKS \{ color = 5; 5.0|0.5; 5.0|9.5; \} \\
      \tt    \> \tt  gate RECHTS \{ color = 5; 45.0|0.5; 45.0|9.5; \} \\ 
      \tt    \> \tt  offset = 0.5; \\
      \tt    \> \tt  door MITTE \{ color = 5; 25.0|5.75; 25.0|4.25; \} \\
      \\
      \tt    \> \tt  p = 2.1; sig = 0.3; lam = 10.0; mu = 0.1; vfaktor = 1.2; \\
      \tt    \> \tt  v0 = norm(1.34,0.26); t = unif(0.0,50.0); \\
      \tt    \> \tt  pedestrian A $[$30$]$ \{ \\
      \tt    \> \tt  \> \tt  color = 2; LINKS;  MITTE; RECHTS; GangTuer; clan = 0; \\
      \tt    \> \tt  \} \\
      \tt    \> \tt  pedestrian B $[$30$]$ \{ \\
      \tt    \> \tt  \> \tt  color = 4; RECHTS;  MITTE; LINKS; GangTuer; clan = 1; \\
      \tt    \> \tt  \} \\
      \tt    \} \\
    \end{tabbing}
    \mycaption{Beispiel einer Modellspezifikation}
  {Beispiel einer Modellspezifikation. Das Objekt {\tt GangTuer}
    bildet die Umgebung f"ur die {\fussgaenger}. Darin werden die W"ande mit
    ihren Wechselwirkungsparametern definiert. Das Objekt {\tt F} umfa"st die
    Definition der Durchg"ange {\tt LINKS}, {\tt RECHTS} und {\tt MITTE} sowie
    die der {\fussgaenger}populationen {\tt A} und {\tt B} mit jeweils 30
    Mitgliedern. Die Reihenfolge der Deklaration der Durchg"ange im
    Geltungsbereich der {\fussgaenger} gibt deren Weg an. Gemeinsame Parameter
    werden im "ubergeordneten Geltungsbereich zugewiesen.}
  \label{fig:modellspezifikation}
\end{figure}

\chapter*{Symbolverzeichnis}
\label{cha:symbol}
\addcontentsline{toc}{chapter}{Symbolverzeichnis}
\markboth{SYMBOLVERZEICHNIS}{SYMBOLVERZEICHNIS}
\newcommand{\mySym}[1]{\rule{0pt}{16pt}\parbox[t]{4cm}{#1}}
\newcommand{\myDef}[1]{\parbox[t]{12cm}{#1}}

\begin{tabbing}
  \rule{4cm}{0cm}\= \kill
  17\,0123.987 \> \myDef{Schreibweise der Zahlen mit Dezimalpunkt und Leerraum zwischen den Tausendern} \\
  ${\rm I\!R}$ \> \myDef{Menge der reellen Zahlen} \\
  ${\rm I\!R}^n$ \> \myDef{Menge reellwertige  Vektoren, n-dimensional} \\
  ${\rm Z\!\!Z}$ \> \myDef{Menge der ganzen Zahlen} \\ 
  \parbox{3cm}{$x_{i}$, $x^i$, $x_{ij}$, $x_i^j$, $x_{ij}^k$, $x^1$, $x^2$, $x^3$} \>
  \myDef{$i$, $j$ und $k$ sind Indizes 
    der Gr"o"se $x$ und werden sowohl oben, als auch unten angeordnet, ohne
    weitere Angaben stellt die Bezeichnung keine Potenzschreibweise
    dar} \\
  $x^\prime$, $x^{\prime\prime}$ \> \myDef{Striche dienen zur Unterscheidung von  Gr"o"sen, keine Ableitung} \\
  $\sq{x}$, $x^{\bf 3}$ \> \myDef{Quadrat und und dritte Potenz, der
    Exponent ist fett gedruckt} \\  
  $\vec{p}$ \> \myDef{Vektor mit den Komponenten $p_1\dots p_n$, in der Regel
    zweidimensional ($n= 2$), auch $p_x$, $p_y$} \\
  $\vp{\vec{p}}{\vec{q}}$ \> \myDef{Skalarprodukt zweier Vektoren $\vec{p}$ und
    $\vec{q}$, ${\rm I\!R^n,I\!R^n} \rightarrow {\rm I\!R}$ } \\
  ${\N}$, $\N(x,\sigma)$ \> \myDef{normalverteilte Zufallszahl mit der
    Verteilungsfunktion 
    $P(\N(x,\sigma), x^\prime) =
    \frac{1}{\sqrt{2\pi\sigma}} e^{-\sq{(x^\prime-x)}/2\sigma}$,
    $x$ gibt dabei den Mittelwert und $\sigma$ die Standardabweichung an, ohne
    weitere Angaben wird $x=0$ und $\sigma=1$ angenommen } \\
  $\Z$, $\Z(a,b)$ \> \myDef{uniformverteilte Zufallszahl, alle Werte im
    Intervall $\left[a,b\right]$ treten mit gleicher Wahrscheinlichkeit auf,
    ohne Angaben  wird $a=0$ und $b=1$ angenommen}  \\ 
  \\
  {\bf Einheiten} \\
  \\
  m \> \myDef{Meter (L"ange)}\\
  m$^{\bf 2}$ \>  \myDef{Quadratmeter (Fl"ache)} \\
  m/s \> \myDef{Meter pro Sekunde (Geschwindigkeit)}   \\
  P/m$^{\bf 2}$ \> \myDef{Personen pro Quadratmeter ({\fussgaenger}dichte)} \\
  P/ms \> \myDef{Personen pro Meter und Sekunde (spezifischer Flu"s)} \\
  $^{\circ}$\,C \> \myDef{Grad Celsius (Temperatur)} \\ 
  \\
  {\bf Funktionen und Operatoren} \\
  \\
  $\displaystyle \sum$, $\displaystyle \sum_i$, $\displaystyle \sum_{i=1}^N$ \>
  \myDef{Summe, mit Laufindex $i$, mit 
    Indexbereich $1\dots N$, die Summation bezieht sich auf alle nachfolgenden
    Ausdr"ucke bis zum n"achsten $+$ oder $-$Operator} \\
  $\displaystyle\int dt$, $\displaystyle\int\limits_{t_0}^{t_0+T} dt$ \>
  \myDef{Integral mit 
    Integrationsvariable $t$, Integrationsgrenzen $t_0\dots t_0+T$, die
    Integration bezieht sich auf alle nachfolgenden 
    Ausdr"ucke bis zum n"achsten $+$ oder $-$Operator} \\
  $\displaystyle \frac{\displaystyle d}{\displaystyle dt}$ \> \myDef{Ableitung
    nach $t$}\\[4pt] 
  $\displaystyle \frac{\displaystyle \partial}{\displaystyle \partial x_i}$ \> \myDef{partielle Ableitung nach der
    Komponenete $x_i$}\\
  $\nabla$ \> \myDef{Nabla-Operator, Vektor der komponentenweisen Ableitungen}   \\
  $\exp(x)$, $e^x$ \> \myDef{Exponentialfunktion} \\
  $\ln x, \ln(x)$ \>  \myDef{nat"urlicher Logarithmus, bezieht sich auf
    alle nachfolgenden 
    Ausdr"ucke bis zum n"achsten $+$ oder $-$Operator} \\
  $\cos x, \cos (x)$ \> \myDef{Cosinus-Funktion, bezieht sich auf
    alle nachfolgenden 
    Ausdr"ucke bis zum n"achsten $+$ oder $-$Operator, }   \\
  $\delta_{ij} $ \> \myDef{Kronecker-Symbol $\delta_{ij} = 1$ f"ur $i=j$ und
    $\delta_{ij} = 0$ f"ur $i\ne j$}   \\  
  $\ave{x}$ \> \myDef{Mittelwert zur Gr"o"se x mit $$\ave{x} =
    \frac{1}{N}\sum_{i=1}^N x_i$$ }   \\ 
  $\sq{\sigma}(x)$ \> \myDef{ Varianz $$\sq{\sigma}(x) =
    \frac{1}{N}\sum_{i=1}^N \sq{x_i} - \frac{1}{\sq{N}}\sum_{i=1}^N x_i $$ }  \\
  $\sq{\chi}$ \> \myDef{Quadratischer Fehler aus dem Verfahren des
    Minimalen-Fehlerquadrats}   \\
   \\
  {\bf Bemessungsgrundlagen von Fu"sg"angeranlagen} \\
  \\
  $ N $ \> \myDef{ Anzahl von Personen, 
    die pro Stunde einen Querschnitt passieren  } \\
  $\rho $ \> \myDef{{\fussgaenger}dichte  } \\
  $ v_h $ \> \myDef{ Horizontalgeschwindigkeit der {\fussgaenger}  } \\
  $ b_n$ \> \myDef{ nutzbare Breite des Fu"sweges  } \\
  $T $ \> \myDef{ Beobachtungszeitraum } \\
  $  L $ \> \myDef{ Leistungsf"ahigkeit einer {\fussgaenger}anlage  } \\
  $  \hat{L} $ \> \myDef{spezifische Leistungsf"ahigkeit, bezogen auf die
    Breite von \meter{1} } \\
  $v_h^0 $ \> \myDef{maximal zul"assige Geschwindigkeit auf freier Fl"ache } \\
  \\
  {\bf Sozial-psychologische Modelle} \\ 
  \\
  $ \hat{\i} $ \> \myDef{ Soziale Wirkung (social impact) } \\
  $f(SIN) $ \> \myDef{ Soziale Kraft} \\
  $S $ \> \myDef{ St"arke der Beeinflussung (strength)  } \\
  $I $ \> \myDef{ Direktheit, Unmittelbarkeit (immediacy) } \\
  $N $ \> \myDef{ Anzahl der beeinflussenden Individuen } \\
  $\Psi $ \> \myDef{subjektive Wahrnehmungsintensit"at  } \\
  $\Phi $ \> \myDef{objektiver physikalischer Reiz   } \\
  $\beta $ \> \myDef{Potenz  } \\
  $\kappa $ \> \myDef{Proportionalit"atsfaktor  } \\
  \\
  {\bf Soziale-Kr"afte-Modell} \\
  \\
  $\alpha$, $\beta$ \> \myDef{Index f"ur Individuen ({\fussgaenger}), $\alpha$
    bezeichnet die Person, auf die Kr"afte ausge"ubt wird, $\beta$ steht f"ur
    den Verursacher} \\
  $B$ \> \myDef{Index f"ur Hindernisse und Begrenzungen} \\
  $i$ \> \myDef{Index f"ur Attraktionen} \\
  $\alpha^\prime$ \> \myDef{Gruppenmitglied das zu $\alpha$ geh"ort} \\

  $U_{(\cdot)}$ \> \myDef{Potential, Soziales Feld, Die Indizes $(\cdot)$ geben
    die Wechselwirkungspartner an} \\
  $\vec{f}_{(\cdot)} $ \> \myDef{ Soziale Kraft, falls es ein Potential gibt,
    gilt $\vec{f} = - \nabla U$ } \\
  $\vec{\cal F}$ \> \myDef{Fluktuationsterm, vektorielle Zufallsgr"o"se}  \\ 
  $t$, $\Delta t$ \> \myDef{Zeit, Zeitschritt} \\
  $\Ra$, $\Rb$  \>  \myDef{momentante Position der {\fussgaenger}}  \\
  $\Va$, $\Vb$  \>  \myDef{momentane Geschwindigkeit}  \\
  $\Eanull$  \>  \myDef{ Einheitsvektor in momentaner Zielrichtung} \\
  $\Esanull$ \> \myDef{ Einheitsvektor senkrecht zur momentanen Zielrichtung,
    es gilt $\vp{\Esanull}{\Eanull} = 0$, beide Richtungsm"oglichkeiten f"uhren
    im Modell zum selben Ergebnis} \\  
  $\Fanull$, $\Fbnull$  \>  \myDef{Antriebskraft} \\
  $\vanull$  \> \myDef{(Betrag der) Wunschgeschwindigkeit} \\ 
  $\vamax$  \>  \myDef{Betrag der Maximalgeschwindigkeit, mit der sich ein
    {\fussgaenger} fortbewegen kann} \\ 
  $\tau_\alpha$ \> \myDef{Relaxationszeit} \\ 
  $a_{\alpha i}$, $a_{\alpha i}^0$ \> \myDef{momentane  Attraktionsst"arke,
    Anfangsinteresse} \\
  $\Utr $ \> \myDef{Trampelpfadepotenital } \\
  $\Ftr $ \> \myDef{ Anziehungskraft existierender Trampelpfade} \\
  $Q_\beta $ \> \myDef{Potential eines Fu"sabdruckes } \\
  $T $ \> \myDef{Zerfallszeit des Trampelpfadepotentials} \\
  \\
  {\bf Leistungsma"se} \\
  \\
  $Y_\alpha$, $Y$ \> \myDef{Individuelles Leistungsma"s, das ein {\fussgaenger}
    $\alpha$ bestimmt hat, und Leistungsma"s eines Abschnittes der Anlage } \\
  $T$ \> \myDef{ Reisezeit durch den Abschnitt der {\fussgaenger}anlage } \\
  $Y^1$ \> \myDef{Effizienzma"s } \\
  $Y^2$ \> \myDef{H"aufigkeit erzwungener Geschwindigkeitswechsel } \\
  $Y^3$ \> \myDef{St"arke der Beeinflussung, Wohlbefinden } \\
  $Y^4$ \> \myDef{Ma"s f"ur das Zusammenbleiben von Gruppen } \\
  $Y^5$ \> \myDef{Grade der Segregation verschiedener Subpopulationen in
    einer {\fussgaenger}menge } \\
  $  \lanull $ \> \myDef{ Effiziente L"ange eines Anlagenabschnittes} \\
  $N_l^{+}$ ,$N_l^{-}$ \> \myDef{Anzahl der Personen, die eine (Teil-)Anlage
    in einem gewissen Zeitraum betreten, und die Anzahl, die diese verlassen
    } \\ 
  $\Phi_l^{\pm} $ \> \myDef{Flu"sdichte } \\
  $\rho_l $ \> \myDef{{\fussgaenger}dichte im Abschnitt $l$ } \\
  $v^{\pm} $ \> \myDef{mittlere Geschwindigkeit der hinein und hinaus
    str"omenden {\fussgaenger} } \\
  $ b_l $ \> \myDef{ Breite des Abschnitts $l$} \\
  $ L $ \> \myDef{ Leistungsf"ahigkeit eines Teilst"uckes  } \\
  $\bf C $ \> \myDef{Subpopulation von Individuen mit gemeinsamen
    Eigenschaften } \\ 
  $ \Pac $ \> \myDef{ anteiliger Einflu"s einer Subpoulation auf das
    Individuum $\alpha$ } \\
  $S_\alpha $ \> \myDef{ Individuelle, momentane Ordnung } \\
  $S  $ \> \myDef{ momentane Entropie eines Systems mit N Individuen } \\
  $S_T $ \> \myDef{ Entropie eines Systems "uber die Periode $T$ } \\
  \\
  {\bf Evolution"are Methoden} \\
  \\
  $x, x^\prime, x^{\prime\prime}, x^1, x^2$ \> \myDef{Genetsiche Darstellung
    einer potentiellen L"osung durch Bin"arzahlen oder reellwertige Vektoren
    der Dimension $n$,
    $x = (x_1,\dots,x_n)$ } \\
  $\mu $ \> \myDef{Populationsgr"o"se } \\
  $\lambda $ \> \myDef{Gr"o"se der Zwischenpopulation } \\
  $p_k $ \> \myDef{Reproduktionswahrscheinlichkeit eines Individuums $k$ } \\
  $W $ \> \myDef{Erfolgswahrscheinlichkeit eines Evolutionsschrittes  } \\
  $\sigma $ \> \myDef{Standardabweichung des Mutationsoperators } \\
  $(x,\sigma) $ \> \myDef{Darstellung eines Individuums in der
    Evolutionstrategie mit anpassungsf"ahiger Schrittweite  } \\ 
  $p_c$, $p_m$ \> \myDef{Anwendungswahrscheinlichkeit des Crossover- und des
    Mutationsoperators} \\
  $c_{\alpha k}$  \> \myDef{"Ubereinstimmung der Eigenschaften eines {\fussgaenger}s
    $\alpha$ und einer potentiellen L"osung $k$ } \\
  $p_{\alpha k} $ \> \myDef{von Eigenschaften abh"angige
    Reproduktionswahrscheinlichkeit } \\
  \\
  {\bf Entscheidungsmodell} \\
  \\
  $i$, $j$ \> \myDef{Index einer Entscheidungsm"oglichkeit} \\
  $  p_{j\leftarrow i} $ \> \myDef{"Ubergangswahrscheinlichkeit } \\
  $ U_i$, $U_j$ \> \myDef{Nutzen der Alternativen $i$ und $j$ } \\
  $ S_{j\leftarrow i}$ \> \myDef{bei einem Wechsel entstehender Verlust } \\
  $\xi $ \> \myDef{Bereitschaft zur Akzeptanz schlechterer Alternativen } \\
  \\
  {\bf Wegenetze} \\
  \\
  $G(V,E) $ \> \myDef{ Graph mit $V$-Knoten und $E$-Kanten} \\
  $(M_{ab}) $ \> \myDef{Entfernungsmatrix, $M_{ab}$ gibt die L"ange der
    k"urzesten Strecke von $a$ nach $b$ an} \\
  $(R_{ab}) $ \> \myDef{Routenmatrix, $R_{ab}$ gibt den n"achsten
    (Zwischen-)Knoten der k"urzesten Strecke von $a$ nach $b$ an} \\
  $F_i $ \> \myDef{Benutzungsfrequenz der Kante $i$ } \\
  $z_i $ \> \myDef{Ordnun gszahl der Kante $i$, sortiert nach
    Benutzungsfrequenz, $z_1$ ist die am h"aufigsten belaufene Kante } \\
  $\epsilon_t $ \> \myDef{Fehler nach $t$ Iterationen des
    Random-Warshall-Floyd-Algorithmus } \\
  \\
  {\bf Simulationstheorie} \\
  \\
  $S_i $ \> \myDef{Simulationsobjekt $i$ } \\
  $q_i $ \> \myDef{Z"ahler des Objektes $i$ } \\
  $Dq $ \> \myDef{Jacobi-Matrix der Z"ahler } \\
  $z_i$ \> \myDef{Zustand von $i$} \\
  $u_i $ \> \myDef{Aktualisierungsfunktion von $i$} \\
  $t $ \> \myDef{Zeit } \\
  $k$, $k_0 $ \> \myDef{Takt, Anfangstakt } \\
  $\Delta t$, $\Delta T$ \> \myDef{Zeitschritt } \\
  $Z$ \> \myDef{Zahl der ben"otigten Vergleichsoperationen } \\
\end{tabbing}
\clearpage

\addcontentsline{toc}{chapter}{Literaturverzeichnis}
\bibliographystyle{alpha}
\bibliography{phd}

\end{document}